\documentclass[3p,sort&compress]{elsarticle}
\usepackage{hyperref}

\usepackage{fleqn}
\usepackage{soul}

\usepackage{graphicx}
\usepackage{amssymb,latexsym,amsthm,amsmath,dsfont,mathtools}
\usepackage{color,linegoal}
\usepackage[utf8x]{inputenc}
\usepackage[T1]{fontenc}
\usepackage[polutonikogreek,english]{babel}
\usepackage{mathdots}
\usepackage{multirow}
\usepackage{hyperref}

\newcommand{\la}{\left\langle}
\newcommand{\ra}{\right\rangle}
\newcommand{\greek}[1]{{\selectlanguage{polutonikogreek}#1}}
\newcommand{\ddt}{\frac{\mathrm d}{\mathrm {dt}}}
\newcommand{\dds}{\frac{\mathrm d}{\mathrm {ds}}}

\usepackage{bm} 
\newcommand{\be}{\begin{equation}}
\newcommand{\ee}{\end{equation}}
\newcommand{\ba}{\begin{eqnarray}}
\newcommand{\ea}{\end{eqnarray}}
\def\bea{\begin{eqnarray}}
\def\eea{\end{eqnarray}}

\newcommand{\Avg}[1]{\left\langle{#1}\right\rangle}
\newcommand{\avg}[1]{\langle{#1}\rangle}

\journal{Physics Reports}

\begin{document}

\begin{frontmatter}

\title{The structure and dynamics of multilayer networks}

\author[SB,SB2]{S. Boccaletti\corref{corrauthor}}
\cortext[corrauthor]{Corresponding author}
\ead{stefano.boccaletti@gmail.com}
\address[SB]{CNR- Institute of Complex Systems, Via Madonna del Piano,
  10, 50019 Sesto Fiorentino, Florence, Italy}
\address[SB2]{The Italian Embassy in Israel, 25 Hamered st., 68125 Tel Aviv, Israel}

\author[GB]{G. Bianconi}
\address[GB]{School of Mathematical Sciences, Queen Mary University
  of London, London, United Kingdom}

\author[RC,CTB]{R. Criado}
\address[RC]{Departamento de Matem\'atica Aplicada, Universidad  Rey Juan Carlos,  28933 M\'ostoles, Madrid, Spain}
\address[CTB]{Center for Biomedical Technology, Universidad
  Polit\'ecnica de Madrid,  28223 Pozuelo de Alarc\'on, Madrid, Spain}

\author[CG1,CG2,CG3]{C.I. del Genio}
\address[CG1]{Warwick Mathematics Institute, University of Warwick,
  Gibbet Hill Road, Coventry CV4 7AL, United Kingdom}
\address[CG2]{Centre for Complexity Science, University of Warwick,
  Gibbet Hill Road, Coventry CV4 7AL, United Kingdom}
\address[CG3]{Warwick Infectious Disease Epidemiology Research (WIDER) Centre, \\ University of Warwick,
  Gibbet Hill Road, Coventry CV4 7AL, United Kingdom}

\author[JGG]{J. G\'omez-Garde\~nes}
\address[JGG]{Institute for Biocomputation and Physics of Complex
  Systems,  University of Zaragoza, Zaragoza, Spain}

\author[RC,CTB]{M. Romance}

\author[ISN,CTB]{I. Sendi\~na-Nadal}
\address[ISN]{Complex Systems Group, Universidad Rey Juan Carlos,
  28933 M\'ostoles, Madrid, Spain}

\author[ZW,ZW2]{Z. Wang}
\address[ZW]{Department of Physics, Hong Kong Baptist University,  Kowloon Tong, Hong Kong SRA, China}
\address[ZW2]{Center for Nonlinear Studies, Beijing-Hong
  Kong-Singapore Joint Center for Nonlinear and \\Complex Systems (Hong
  Kong) and  Institute of Computational and Theoretical Studies, \\Hong Kong Baptist University, Kowloon Tong, Hong Kong SRA, China
}

\author[MZ,MZ2]{M. Zanin}
\address[MZ]{Innaxis Foundation \& Research Institute,  Jos\'e Ortega y Gasset 20, 28006 Madrid, Spain}
\address[MZ2]{Faculdade de Ci\^encias e Tecnologia, Departamento de Engenharia Electrot\'ecnica, \\Universidade Nova de Lisboa, 2829-516 Caparica, Portugal}

\begin{abstract}
In the past years, network theory has successfully characterized the interaction among
the constituents of a variety of complex systems, ranging from biological to technological, and social systems.
 However, up until recently, attention was almost exclusively given to networks in which all components were treated on
equivalent footing, while neglecting all the extra information about the temporal- or context-related properties of the
interactions under study. Only in the last years, taking advantage of the enhanced resolution in real data sets,
network scientists have directed their interest to the multiplex character of real-world systems, and explicitly
considered the time-varying and multilayer nature of networks. We offer here a comprehensive review on both structural and
dynamical organization of graphs made of diverse relationships (layers) between its constituents, and cover several
relevant issues, from a full redefinition of the basic structural measures, to understanding how the multilayer nature of
the network affects processes and dynamics.
\end{abstract}


\end{frontmatter}
\tableofcontents


\section{Introduction}\label{intro}

\subsection{The multilayer network approach to nature}

If we just turn our eyes to the immense majority of phenomena that occur around us
(from those influencing our social relationships, to those transforming the overall environment where we live, to even those affecting our own biological functioning), we realize immediately that they are nothing but the result of the emergent dynamical organization of systems that, on their turn, involve a multitude of basic constituents (or entities) interacting with each other via somehow complicated patterns.

One of the major effort of modern physics is then providing proper and suitable representations of these systems, where constituents are considered as nodes (or units) of a network, and interactions are modeled by links of that same network.
Indeed, having such a representation in one hand and the arsenal of mathematical tools for extracting information in the other (as inherited by several gifted centuries of thoughts,
concepts and activities in applied mathematics and statistical mechanics) is the only suitable way through which we can even dare to {\it understand} the observed phenomena, identify the rules and mechanisms that are lying behind them, and possibly control and manipulate them conveniently.

The last fifteen years have seen the birth of a movement in science, nowadays very well known under the name of {\it complex networks theory}. It involved the interdisciplinary effort of some of our best scientists in the aim of
exploiting the current availability of {\it big data} in order to extract the ultimate and optimal representation of the underlying complex systems and mechanisms. The main goals were {\it i)} the extraction of unifying principles that could encompass and describe (under some generic and universal rules) the structural accommodation that is being detected ubiquitously, and {\it ii)} the modeling of the resulting emergent dynamics to explain what we actually {\it see} and experience from the observation of such systems.

It would look like even pleonastic to report here on each and every original work that was carried out in specific contexts under study (a reader, indeed, will find, along this review, all the relevant literature that was produced so far on this subject, properly addressed and organized in the different sections of the report).
At this initial stage, instead, and together with the pioneering articles \cite{Strogatz2001,AlbertBarabasi2001,DorogovtsevMendes2002,Newman_SIAM} and classical books \cite{Wattsbook,Schuster2003,mendesbook2003,vespbook08} on complex networks, we address the interested reader to some other reports \cite{Boccaletti2006,szabo2007evolutionary,ArenasPR2008,Fortunatoregino2010Physrep,BarthelemyPR2011} that were published recently on this same Journal, which we believe may constitute very good guides to find orientation into the immensely vast literature on the subject.
In particular, Ref.~\cite{Boccaletti2006} is a complete {\it compendium} of the ideas and concepts involved in both structural and dynamical properties of complex networks, whereas Refs.~\cite{Fortunatoregino2010Physrep,BarthelemyPR2011} have the merit of accompanying and orienting the reader through the relevant literature discussing modular networks \cite{Fortunatoregino2010Physrep}, and space-embedded networks \cite{BarthelemyPR2011}. Finally, Refs.~\cite{szabo2007evolutionary,ArenasPR2008} constitute important accounts on the state of the art for what concerns the study of synchronous organization of networking systems \cite{ArenasPR2008},
and processes like evolutionary games on networks \cite{szabo2007evolutionary}.

The traditional complex network approach to nature has mostly been concentrated to the case in which each system's constituent (or elementary unit) is charted into a network node, and each unit-unit interaction is represented as being a (in general real) number quantifying the weight of the corresponding graph's connection (or link). However, it is easy to realize that treating all the network's links on such an equivalent footing is too big a constraint, and may occasionally result in not fully capturing  the details present in some real-life problems, leading even to incorrect descriptions of some phenomena that are taking place on real-world networks.

The following three examples are representative, in our opinion, of the major limitation of that approach.

The first example is borrowed from sociology. Social networks analysis
is one of the most used paradigms in behavioral sciences, as well as
in economics, marketing, and industrial engineering
\cite{wasserman94}, but some questions related to the real structure
of social networks have been not properly understood. A social network
can be described as a set of people (or groups of people) with some
pattern of contacts or interactions between them
\cite{wasserman94,Scott}. At a first glance, it seems natural to
assume that all the connections or social relationships between the
members of the network take place at the same level. But the real
situation is far different. The actual relationships amongst the
members of a social network take place (mostly) inside of different
groups (levels or \emph{layers}), and therefore they cannot be
properly modeled if only the natural local-scale point of view used in
classic complex network models is taken into account. Let us for a moment think to the problem of spreading of information, or rumors, on top of a social network like {\sf Facebook}. There, all users can be seen as nodes of a graph, and all the friendships that users create may be considered as the network's links. However, friendships in {\sf Facebook} may result from relationships of very different origins: two {\sf Facebook}'s users may share a friendship because they are colleagues in their daily occupations, or because they are fans of the same football team, or because they occasionally met during their vacation time in some resort, or for any other possible social reason. Now, suppose that a given user gets aware of a given information and wants to spread it to its {\sf Facebook}'s neighborhood of friends. It is evident that the user will first select that subgroup of friends that he/she believes might be potentially interested to the specific content of the information, and only after will proceed with spreading it to that subgroup. As a consequence, representing {\sf Facebook} as a {\it unique} network of acquaintances (and simulating there a classical diffusion model) would result in drawing incorrect conclusions and predictions of the real dynamics of the system. Rather, the correct way to proceed is instead to chart each social group into a different layer of interactions, and operating the spreading process separately on each layer. As we will see in Section~\ref{sec:spreading}, such a multilayer approach leads actually to a series of dramatic and important consequences.

A second paradigmatic example of intrinsically multirelational (or
multiplex) systems is the situation that one has to tackle when trying
to describe transportation networks, as for instance the Air
Transportation Network (ATN)  \cite{CardilloEPJST13} or subway
networks \cite{CRV10,CHR07}. In particular, the traditional study of
ATN is by representing it as a single-layer network, where nodes
represent airports, while links stand for direct flights between two
airports. On the other hand, it is clear that a more accurate mapping
is considering that each commercial airline corresponds instead to a
different layer, containing all the connections operated by the same
company. Indeed, let us suppose for a moment that one wants to make predictions on the propagation of the delays in the flight scheduling through the system, or the effects of such a dynamics on the movement of passengers \cite{zanin2013}. In particular, Ref.~\cite{CardilloEPJST13} considered the problem of passenger rescheduling when a fraction of the ATN links, i.e. of flights, is randomly removed. It is well known that each commercial airline incurs in a rather high cost whenever a passenger needs to be rescheduled on a flight of another company, and therefore it tries first to reschedule the passenger by the use of the rest of its own network. As we will see in full details
in Section~\ref{sec:applications}, the proper framework where predictions about such dynamical processes can be made suitably is, in fact, considering the ATN as a fully multilayer structure.

Moving on to biology, the third example is the effort of scientists in trying to rank the importance of a specific component in a biological system. For instance, the {\it Caenorhabditis elegans} or {\it C. elegans} is a small nematode, and it
is actually the first ever organism for which the entire genome was sequenced. Recently, biologists
were even able to get a full mapping of the {\it C. elegans}' neural network, which actually consists of 281 neurons and around two thousands of connections. On their turn, neurons can be connected either by a chemical link, or by an ionic channel, and the two types of connections have completely different dynamics. As a consequence, the only proper way to describe such a network is a multiplex graph with 281 nodes and two layers, one for the chemical synaptic links and a separate one for the gap junctions' interactions.
The most important consequence is that each neuron can play a very different role in the two layers, and a proper ranking should be then able to distinguish those cases in which a node of high centrality in a layer is just marginally central in the other.

These three examples, along with the many others that the reader will be presented with throughout the rest of this report,
well explain why the last years of research in network science have been characterized by more and more attempts to generalize the traditional network theory by developing (and validating) a novel framework for the study of multilayer networks, i.e. graphs where constituents of a system are the nodes, and several different layers of connections have to be taken into account to accurately describe the network's unit-unit interactions, and/or the overall system's parallel functioning.

\emph{Multilayer networks} explicitly incorporate multiple channels of connectivity and constitute the natural environment to describe systems interconnected through different categories of connections: each channel (relationship, activity, category) is represented by a layer and the same node or entity may have different kinds of interactions (different set of neighbors in each layer). For instance, in social networks, one can consider several types of different actors' relationships:  friendship, vicinity, kinship, membership of the same cultural society, partnership or coworker-ship, etc.  

Such a change of paradigm, that was termed in disparate ways (multiplex networks, networks of networks, interdependent networks, hypergraphs, and many others), led already to a series of very relevant and unexpected results, and we firmly believe that: {\it i)} it actually constitutes the new frontier in many areas of science, and {\it ii)} it will rapidly expand and attract more and more attention in the years to come, stimulating a new movement of interdisciplinary research.

Therefore, in our intentions, the present report would like to constitute a survey and a summary of the current state of the art, together with a weighted and meditated outlook to the still open questions to be addressed in the future.

\subsection{Outline of the report}

Together with this introductory Preamble, the Report is organized along other 8 sections.

In the next section, we start by offering the overall mathematical definitions that will accompany the rest of our discussion. Section 2 contains several parts that are necessarily rather formal, in order to properly introduce the quantities (and their mathematical properties and representation) that characterize the structure of multilayer networks. The reader will find there the attempt of defining an overall mathematical framework encompassing the different situations and systems that will be later extensively treated. If more interested instead in the modeling or physical applications of multilayer networks, the reader is advised, however, to take Section 2 as a {\it vocabulary} that will help and guide him/her in the rest of the paper.

Section 3 summarizes the various (growing and not growing) models that have been introduced so far for generating artificial multilayer networks with specific structural features. In particular, we extensively review the available literature for the two classes of models considered so far: that of growing multiplex networks (in which the number of  nodes grows in time
and fundamental rules are imposed to govern the dynamics of the network structure), and that of multiplex network ensembles
(which are ultimately ensembles of networks satisfying a
certain set of structural constraints).

Section 4  discusses the concepts, ideas, and available results related to multilayer networks' robustness and resilience, together with the process of percolation on multilayer networks, which indeed has attracted a huge attention in recent years. In particular, Section 4 will manifest the important differences, as far as these processes are concerned, between the traditional approach of single-layer networks, and the case where the structure of the network has a multilayer nature.

In Section 5, we summarize the state of current understanding of spreading processes taking place on top of a multilayer structure, and discuss explicitly linear diffusion, random walks, routing and congestion phenomena and spreading of information and disease. The section includes also a complete account on evolutionary games taking place on a multilayer network.

Section 6 is devoted to synchronization, and we there describe the cases of both alternating and coexisting layers. In the former, the different layers correspond to different connectivity configurations that alternate to
define a time-dependent structure of coupling amongst a given set of dynamical units. In the latter, the different layers are simultaneously responsible for the coupling of the network's units, with explicit additional layer-layer interactions taken into account.

Section 7 is a large review of applications, especially those that were studied in social sciences, technology, economy, climatology, ecology and biomedicine.

Finally, Section 8 presents our conclusive remarks and perspective ideas.


\section{The structure of multilayer networks}
\label{sec:structuremultilayer}


Complexity science is the study of systems with many interdependent components, which, in turn, may interact through many different channels. Such systems -- and the self-organization and emergent phenomena they manifest -- lie at the heart of many challenges of global importance for the future of the Worldwide Knowledge Society. The development of this science is providing radical new ways of understanding many different mechanisms and processes from physical, social, engineering, information and biological sciences. Most complex systems include multiple subsystems and layers of connectivity, and they are often open, value-laden, directed, multilevel, multicomponent, reconfigurable systems of systems, and placed within unstable and changing environments. They evolve, adapt and transform through internal and external dynamic interactions affecting the subsystems and components at both local and global scale. They are the source of very difficult scientific challenges for observing, understanding, reconstructing and predicting their multiscale and multicomponent dynamics. The issues posed by the multiscale modeling of both natural and artificial complex systems call for a generalization of the ``traditional'' network theory, by developing a solid foundation and the consequent new associated tools to study multilayer and multicomponent systems in a comprehensive fashion. A lot of work has been done during the last years to understand the structure and dynamics of these kind of systems \cite{Belingerio01,Domenico2013,Kivela2013,Batiston2013,Donges2011}. Related
notions, such as networks of networks
{\cite{Buldyrev10,Gao2012GinestraNatPhys}}, multidimensional networks
\cite{Belingerio01}, multilevel networks, multiplex networks,
interacting networks, interdependent networks, and many others have
been introduced, and even different mathematical approaches, based on
tensor representation ~\cite{Domenico2013,Kivela2013} or otherwise~\cite{Batiston2013,Donges2011}, have been proposed. It is the purpose of this section to survey and discuss a general framework for multilayer networks and review some attempts to extend the notions and models from single layer to multilayer networks. As we will see, this framework includes the great majority of the different approaches addressed so far in the literature.

\subsection{Definitions and notations}

\subsubsection{The formal basic definitions}\label{cap2:basicdef}
A \emph{multilayer network }is a pair $\mathcal{M}=(\mathcal{G},\mathcal{C})$ where $\mathcal{G}=\{G_\alpha;\enspace \alpha\in \{1,\cdots,M\} \}$ is a family of (directed or undirected, weighted or unweighted) graphs $G_\alpha=(X_\alpha,E_\alpha)$ (called layers of $\mathcal{M}$) and
\begin{equation}
\mathcal{C}=\{E_{\alpha\beta}\subseteq X_\alpha \times X_\beta;\enspace \alpha,\beta\in \{1,\cdots,M\},\enspace\alpha\neq \beta \}
\end{equation}
is the set of interconnections between nodes of different layers $G_\alpha$ and $G_\beta$ with $\alpha\neq \beta$. The elements of $\mathcal{C}$ are called \emph{crossed layers}, and the elements of each $E_\alpha$ are called
\emph{intralayer} connections of $\mathcal{M}$ in contrast with the elements of each $E_{\alpha\beta}$ ($\alpha\ne \beta$) that are called \emph{interlayer} connections.

In the remainder, we will use Greek subscripts and superscripts to denote the layer index. The set of nodes of the layer $G_\alpha$ will be denoted by $X_{\alpha}=\{x^{\alpha}_1,\cdots,x^{\alpha}_{N_\alpha}\}$ and the adjacency
matrix of each layer $G_\alpha$ will be denoted by $A^{[\alpha]}=(a^{\alpha}_{ij})\in \mathbb{R}^{N_{\alpha} \times N_{\alpha}}$, where
\begin{equation}
a^{\alpha}_{ij}=\left\{
\begin{tabular}{ll}
1  & \text{if $(x^{\alpha}_i,x^{\alpha}_j)\in E_\alpha,$} \\
0  & \text{otherwise,}
\end{tabular}
\right.
\end{equation}
for $1\le i,j\le N_{\alpha}$ and $1\le \alpha\le M$. The {\sl interlayer adjacency matrix} corresponding to $E_{\alpha\beta}$ is the matrix $A^{[\alpha,\beta]}=(a^{\alpha\beta}_{ij})\in\mathbb{R}^{N_\alpha\times N_\beta}$ given by:
\begin{equation}
a^{\alpha\beta}_{ij} =\left\{
\begin{tabular}{ll}
1  & \text{if $(x^{\alpha}_i,x^{\beta}_j)\in E_{\alpha\beta},$} \\
0  & \text{otherwise.}
\end{tabular}
\right.
\end{equation}

The projection network of  $\mathcal{M}$ is the graph $proj(\mathcal{M})=(X_\mathcal{M},E_\mathcal{M})$ where
\begin{equation}
X_\mathcal{M}=\bigcup_{\alpha=1}^M X_\alpha,
 \qquad\qquad E_\mathcal{M}= \left(\bigcup_{\alpha=1}^M E_\alpha\right)\bigcup
\left( \bigcup_{\substack{\alpha,\beta=1\\\alpha\neq\beta}}^M E_{\alpha\beta}\right).
\end{equation}

We will denote the adjacency matrix of $proj(\mathcal{M})=(X_\mathcal{M},E_\mathcal{M})$ by $\overline{A_\mathcal{M}}$.

This mathematical model is well suited to describe phenomena in social systems, as well as many other complex systems. An example is the dissemination of culture in social networks in the Axelrod Model~\cite{axelrod}, since each social group can be understood as a {\sl layer} within a {\sl multilayer network}. By using this representation we simultaneously take into account:
 \begin{itemize}
 \item[{\it (i)}] the links inside the different groups,
 \item[{\it (ii)}] the nature of the links and the relationships between elements that (possibly) belong to different layers,
 \item[{\it (iii)}] the specific nodes belonging to each layer involved.
\end{itemize}

A \emph{multiplex network} \cite{SRCFGB13} is a special type of multilayer network in which $X_{1}=X_{2}=\cdots=X_{M}=X$ and the only possible type of interlayer connections are those in which a given node is only connected to its counterpart nodes in the rest of layers, i.e., $E_{\alpha\beta}=\{(x,x);\enspace x\in X\}$ for every $\alpha,\beta\in \{1,\cdots,M\},\alpha\neq \beta $. In other words, multiplex networks consist of a fixed set of nodes connected by different types of links. The paradigm of multiplex networks is social systems, since these systems can be seen as a superposition of a multitude of complex social networks, where nodes represent individuals and links capture a variety of different social relations.

A given multiplex network $\mathcal{M}$, can be associated to several (monolayer) networks providing valuable information about it. A specific example is the {\sl projection network} $proj(\mathcal{M})=(X_\mathcal{M},E_\mathcal{M})$. Its adjacency matrix $\overline{A_\mathcal{M}}$ has elements
\begin{equation}
\overline{a_{ij}} =\left\{
\begin{tabular}{ll}
1  & \text{if $a^{\alpha}_{ij}=1$ for some $1\le\alpha\le M$} \\
0  & \text{otherwise.}
\end{tabular}
\right.
\end{equation}

A first approach to the concept of multiplex networks could suggest that these new objects are actually (monolayer) networks with some (modular) structure in the mesoscale. It is clear that if we take a multiplex  $\mathcal{M}$, we can associate to it a (monolayer) network $\tilde{\mathcal{M}}=(\tilde X,\tilde E)$, where $\tilde X$ is the {\sl disjoint} union of all the nodes of $G_1,\cdots,G_M$, i.e.
\begin{equation}
\tilde X=\bigsqcup_{1\le \alpha \le M} X_\alpha=\left\{x^\alpha;\enspace x\in
X_\alpha\right\}
\end{equation}
and $\tilde E$ is given by
\begin{equation}
\left( \bigcup_{\alpha=1}^M\left\lbrace\left(x_i^\alpha, x_j^\alpha\right);\ \left(x_i^\alpha,x_j^\alpha\right)\in E_\alpha\right\rbrace\right) \bigcup \left(\bigcup_{\substack{\alpha,\beta=1\\\alpha\neq\beta}}^M\left\lbrace\left(x_i^\alpha,x_i^\beta\right); \enspace x_i\in X\right\rbrace\right)\:.
\end{equation}
Note that $\tilde{\mathcal{M}}$ is a (monolayer) graph with $N\times M$ nodes whose adjacency matrix, called {\sl supra-adjacency} matrix of $\mathcal{M}$, can be written as a block matrix
\begin{equation}
\tilde A=\left(
\begin{array}{c|c|c|c}
A_1    & I_N & \cdots & I_N \\ \hline
I_N & A_2    & \cdots & I_N \\ \hline
\vdots & \vdots & \ddots & \vdots \\ \hline
I_N & I_N & \cdots & A_M
\end{array}
\right)\in \mathbb{R}^{NM\times NM},
\end{equation}
where $I_N$ is the $N$-dimensional identity matrix.

The procedure of assigning a matrix to a multilayer network is often called $flattening$, $unfolding$ or $matricization$. It is important to remark that the behaviors of $\mathcal{M}$ and $\tilde{\mathcal{M}}$ are related but different, since a single node of $\mathcal{M}$ corresponds to different nodes in $\tilde{\mathcal{M}}$. Therefore, the properties and behavior of a multiplex $\mathcal{M}$ can be understood as a type of non-linear {\sl quotient} of the properties of the corresponding (monolayer) network $\tilde{\mathcal{M}}$.

It is important to remark that the concept of multilayer network extends that of other mathematical objects, such as:
\begin{enumerate}
 \item {\sl Multiplex networks}. As we stated before, a multiplex network \cite{SRCFGB13} $\mathcal{M}$, with $M$ layers is a set of layers $\{G_\alpha;\enspace \alpha\in \{1,\cdots,M\} \}$, where each layer is a (directed or undirected, weighted or unweighted) graph $G_\alpha=(X_\alpha,E_\alpha)$, with $X_\alpha=\{x_1,\cdots,x_N\}$. As all layers have the same nodes, this can be thought of as a multilayer network by taking $X_1=\cdots =X_M=X$ and $E_{\alpha\beta}=\{(x,x);\enspace x\in X\}$ for every $1\le \alpha\ne \beta\le M$.
 \item {\sl Temporal networks} \cite{Holmeregino2012physicsreport}. A temporal network $(G(t))_{t=1}^T$ can be represented as a multilayer network with a set of layers $\{G_1,\cdots, G_{T}\}$ where $G_t=G(t)$, $E_{\alpha\beta}=\emptyset$ if $\beta\ne \alpha+1$, while
     \begin{equation}
     E_{\alpha,\alpha+1}=\{(x,x);\enspace x\in X_\alpha\cap X_{\alpha+1}\}
     \end{equation}
     (see the schematic illustration of Fig.~\ref{ftemporal}). Notice that here $t$ is an integer, and not a continuous parameter as it will be used later on in Sec.~\ref{master-temporal}.
\begin{figure}
\centering
\includegraphics[width=0.8\textwidth]{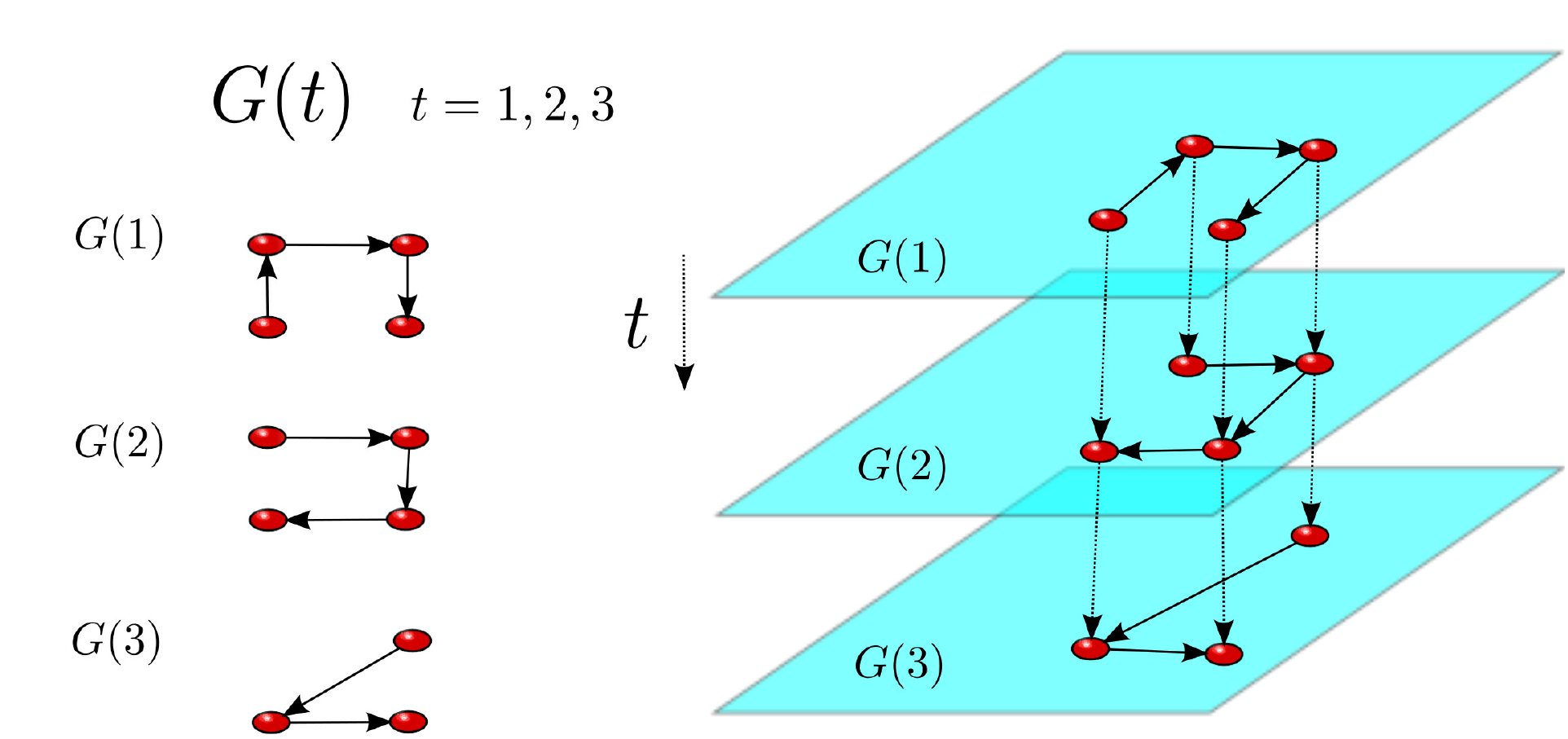}
\caption{(Color online). Schematic illustration of the mapping of a temporal network into a multilayer network. (Left side) At each time $t=1,2,3$, a different graph characterizes the structure of interactions between the system's constituents. (Right side) The corresponding multilayer network representation where each time instant is mapped into a different layer.}
\label{ftemporal}
\end{figure}
 \item {\sl Interacting} or {\sl interconnected networks} \cite{Donges2011}. If we consider a family of networks $\{G_1,\cdots, G_{L}\}$ that interact, they can be modeled as a multilayer network of layers $\{G_1,\cdots,G_{L}\}$ and whose crossed layers $E_{\alpha\beta}$ correspond to the interactions between network $G_\alpha$ and $G_\beta$ (see Fig.~\ref{finteracting}).
 \item {\sl Multidimensional networks} \cite{Belingerio01,Coscia,BerCosGi12,BerPiCa13,CosRoPe13}. Formally, an \emph{edge-labeled multigraph} (multidimensional network) \cite{Coscia} is a triple $G=(V,E,D)$ where $V$ is a set of nodes, $D$ is a set of labels representing the dimensions, and $E$ is a set of labeled edges, i.e. it is a set of triples $E=\{(u,v,d);\ u,v\in V, d\in D\}$.

It is assumed that given a pair of nodes $u,v \in V$ and a label $d \in D$, there may exist only one edge $(u,v,d)$. Moreover, if the model considered is a directed graph, the edges $(u,v,d)$ and $(v,u,d)$ are distinct. Thus, given $|D|=m$, each pair of nodes in $G$ can be connected by at most $m$ possible edges. When needed, it is possible to consider weights, so that the edges are no longer triplets, but quadruplets $(u,v,d,w)$, where $w$ is a real number representing the weight of the relation between nodes $u,v \in V$ and labeled with $d \in D$. A multidimensional network  $G=(V,E,D)$ can be modeled as a multiplex network (and therefore, as a multilayer network) by mapping each label to a layer. Specifically, $G$ can be associated to a multilayer network of layers $\{G_1,\cdots, G_{|D|}\}$  where for every $\alpha\in D$, $G_\alpha=(X_\alpha,E_\alpha)$, $X_\alpha=V$,
\begin{equation}
E_\alpha=\{(u,v) \in V\times V;\enspace (u,v,d) \in E\enspace{\rm{and}}\enspace d=\alpha\}
\end{equation}
and $E_{\alpha\beta}=\{(x,x);\enspace x\in V\}$ for every $1\le \alpha\ne \beta\le |D|$.
\item {\sl Interdependent (or layered) networks} \cite{Kuran-Thiran06,Buldyrev10,Parshani10}. An interdependent (or layered) network is a collection of different networks, the layers, whose nodes are interdependent to each other. In practice, nodes from one layer of the network depend on control nodes in a different layer. In this kind of representation, the dependencies are additional edges connecting the different layers. This structure, in between the network layers, is often called  \emph{mesostructure}. A preliminary study about interdependent networks  was presented in Ref.~\cite{Kuran-Thiran06} (there called layered networks). Similarly to the previous case of multidimensional networks, we can consider an interdependent (or layered) network as a multilayer network by identifying each network with a layer.
\begin{figure}[t!]
\centering
\includegraphics[width=0.8\textwidth]{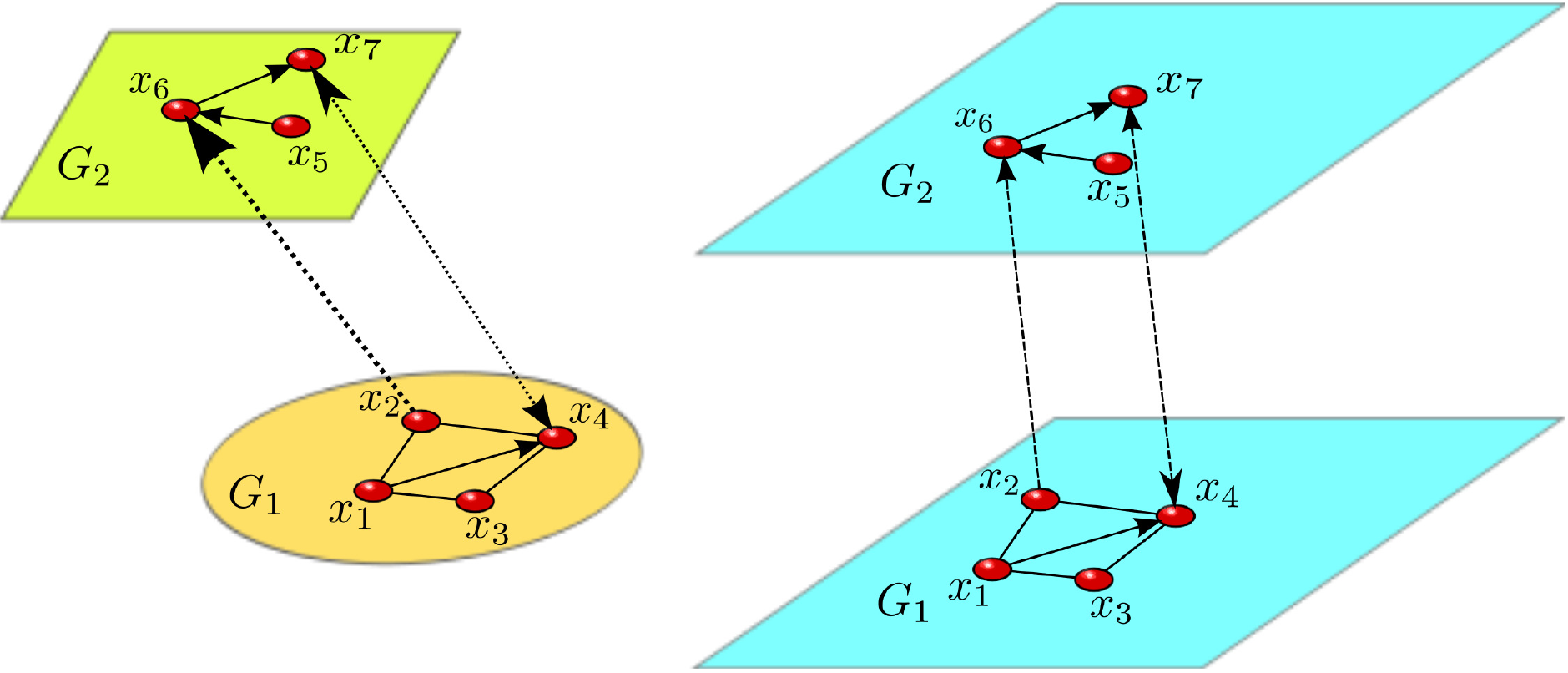}
\caption{(Color online). Schematic illustration of the rather straightforward mapping of interacting networks into multilayer networks. Each different colored network on the left side corresponds to a different blue layer on the right side.}
\label{finteracting}
\end{figure}

\item {\sl Multilevel networks} \cite{CFGGR12}. The formal definition of a multilevel network is the following. Let $G=(X,E)$ be a network. A {\sl multilevel network} is a triple $\mathbf{M}=(X,E,\mathcal{S})$, where $\mathcal{S}=\{S_1,\ldots,S_p\}$ is a family of subgraphs $S_j=(X_j,E_j), j=1,\ldots,p$ of $G$ such that
\begin{equation}
X=\bigcup_{j=1}^pX_j,\qquad
E=\bigcup_{j=1}^pE_j.
\end{equation}
The network $G$ is the {\sl projection network} of $\mathbf{M}$ and each subgraph $S_j\in \mathcal{S}$ is called a {\sl slice} of the multilevel network $\mathbf{M}$. Obviously, every multilevel network $\mathbf{M}=(X,E,\mathcal{S})$ can be understood as a multilayer network with layers $\{S_1,\ldots,S_p\}$ and crossed layers  $E_{\alpha\beta}=\{(x,x);\enspace x\in X_\alpha\cap X_\beta\}$ for every $1\le \alpha\ne \beta\le p$. It is straightforward to check that every multiplex network is a multilevel network, and a multilevel network $\mathbf{M}=(X,E,\mathcal{S})$ is a multiplex network if and only if $X_\alpha=X_\beta$ for all $1\le \alpha,\beta\le p$.

\item {\sl Hypernetworks} (or hypergraphs) \cite{Berge1989,Duchetregino1995,CRV10,Johnsonregino2006,Estradaregino2006PhysA,Konstantinovaregino2001DM,Karonskiregino1996,Rodriguezregino2003LMA}. A hypergraph is a pair $\mathcal{H}=(X,H)$ where $X$ is a non-empty set of nodes and $H=\{H_1,\ldots, H_p\}$ is a family of non-empty subsets of $X$, each of them called a {\sl hyperlink} of $\mathcal{H}$. Now, if $G=(X,E)$ is a graph, a hyperstructure $S=(X,E,H)$ is a triple formed by the vertex set $X$, the edge set $E$, and the hyper-edge set $H$. If $\mathcal{H}=(X,H)$ is a hypernetwork (or hypergraph), then it can be modeled as a multilayer network, such that for every hyperlink $h=(x_1,\cdots, x_k)\in H$ we define a layer $G_h$ which is a complete graph of nodes $x_1,\cdots, x_k$, and the crossed layers are $E_{\alpha\beta}=\{(x,x);\enspace x\in X_\alpha\cap X_\beta\}$ (see Fig.~\ref{fhypergraph}).
\end{enumerate}
\begin{figure}
\centering
\includegraphics[width=0.8\textwidth]{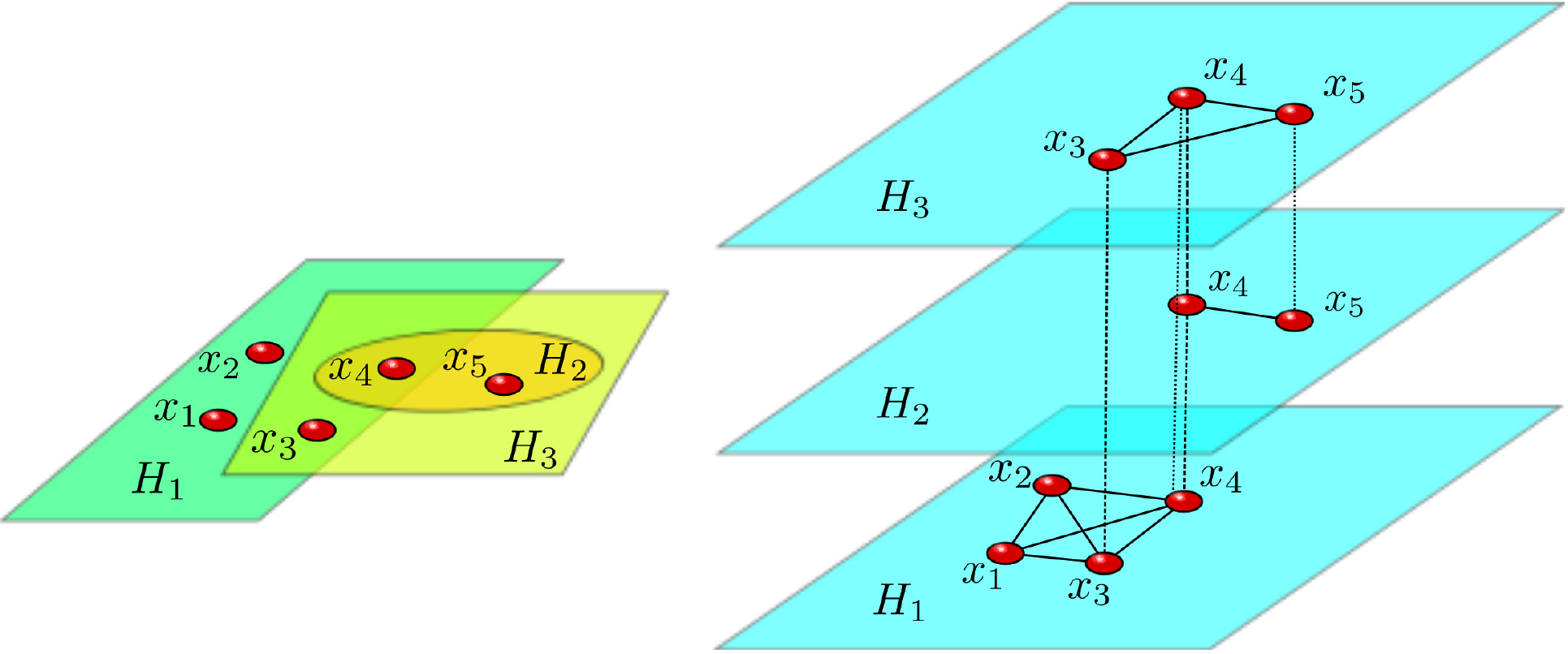}
\caption{(Color online). Schematic illustration of the transformation of a hypergraph into a multilayer network. The set of red nodes on the left side defines three hyperlinks ($H_1$, $H_2$, and $H_3$), each of which is mapped to a layer consisting of a complete graph of its nodes. }
\label{fhypergraph}
\end{figure}

\subsubsection{A small (and less formal) handbook}
A large amount of studies has shown how representing the elements of a complex system and their interactions with nodes and links can help us to provide insights into the system's structure, dynamics, and function. However, except for a number of complex systems, the simple abstraction of their organization into a single layer of nodes and links is not sufficient. As we have discussed in the previous section, several extensions of complex networks to \emph{multistructure} or \emph{multirelational} networks have been developed in recent years~\cite{Belingerio01,Domenico2013,Kivela2013,Donges2011,SRCFGB13,Buldyrev10,Parshani10,Guerra,GoDi13,Dagostino2014,Cozzo2013,MinGo2013,Bianconi13,BruLee12,CellaiLo13,CoAre12,CoBa13,Halu13,KimGo13,LeeKim12,NiBiLa13,CardilloEPJST13,Nicosiaregino2012SR,SoDo13,YaGli12}.
 In the following, we briefly describe the characteristics of the main ones.

Amongst the mathematical models that consider non-local structures are
the hypergraphs (or hypernetworks) \cite{Berge1989} and the
hyperstructures \cite{CRV10}. 
However, these models are not able to combine the local scale with the
global and the mesoscale structures of the system. 
For instance, if one wants to model how a rumor spreads within a
social network, it is necessary to have in mind not only that
different groups are linked through some of their members, 
but also that two people who know the same person do not necessarily
know each other. 
In fact, these two people may very well belong to entirely different groups or levels. If one tries to model this situation with
hypernetworks, then one only takes into account the social groups, and
not the actual relationships between their members. 
In contrast, if one uses the hyperstructure model, then one cannot determine to which social group each contact between two nodes belongs. The key-point that makes hypernetworks and hyperstructures not the best mathematical model for systems with mesoscale structures is that they  are both node-based models, while many real systems combine a node-based point of view with a link-based perspective.

Following on with the social network example, when one considers a relationship between two members of a social group, one has to take into account not only the social groups that hold the members, but also the social groups that hold the relationship itself. In other words, if there is a relationship between two people that share two distinct social groups, such as a work and a sporting environment, one has to specify if the relationship arises in the work place, or if it has a sport nature. A similar situation characterizes also public transport systems, where a link between two stations belonging to several transport lines can occur as a part of different lines.

Nevertheless, the use of hypergraph analysis is far from being inadequate. In Ref.~\cite{Ghos-Zlatic2009}, the Authors investigate the theory of random tripartite hypergraphs with given degree distributions, providing an extension of the configuration model for traditional complex networks, and a theoretical framework in which analytical properties of these random graphs are investigated. In particular, the conditions that allow the emergence of giant components in tripartite hypergraphs are analyzed, along with classical percolation events in these structures. Moreover, in Ref.~\cite{Zlatic2009} a collection of useful metrics on tripartite hypergraphs is provided, making this model very robust for basic analysis.

Another mathematical model capturing  multiple different relations that act at the same time is that of multidimensional networks \cite{Belingerio01,Coscia,BerCosGi12,BerPiCa13,CosRoPe13,wasserman94,Szell2010,Barret2012,BerCosGi11,KaMuKa11,KaMuKu11}. In a \emph{multidimensional network}, a pair of entities may be linked by different kinds of links. For example, two people may be linked because they are acquaintances, friends, relatives or because they communicate to each other by phone, email, or other means. Each possible type of relation between two entities is considered as a particular dimension of the network. In the case of a multidimensional network model, a network is a labeled multigraph, that is, a graph where both nodes and edges are labeled and where there can exist two or more edges between two nodes. Obviously, it is possible to consider edge-only labeled multigraphs as particular models of multidimensional networks.

When edge-labeled multigraphs are used to model a multidimensional network, the set of nodes represents the set of entities or actors in the networks, the edges represent the interactions and relations between them, and the edge labels describe the nature of the relations, i.e. the dimensions of the network. Given the strong correlation between labels and dimensions, often the two terms are used interchangeably. In this model, a node $u$ belongs to (or appears in) a given dimension $d$ if $u$ has at least one edge labeled with $d$. So, given a node $u \in V$, $n_u$ is the number of dimensions in which $u$ appears, i.e.
\begin{equation}
n_u=\left|\{d \in D;\enspace {\rm{there\enspace is}}\enspace v \in V \enspace {\rm {s.t.}} \enspace (u,v,d)\in E\}\right|.
\end{equation}
Similarly, given a pair of nodes $u,v \in V$, $n_{uv}$ is the number of dimensions which label the edges between $u$ and $v$, i.e. $n_{uv}=|\{d \in D;\ (u,v,d)\in E\}|$. A multidimensional network can be considered as a multilayer network by making every label equivalent to a layer.

Other alternative models focus on the study of networks with multiple kinds of interactions in the broadest as possible sense. For example, interdependent networks were used to study the interdependence of several real world infrastructure networks in Refs.~\cite{Buldyrev10} and \cite{Parshani10}, where the Authors also explore several properties of these structures such as cascading failures and percolation. Related to this concept, in
Ref.~\cite{Parshani10} the Authors report the presence of a critical threshold in percolation processes, creating an analogy between interdependent networks and ideal gases. The main advantage of interdependent networks is their ability of mapping a node in one relation with many different nodes in another relation. Thus, the mesostructure can connect a single node in a network $N_{1}$ to several different nodes in a network $N_{2}$. This is not possible in other models where there is no explicitly defined mesostructure, and therefore a node is a single, not divisible entity.

Multilevel networks \cite{CFGGR12} lie in between the multidimensional and the interdependent networks, as they extend both the classic complex network model and the hypergraph model \cite{Berge1989}. Multilevel networks are completely equivalent (isomorphic) to multidimensional networks by considering a dimension $d$ in conjunction with the equivalent slice $S_d$. The set $D$ corresponding to dimension $d$ is the collection of edges labeled with the label $d$, i.e. $D=\{(u,v,x);\ x=d\}$, while $S_d$ is the collection of nodes and edges of relation $d$, i.e. all the $u$ and $v$ that are present in at least one edge labeled with $d$. In order to perform more advanced studies such as the shortest path detection, in Ref.~\cite{CFGGR12} Authors introduced also a mesostructure called auxiliary graph. Every vertex of the multilevel network $\mathbf{M}$ is represented by a vertex in the auxiliary graph and, if a vertex in $\mathbf{M}$ belongs to two or more slice graphs in $\mathbf{M}$, then it is duplicated as many times as the number of slice graphs it belongs to. Every edge of $E$ is an edge in the auxiliary graph and there is one more (weighted) edge for each vertex duplication between the duplicated vertex and the original one. This operation breaks the isomorphism with multidimensional networks and brings this representation very close to a layered network \cite{Coscia}. However, these two models are not completely equivalent, since in the multilevel network the one-to-one correspondence of nodes in different slices is strict, while this condition does not hold for layered networks. In Ref.~\cite{CFGGR12}, the Authors also define some extensions of classical network measures for multilevel networks, such as the slice clustering coefficient and the efficiency, as well as a collection of network random generators for multilevel structures.

In Ref.~\cite{Mucha10} the Authors introduce the concept of \emph{multiplex networks} by extending the popular modularity function for community detection, and by adapting its implicit null model to fit a layered network. The main idea is to represent each layer with a slice. Each slice has an adjacency matrix describing connections between nodes belonging to the previously considered slice. This concept also includes a mesostructure, called \emph{interslice couplings} which connects a node of a specific slice $S_{\alpha}$ to its copy in another slice $S_{\beta}$. The mathematical formulation of multiplex networks has been recently developed through many works
\cite{Batiston2013,SRCFGB13,GoDi13,Cozzo2013,MinGo2013,Bianconi13,BruLee12,CellaiLo13,CoAre12,CoBa13,Halu13,KimGo13,LeeKim12,NiBiLa13,SoDo13,YaGli12}. For instance, in Ref.~\cite{Batiston2013} a comprehensive formalism to deal with multiplex systems is proposed, and a number of metrics to characterize multiplex systems with respect to node degree, edge overlap, node participation to different layers, clustering coefficient, reachability and eigenvector centrality is provided.

Other recent extensions include multivariate networks \cite{Pattison99}, multinetworks \cite{Barigozzi2010,Barigozzi2011}, multislice networks \cite{Mucha10,BaPor13,CarLon11,MuchaPor10}, multitype networks \cite{AllNo09,HinSin13,Vazquez06}, multilayer networks \cite{Domenico2013,Kivela2013,CardilloSR13,Ruben2014}, interacting networks \cite{Donges2011,LeiSo09,BruSo12}, and networks of networks \cite{Dagostino2014},
most of which can be considered particular cases of the definition of multilayer networks given in Sec.~\ref{cap2:basicdef}. Finally, it is important to remark that the terminology referring to networks with multiple different relations has not yet reached a consensus. In fact, different scientists from various fields still use similar terminologies to refer to different models, or distinct names for the same model. It is thus clear that the introduction of a sharp mathematical model that fits these new structures is crucial to properly analyze the dynamics that takes place in these complex systems which even nowadays are far from being completely understood.

The notation proposed in Sec.~\ref{cap2:basicdef} is just one of the possible ways of dealing with multilayer networks, and indeed there have been other recent attempts to define alternative frameworks. In particular, the tensor formalism proposed in Refs.~\cite{Domenico2013} and~\cite{Kivela2013}, which we will extensively review in Sec.~\ref{cap2:tensor}, seems promising, since it allows the synthetic and compact expression of multiplex metrics. Nevertheless, we believe that the notation we propose here is somehow more immediate to understand and easier to use for the study of real-world systems.

In the following, we describe the extension to the context of multilayer networks of the parameters that are traditionally used to characterize the structural properties of a monolayer graph.

\subsection{Characterizing the structure of multilayer networks}

\subsubsection{Centrality and ranking of nodes}\label{subsec:centrality}

The problem of identifying the nodes that play a central structural role is one of the main topics in the traditional analysis of complex networks. In monolayer networks, there are many well-known parameters that measure the structural relevance of each node, including the node degree, the closeness, the betweenness, eigenvector-like centralities and PageRank centrality. In the following, we discuss the extension of these measures to multilayer networks.

One of the main centrality measures is the {\sl degree} of each node: the more links a node has, the more relevant it is. The degree of a node $i\in X$ of a multiplex network $\mathcal{M}=(\mathcal{G},\mathcal{C})$ is the vector
\cite{Belingerio01,Batiston2013}
\begin{equation}
{\mathbf k}_i=(k_i^{[1]},\cdots, k_i^{[M]}),
\end{equation}
where $k_i^{[\alpha]}$ is the degree of the node $i$ in the layer $\alpha$, i.e. $k_i^{[\alpha]}=\sum_{j}a_{ij}^{[\alpha]}$. This vector-type node degree is the natural extension of the established definition of the node degree in a monolayer network.

One of the main goals of any centrality measure is ranking the nodes
to produce an ordered list of the vertices according to their
relevance in the structure. However, since the node degree in a
multiplex network is a vector, there is not a clear ordering in $\mathbb{R}^M$ that could produce such a ranking. In fact, one can define many complete orders in $\mathbb{R}^M$, and therefore we should clarify which of these are relevant. Once one has computed the vector-type degree of the nodes, one can aggregate this information and define the {\sl overlapping degree} \cite{Batiston2013}  of the node $i\in X$, as
\begin{equation}
o_i=\sum_{\alpha=1}^M k_i^{[\alpha]},
\end{equation}
i.e. $o_i=\|{\mathbf k}_i\|_1$. In fact, many other aggregation measures $f({\mathbf k}_i)$ could be alternatively used to compute the degree centrality, such as a convex combination of $k_i^{[1]},\cdots, k_i^{[M]}$, or any norm of ${\mathbf k}_i$.

Other centrality measures, such as closeness and betweenness centrality, are based on the metric structure of the network. These measures can be easily extended to multilayer networks, once the metric and geodesic structure are defined. In Sec.~\ref{sec:metric}, the reader will find a complete discussion of the metric structures in multilayer networks, which allow straightforward extensions of this kind of centrality measures.

A different approach to measure centrality employs the spectral properties of the adjacency matrix.
In particular, the {\sl eigenvector centrality} considers not only the number of links of each node but also the quality
of such connections \cite{bonacichAmJSociol87}. There are several different ways to extend this idea to multilayer networks,
as discussed in Ref.~\cite{SRCFGB13}, or in Ref.~\cite{Aguirre2013IreneNatPhys} where eigenvector centrality is used to optimize the outcome of interacting competing networks. In the former reference, several definitions of eigenvector-like centrality measures for multiplex networks are presented, along with studies of their existence and uniqueness.

The simplest way to calculate eigenvector-like centralities in multiplex networks is to consider the eigenvector centrality ${\mathbf c}_{\alpha}=(c_1^{[\alpha]},\cdots,c_{N}^{[\alpha]})$ in each layer $1\le\alpha\le M$ separately. With this approach, for every node $i\in X$ the eigenvector centrality ${\mathbf c}_i$ is another vector
\begin{equation}
{\mathbf c}_i=(c_i^{[1]},\cdots,c_i^{[M]})\in \mathbb{R}^M,
\end{equation}
where each coordinate is the centrality in the corresponding layer. Once all the eigenvector centralities have been computed, the {\sl independent layer} eigenvector-like centrality \cite{SRCFGB13} of $\mathcal{M}$ is the
matrix
\begin{equation}
C=
\left(
      \begin{array}{c|c|c|c}
            {\mathbf c}_1^T & {\mathbf c}_2^T & \dots & {\mathbf c}_M^T
      \end{array}
\right)
\in \mathbb{R}^{N\times M}.
\end{equation}
Notice that $C$ is column stochastic, since all components of ${\mathbf c}_{\alpha}$ are semipositive definite and $\|{\mathbf c}_{\alpha}\|_1=1$ for every $1\le \alpha \le M$, and the centrality ${\mathbf c}_i$ of each node $i\in X$ is the $i^{th}$ row of $C$. As for the degree-type indicators, a numeric centrality measure of each node can be obtained by using an aggregation measure $f({\mathbf c}_i)$ such as the sum, the maximum, or the $\ell_p$-norm. The main limitation of this parameter is that it does not fully consider the multilevel interactions between layers and its influence in the centrality of each node.

If one now bears in mind that the centrality  of a node must be proportional to the centrality of its neighbors (that are distributed among all the layers), and if one considers that all the layers have the same importance, one has that
\begin{equation}
\forall x_{i}^{\alpha},x_{j}^{\alpha}\in X_\alpha,\,\,\, c(x_{i}^{\alpha})
\varpropto c(x_{j}^{\alpha})\,\,\, \mbox{ if }(x_{j}^{\alpha}\to
x_{i}^{\alpha})\in G_\alpha,\,\,\, \alpha \in\{1,\dots,M\} ,
\end{equation}
so the {\sl uniform eigenvector-like centrality}  is defined \cite{SRCFGB13} as the positive and normalized eigenvector $\widetilde c$
(if it exists) of the matrix $\widetilde A$ given by
\begin{equation}
\widetilde A=\sum_{\alpha=1}^M (A^{[\alpha]})^\mathrm{T},
\end{equation}
where $(A^{[\alpha]})^\mathrm{T}$ is the transpose of the adjacency matrix of layer $\alpha$. This situation happens, for instance, in social networks, where different people may have different relationships with other people, while one is generically interested to measure the centrality of the network of acquaintances.

A more complex approach is to consider different degrees of importance (or influence) in different layers of the network, and to include this information in the definition of a matrix that defines the mutual influence between the layers. Thus, to calculate the importance  of a node within a specific layer, one must take into account also all the other layers, as some of them may be highly relevant for the calculation. Consider, for instance, the case of a boss living in the same block of flats as one of his employees: the relationship between the two fellows within the condominium layer formed by all the neighbors has a totally different nature from that occurring inside the office layer, but the role of the boss (i.e. his centrality) in this case can be even bigger than if he was the only person from the office living in that block of flats. In other words, one needs to consider the situation where the influence amongst layers is heterogeneous.

To this purpose, one can introduce an {\sl influence matrix} $W=(w_{\alpha \beta})\in \mathbb{R}^{M\times M}$, defined as a non-negative matrix $W\ge 0$ such that $w_{\alpha \beta}$ measures the influence on the layer $G_\alpha$ given by the layer $G_\beta$. Once $\mathcal{G}$ and $W=(w_{\alpha \beta})$ have been fixed, the {\sl local heterogeneous eigenvector-like centrality} of  $\mathcal{G}$ on each layer $G_\alpha$ is defined \cite{SRCFGB13} as a positive and normalized eigenvector
$c^{\star}_\alpha\in \mathbb{R}^N$ (if it exists) of the matrix
\begin{equation}
A^{\star}_\alpha=\sum_{\beta=1}^M w_{\alpha \beta}(A^{[\beta]})^\mathrm{T}.
\end{equation}
So, the {\sl local heterogeneous eigenvector-like centrality} matrix can be defined as
\begin{equation}
C^{\star}=
\left(
      \begin{array}{c|c|c|c}
        {\mathbf c}^{\star}_1 & {\mathbf c}^{\star}_2 & \dots & {\mathbf
c}^{\star}_M
      \end{array}
\right)
\in \mathbb{R}^{N\times M}.
\end{equation}
A similar approach was introduced in Ref.~\cite{Batiston2013}, but using a matrix
\begin{equation}
A^{\star}_\alpha=\sum_{\beta=1}^M b_{\beta}(A^{[\beta]})^\mathrm{T},
\end{equation}
where $b_{\beta}\ge 0$ and $\sum_\beta b_{\beta}=1$.

Another important issue is that, in general, the centrality of a node $x_{i}^{\alpha}$ within a specific layer $\alpha$ may depend not only on the neighbors that are linked to $x_{i}^{\alpha}$ within that layer, but also on all other neighbors of $x_{i}^{\alpha}$ that belong to the other layers. Consider the case of scientific citations in different areas of knowledge. For example, there could be two scientists working in different subject areas (a chemist and a physicist) with one of them awarded the Nobel Prize: the importance of the other scientist will increase even though the Nobel prize laureate had few citations within the area of the other researcher. This argument leads to the introduction of another concept of centrality \cite{SRCFGB13}. Given a multiplex network $\mathcal{M}$ and an influence matrix $W=(w_{\alpha \beta})$, the {\sl global heterogeneous eigenvector-like centrality} of  $\mathcal{M}$ is defined as a positive and normalized eigenvector
$c^{\otimes}\in \mathbb{R}^{N M}$ (if it exists) of the matrix
\begin{equation}
A^{\otimes}=
\left(
\begin{array}{c|c|c|c}
  w_{11}(A^{[1]})^\mathrm{T} & w_{12}(A^{[2]})^\mathrm{T} & \cdots & w_{1M}(A^{[M]})^\mathrm{T} \\ \hline
  w_{21}(A^{[1]})^\mathrm{T} & w_{22}(A^{[2]})^\mathrm{T} & \cdots & w_{2M}(A^{[M]})^\mathrm{T} \\ \hline
  \vdots    & \vdots    & \ddots & \vdots    \\ \hline
  w_{L1}(A^{[1]})^\mathrm{T} & w_{L2}(A^{[2]})^\mathrm{T} & \cdots & w_{MM}(A^{[M]})^\mathrm{T}
\end{array}
\right) \in \mathbb{R}^{(N M)\times (N M)}.
\end{equation}
Note that $A^{\otimes}$ is the Khatri-Rao product of the matrices
\begin{equation}
W=
\left(
  \begin{array}{c|c|c}
    w_{11} & \cdots & w_{1M} \\ \hline
    \vdots & \ddots & \vdots \\ \hline
    w_{M1} & \cdots & w_{MM}
  \end{array}
\right)
\text{ and }
\left(
\begin{array}{c|c|c|c}
(A^{[1]})^\mathrm{T} & (A^{[2]})^\mathrm{T} & \cdots & (A^{[M]})^\mathrm{T}
\end{array}
\right).
\end{equation}
In analogy with what done before, one can introduce the notation
\begin{equation}
c^{\otimes}=
\left(
\begin{array}{c}
  {\mathbf c}^{\otimes}_1 \\ \hline
  {\mathbf c}^{\otimes}_2 \\ \hline
  \vdots        \\ \hline
  {\mathbf c}^{\otimes}_M
\end{array}
\right),
\end{equation}
where ${\mathbf c}^{\otimes}_1,\cdots, {\mathbf c}^{\otimes}_M\in \mathbb{R}^N$. Then, the {\sl global heterogeneous eigenvector-like centrality matrix} of $\mathcal{M}$ is defined as the matrix given by
\begin{equation}
C^{\otimes}=
\left(
    \begin{array}{c|c|c|c}
        {\mathbf c}^{\otimes}_1 & {\mathbf c}^{\otimes}_2 & \dots & {\mathbf
c}^{\otimes}_M
    \end{array}
\right)
\in \mathbb{R}^{N \times M}.
\end{equation}
Note that, in general $C^{\otimes}$ is neither column stochastic nor row stochastic, but the sum of all the entries of $C^{\otimes}$ is 1.

Other spectral centrality measures include parameters based on the stationary distribution of random walkers with additional random jumps, such as the {\sl PageRank centrality} \cite{BrinPage1998}. The extension of these measures and the analysis of random walkers in multiplex networks will be discussed in Sec.~\ref{sec:spreading}.

\subsubsection{Clustering}\label{subsec:clustering}
The graph clustering coefficient introduced by Watts and Strogatz in Ref.~\cite{ws98} can be extended to multilayer networks in many ways. This coefficient quantifies the tendency of nodes to form triangles, following the popular saying ``the friend of your friend is my friend''.

Recall that given a network $\mathcal{G} = (X,E)$ the clustering coefficient of a given node $i$ is defined as
\begin{equation}
c_{\mathcal{G}}(i)=\frac{\text{\# of links between the neighbors of
$i$}}{\text{largest possible \# of links between the neighbors
of $i$}} \,.
\end{equation}
If we think of three people $i$, $j$ and~$k$ with mutual relations between $i$ and $j$ as well as between $i$ and $k$, the clustering coefficient of $i$ represents the likelihood that $j$ and $k$ are also related to each other.
The \emph{global clustering coefficient} of $\mathcal{G}$ is further defined as the average of the clustering coefficients of all nodes. Obviously, the local clustering coefficient is a measure of transitivity \cite{Luce49}, and it can be interpreted as the density of the local node's neighborhood.

Notice that the global clustering coefficient is sometimes defined differently, with an expression that relates it directly to the global features of a network. This alternative definition is not equivalent to the previous one, and it is commonly used in the social sciences~\cite{lamar03}. Its expression, sometimes called network transitivity~\cite{wasserman94}, is
\begin{equation}
T=\frac{\text{\# of triangles in the network}}{\text{\# of triads in the
network}} \,.
\end{equation}

In order to extend the concept of clustering to the context of multilayer networks, it is necessary to consider not only the intralayer links, but also the interlayer links. In Ref.~\cite{CFGGR12}, the Authors establish some relations between the clustering coefficient of a multilevel network, the clustering coefficient of its layers and the clustering coefficient of its projection network. This generalization is obtained straightforwardly by identifying each slice $S_q$ of a multilevel network $\mathcal{M}$ with a layer $G_{\alpha}$ of the corresponding  multiplex network $\mathcal{M}=(\mathcal{G},\mathcal{C})$. Before giving a definition of the clustering coefficient of a node $i \in X$ within a multilevel network $\mathcal{M}$, we need to introduce some notation.

For every node $i\in X$ let $\mathcal{N}(i)$ be the set of all neighbors of $i$ in the projection network $proj(\mathcal{M})$.  For every $\alpha\in \{1,\cdots,M\}$ let $\mathcal{N}_\alpha(i)=\mathcal{N}(i)\cap X_\alpha$ and $\overline{S_\alpha}(i)$ be the subgraph of the layer $G_\alpha$ induced by $\mathcal{N}_\alpha(i)$, i.e. $\overline{S_\alpha}(i)=(\mathcal{N}_\alpha(i), \overline{E_\alpha}(i))$, where
\begin{equation}
\overline{E_\alpha}(i)=\big\{(k,j)\in E_\alpha;\, k,j\in
\mathcal{N}_\alpha(i)\big\}.
\end{equation}
Similarly, we will define $\overline{S}(i)$ as  the subgraph of the projection network $proj(\mathcal{M})$ induced by $\mathcal{N}(i)$. In addition, the complete graph generated by $\mathcal{N}_\alpha(i)$ will be denoted  by $K_{\mathcal{N}_\alpha(i)}$, and the number of links in $K_{\mathcal{N}_\alpha(i)}$ by $|\overline{E_\alpha}(i)|$. With this notation we can define the {\sl clustering coefficient of a given node} $i$ in $\mathcal{M}$ as
\begin{equation}
{\textbf{C}_\mathcal{M}}(i)=\frac{\displaystyle 2\sum_{\alpha=1}^M
|\overline{E_\alpha}(i)|}{\displaystyle\sum_{\alpha=1}^M|\mathcal{N}
_\alpha(i)|(|\mathcal{N}_\alpha(i)|-1)} \,.
\end{equation}
Then, the {\sl clustering coefficient} of $\mathcal{M}$ can be defined as the average of all ${\textbf{C}_\mathcal{M}}(i)$.

\begin{figure}
\centering
\label{excluster}
\includegraphics[width=0.3\textwidth]{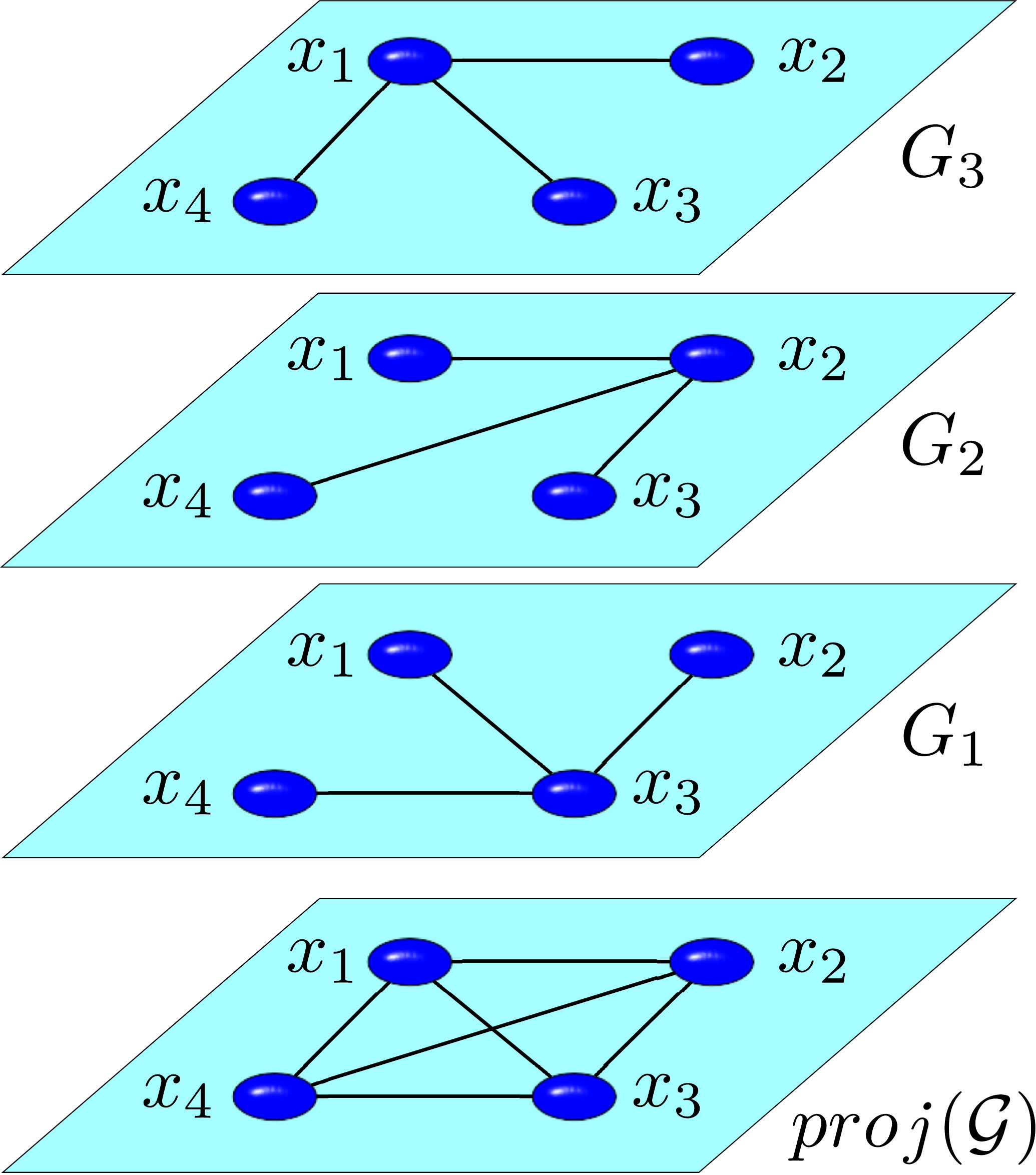}
\caption{(Color online). An example of the difference between clustering coefficients.
The local clustering coefficient of all nodes is~0 in each single layer,
but~1 in the projection network.}\label{clucoe}
\end{figure}

Once again, the clustering coefficient may be defined in several different ways. For instance, we may consider the average of the clustering coefficients of each layer $G_\alpha$. However, it is natural to opt for a definition that considers the possibility that a given node $i$ has two neighbors $k$ and $j$ with $(i,k)\in E_\alpha$, $(i,j)\in E_{\beta}$ with $\alpha \neq \beta$, and $(j,k)\in E_{\gamma}$ with $\gamma \neq \alpha,\beta$. This is a situation occurring in social networks: one person $i$ may know $j$ from the aerobic class and $k$ from a reading club, while $j$ and $k$ know each other from the supermarket. Averaging the clustering coefficients of the layers does not help in describing such situations, and an approach based on the projection network seems more relevant, as evidenced by the following example: consider the multiplex network $\mathcal{M}$ with layers $\{G_1,G_2,G_3\}$ and $X=\{x_1,x_2,x_3,x_4\}$ (Fig.~\ref{clucoe}); it is easy to check that $c_{proj(\mathcal{M})}(x_i)=1$ for all $x_i \in X$ but $c_{G_{\alpha}}(x_i)=0$ for all $x_i$ and $\alpha$.
In order to establish a relation between the clustering coefficients of the nodes in the multiplex network, $\textbf{C}_\mathcal{M}(i)$, and the clustering coefficients of the nodes in the projected network, $c_{proj(\mathcal{M})}(i)$, we can use the same arguments employed in Ref.~\cite{CFGGR12} for multilevel networks. Define $\theta(i)$ as the number of layers in which $i$ has less than two neighbors. Then,
\begin{equation}\label{E:multiclustering_homogeneo}
\frac{1}{M-\theta(i)}\, c_{proj(\mathcal{M})}(i)\le
{\textbf{C}_\mathcal{M}}(i)\le c_{proj(\mathcal{M})}(i).
\end{equation}

Note that Eq.~(\ref{E:multiclustering_homogeneo}) shows that the range of ${\textbf{C}_\mathcal{M}}(i)$ increases with $\theta(i)$.

While the previous example (Fig.~\ref{clucoe}) shows the implicit limitations in defining the clustering coefficient of nodes in a multiplex from that of individual layers, still it makes sense to provide an alternative {\sl ``layers''} definition. Similarly to the previous case, let $\mathcal{N}_\alpha^*(i)=\{j\in X \,;\, j\  \text{is a neighbor of}\  i\ \text{in}\ G_\alpha\}$ and ${S_\alpha}(i)=(\mathcal{N}_\alpha^*(i),E_\alpha(i))$, where
$E_\alpha(i)=\left\{(k,j)\in E_\alpha \,;\, k,j\in\mathcal{N}_\alpha^*(i)\right\}$. Note that ${S_\alpha}(i)$ is the subgraph of the layer $G_\alpha$ induced by $\mathcal{N}_\alpha^*(i)$. Then, the \textit{layer clustering} coefficient of node $i$  is defined as
\begin{equation}
\textbf{C}_\mathcal{M}^{ly}(i)=\frac{2\displaystyle\sum_{\alpha=1}^M
|E_\alpha(i)|}{\displaystyle\sum_{\alpha=1}^M|\mathcal{N}_\alpha^*(i)|(|\mathcal
{N}_\alpha^*(i)|-1)}.
\end{equation}

Notice that $S_{\alpha}(i)$ is a subgraph of $\overline{S_{\alpha}}(i)$. Accordingly, $\mathcal{N}_\alpha^*(i)\subseteq\mathcal{N}_\alpha(i)$ and so the  largest possible number of links between neighbors of $i$ in layer $\alpha$ cannot exceed the corresponding largest possible number of links between neighbors of $i$ in $\mathcal{N}_\alpha(i)$. Also, we have the relation
\begin{equation}
\frac{|\mathcal{N}(i)|(|\mathcal{N}(i)|-1)}{2}\le \frac 12\sum_{\alpha=1}^M
|\mathcal{N}_\alpha^*(i)|(|\mathcal{N}_\alpha^*(i)|-1)+\sum_{k<j}|\mathcal{N}
_k^*(i)||\mathcal{N}_j^*(i)|,
\end{equation}
where the last sum has $\binom{M}{2}$ terms. We can further assume that the nodes can be rearranged so that $|\mathcal{N}_\alpha^*(i)|\le|\mathcal{N}_{\alpha+1}^*(i)|$ for all $1\le \alpha \le M$. Thus, using the method described
in Ref.~\cite{CFGGR12}, we can obtain a relation between the clustering coefficient in the projected network $c_{proj(\mathcal{M})}(i)$ and the layer clustering coefficient $\textbf{C}_\mathcal{M}^{ly}(i)$:
\begin{equation}
\textbf{C}_\mathcal{M}^{ly}(i)\le M\cdot
\,c_{proj(\mathcal{G})}(i)\left[1+(M-1)\left(4+ \frac
{\theta(i)}{M-\theta(i)}\right)\right].
\end{equation}

Another possible  definition of the layer clustering coefficient of a node $i$ is the  average over the clustering coefficients $c_{G_{\alpha}}(i)$ of the slices
\begin{equation}
\textbf{C}_{proj(\mathcal{M})}^{\overline{ly}}(i)=\frac{1}{M}\sum_{\alpha=1}^M
c_{G_{\alpha}} \,.
\end{equation}

The following relationship between both clustering coefficients holds ~\cite{CFGGR12}:
\begin{equation}
M\cdot \min_{1 \leq \alpha \leq M}
\frac{|\mathcal{N}_\alpha^*(i)|(|\mathcal{N}_\alpha^*(i)|-1)}{2} \,
\textbf{C}_{proj(\mathcal{M})}^{\overline{ly}}(i) \leq
\textbf{C}_{proj(\mathcal{M})}^{ly}(i) \leq
M \cdot \max_{1 \leq \alpha \leq M}
\frac{|\mathcal{N}_\alpha^*(i)|(|\mathcal{N}_\alpha^*(i)|-1)}{2}
\,\textbf{C}_{proj(\mathcal{M})}^{\overline{ly}}(i).
\end{equation}

Further generalizations of the notion of clustering coefficient to multilayer networks have been proposed
in Refs. ~\cite{Batiston2013} and~\cite{Cozzo2013}. In Ref.~\cite{Batiston2013}, the Authors point out the necessity of extending the notion of triangle to take into account the richness added by the presence of more than one layer. They define a 2-triangle as a triangle formed by an edge belonging to one layer and two edges belonging to a second layer. Similarly,  a 3-triangle is a triangle which is composed by three edges all lying in different layers. In order to quantify the added value provided by the multiplex structure in terms of clustering, they consider two parameters of clustering interdependence, $I_1$ and $I_2$. $I_1$ ($I_2$) is the ratio between the number of triangles in the multiplex which can be obtained only as 2-triangles (3-triangles), and the number of triangles in the aggregated system.  Then, $I=I_1+I_2$ is the total fraction of triangles of the aggregated  network which cannot be found entirely in one of the layers. They also define a $1$-triad centered at node $i$, for instance $j-i-k$, as a triad in which both edge $(j-i)$ and edge $(i-k)$ are on the same layer. Similarly, a $2$-triad is a triad whose two links belong to two different layers of the systems. This way, they establish two further definitions of clustering coefficient for multiplex networks. For each node $i$ the clustering coefficient $C_{i1}$ is the ratio between the number of $2$-triangles with a vertex in $i$ and the number of $1$-triads centered in $i$. A second clustering coefficient $C_{i2}$ is defined as the ratio between the number of $3$-triangles with node $i$ as a vertex, and the number of $2$-triads centered in $i$. As the Authors point out, while $C_{i1}$ is a suitable definition for multiplexes with $M\geq 2$, $C_{i2}$ can only be defined for systems composed of at least three layers, and both coefficients are poorly correlated, so it is necessary to use both clustering coefficients in order to properly quantify the abundance of triangles in multilayer networks. Averaging over all the nodes of the system, they obtain the network clustering coefficients $C_{1}$ and $C_{2}$.

In Ref.~\cite{Batiston2013} the Authors also generalize the definition of transitivity. They propose two measures of transitivity: $T_1$ as the ratio between the number of $2$-triangles and the number of $1$-triads, and $T_2$ as the ratio between the number of $3$-triangles and the number of $2$-triads. As it is stressed by the Authors, clustering interdependencies $I_1$ and $I_2$, average multiplex clustering coefficients $C_1$ and $C_2$, and multiplex transitivities $T_1$ and $T_2$ are all global network variables which give a different perspective on the multilayer patterns of clustering and triadic closure with respect to the clustering coefficient and the transitivity computed for each layer of the network.

In Ref.~\cite{Cozzo2013}, the Authors derive measurements of transitivity for multiplex networks by developing several multiplex generalizations of the clustering coefficient, and provide a comparison between some different formulations of multiplex clustering coefficients. For instance, Authors point out that the balance between intralayer versus interlayer clustering is different in social versus transportation networks,
reflecting the fact that transitivity emerges from different mechanisms in these cases.
Such differences are rooted in the new degrees of freedom that arise from interlayer connections,
and are invisible to calculations of clustering coefficients on single-layer networks obtained via aggregation.
Generalizing clustering coefficients for multiplex networks thus makes it possible to explore such phenomena and to gain deeper insights into different types of transitivity in networks.
Further multiplex clustering coefficients are defined in Refs.~\cite{Barret2012,Brodka2010,Brodka2012}.

\subsubsection{Metric structures: Shortest paths and distances\label{sec:metric}}
The metric structure of a complex network is related to the topological distance between nodes, written in terms of walks and paths in the graph. So, in order to extend the classical metric concepts to the context of multilayer networks, it is necessary to establish first the notions of {\sl path}, {\sl walk} and {\sl length}. In order to introduce all these concepts, we will follow a similar scheme to that used in Ref.~\cite{Fontouraregino2007AdvancesinPhysics}. Given a multilayer network $\mathcal{M}=(\mathcal{G},\mathcal{C})$, we consider the set
\begin{equation}
 E(\mathcal{M})=\{E_1,\cdots,E_M\} \bigcup \mathcal{C}.
\end{equation}

A {\sl walk} (of length $q-1$) in $\mathcal{M}$ is a non-empty alternating sequence
\begin{equation}
\{x^{\alpha_1}_1,\ell_1,x^{\alpha_2}_2,\ell_2,\cdots,\ell_{q-1},x^{\alpha_q}_q\},
\end{equation}
of nodes and edges with $\alpha_1,\alpha_2,\cdots,\alpha_q \in \{1,\cdots,M\} $, such that for all $r < q$ there exists an $\mathcal{E}\in E(\mathcal{M})$ with
\begin{equation}
 \ell_r=\left({x^{\alpha_r}_r, x^{\alpha_{r+1}}_{r+1}}\right) \in \mathcal{E}.
\end{equation}

If the edges $\ell_1,\ell_2,\ell_{q-1}$ are weighted, the length of the walk can be defined as the sum of the inverse of the corresponding weights. If $x^{\alpha_1}_1 = x^{\alpha_q}_q$, the  walk is said to be {\sl closed}.

A {\sl path} $\omega=\{x^{\alpha_1}_1,x^{\alpha_2}_2,\cdots,x^{\alpha_q}_q\},$ between two nodes $x^{\alpha_1}_1 $ and $x^{\alpha_q}_q$ in $\mathcal{M}$ is a walk through the nodes of $\mathcal{M}$ in which each node is visited only once. A {\sl cycle} is a closed path starting and ending at the same node. If it is possible to find a path between
any pair of its nodes, a multilayer network $\mathcal{M}$ is referred to as {\sl connected}; otherwise it is called disconnected. However, different types of reachability may be considered \cite{Belingerio01,Batiston2013} depending, for example, on whether we are considering only the edges included in some layers.

The {\sl length} of a path is the number of edges of that path. Of course, in a multilayer network there are at least two types of edges, namely intralayer and interlayer edges. Thus, this definition changes depending on whether we consider
interlayer and intralayer edges to be equivalent. Other metric definitions may be easily generalized from monolayer  to multilayer networks. So, a {\sl geodesic} between two nodes $u$ and $v$ in $\mathcal{M}$ is one of the shortest path that connects  $u$ and $v$. The {\sl distance} $d_{uv}$ between $u$ and $v$  is the length of any geodesic between $u$ and $v$.  The maximum distance $D(\mathcal{M})$ between any two vertices in $\mathcal{M}$
is called the {\sl diameter} of $\mathcal{M}$. By $n_{uv}$ we will denote the number of different geodesics that join $u$ and $v$. If $x$ is a node and $\ell$ is a link, then $n_{uv}(x)$ and $n_{uv}(\ell)$ will denote the number of geodesics that join the nodes $u$ and $v$ passing through $x$ and $\ell$ respectively.

A multilayer network $\mathcal{N}=(\mathcal{G'},\mathcal{C'})$ is a subnetwork of $\mathcal{M}=(\mathcal{G},\mathcal{C})$ if for every $\gamma,\delta$ with $\gamma\neq \delta$, there exist $\alpha,\beta$ with $\alpha\neq \beta$ such that $X'_\gamma\subseteq X_\alpha$, $E'_\gamma\subseteq E_\alpha$ and $E'_{\gamma\delta}\subseteq E_{\alpha\beta}$.
A {\sl connected component} of $\mathcal{M}$  is a maximal connected subnetwork of $\mathcal{M}$. Two paths connecting the same pair of vertices in a multilayer network are said to be vertex-independent if they share no vertices other than the starting and ending ones. A {\sl $k$-component} is a maximal subset of the vertices such that every vertex in the subset is connected to every other by $k$ independent paths. For the special cases $k=2$, $k=3$, the $k$-components are called {\sl bi-components} and {\sl three-components} of the multilayer network. For any given network, the $k$-components are nested: every three-component is a subset of a bi-component, and so forth.

The {\sl characteristic path length} is defined as
\begin{equation}
L(\mathcal{M})=\frac 1{N(N-1)} \sum_{\substack{u,v \in X_\mathcal{M} \\ u \ne
v}} d_{uv},
\end{equation}
where $|X_\mathcal{M}|=N$, and is also a way of measuring the performance of a graph.

The concept of {\sl efficiency}, introduced by Latora and Marchiori in Ref.~\cite{lm01} may also be extended to multilayer networks in a
similar way to the one used in Ref.~\cite{CFGGR12}. With the same notation, the efficiency of a multilayer network $\mathcal{M}$ is defined as
\begin{equation}
E(\mathcal{M})= \frac{1}{N(N-1)} \sum_{\substack{u,v \in X_\mathcal{M} \\ u \ne
v}} \frac{1}{d_{uv}}.
\end{equation}

It is quite natural to try to establish comparisons between the efficiency $E(\mathcal{M})$ of a multilayer network $\mathcal{M}$, the efficiency $E(proj(\mathcal{M}))$ of its projection and the efficiencies of the different layers $E(G_\alpha)$. In Ref.~\cite{CFGGR12} there are some analytical results that establish some relationships
 between these parameters.

If we consider  interlayer and intralayer edges not to be equivalent, we can give alternative definitions of the metric quantities. Let $\mathcal{M}=(\mathcal{G},\mathcal{C})$ be a multilayer network, and
\begin{equation}
\omega=\{x^{\alpha_1}_1,\ell_1,x^{\alpha_2}_2,\ell_2,\cdots,\ell_{q},x^{\alpha_{q+1}}_{q+1}\},
\end{equation}
be a path in $\mathcal{M}$. The {\sl length} of $\omega$ can be defined as the non-negative value
\begin{equation}
 \ell(\omega)=q+\beta\sum_{j=2}^q \Delta(j),
\end{equation}
where
\begin{equation}
\Delta(j)= \left\{
  \begin{array}{ll}
    1 & \mbox{if\enspace $\ell_j \in \mathcal{C}$,\quad \text{(i.e. if $\ell_j$
is a crossed layer)} } \\
    0 & \mbox{otherwise,}
  \end{array}
\right.
\end{equation}
and $\beta$ is an arbitrarily chosen non-negative parameter.

In this case, the {\sl distance} in  $\mathcal{M}$ between two nodes  $i$ and $j$ is the minimal length among all possible paths in from $i$ to $j$.

Notice that for $\beta=0$, the previous definition reduces to the natural metric in the projection network $proj(\mathcal{M})$, while positive values of $\beta$ correspond to metrics that take into account also the interplay between the different layers.

One can generalize the definition of path length even further by replacing the {\sl jumping weight} $\beta$ with an $M\times M$ non-negative matrix $\Theta=(\beta (G_\lambda,G_\mu))$, to account for different distances between layers. Thus, the length of a path $\omega$ becomes
\begin{equation}
 \tilde\ell(\omega)=q+\sum_{j=2}^q \tilde\Delta(j),
\end{equation}
where
\begin{equation}
\tilde\Delta(j)= \left\{
  \begin{array}{ll}
    \beta(G_{\sigma(j-1)},G_{\sigma(j)}) & \mbox{if\enspace $G_{\sigma(j)}\ne
G_{\sigma(j-1)}$,} \\
    0 & \mbox{otherwise.}
  \end{array}
\right.
\end{equation}
The {\sl jumping weights} can be further extended to include a dependence not only on the two layers involved in the layer jump, but also on the node from which the jump starts.

For a multiplex network, it is also important to quantify the participation of single nodes to the structure of each layer in terms of node reachability \cite{Batiston2013}. Reachability is an important feature in networked systems. In single-layer networks it depends on the existence and length of shortest paths connecting pairs of nodes. In multilevel systems, shortest paths may significantly differ between different layers, as well as between
each layer and the aggregated topological networks. To address this, the so-called node interdependence
was introduced in Refs.~\cite{NiBiLa13} and ~\cite{MorBar12}. The interdependence $\lambda_i$ of a node $i$ is defined as:
\begin{equation}
\lambda_i=\sum_{j \ne i}\frac {\psi_{ij}}{\sigma_{ij}},
\end{equation}
where $\sigma_{ij}$  is the total number of shortest paths between node $i$ and node $j$ on the multiplex network, and $\psi_{ij}$ is the number of shortest paths between node $i$ and node $j$ which make use of links in two or more layers. Therefore, the node interdependence is equal to $1$ when all shortest paths make use of edges laying on at least two layers, and equal to $0$ when all shortest paths use only one layer of the system. Averaging $\lambda_i$ over all nodes, we obtain the network interdependence.

The interdependence is a genuine multiplex measure that provides information in terms of reachability. It is slightly anti-correlated to measures of degree such as the overlapping degree. In fact, a node with high overlapping degree has a higher number of links that can be the first step in a path toward other nodes; as such, it is likely to have a low $\lambda_i$. Conversely, a node with low overlapping degree is likely to have a high value of $\lambda_i$, since its shortest paths are constrained to a smaller set of edges and layers as first step. This measure is validated in Ref.~\cite{Batiston2013} on the data set of Indonesian terrorists, where information among 78 individuals are recorded with respect to mutual trust, common operations, exchanged communications and business relationships.

\subsubsection{Matrices and spectral properties}\label{cap2:spectralp}
It is well known that the spectral properties of the adjacency and Laplacian matrices of a network provide insights
into its structure and dynamics ~\cite{VanMie12, BroHa12,Boccaletti2006}. A similar situation is possible for multilayer networks, if proper matrix representations are introduced.

Given a multilayer network ${\mathcal{M}}$, several {\sl adjacency} matrices can give information about its structure: amongst others, the most used are the adjacency matrix $A^{[\alpha]}$ of each layer $G_\alpha$, the adjacency matrix $\overline{A_\mathcal{M}}$ of the projection network $proj(\mathcal{M})$ and the supra-adjacency matrix $A_{\mathcal{M}}$. The spectrum of the supra-adjacency matrix is directly related to several dynamical processes that take place on a multilayer network. Using some results from interlacing of eigenvalues of quotients of matrices \cite{VanMie12,BroHa12,HaemersMiguel1995LAA}, in Ref.~\cite{Ruben2014} it was proven that if $\lambda_1\le\cdots\le\lambda_N$ is the spectrum of the supra-adjacency matrix $A_{\mathcal{M}}$ of an undirected multilayer network and $\mu_1\le\cdots\le \mu_{n_\alpha}$ is the spectrum of the adjacency matrix $A^{[\alpha]}$ of layer $G_\alpha$, then for every $1\le k\le n_\alpha$
\begin{equation}
\lambda_k\le \mu_k\le \lambda_{k+N-n_\alpha}.
\end{equation}

A similar situation occurs for the Laplacian matrix of a multiplex network.  The Laplacian matrix  ${\cal L}_\mathcal{M}={\cal L}$ (also called {\sl supra-Laplacian matrix}) of a multiplex  $\mathcal{M}$ is defined as the
$M N\times M N$ matrix of the form:
\begin{equation}
{\cal L}=\left(
  \begin{array}{cccc}
    D_1 {\bf L^{1}} & 0 & \dots & 0\\
    0 & D_2{\bf L^{2}} & \dots & 0\\
    \vdots & \vdots & \ddots & \vdots\\
    0 & 0 & \dots & D_M {\bf L^{M}}\\
  \end{array}
  \right) +\left(
\begin{array}{cccc}
   \sum_{\beta} D_{1\beta} {\bf I} & -D_{12}{\bf I} & \dots & -D_{1M} {\bf I}\\
    -D_{21}{\bf I} & \sum_{\beta} D_{2\beta} {\bf I}  & \dots & -D_{2M} {\bf
I}\\
    \vdots & \vdots & \ddots & \vdots\\
    -D_{M1}{\bf I}  & -D_{M2}{\bf I}  & \dots & \sum_{\beta}  D_{M\beta} {\bf I}
\\
  \end{array}\right) \;.
  \label{supraLaplacian}
\end{equation}
In the equation above, {\bf I} is the $N\times N$ identity matrix and ${\bf L^{\alpha}}$ is the usual $N\times N$ Laplacian matrix of the network layer $\alpha$ whose elements are $L^{\alpha}_{ij}=s^{\alpha}_i\delta_{ij}-w_{ij}^{\alpha}$ where $s^{\alpha}_i$ is the strength of node $i$ in layer $\alpha$, $s^{\alpha}_i=\sum_j w_{ij}^{\alpha}$. We will see in
Sec.~\ref{sec:spreading} that the diffusion dynamics on a multiplex network is strongly related to the spectral properties of ${\cal L}$. In Ref.~\cite{Ruben2014}, the Authors prove that if $\lambda_1\le \cdots\le \lambda_N$ is the spectrum of the Laplacian matrix ${\cal L}$ of an undirected multilayer network and $\mu_1\le\cdots\le \mu_{n_\alpha}$ is the spectrum of the Laplacian matrix $\bf L^{\alpha}$ of layer $G_\alpha$, then for every $1\le k\le n_\alpha$
\begin{equation}
\mu_k\le \lambda_{k+N-n_\alpha}.
\end{equation}
Similar results can be found using perturbative analysis~\cite{SoDo13,radicchi}.

In addition to the spectra of the adjacency and Laplacian matrices, other types of spectral properties have been studied for multiplex network, such as the irreducibility \cite{CriadoMiguel2014Pre}. Several problems in network theory involve the analysis not only of the eigenvalues, but also of the eigenvectors of a matrix~\cite{Boccaletti2006}. This analysis typically includes the study of the existence and uniqueness of a positive and normalized eigenvector (Perron vector), whose existence is guaranteed if the corresponding matrix is irreducible (by using the classic Perron-Frobenius theorem). As for the spectral properties, it is possible to relate the irreducibility of such a matrix with the irreducibility in each layer and on the projection network. In Refs.~\cite{SRCFGB13, CriadoMiguel2014Pre} it was shown that if $w_{ij}>0$ and the adjacency matrix $\overline{A_\mathcal{M}}$ of the projection network $proj(\mathcal{M})$ is irreducible, then the matrix $A^{\otimes}$ defining the global heterogeneous eigenvector-like centrality is irreducible. A similar situation occurs when we consider random walkers in multiplex networks~\cite{Domenico13}. In this case, the uniqueness of a stationary state of the Markov process is guaranteed by the irreducibility of a matrix of the form
\begin{equation}
{\mathbb {B}}=
\left(
\begin{array}{c|c|c|c}
  P_{11} & P_{12} & \cdots & P_{1M} \\ \hline
  P_{21} & P_{22} & \cdots & P_{2M} \\ \hline
  \vdots    & \vdots    & \ddots & \vdots    \\ \hline
  P_{M1} & P_{M2} & \cdots & P_{MM}
\end{array}
\right) \in \mathbb{R}^{(N M)\times (N M)},
\end{equation}
where $P_{\alpha\beta}=W_{\alpha\beta}\odot A^{[\beta]}+v_{\alpha\beta}{\bf I}$, $W_{\alpha\beta}\in\mathbb{R}^{N\times N}$, $v_{\alpha\beta}\in {\mathbb{R}}^N$, and $W_{\alpha\beta}\odot A^{[\beta]}$ is the Hadamard product of matrices $W_{\alpha\beta}$ and $A^{[\beta]}$ (see Ref.~\cite{CriadoMiguel2014Pre}). It can be proven that, under some hypotheses, if the adjacency matrix $\overline{A_\mathcal{M}}$ of the projection network $proj(\mathcal{M})$ is irreducible, then ${\mathbb {B}}$ is irreducible and hence the random walkers proposed in Ref.~\cite{Domenico13} have a unique stationary state.

\subsubsection{Mesoscales: Motifs and modular structures}
An essential method of network analysis is the detection of mesoscopic structures known as communities (or cohesive groups). These communities are disjoint groups (sets) of nodes which are more densely connected to each other than they are to the rest of the network \cite{Fortunatoregino2010Physrep,Porterregino2009noticesoftheAMS,Newmanregino2004PNAS}. Modularity is a scalar that can be calculated for any partition of a network into disjoint sets of nodes. Effectively, modularity is a quality function that counts intracommunity edges compared to what one would expect at random. Thus, one tries to determine a partition that maximizes modularity to identify communities within a network.

In Ref.~\cite{Domenico2013}, there is a definition of modularity for multilayer networks in which the Authors introduce a tensor that encodes the random connections defining the null model. Ref.~\cite{Mucha10} provides a  framework for the study of community structure in a very general setting, covering networks that evolve over time, have multiple types of links (multiplexity), and have multiple scales. The Authors generalize the determination of community structure via quality functions to multislice networks that are defined by coupling multiple adjacency matrices.

As described above, communities are mesoscale structures mainly defined at a global level. One may ask whether more local mesoscales, i.e. those defined slightly above the single node level, can also be extended to multilayer networks. Without doubts, the most important of them are {\it motifs},  i.e. small sub-graphs recurring within a network with a frequency higher than expected in random graphs \cite{milo2002}. Their importance resides in the fact that they can be understood as basic building blocks, each associated with specific functions within the global system \cite{shen2002}.

Little attention so far has been devoted to understanding the meaning of motifs in multilayer networks. An important exception is presented in Ref.~\cite{Szell2010} in the framework of  the analysis of the Pardus social network, an on-line community that comprises $300.000$ players interacting in a virtual universe. When links are grouped in two layers, corresponding to positively and negatively connoted interactions, some resulting structures, which can be seen as multilayered motifs, appear with a frequency much higher (e.g. triplets of users sharing positive relations) or lower (e.g. two enemy users that both have a positive relation with a third) than expected in random networks.


\begin{figure}[t!]
\centering
\includegraphics[width=0.3\textwidth]{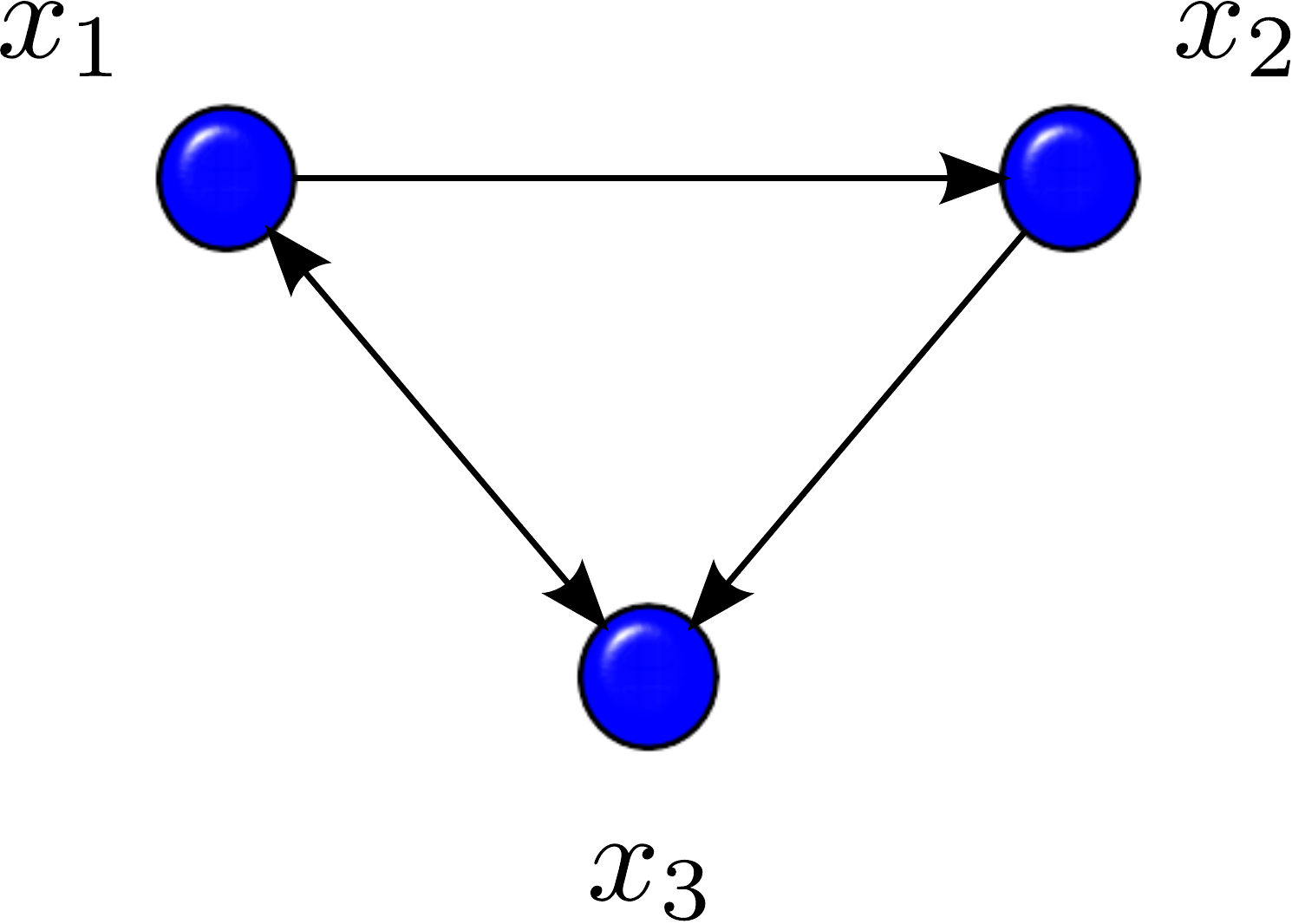}
\caption{Schematic illustration of the monoplex network discussed in Sec.~\ref{cap2:tensor} and whose adjacency matrix
is given in Table~\ref{table1}.}
\label{fsimple}
\end{figure}

\subsection{Matrices and tensor representation: Spectral properties}\label{cap2:tensor}
There have been some attempts in the literature for modeling multilayer networks properly by using the concept of tensors \cite{Domenico2013,Kivela2013,Cozzo2013}. Before describing these results, we recall some basic tensor analysis concepts.

There are two main ways to think about tensors:
\begin{itemize}
\item[(1)] tensors as multilinear maps;
\item[(2)] tensors as elements of a tensor product of two or more vector spaces.
\end{itemize}
The former is more applied. The latter is more abstract but more powerful. The tensor product of two real vector spaces $\mathcal{V}$ and $\mathcal{L}$, denoted by $\mathcal{V}\otimes\mathcal{L}$, consists of finite linear combinations of $v\otimes w$, where $v \in \mathcal{V}$ and $w \in \mathcal{L}$.

The dual vector space of a real vector space $\mathcal{V}$ is the vector space of linear functions $f:\mathcal{V}\longrightarrow\mathbb{R}$, indicated by $\mathcal{V}^*$. Denoting by $Hom(\mathcal{V}^*,\mathcal{L})$ the set of linear functions from $\mathcal{V}^*$ to $\mathcal{L}$, there is a natural isomorphism between the linear spaces $\mathcal{V}\otimes \mathcal{L}$ and $Hom(\mathcal{V}^*,\mathcal{L})$. Moreover, if $\mathcal{V}$ is a finite dimensional vector space, then there exists a natural isomorphism (depending on the bases considered) between the linear spaces $\mathcal{V}$ and $\mathcal{V}^*$. In fact, if $\mathcal{V}$ is finite-dimensional, the relationship between $\mathcal{V}$ and $\mathcal{V}^*$ reflects in an abstract way the relation between the $(1\times N)$ row-vectors  and the $(N\times 1)$ column-vectors of a $(N\times N)$ matrix. However, if $\mathcal{V}$ and $\mathcal{L}$ are finite-dimensional, then we can identify the linear spaces $\mathcal{V}\otimes \mathcal{L}$ with $Hom(\mathcal{V},\mathcal{L})$. Thus, a tensor $\sigma \in \mathcal{V}\otimes \mathcal{L}$ can be understood as a linear function $\sigma:\mathcal{V}\longrightarrow \mathcal{L}$, and therefore, once two bases of the corresponding vector spaces have been fixed, a tensor may be identified with a specific matrix.

Now, let us consider a multiplex network $\mathcal{M}$ made of $M$ layers $\{G_\alpha;\enspace 1\le\alpha\le M\}$,  $G_\alpha=(X_\alpha,E_\alpha)$, with $X_1= \cdots =X_M=\{x_1,\cdots,x_N\}=X$, in order to describe it as a tensor product
\begin{equation}
\mathcal{M}=\left\langle X\right\rangle \otimes \left\langle\{\ell_1, \cdots, \ell_M\}\right\rangle,
\end{equation}
where $\ell_i$ represents the layer $G_i$.

Linear combinations of a family of vectors are obtained by multiplying a finite number among them by nonzero scalars (usually real numbers) and adding up the results. The span of a family of vectors is the set of all their linear combinations. In the following, we will denote the span of a family of vectors $\{v_1,\cdots,v_m\}$ by
\begin{equation}
span (\{v_1,\cdots,v_m\})=\left\langle\{v_1,\cdots,v_m\}\right\rangle\:.
\end{equation}
Given the sets $\{x_1,\cdots,x_N\}$ and $\{\ell_1, \cdots, \ell_M\}$, we consider the formal vector spaces on $\mathbb{R}$ of all their linear combinations as follows:
\begin{equation}
\mathcal{V}=\left\langle\{x_1,\cdots,x_N\}\right\rangle,
\end{equation}
\begin{equation}
\mathcal{L}=\left\langle\{\ell_1, \cdots, \ell_M\}\right\rangle.
\end{equation}

It is straightforward to check that $\{x_1,\cdots,x_N\}$ and $\{\ell_1, \cdots, \ell_M\}$ are basis of $\mathcal{V}$ and $\mathcal{L}$ respectively, and therefore
\begin{equation}
\{x_i\otimes\ell_\alpha;\enspace 1\le i\le N,\enspace 1\le\alpha\le M\}
\end{equation}
is a basis of $\mathcal{V}\otimes \mathcal{L}$. So, we may give $x_i\otimes\ell_\alpha$ the intuitive meaning of ``being in the node $i$ and in the layer $\ell_\alpha $''.

\begin{figure}[t!]
\centering
\includegraphics[width=0.35\textwidth]{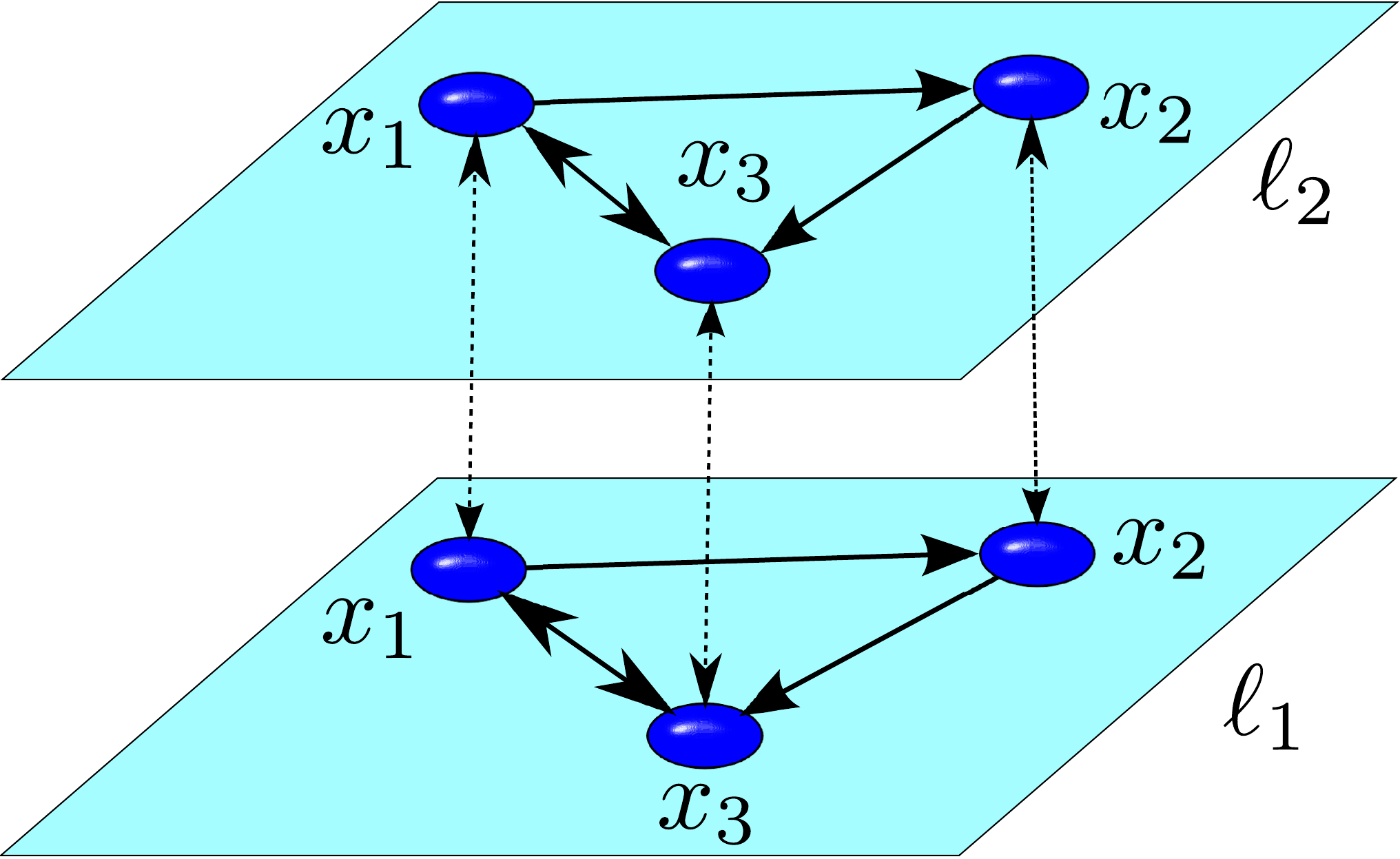}
\caption{(Color online). Schematic illustration of a two-layer multiplex network obtained by duplicating the network of Fig.~\ref{fsimple}, and connecting each node with its counterpart in the other layer. }\label{twolmult}
\end{figure}

\begin{table}[b!]
\begin{center}
\begin{tabular}{|c|c|c|c|}
  \hline
   &  $x_1$& $x_2$ &$x_3 $\\
  \hline
  $x_1$  & 0  & 1  & 1  \\
  \hline
  $x_2$  & 0  & 0  & 1  \\
  \hline
  $x_3$  & 1  & 0  & 0  \\
  \hline
\end{tabular}
\end{center}
\caption{The adjacency matrix of the monoplex in Fig.~\ref{fsimple}}\label{table1}
 \end{table}

Now, if we consider a monolayer network $G=(X,E)$ of $N$ nodes, with $V=\left\langle\{x_1,\cdots,x_N\}\right\rangle$, we can identify $G$ with the linear transformation $\sigma:V\longrightarrow V$ such that the matrix associated to $\sigma$ with respect to the basis $X=\{x_1,\cdots,x_N\}$ is the adjacency matrix of $G$.

Thus, if the adjacency matrix of $G$ is $A$, we can find the set of nodes accessible from any node $i$ in one step by multiplying $A$ on the left by the $i^\mathrm{th}$ canonical basis row-vector. For example, the adjacency matrix of the graph in Fig.~\ref{fsimple} is
\begin{equation}
\left(
\begin{array}{ccc}
0 & 1 & 1 \\
0 & 0& 1 \\
1 & 0& 0
\end{array}
\right)\:.
\end{equation}
From node~1, we can access in one step nodes~2 and~3, since
\begin{equation}
\left(
 \begin{array}{ccc}
  1&0&0
 \end{array}
\right)
\left(
\begin{array}{ccc}
0 & 1 & 1 \\
0 & 0& 1 \\
1 & 0& 0
\end{array}
\right) = \left(\begin{array}{c}
0\\
1\\
1
\end{array}
\right)\:.
\end{equation}

In the same way, a multilayer network $\mathcal{M}$ of $N$ nodes and with layers $\{\ell_1, \cdots, \ell_M\}$ can be identified with a linear transformation $\sigma:\mathcal{V} \otimes \mathcal{L}\longrightarrow \mathcal{V}\otimes \mathcal{L}$ where $X_\mathcal{M}= \{x_1,\cdots,x_N\}$,
\begin{equation}
\mathcal{V}= \left\langle\{x_1,\cdots,x_N\}\right\rangle,
\end{equation}
\begin{equation}
\mathcal{L}= \left\langle\{\ell_1, \cdots, \ell_M\}\right\rangle.
\end{equation}

\begin{figure}[t!]
\begin{center}
\includegraphics[width=0.35\textwidth]{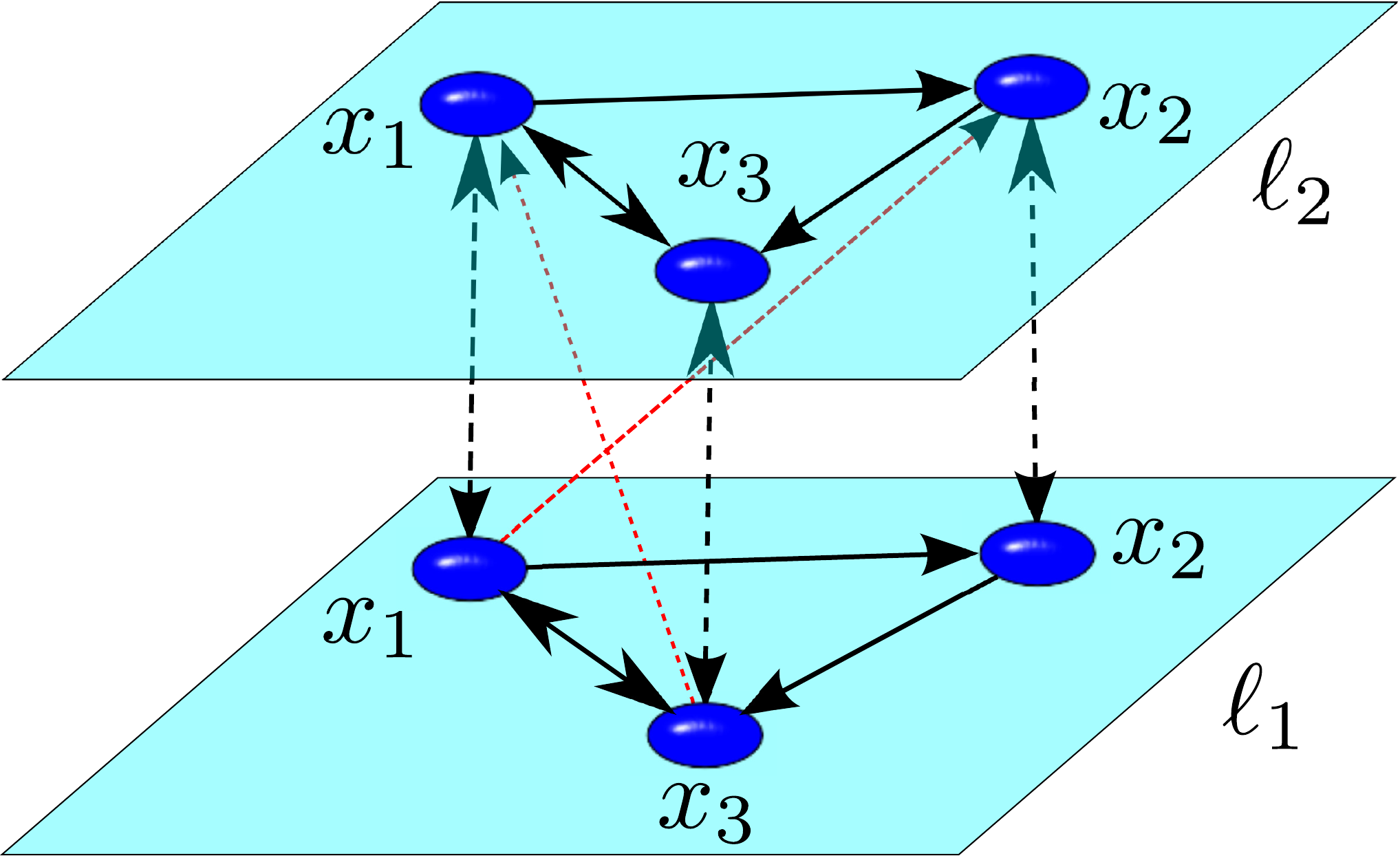}
\end{center}
\caption{(Color online). Schematic illustration of a generic two-layer network with interlayer connections.}\label{twolintcon}
\end{figure}

It is easy to check that $\mathcal{M}$ is completely determined by the matrix associated to $\sigma$ with respect to the basis
\begin{equation}
\{x_i\otimes\ell_\alpha;\ 1\le i\le N,\enspace 1\le \alpha \le M\},
\end{equation}

As an example, consider the monolayer network $G=(X,E)$, with $X=\{x_1,x_2,x_3\}$, whose adjacency matrix $A$ is given in Table~\ref{table1}.

\begin{table}[b!]
 \begin{center} \begin{tabular}{c|c|c|c|c|c|c|c|}
\multicolumn{2}{c|}{}&\multicolumn{2}{|c|}{$x_1$}&\multicolumn{2}{|c|}{$x_2$}
&\multicolumn{2}{|c|}{$x_3$}\\
     \cline{3-8}

\multicolumn{2}{c|}{}
&$x_1\otimes\ell_1$&$x_1\otimes\ell_2$&$x_2\otimes\ell_1$&$x_2\otimes\ell_2$&$x_
3\otimes\ell_1$&$x_3\otimes\ell_2$\\
     \hline
     \multirow{2}{*}{$x_1$}&$x_1\otimes\ell_1$&0& 1& 1&0& 1&0 \\
     \cline{2-8}
                           &$x_1\otimes\ell_2$&1& 0&0 &1 & 0& 1\\
     \hline
     \multirow{2}{*}{$x_2$}&$x_2\otimes\ell_1$&0&0 &0 & 1&1 &0 \\
     \cline{2-8}
                           &$x_2\otimes\ell_2$&0& 0&1& 0 & 0& 1\\
     \hline
     \multirow{2}{*}{$x_3$}&$x_3\otimes\ell_1$&1&0 & 0& 0& 0& 1\\
     \cline{2-8}
                           &$x_3\otimes\ell_2$&0& 1& 0& 0& 1&0 \\
     \hline
\end{tabular}
\end{center}
\caption{Adjacency matrix of the duplex obtained by duplicating the network of Fig.~\ref{fsimple}, and connecting each node with its counterpart in the other layer.}\label{table2}
\end{table}

Now consider the multiplex network $\mathcal{M}$ with layers $\{\ell_1, \ell_2\}$, and nodes $X_\mathcal{M}= \{x_1,x_2,x_3\}$, obtained by duplicating the previous monoplex network in both layers, and connecting each node with its counterpart in the other layer (Fig.~\ref{twolmult}). The adjacency matrix of $\mathcal{M}$ is given in Table~\ref{table2}.

Of course, the specific matrix representing a tensor depends on the way we order the elements of the basis chosen.
\begin{equation}
\begin{split}
\{x_i\otimes\ell_\alpha;\ i \in \{1,\cdots,N\}, \enspace \alpha \in \{1,\cdots,M\}\},
 \end{split}
\end{equation}
However, if $A$ and $B$ are two different matrices assigned to the same tensor, there exists a permutation matrix $P$ such that $B=P \cdot A \cdot P^{-1}$, and thus both matrices have the same spectral properties (see Sec.~\ref{cap2:spectralp}).

If in the previous example we swap the order of layers and nodes, we get a different adjacency matrix, called \emph{supra-adjacency matrix} of $\mathcal{M}$ (Table~\ref{table3}), which can be written in the usual form
\begin{equation}
\mathcal{A}=
\left(
\begin{array}{c|c}
A_1 & I_3 \\ \hline
I_3 & A_2
\end{array}
\right)\in \mathbb{R}^{6\times 6},
\end{equation}
where $A_i$ is the adjacency matrix of layer $\ell_i$. Note that,
unlike the case of multiplex networks, the blocks in the
supra-adjacency matrix of a general multilayer networks are not
necessarily square matrices. It is worth mentioning the dual roles
between layers and nodes that may be observed by comparing Table~\ref{table2} and
Table~\ref{table3}.

An example of a generic 2-layer network, can be seen in Fig.~\ref{twolintcon}, where we can pass in one step from node $x_1\otimes \ell_1 $ to node $x_2\otimes \ell_2 $, and from node $x_3\otimes \ell_1 $ to node $x_1\otimes \ell_2 $. Its adjacency matrix is listed in Table~\ref{table4}. It is important to stress that while the adjacency matrix of a multilayer network with $M$ layers and $N$ nodes has $N^{2}M^{2}$ entries, the adjacency matrix of a multiplex network with the same layers and nodes is completely determined by only $N^{2}M + M^{2}$ elements. In fact, a multiplex network of $M$ layers is completely determined by the $M$ adjacency matrices corresponding to the layers and its influence matrix $W=(w_{\alpha \beta}) \in \mathbb{R}^{M\times M}$, i.e. the non-negative matrix $W\ge 0$ such that $w_{\alpha \beta}$ is the influence of the layer $G_{\alpha}$ on the layer $G_{\beta}$.

\begin{table}[t!]
 \begin{center} \begin{tabular}{c|c|c|c|c|c|c|c|}

\multicolumn{2}{c|}{}&\multicolumn{3}{|c|}{$\ell_1$}&\multicolumn{3}{|c|}{
$\ell_2$}\\
     \cline{3-8}

\multicolumn{2}{c|}{}
&$x_1\otimes\ell_1$&$x_2\otimes\ell_1$&$x_3\otimes\ell_1$&$x_1\otimes\ell_2$&$x_
2\otimes\ell_2$&$x_3\otimes\ell_2$\\
     \hline
     \multirow{3}{*}{$\ell_1$}&$x_1\otimes\ell_1$&0& 1& 1&1& 0&0 \\
     \cline{2-8}
                           &$x_2\otimes\ell_1$&0& 0&1 &0 & 1& 0\\
     \cline{2-8}
                           &$x_3\otimes\ell_1$&1& 0&0 &0 & 0& 1\\
     \hline
     \multirow{3}{*}{$\ell_2$}&$x_1\otimes\ell_2$&1&0 &0 & 0&1 &1 \\
     \cline{2-8}
                           &$x_2\otimes\ell_2$&0& 1&0& 0 & 0& 1\\
     \cline{2-8}
                           &$x_3\otimes\ell_2$&0&0 & 1& 1& 0& 0\\
     \hline

\end{tabular}
\end{center}
\caption{The supra-adjacency matrix (see text for definition) corresponding to the duplex network in Fig.~\ref{twolmult}.}\label{table3}
\label{tab:tabla2}
\end{table}

\begin{table}[b!]
\begin{center} \begin{tabular}{c|c|c|c|c|c|c|c|}
\multicolumn{2}{c|}{}&\multicolumn{2}{|c|}{$x_1$}&\multicolumn{2}{|c|}{$x_2$}
&\multicolumn{2}{|c|}{$x_3$}\\
     \cline{3-8}

\multicolumn{2}{c|}{}
&$x_1\otimes\ell_1$&$x_1\otimes\ell_2$&$x_2\otimes\ell_1$&$x_2\otimes\ell_2$&$x_
3\otimes\ell_1$&$x_3\otimes\ell_2$\\
     \hline
     \multirow{2}{*}{$x_1$}&$x_1\otimes\ell_1$&0& 1& 1&1& 1&0 \\
     \cline{2-8}
                           &$x_1\otimes\ell_2$&1& 0&0 &1 & 0& 1\\
     \hline
     \multirow{2}{*}{$x_2$}&$x_2\otimes\ell_1$&0&0 &0 & 1&1 &0 \\
     \cline{2-8}
                           &$x_2\otimes\ell_2$&0& 0&1& 0 & 0& 1\\
     \hline
     \multirow{2}{*}{$x_3$}&$x_3\otimes\ell_1$&1&1 & 0& 0& 0& 1\\
     \cline{2-8}
                           &$x_3\otimes\ell_2$&0& 1& 0& 0& 1&0 \\
     \hline
\end{tabular}
\end{center}
\caption{The adjacency matrix of the two-layer network with interlayer connections sketched in Fig.~\ref{twolintcon}.}\label{table4}
\end{table}

Multilayer networks, multidimensional networks, hypergraphs, and some other objects can be
represented with tensors, which represent a convenient formalism to implement different models~\cite{KB09}.
Specifically, tensor-decomposition methods \cite{KB09,DunKolKe11} and multiway data analysis \cite{AcYe2009}
have been used to study various types of networks. These kind of methods are based on representing multilayer networks as adjacency higher-order tensors and then applying generalizations of methods such as Singular Value Decomposition (SVD)~\cite{MarPor12}, and the combination of CANDECOMP (canonical decomposition) by Carroll and Chang~\cite{CaChan70} and PARAFAC (parallel factors) by Harshman~\cite{Harshman70} leading to CANDECOMP/PARAFAC (CP).
These methods have been used to extract communities~\cite{DunKolKe11}, to rank nodes~\cite{KolBa2006,KolBaKe05} and to perform data mining~\cite{KB09,Kolda2008,Acar2009}.

\subsection{Correlations in multiplex networks}
Multiplex networks encode significantly more information than their single layers taken in isolation, since they include correlations between the role  of the nodes in different layers and between statistical properties of the single layers. Discovering the  statistically significant correlations in multilayer networks
is likely to be  one of the major goals of network science for  the next years. Here we outline the major types of correlations explored so far. We distinguish between:
\begin{itemize}
 \item {\it Interlayer degree correlations}\\
Generally speaking these correlations are able to  indicate  if the hubs in one layer are also the hubs, or  they are typically low degree nodes, in the other layer.
\item {\it Overlap and multidegree}\\
The node connectivity patterns can be correlated in two or more layers and these correlations can be captured by the overlap of the links. For example, we usually have a large fraction of friends with which we communicate through multiple means of communications, such as phone, email and instant messaging. This implies that the mobile phone social network has a significant overlap with the one of email communication or the one of  instant messaging.
The overlap of the links can be quantified by the global or local  overlap between two layers, or by the multidegrees of the nodes that  determine the specific overlapping pattern.
\item{\it Multistrengths and inverse multiparticipation ratio of weighted multiplex}\\
The weights of the links in the different layers can be correlated with other structural properties of the multiplex. For example, we tend to cite collaborators differently from other scientists. These types of correlations between structural properties of the multiplex and the weights distribution are captured by the multistrengths and inverse multiparticipation ratio.
\item {\it Node pairwise multiplexity}\\
When the nodes are not all active in all layers two nodes can have correlated activity patterns. For example they can be active on the same, or on different layers. These correlations are captured by the Node Pairwise Multiplexity.
\item {\it Layer pairwise multiplexity}\\
When the nodes are not all active in all layers, two layers  can have correlated activity patterns. For example they can  contain the same active nodes, or different active nodes. These correlations are captured by the Layer
Pairwise Multiplexity.
\end{itemize}

\subsubsection{Degree correlations in multiplex networks}
\label{par_corr}
\begin{figure}[!t]
 \centering
   \includegraphics[width=0.65\textwidth]{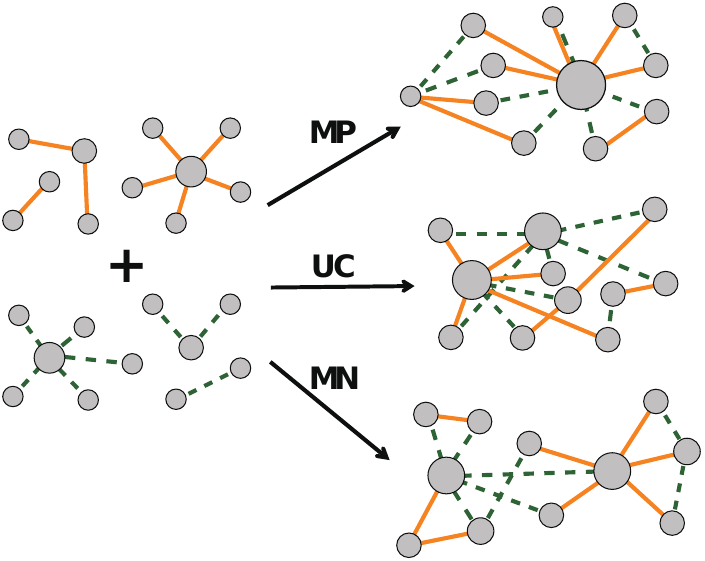}
 \caption{(Color online). Schematic illustration of three kinds of correlated multiplex networks, maximally-positive (MP), uncorrelated (UC), and maximally-negative (MN). Each layer of the networks has different types of links, indicated by
solid and dashed links, respectively. Reprinted figure with permission from Ref.~\cite{MinGo2013}. \copyright\, 2014 by the American Physical
Society.}
    \label{fig:goh_corr}
\end{figure}

Every node $i=1,2,\ldots, N$ of a multiplex has a  degree $k_i^{[\alpha]}$ in each layer $\alpha=1,2,\ldots, M$. The degrees of the same node in different layers can be correlated. For example, a node that is a hub in the scientific collaboration network is likely to be a hub also in the citation network between scientists. Therefore, the degree in the collaboration network is positively correlated with that of the citation network. Negative correlations also exist, when the hubs of one layer are not the hubs of another layer. The methods to evaluate the degree correlations between a layer $\alpha$ and a layer $\beta$  are the following:
\begin{itemize}
\item{\it Full characterization of the matrix $P(k^{\alpha},k^{\beta})$.}\\
One can construct the matrix
\begin{equation}
P(k^{\alpha},k^{\beta})=\frac{N(k^{\alpha},k^{\beta})}{N},
\end{equation}
where $N(k^{\alpha},k^{\beta})$ is the number of nodes that have degree $k^{\alpha}$ in layer $\alpha$ and degree $k^{\beta}$ in layer $\beta$. From this matrix, the full pattern of correlations can be determined.
\item{\it Average degree in layer $\alpha$ conditioned on the degree of the node in layer $\beta$}\\
A more coarse-grained measure of correlation is $\bar{k^{\alpha}}(k^{\beta})$, i.e. the average degree of a node in layer $\alpha$ conditioned to the degree of the same node in layer $\beta$:
\begin{equation}
\bar{k}^{\alpha}(k^{\beta})=\sum_{k^{\alpha}}k^{\alpha}P(k^{\alpha}|k^{\beta}
)=\frac{\sum_{k^{\alpha}}k^{\alpha}P(k^{\alpha},k^{\beta})}{\sum_{k^{\alpha}}
P(k^{\alpha},k^{\beta})}.
\end{equation}
If this function does not depend on $k^{\beta}$, the degrees in the two layers are uncorrelated.
If this function is increasing (decreasing) in $k^{\beta}$, the degrees of the nodes in the two layers are positively (negatively) correlated.
\item{\it Pearson, Spearman and Kendall correlations coefficients}\\
Even more coarse-grained correlation measures are the Pearson, the Spearman and the Kendall's correlations coefficients.

The Pearson correlation coefficient $r_{\alpha\beta}$ is given by
\begin{equation}
r_{\alpha\beta}=\frac{\Avg{k_i^{[\alpha]}k_i^{[\beta]}}-\Avg{k_i^{[\alpha]}}\Avg{k_i^{[\beta]}},
}{\sigma_{\alpha}\sigma_{\beta}}
\end{equation}
where $\sigma_{\alpha}=\sqrt{\Avg{k_i^{[\alpha]}k_i^{[\alpha]}}-\Avg{k_i^{[\alpha]}}^2}$. The Pearson correlation coefficient can be dominated by the correlations of the high degree nodes if the degree distributions are broad.

The Spearman correlation coefficient $\rho_{\alpha\beta}$  between the degree sequences $\{k_i^{[\alpha]}\}$ and $\{k_i^{[\beta]}\}$ in the two layers $\alpha$ and $\beta$ is given by
\begin{equation}
\rho_{\alpha\beta}=\frac{\Avg{x_i^{\alpha}x_i^{\beta}}-\Avg{x_i^{\alpha}}\Avg{x_i^{\beta}}}{\sigma({x^
{\alpha}})\sigma(x_i^{\beta})},
\end{equation}
where $x_i^{\alpha}$ is the rank of the degree $k_i^{[\alpha]}$ in the sequence $\{k_i^{[\alpha]}\}$ and  $x_i^{\beta}$ is the rank of the degree $k_i^{[\beta]}$ in the sequence $\{k_i^{[\beta]}\}$. Moreover, $\sigma({x_i^{\alpha}})=\sqrt{\Avg{(x_i^{\alpha})^2}-\Avg{x_i^{\alpha}}^2}$ and $\sigma(x_i^{\beta})=\sqrt{\Avg{(x_i^{\beta})^2}-\Avg{x_i^{\beta}}^2}$. The Spearman coefficient has the problem that the ranks of the degrees in each
degree sequence are not uniquely defined because degree sequences must always have some degeneracy. Therefore, the Spearman correlation coefficient is not a uniquely defined number.

The Kendall's $\tau $ correlation coefficient between the degree sequences $\{k_i^{[\alpha]}\}$ and $\{k_i^{[\beta]}\}$  is a measure that takes into account the sequence of ranks $\{x_i^{\alpha}\}$ and $\{x_i^{\beta}\}$. A pair of nodes $i$ and $j$ is concordant if their ranks have the same order in the two sequences, i.e. $(x_i^{\alpha}-x_j^{\alpha})(x_i^{\beta}-x_j^{\beta})>0$, and discordant otherwise. The Kendall's $\tau$ is defined in terms of the number $n_c$ of concordant pairs and the number $n_d$ of discordant pairs and is given by
\begin{equation}
\tau=\frac{n_c-n_d}{\sqrt{(n_0-n_1)(n_0-n_2)}}\:.
\end{equation}
In the equation above, $n_0=1/2N(N-1)$ and the terms $n_1$ and $n_2$ account for the degeneracy of the ranks and are given by
\begin{align}
n_1=\frac{1}{2}\sum_n u_n(u_n-1),\nonumber \\
n_2=\frac{1}{2}\sum_n v_n(v_n-1),
\end{align}
where $u_n$  is the number of nodes in the $n^\mathrm{th}$ tied group of the degree sequence $\{k_i^{[\alpha]}\}$, and $v_n$ is the number of nodes in the $n^\mathrm{th}$ tied group of the degree sequence $\{k_i^{[\beta]}\}$.
\end{itemize}
The level of degree correlations in the multiplex networks can be tuned by reassigning new labels to the nodes of the multiplexes \cite{MinGo2013}. Let us for simplicity  consider a duplex where we have two degree sequences
$\{k_i^{[1]}\}$ and $\{k_i^{[2]}\}$ as proposed in Ref.~\cite{MinGo2013}. The label of the nodes in layer 1 and layer 2 can be reassigned in order to generate the desired correlations between the two degree sequences. Assume we have ranked the degree sequences in the two layers. If the labels of the nodes in both sequences are given according to the increasing rank of the degree in each layer, then the sequences are maximally-positive correlated (MP), i.e. the nodes of equal rank in the two sequences are replica nodes of the multiplex. Conversely, if the label of a node in one layer is given according to the increasing rank of the degree in the same layer, and the label of a node in the other layer is given according to the decreasing rank of the degree in the same layer, the two degree sequences are maximally-negative correlated (MN). In Fig.~\ref{fig:goh_corr}, an example is given for the construction of maximally-positive (MP) correlations and maximally-negative (MN) correlations from two given degree sequence in two layers of a multiplex network.

\subsubsection{Overlap and multidegree}
A large variety of data sets, including the multiplex airport network, on-line social games, collaboration and citation networks~\cite{Szell2010,CardilloSR13,Menichetti}, display a significant overlap of the
links. This means that the number of links present at the same time in two different layers is not negligible with respect to the total number of links in the two layers.

The {\em total overlap } $O^{\alpha\beta}$ between two layers $\alpha$ and $\beta$ \cite{Bianconi13,Szell2010} is defined as the total number of links that are in common between layer $\alpha$ and layer $\beta$:
\begin{equation}
O^{\alpha\beta}=\sum_{i<j}a_{ij}^{\alpha}a_{ij}^{\beta},
\end{equation}
where $\alpha\neq \beta$.

It is sometimes useful also to define the {\em local overlap} $o_i^{\alpha\beta}$ between two layers $\alpha$ and $\beta$ \cite{Bianconi13}, defined as the total number of neighbors of node $i$ that are neighbors in  both layer $\alpha$ and layer $\beta$:
\begin{equation}
o_i^{\alpha\beta}=\sum_{j=1}^Na_{ij}^{\alpha}a_{ij}^{\beta}.
\end{equation}
The total and local overlap in multilayer networks can be very significant. Take for example the online social game studied
in Ref.~\cite{Szell2010}. In this data set the friendship and the communication layers have a significant overlap,
similarly to the communication and trade layers.  Even the two ``negative'' layers of  enmity and attack have significant overlap of the links. As a second example of multiplex network with significant overlap, consider the APS data set of citations and collaboration networks \cite{Menichetti}. The two layers in this data set display significant overlap because two co-authors are also usually citing each other in their papers.

\begin{figure}[!t]
 \centering
  \includegraphics[width=0.8\textwidth]{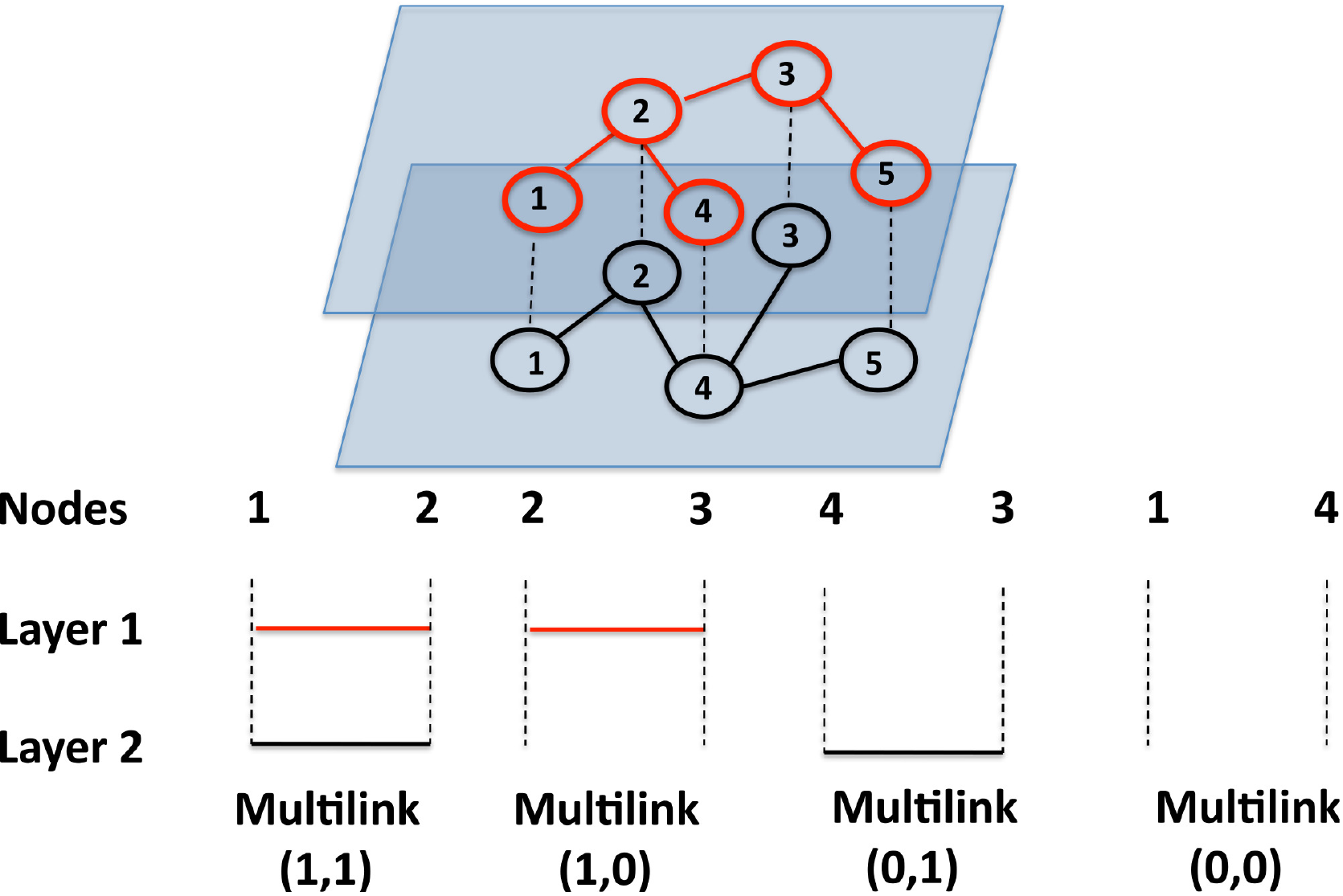}
 \caption{(Color online). Example of all possible multilinks in a
   multiplex network with $M=2$ layers and $N=5$ nodes. Nodes $i$ and
   $j$ are linked by one multilink $\vec{m}=(m_{\alpha},m_{\alpha
     '})$.  Reprinted figure from Ref.~\cite{Menichetti}.}
    \label{fig:multilink}
\end{figure}

One way to characterize the link overlap is by introducing the concept of {\em multilinks} \cite{Bianconi13,Menichetti}. A multilink fully determines all the links present between any given two nodes $i$ and $j$ in the multiplex. Consider for example the multiplex with $M=2$ layers, i.e. the duplex shown in Fig.~~$\ref{fig:multilink}$. Nodes~1 and~2 are connected by one link in the first layer and one link in the second. Thus, we say that the nodes are connected by a multilink $(1,1)$. Similarly, nodes~2 and~3 are connected by one link in the first layer and no link in layer~2. Therefore, they are connected by a multilink $(1,0)$. In general, for a multiplex of $M$ layers we can define {\it multilinks}, and {\it multidegrees} in the following way. Let us consider the  vector $\vec{m}=(m_1,m_2,\ldots,m_{\alpha},\ldots m_M)$ in which every element $m_{\alpha}$ can take only two values $m_{\alpha}=\left\lbrace 0,1\right\rbrace$. We define a {\it multilink} $\vec{m}$ as the set of links connecting a given pair of nodes in the different layers of the multiplex: $m_\alpha=1$ if and only if they are connected in layer $\alpha$. We can therefore introduce the {\it multiadjacency matrices $A^{\vec{m}}$} with elements $A_{ij}^{\vec{m}}$ equal to~1 if there is a multilink $\vec{m}$ between node $i$ and node $j$, and zero otherwise:
\begin{equation}
 A_{ij}^{\vec{m}}=\prod_{\alpha=1}^M
\left[a_{ij}^{\alpha}m_{\alpha}+(1-a_{ij}^{\alpha})(1-m_{\alpha})\right].
 \label{MA}
\end{equation}

Thus, multiadjacency matrices satisfy the condition
\begin{equation}
\sum_{\vec{m}} A_{ij}^{\vec{m}}=1,
\end{equation}
for every fixed pair of nodes $(i,j)$. Multiadjacency matrices are a way to describe the same information encoded in the $M$ adjacency matrices of each layer, and while they do not encode more information, they can be useful tools to characterize partial overlap and correlations. Finally, we define the {\it multidegree} $\vec{m}$  of a node $i$, $k_i^{\vec{m}}$ as the total number of multilinks $\vec{m}$ incident to node $i$, i.e.
\begin{equation}
k_i^{\vec{m}}=\sum_{j=1}^N A_{ij}^{\vec{m}}.
\end{equation}
For a duplex the non trivial multidegrees, i.e. the multidegrees $\vec{m}\neq\vec{0}$, have the explicit expression in terms of the adjacency matrix ${\bf a}^{1}$ of layer 1 and ${\bf a}^2$ of layer 2 given by
\begin{align}
k^{(1,0)}_i&=\sum_{j=1}^N a_{ij}^1(1-a_{ij}^2),\nonumber \\
k^{(0,1)}_i&=\sum_{j=1}^N (1-a_{ij}^1)a_{ij}^2,\nonumber \\
k^{(1,1)}_i&=\sum_{j=1}^Na_{ij}^1a_{ij}^2.
\end{align}
Considering the example of a social duplex, formed by the same set of agents interacting by mobile phone in layer 1 and by email in layer 2, $k_i^{(1,0)}$ indicates the number of acquaintances of node $i$ that only communicate with it via mobile phone, $k^{(0,1)}_i$ indicates the number of acquaintances of node $i$ that only communicate with it via email, and $k^{(1,1)}_i$ indicates the number of acquaintances of node $i$ that communicate with it via both mobile phone and email.

When the number of layers is very large, it is difficult to keep track of all the different types of multilinks. In this case one can consider the {\em multiplicity of the overlap} $\nu_{ij}$ between node $i$ and node $j$ indicating the total number of layers in which the two nodes are connected, i.e.
\begin{equation}
\nu_{ij}=\sum_{\alpha=1}^M a_{ij}^{\alpha}=\sum_{\alpha=1}^M m_{ij}^{\alpha},
\end{equation}
where the nodes $i$ and $j$ are linked by a multilink $\vec{m}=\vec{m}_{ij}$.

\begin{figure}[!t]
 \centering
   \includegraphics[width=0.75\textwidth]{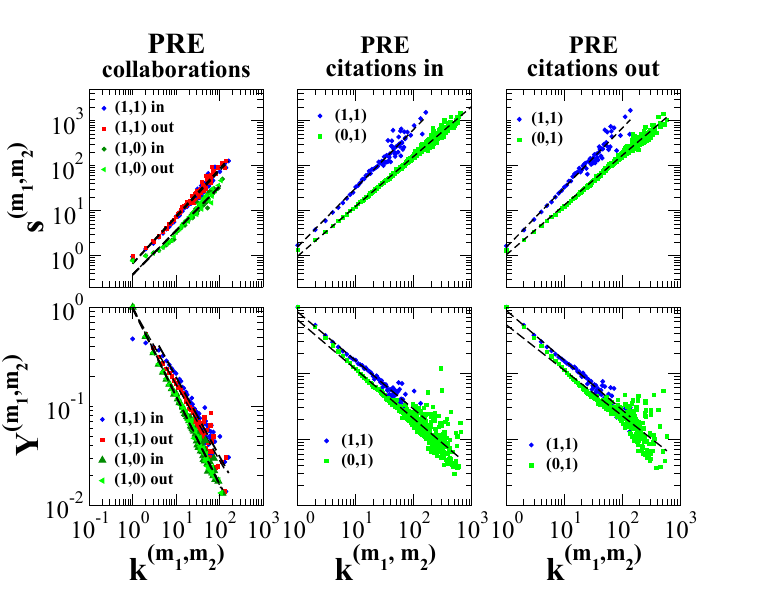}
 \caption{(Color online). Properties of multilinks in the weighted
   collaboration (layer 1)/citation (layer 2) multiplex network based
   on the PRE publications   analyzed in Ref.~\cite{Menichetti}. In
   the case of the collaboration network, the distributions of
   multistrengths versus multidegrees always have the same exponent,
   but the average weight of multilinks $(1,1)$ is larger than the
   average weight of multilinks $(1,0)$. Moreover, the exponents
   $\lambda_{(1,0),col,in}$, $\lambda_{(1,0),col,out}$ are larger than
   exponents $\lambda_{(1,1),col,in},\lambda_{(1,1),col,out}$. In the
   case of the citation layer, both the incoming multistrengths and
   the outgoing multistrengths have a functional behavior that varies
   depending on the type of multilink. Conversely, the average inverse
   multiparticipation ratio in the citation layer does not show any
   significant change of behavior when compared across different
   multilinks. Reprinted figure from Ref.~\cite{Menichetti}.  }
    \label{fig:weighted_multi}
\end{figure}

\subsubsection{Multistrength and inverse multiparticipation ratio in  weighted
multiplex networks}
In weighted multiplex networks the weights might be correlated with the multiplex network structure in a non trivial way. To study weighted multiplex networks,  two new measures, the {\it multistrengths } and the {\it  inverse multiparticipation ratio} have been introduced in Ref.~\cite{Menichetti}. For layer $\alpha$ associated to multilinks $\vec{m}$, such that $m_{\alpha}>0$, we define the multistrength $s^{\vec{m}}_{i,\alpha}$ and the inverse multiparticipation ratio $Y^{\vec{m}}_{i,\alpha}$ of node $i$, as
\begin{equation}
s^{\vec{m}}_{i,\alpha}=\sum_{j=1}^N a_{ij}^{\alpha}A_{ij}^{\vec{m}},
\end{equation}
\begin{equation}
Y^{\vec{m}}_{i,\alpha}=\sum_{j=1}^N
\left(\frac{a_{ij}^{\alpha}A_{ij}^{\vec{m}}}{\sum_r a_{ir}^{\alpha}
A_{ir}^{\vec{m}}}\right)^2,
\end{equation}
respectively. The multistrength $s^{\vec{m}}_{i,\alpha}$ measures the total weights of the links incident to node $i$ in layer $\alpha$ that form a multilink of type $\vec{m}$. For example, in a social multiplex network where each layer is either mobile phone communication (layer 1) or email communication (layer 2), $s_{i,1}^{(1,0)}$ is the total strength of phone calls of node $i$ with people that do not communicate with it by email, $s_{i,2}^{(0,1)}$ is the total strength of email contacts of node $i$ with people that do not communicate with it via mobile phone, $s_{i,1}^{(1,1)}$ is the total strength of phone calls of node $i$ with people that also communicate with it via email, and $s_{i,2}^{(1,1)}$ is the total strength of email communication of node $i$ with people that also communicate with it via mobile phone. The inverse multiparticipation ratio $Y^{\vec{m}}_{i,\alpha}$ is a measure of the inhomogeneity of the weights of the nodes that are incident to node $i$ in layer $\alpha$ and at the same time are part of a multilink $\vec{m}$. Since for every layer $\alpha$ and node $i$  the non trivial multistrengths must
involve multilinks $\vec{m}$ with $m_{\alpha}=1$, the number of non trivial multistrengths $s^{\vec{m}}_{i,\alpha}$ is given by $2^{M-1}$. Therefore, for each node, the number of multistrengths that can be defined in a multiplex
network of $M$ layers is $M2^{M-1}$. Similarly to what happens for single layers \cite{BarratGinestraPNAS,BarthelemyGinestraPA,AalmasGinestraPA}, the average multistrength of nodes with a given multidegree, i.e. $s^{\vec{m},\alpha}(k^{\vec{m}})=\Avg{s_i^{\vec{m},\alpha}\delta(k_i^{\vec{m}}, k^{\vec{m}})}$, and the average inverse multiparticipation ratio of nodes with a given multidegree, $Y^{\vec{m},\alpha}(k^{\vec{m}})=\Avg{Y_i^{\vec{m},\alpha}\delta(k_i^{\vec{m}}, k^{\vec{m}})}$, are expected to scale as
\begin{align}
s^{\vec{m},\alpha}(k^{\vec{m}}) &\propto (k^{\vec{m}})^{\beta_{\vec{m},\alpha}}
\nonumber \\
Y^{\vec{m},\alpha}(k^{\vec{m}})&\propto \frac{1}{(k^{\vec{m}})^{\lambda_{\vec{m},\alpha}}},
\label{SkYk}
\end{align}
with exponents $\beta_{\vec{m},\alpha} \geq 1$ and $\lambda_{\vec{m},\alpha}\leq 1$. In Ref.~\cite{Menichetti} the weighted multilink properties of multiplex networks between scientist collaborating (in layer 1) and citing each other (in layer 2) have been analyzed starting from the APS data set. In this case the citation network is also directed, and the Authors find that the exponents $\lambda$ and $\beta$ depend on the type of multilink, i.e. they are significantly different  for links without overlap and for links with overlap. It is clear from this analysis that many weighted multiplex networks have a distribution of the weights that depends on the non-trivial correlations existing in their structure. In Fig.~$\ref{fig:weighted_multi}$, the multistrength and inverse multiparticipation ratio for the collaboration and citation networks of PRE Authors are shown.

\subsubsection{The activities of the nodes and the pairwise multiplexity}
\begin{figure}
 \centering
  \includegraphics[width=0.9\textwidth]{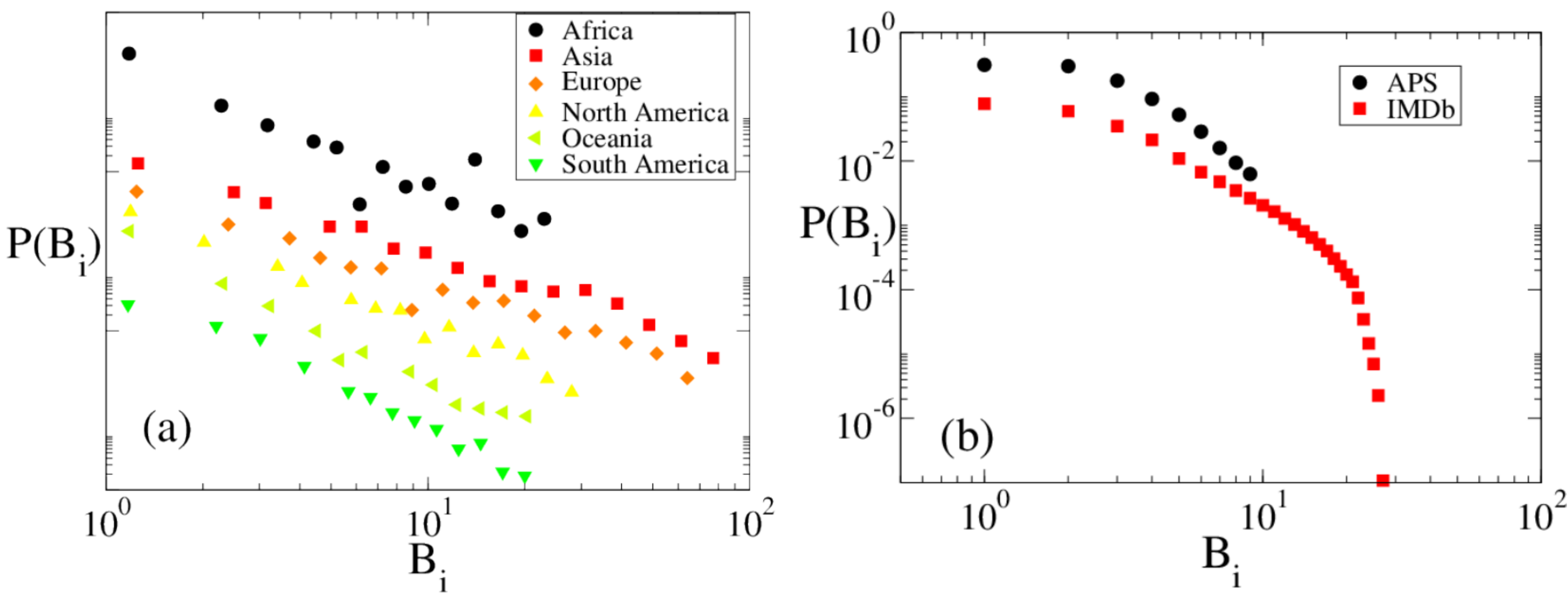}
 \caption{(Color online). Distributions $P(B_i)$ of the node-activity
   $B_i$ for (a) the six continental airline multiplex networks and
   for (b) APS and IMDb. Note that these distributions are broad and
   can be fitted by power-law exponents between $1.5$ and $3.0$
   indicating that the typical number of layers in which a node is
   active is subject to unbound fluctuations. Reprinted figure from
   Ref.~\cite{VitoGinestraAr}. Courtesy of V. Latora.}
\label{fig:activities}
\end{figure}
{From} the $M$ adjacency matrices ${\bf a}^{\alpha}$ of each layer $\alpha=1,2\ldots M$ of the network it is possible to construct an $N\times M$ activity matrix  ${\bf b}$ of elements $b_{i,\alpha}$ indicating if node $i$ is present in layer $\alpha$. This matrix can be seen as an adjacency matrix between nodes and layers indicating which node is active in each layer of the multiplex. For a undirected multiplex network, node $i$ is active in layer $\alpha$  if it  has a positive degree in layer $\alpha$, i.e. $k_i^{[\alpha]}>0$. Therefore  we have
\begin{equation}
b_{i,\alpha}=1-\delta_{0,k_i^{[\alpha]}}=1-\delta_{0,\sum_{j=1}^N a_{ij}^{\alpha}
},
\end{equation}
where $\delta_{x,y}$ indicates the Kronecker delta. For a directed multiplex network, node $i$ is inactive in layer $\alpha$ if both its in-degree and its out-degree in layer $\alpha$ are zero. Therefore we can define the matrix elements $b_{i,\alpha}$ as
\begin{equation}
b_{i,\alpha}=1-\delta_{0,\sum_{j=1}^N a_{ij}^{\alpha} }\delta_{0,\sum_{j=1}^N
a_{ji}^{\alpha} }.
\end{equation}
The activity $B_i$ of a node $i$ has been defined in Ref.~\cite{VitoGinestraAr} and is given by the number of layers in which node $i$ is active:
\begin{equation}
B_i=\sum_{\alpha=1}^Mb_{i,\alpha}.
\end{equation}
The layer activity $N_{\alpha} $ has been defined in Ref.~\cite{VitoGinestraAr} and is given by the number of nodes active in layer $\alpha$:
\begin{equation}
N_{\alpha}=\sum_{i=1}^Nb_{i,\alpha}.
\end{equation}
By analyzing a large set of multiplex networks, including a multiplex formed by a very large number of layers, Nicosia and Latora in Ref.~\cite{VitoGinestraAr} have shown that the activity distribution  $P(B_i)$ is typically broad and can be fitted by a power-law $P(B_i)\simeq B_i^{-\delta}$ with $\delta\in [1.5,3.0]$. This implies that for some multiplex networks the bipartite network between nodes and layers described by the activity adjacency matrix can be either dense ($\delta\leq 2$) or sparse ($\delta>2$) but the typical number of layers in which a node is active is always subject to unbound fluctuations (see Fig.~$\ref{fig:activities}$). Moreover, in Ref.~\cite{VitoGinestraAr} a broad distribution $P(N_{\alpha})$ of layer activities has been reported.

In Ref.~\cite{VitoGinestraAr} the Authors studied the {\em layer pairwise multiplexity} $Q_{\alpha\beta}$ measuring the  correlation between the layers. The layer pairwise multiplexity is defined as
\begin{equation}
Q_{\alpha\beta}=\frac{1}{N}\sum_{i=1}^N b_{i,\alpha}b_{i,\beta}\:,
\end{equation}
and quantifies the fraction of nodes that are active in layer $\alpha$ as well as in layer $\beta$. Similarly, one can use the {\em node pairwise multiplexity} $Q_{ij}$, measuring the correlation of activities between two nodes. The node pairwise multiplexity, introduced in Ref.~\cite{CFGGR12}, is defined as
\begin{equation}
Q_{ij}=\frac{1}{M}\sum_{\alpha=1}^M b_{i,\alpha}b_{i,\beta},
\end{equation}
and quantifies the fraction of layers in which both node $i$ and node $j$ are active.
\begin{figure}
 \centering
   \includegraphics[width=0.7\textwidth]{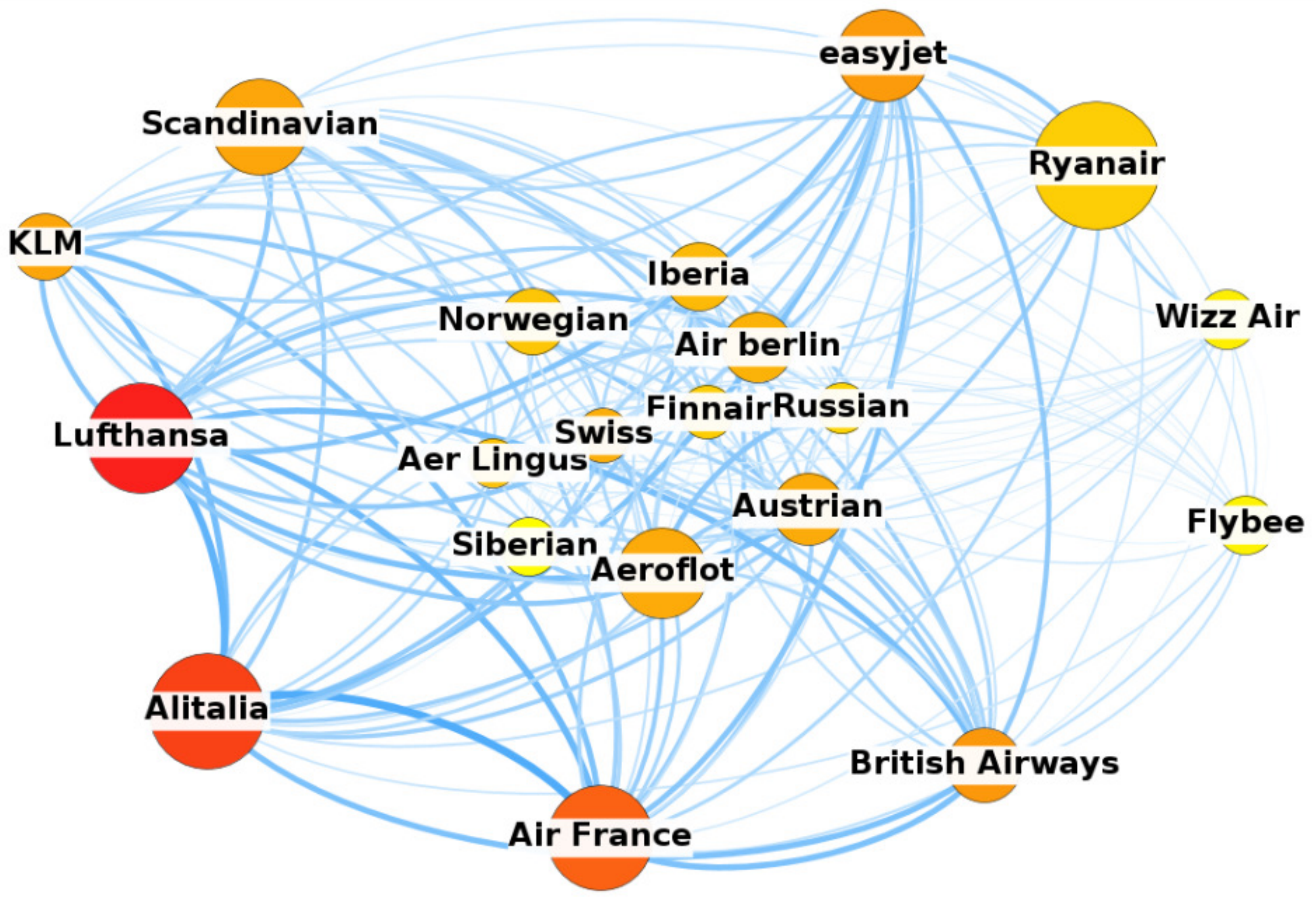}
 \caption{The network between 20 European airline companies in which
   each link is weighted with their layer pairwise multiplexity. This
   method can reveal non trivial correlations between the activities
   of the nodes in the different layers. Reprinted figure from
   Ref.~\cite{VitoGinestraAr}. Courtesy of V. Latora.}
    \label{fig:air_vito}
\end{figure}

In Fig.~\ref{fig:air_vito} the network of European airline companies is plotted,  where the links between two different companies is weighted with the layer pairwise multiplexity.
This method can reveal non trivial correlations between the activities of the nodes in different layers.

\section{Generative models}
\label{sec:generativemodels}

In the last fifteen years large attention has been devoted to unveil the basic mechanisms
for the generation of monolayer networks with specific structural properties.
The discussion of all generative models for single-layer networks is beyond the scope of
this report, and we refer the interested reader to the existing
reviews
~\cite{Boccaletti2006,AlbertBarabasi2001,DorogovtsevMendes2002,
  DorogovtsevMendes2003, Newman_SIAM} 
and to more recent works \cite{Del10,Kim12} on the subject.
Here we mention only that the models introduced so far in the Literature can be cast in two major categories:
\begin{itemize}
\item
{\it the growing networks models}, including the mechanism of {\em preferential attachment} introduced in the
Barab\'asi-Albert (BA) model  to explain the emergence of power-law
distributions, and
\item
{\it the static models}, describing networks ensembles with given structural properties.
Examples of such structural properties are: the degree sequence (the configuration model),
the expected degree sequence (the exponential random graph, or the
hidden-variable model),  the degree sequence and the block structure or the expected degree sequence and the block structure.
These network ensembles can be considered as a generalization of the random graphs to more complex network topologies.
\end{itemize}
Similarly, most of the  generative models for multilayer networks can be divided also into two classes:
\begin{itemize}
\item{\em growing multilayer networks models}, in which the number of the nodes grows, and there is a {\em generalized preferential attachment} rule \cite{KimGo13,NiBiLa13,Nonlinear,magnani2011,magnani2013}. These models explain
multilayer network evolution starting from simple, and fundamental rules for their dynamics.
\item {\em multilayer network ensembles}, which are ensembles of networks with $N$ nodes in each layer satisfying
a certain set of structural constraints \cite{Bianconi13,Pattison99,Menichetti,WangGinestraSN}. These ensembles are able to generate multilayer networks with fully controlled set of degree-degree correlations and of overlap.
\end{itemize}
In the case of multilayer networks the modeling framework is still in its infancy. Yet, different ways to construct multiplex networks with different topological features have been proposed, and summarizing the main results and concepts coming from these works is the object of the present section.
\subsection{Multiplex networks models}

\subsubsection{Growing multiplex networks models}

Growing multiplex networks' models start from the consideration that
many real networks are the result of a growing process, such as, for
instance, the case of the airport network formed by airline
connections of different airline companies \cite{CardilloSR13,VitoGinestraAr}, or that of the social network between scientists collaborating with each other and citing each other \cite{Menichetti}.
Similarly to what happens for single-layer networks, the way the links of the new nodes are attached to the existing nodes
{\it is not random}, therefore it is important to explore the consequences on the resulting topology
of assuming different attachment kernels.

In two recent papers \cite{KimGo13,NiBiLa13}, a similar growing multiplex model has been proposed.
There, the network has a dynamics dictated by growth, and generalized preferential attachment.
Starting at time $t=0$ from a duplex network with $n_0$ nodes (with a replica in each of the two layers)
connected by  $m_0>m$ links in each layer, the model proceeds as follows:
\begin{itemize}
\item{\it Growth:}
At each time $t\geq 1$ a node with a replica node in each of the two layers is added to the multiplex.
Each newly added replica node is  connected to the other nodes of  the same layer by  $m$ links.
\item{\it Generalized preferential attachment: }
The new link in layers $\alpha=1,2$ is attached to node $i$ with probability  $\Pi_i^{\alpha}$
proportional to a linear combination of the degree $k_i^{[1]}$ of node $i$ in layer 1 and $k_i^{[2]}$ of node $i$ in layer 2, i.e.,
\begin{align}
\Pi_i^{[1]} \propto c_{1,1}k_i^{[1]}+c_{1,2}k_i^{[2]},\nonumber \\
\Pi_i^{[2]} \propto c_{2,1}k_i^{[1]}+c_{2,2}k_i^{[2]},
\end{align}
where $c_{\alpha,\beta}\in [0,1]$ with $c_{1,1}+c_{1,2}=c_{2,1}+c_{2,2}=1$.
\end{itemize}
In the case in which $c_{1,1}=c_{2,2}=1$ the model reduces to two apparently decoupled BA models. Nevertheless, in the multiplex network  two replica nodes have the same age, therefore this characteristic of the model is responsible for generating degree-degree correlations in the layers of the duplex.
The model for  $c_{1,1}=c_{2,2}=1$  has been solved by the use of the master equation approach \cite{NiBiLa13} giving
the joined degree distribution $P(k,q)$ of having a node with degree $k^{[1]}=k$ in layer 1  and degree $k^{[2]}=q$ in layer 2,
\begin{equation}
  P(k,q)=\frac{2\Gamma(2+2m)\Gamma(k)\Gamma(q)\Gamma(k+q-2m+1)}
  {\Gamma(m)\Gamma(m)\Gamma(k+q+3)\Gamma(k-m+1)\Gamma(q-m+1)}.
\end{equation}
The average degree $\bar{k}(q)$ defined in Sec.~\ref{par_corr} in one layer conditioned to the degree in the other layer \cite{NiBiLa13} is given by
\begin{equation}
\bar{k}(q) =\sum_{k^{[1]}} k^{[1]} P(k^{[1]}|k^{[2]}=q)=\frac{m(q+2)}{1+m}.
\end{equation}
Therefore, the two layer have positive degree correlations.

\begin{figure}[!t]
 \centering
   \includegraphics[width=0.9\textwidth]{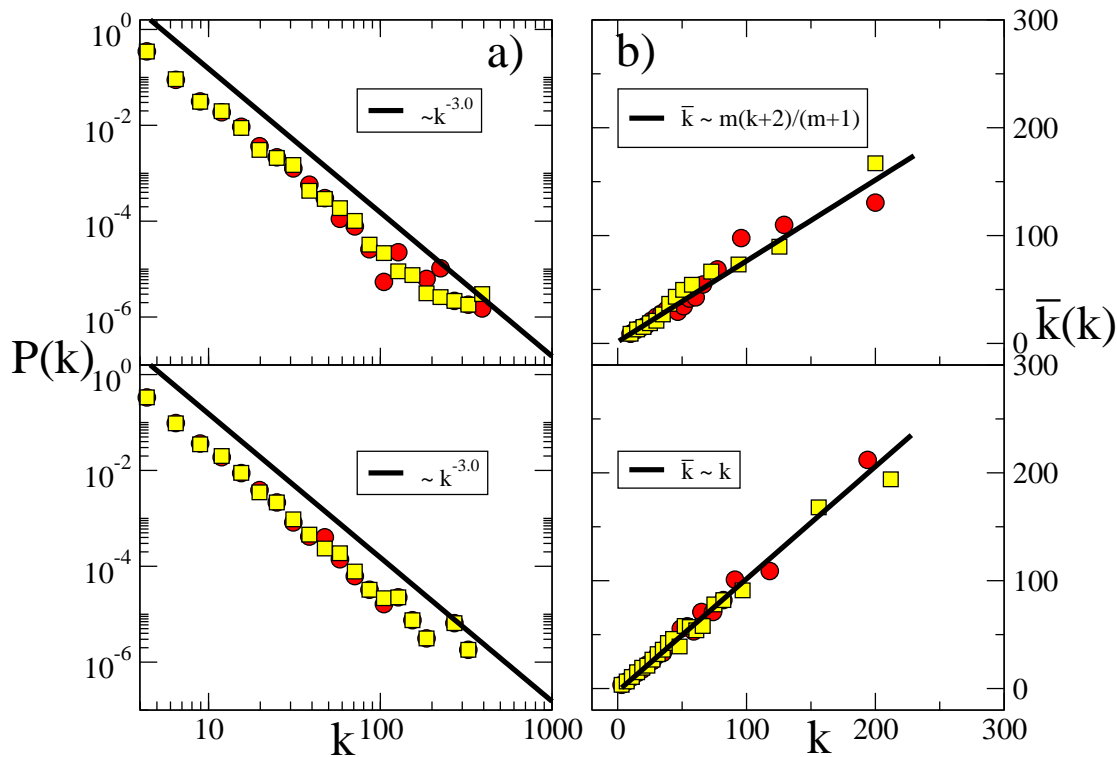}
 \caption{(Color online). Panels (a)-(b): Degree distribution $P(k)$ in each layer (left),
and the interlayer degree-degree correlations $\bar{k}(k)$ (right) for the case $c_{1,1}=1, c_{1,2}=0, c_{2,1}=0, c_{2,2}=1$ (top) and the case $c_{1,1}=0, c_{1,2}=1, c_{2,1}=1, c_{2,2}=0$
(bottom) attachment kernels (red circles are for the first layer,
yellow squares for the second layer). Reprinted figure with permission from
   Ref.~\cite{NiBiLa13}. \copyright\, 2013 by the American Physical Society.  }
    \label{fig:growth}
\end{figure}

In this case, the degree distribution in each layer $P(k)$ is given by the degree distribution of the BA network model.
In Ref.~\cite{NiBiLa13} it has been shown by mean-field calculations and simulations (see Fig.~\ref{fig:growth}) that also for other values of the parameters $\{c_{\alpha,\beta}\}$ the average degree $\bar{k}(q)$ in one layer conditioned to the degree of the other layer remains linear, because older nodes of the multiplex are more connected than younger nodes in both layers. Moreover, in Ref.~\cite{KimGo13} it has been shown that for $c_{1,1}<1$ and $c_{2,2}<1$
the degree correlations in the two layers become more significant.
In fact, the Pearson correlation coefficient $r$ defined in Sec.~\ref{par_corr} between the degrees in the two layers has been characterized for the model with $c_{1,1}=c_{2,2}=1-\epsilon$ and $c_{1,2}=c_{2,1}=\epsilon$, finding in the limit $t\to \infty$
\begin{align}
r=\frac{\Avg{(k-\avg{k})(q-\avg{q})}}{\sigma_k\sigma_q}=\left\{\begin{array}{ccc} \frac{1}{2} &\mbox{for} &\epsilon=0 \\
1& \mbox{for} & \epsilon>0,\end{array}\right.
\end{align}
where $k=k^{[1]}$ indicates the degree in layer 1 and $q=k^{[2]}$ indicates the degree in layer two.
Finally, both Refs.~\cite{NiBiLa13} and ~\cite{KimGo13} explore some variations of this model.
In Ref.~\cite{NiBiLa13} the case of a semilinear attachment kernel (giving rise to networks with different
degree distributions) and that of time delay of the arrival of one replica node in one layer
have been discussed.
In Ref.~\cite{KimGo13} an initial attractiveness has been added to the generalized preferential attachment,
giving rise to two layers formed by scale-free networks with a tunable value of the power-law exponent $\gamma>2$.

A similar growing multiplex model has been also proposed by Magnani and Rossi in Ref.~\cite{magnani2011,magnani2013}.
In their model, the multiplex layers grow by the addition of new links and new nodes that do not arrive
in the different layers at the same time.
Also in their case, the way the new links are attached to the rest of the network follows the preferential attachment rule.
The model displays layers that have either uncorrelated or positively correlated degree features.

\begin{figure}
 \centering
   \includegraphics[width=\textwidth]{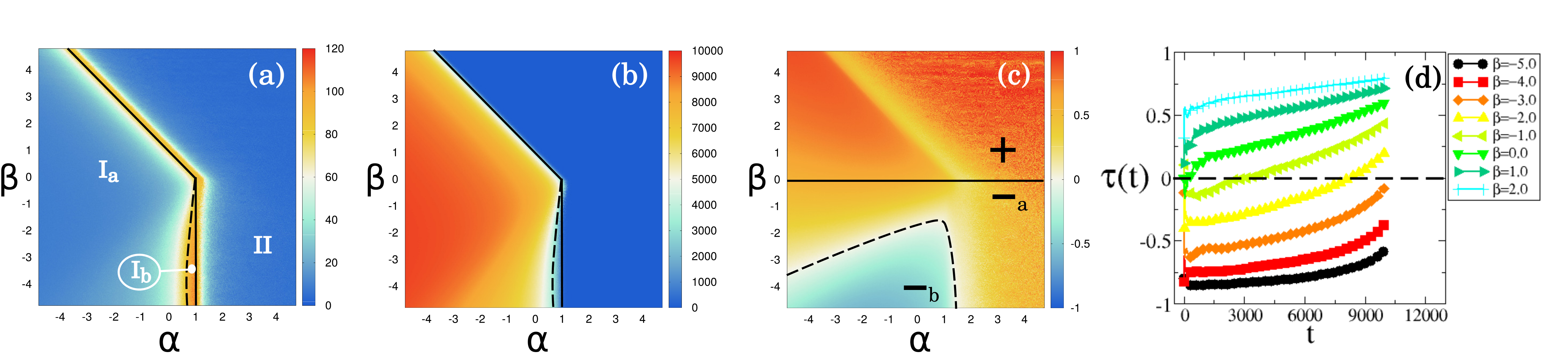}
 \caption{(Color online). Phase diagrams of the non-linear multiplex network growing model. By means of a color code,
the following quantities are plotted (see Ref.~\cite{Nonlinear} for the definition of the plotted measures): (a) the number of distinct degree classes $|k|$, (b) the participation ratio $Y_2^{-1}$, and (c) the Kendall's $\tau$ correlation coefficient. The solid black lines in panel (a) and (b) separate the non-condensed (region I) from the condensed phase (region II, small $|k|$, small $Y_2^{-1}$ ). In region
I we can have either homogeneous (region $I_a$) or heterogeneous
degree distributions (region $I_b$). The solid black line in panel (c)
separates the two regions corresponding to positive (region $+, \tau>
0$) or negative interlayer degree correlations (region
$-,\tau<0$). The value of $\tau$ for the whole multiplex is negative
only in region ${-}_b$. In panel (d) the plot of $\tau(t)$, which is
the Kendall's  restricted to the nodes arrived up to time $t$ is
shown. The dashed black line corresponds to $\tau= 0$ and is reported
for visual reference. For $\beta<0$ the interlayer correlations for
older nodes are disassortative ($\tau(t)<0$ for small $t$), even if
the  value computed on all the nodes might be positive due to the
prevalence of young nodes. Reprinted figure from Ref.~\cite{Nonlinear}.}
    \label{fig:nonlinear}
\end{figure}

In Ref.~\cite{Nonlinear} it has been shown that growing multiplex networks models can also generate
negative degree correlations between the layers. In particular, the attachment kernel used in Ref.~\cite{Nonlinear}
is a non-linear kernel. The model includes growth and non-linear preferential attachment depending on the degree
of the  same node in the different layers.
For a duplex, the model is described by the following algorithm. Starting at time $t=0$ from a duplex network
with $n_0$ nodes (with a replica in each of the two layers) connected by  $m_0>m$ links in each layer, the model
proceed as follows.
\begin{itemize}
\item{\it Growth:}
At each time $t\geq 1$, a node with a replica node in each of the two layers is added to the multiplex.
Each newly added replica node is connected to the other nodes of  the same layer by  $m$ links.
\item{\it Generalized non-linear preferential attachment: }
The new links  are attached to node $i$ with probability  $\Pi_i^{[1]}$ in layer
$1$ and with probability  $\Pi_i^{[2]}$ in layer $2$ with
\begin{align}
\Pi_i^{[1]}&\propto k_i^{[\alpha]}q_i^{[\beta]},\nonumber \\
\Pi_i^{[2]} &\propto q_i^{[\alpha]}k_i^{[\beta]}
\end{align}
where $k_i$ is the degree of node $i$ in layer 1, and $q_i$ is the degree of node $i$ in layer 2. Notice that here  $\alpha$ and $\beta$ are two parameters of the model taking real values.
\end{itemize}
This model displays a very interesting physics, generating multiplex networks with different types of degree distributions,
degree-degree correlations, and with a condensation of the links for some parameter values.
In Ref.~\cite{Nonlinear}, the phase diagram of the model (see Fig.~\ref{fig:nonlinear}) has been completely characterized.
The number of distinct degree classes $|k|$ shown in panel $(a)$ of Fig.~\ref{fig:nonlinear} allows to define  a region of parameter  space corresponding to homogeneous degree distributions (region $I_a$), and another region corresponding to heterogeneous degree distributions (region $I_b$).
The inverse partition ratio $Y_2^{-1} $ of the degree distribution in a given layer shown in panel (b) of Fig.~\ref{fig:nonlinear}  provides evidence of a condensation occurring for $\alpha+\beta>1$ when $\beta\geq 0$, or for $\alpha>1$ when $\beta<0$,  in agreement with the theoretical derivations discussed in Ref.~\cite{Nonlinear}.
Finally, the Kendall $\tau$ correlation coefficient between the degrees of corresponding replica nodes
shows that the model can display at the same time positive and negative degree-degree correlations in the two layers.
Panel (d) of Fig.~\ref{fig:nonlinear} shows the Kendall correlation coefficient $\tau(t)$ between
the degree of corresponding replica nodes, calculated only over the nodes arrived at time $t^{\prime}<t$.
This last panel shows that for every $\beta<0$ the older nodes have always negatively correlated degrees
in the two different layers. Therefore, in the region in which $\tau>0$ and $\beta<0$  shown in panel (c) of
Fig.~\ref{fig:nonlinear}, only the recently added nodes in the multiplex have positively correlated degrees,
due to the  stochastic nature  in the model.

Other growing multiplex network models have been proposed \cite{CFGGR12,VitoGinestraAr},
 where the multiplex network grows by the addition of an entire new layer at each time step.
In Ref.~\cite{CFGGR12}, two nodes $i$ and $j$ in the new layer are linked with a probability $p_{ij}$ that  depends on the
quantity  there called {\it node multiplexity} $Q_{ij}$. In particular, $p_{ij}$ can be either positively correlated with $Q_{ij}$, or negatively
correlated to it. In the first case two nodes that are active at the same time in many layers are more likely to be
connected in the new layer, in the second case two nodes that are active at the same time in many layers have small probability to be connected in the new layer.
In Ref.~\cite{VitoGinestraAr}, instead, every node $i$ of the new layer will be active with a probability $P_i$ proportional to the activity of the node $B_i$, i.e.,
\begin{align}
P_i\propto A+B_i(t),
\end{align}
where $A$ is a parameter of the model. This model  enforces a sort of ``preferential attachment'' of the new layers to nodes of high activity $B_i$, and a power-law distribution $P(B_i)$ of the activities of the nodes.
\subsubsection{Simple static models}
The simplest way to obtain a static generative model for multiplex networks is to generalize the existing
methods for single-layer ones. Fixing the degree sequence in each layer, one can use a configuration
model to obtain a particular realization of the given set of connectivities. Of course, the full
structural information in a multiplex includes
also the connections between layers. In Refs.~\cite{MinGo2013}
and~\cite{manliocent}, the Authors have made the
choice to add interlayer links arbitrarily. A
different approach is to keep using a configuration
model, but to specify the edges between the layers
by means of a joint-degree distribution~\cite{LeeKim12,funkcomp,marceaucomp}.
Note that, while the terms ``joint-degree sequence'',
``joint-degree matrix'' and ``joint-degree distribution''
are normally used in literature to indicate quantities
related to the degrees of the nodes at the end
of the links in a simple network, here they  explicitly refer, instead, to multilayer connections ~\cite{Pat76}.

A similar method is to specify the degree sequences
together with a probability matrix whose element $\left(i,j\right)$
is the fraction of interlayer links between layers $i$
and $j$. The actual link placement is still achieved
via uniform random choice~\cite{Soe02,Soe03,Soe03_2,Soe03_3,Bar11}.
A generalization of this approach has been proposed
in Ref.~\cite{Mel14}. There, the Authors impose the
degree correlations within and between layers by means
of a set of matrices that specify the fraction of edges
between nodes of given degrees in given layers.

Further models of multilayer networks can be created
by defining classes of nodes that share similar structural
characteristics. While the definitions of such classes
are somewhat arbitrary, these models find application
in the detection of higher-level features in real-world
networks, mostly related to community structure~\cite{Hol83,Fie85,Wan87,New07}.

The choice of any network model for a particular study results in the imposition
of constraints of some sort. Each network model corresponds to a network
ensemble, i.e., to the set of networks that satisfy the
constraints implicitly imposed by the model. Thus, it is
important to study the properties of network ensembles,
which we discuss in the following from a statistical physics
perspective.

\subsubsection{Ensembles of multiplex networks}
\paragraph{Network ensembles and their entropy}
A given set of constraints can
give rise to a {\it microcanonical network ensemble}, satisfying the hard constraints, or to a
{\it canonical network ensemble} in which the constraints are satisfied in average over the ensemble. The canonical network ensembles have been widely studied in the sociological literature under the name of exponential random graphs or $p^{\star}$ models \cite{RobinsGinestraSN,Robins2GinestraSN,StraussGinestraJASA,LusherGinestra13}, and recently studied by physicists \cite{ParkGinestraPRE,BianconiGinestraEPL,SquartiniGinestraNJP,AnandGinestraPRE,BianconiGinestraPNAS}. The most popular examples of microcanonical network ensembles  are the $G(N,L)$ ensemble and the configuration model. 
The distinction between the canonical and microcanonical ensembles allows to draw a well defined  parallelism \cite{AnandGinestraPRE} between such network ensembles and the classical ensembles
in statistical mechanics where one considers system configurations compatible either with a fixed value of the energy
(microcanonical ensembles) or with a fixed average of the energy determined by the thermal
bath (canonical ensembles). For example, the $G(N,L)$ random graphs formed by networks of
$N$ nodes and $L$ links is an example of a microcanonical network ensemble, while the $G(N,p)$
ensemble (where each pair of links is connected with probability $p$) is an example of
canonical network ensemble since the number of links can fluctuate but has a fixed average
given by $\avg{L}=pN(N-1)/2$.
Despite this similarity, an ensemble that enforces an extensive number of hard constraints on its networks
(such as the degree sequence) is not equivalent in the thermodynamic limit to the conjugated network ensemble
enforcing the same constraints in average (such as the sequence of expected degrees). For instance,
the ensemble of regular networks with degrees $k_i=c\ \forall i=1,2,\ldots, N$ is clearly not equivalent to the
Poisson network in which every node has the same expected degree $\bar{k}_i=c$. The non-equivalence of microcanonical and canonical network ensembles with an extensive number of constraints needs also to be taken into account when
characterizing dynamical processes on networks, because it can have relevant effects on their critical behavior.

The entropy of network ensembles is given by the logarithm of the number of typical networks in the ensemble, and it
quantifies the complexity of the ensemble. In particular we have that the smaller is
the entropy of the ensemble, the smaller is the number of networks satisfying the
corresponding constraints, implying that these networks are more optimized.
Both the network ensembles and their entropy can be used on several inference problems to
extract information from a given network \cite{BianconiGinestraPNAS,Infomap}.

\paragraph{Multiplex network ensembles}
The statistical mechanics of randomized network ensembles (exponential random graphs)
 has been extended to describe multiplex ensembles \cite{Bianconi13,Pattison99,Menichetti,WangGinestraSN,halu2013}, and applied to analyze different multiplex data sets \cite{Menichetti,LazegaGinestraSN,HeaneyGinestraSN}.
 Consider a multiplex  formed by $N$ labeled nodes $i=1,2,\ldots, N$
and $M$ layers. We can represent the multiplex as described for example in Ref.~\cite{Mucha10}. To this end,
we indicate by $\vec{G}=(G^1,G^2,\ldots,G^M)$ the set of all the networks $G^{\alpha}$ at layer $\alpha=1,2,\ldots, M $ forming the multiplex. Each of  these networks has an  adjacency matrix
with matrix elements $a_{ij}^{\alpha}=1$ if there is a link between node $i$ and node $j$ in layer $\alpha=1,2,\ldots, M $, and zero otherwise.
A multiplex ensemble is specified when the probability $P(\vec{G})$ for each possible multiplex is given.
In a multiplex ensemble, if the probability of a multiplex is given by $P(\vec{G})$, the entropy of the multiplex $S$  is defined as
\begin{equation}
S=-\sum_{\vec{G}}P(\vec{G})\log P(\vec{G}).
\label{entropy}
\end{equation}
and measures the logarithm of the typical number of multiplexes in the ensemble.
One can distinguish between microcanonical and canonical multiplex network ensembles, and between uncorrelated and correlated multiplex network ensembles \cite{Bianconi13}.

\paragraph{Uncorrelated multiplex networks ensembles}
The uncorrelated multiplex ensembles are ensembles of networks in which the links in different layers are not correlated.
Note that the degree in the different layers can nevertheless be eventually correlated, as described below.
The probability $P(\vec{G})$ of  an uncorrelated multiplex ensemble is given by
\begin{equation}
P(\vec{G})=\prod_{\alpha=1}^M P_{\alpha}(G^{\alpha}),
\label{uc0}
\end{equation}
where $P_{\alpha} (G^{\alpha})$ is the probability of network $G^{\alpha}$
on layer $\alpha$. In this case,  the entropy of the multiplex ensemble is the sum of the entropies
of the network in the single layers, i.e.,
\begin{equation}
S=\sum_{\alpha=1}^M S^{\alpha}=-\sum_{\alpha=1}^M P_{\alpha}(G^{\alpha})\log P_{\alpha}(G^{\alpha}).
\end{equation}
 In a uncorrelated multiplex network, the links in two different layers   $\alpha$ and $\alpha'\neq \alpha$  are uncorrelated. In fact, from Eq.~(\ref{uc0}) it immediately follows that
\begin{equation}
\avg{a_{ij}^{\alpha}a_{ij}^{\alpha'}}=\avg{a_{ij}^{\alpha}}\avg{a_{ij}^{\alpha'}}
\label{ucorr}
\end{equation}
for every choice of pair of nodes $i,j$.

Examples of uncorrelated multiplex ensembles are multiplex layers with
given degree sequence, or given degree sequence and community
structure, directed networks with given in-out degree, weighted
networks with given degree and strength of the links, etc.

A pivotal role is played by the ensemble in which we fix either the expected or the exact degree sequence.
These degree sequences can be eventually correlated.
Let us consider here a simple case of a duplex  $M=2$ in which we either fix {\em i)} the expected degree sequence (canonical multiplex network ensemble) or {\em ii)} the exact degree sequence (microcanonical multiplex network ensemble).
\begin{itemize}
\item{\it i) Fixed expected degree sequence }\\
In the first case, when we fix the expected degree sequence $\{\bar{k}_i^{\alpha}\}$, the probability of the multiplex $P_C(\vec{G})$ is given by
\begin{align}
P_C(\vec{G})=\prod_{\alpha=1}^M \prod_{i<j} \left[p_{ij}^{\alpha}a_{ij}^{\alpha}+(1-p_{ij}^{\alpha})(1-a_{ij}^{\alpha})\right]
\end{align}
where $p_{ij}^{\alpha}$ is the probability of the link $(i,j)$ in layer $\alpha$  and can be expressed in terms of the Lagrangian multipliers $\lambda_i^{\alpha}$ as
\begin{align}
p_{ij}^{\alpha}=\frac{e^{-(\lambda_i^{\alpha}+\lambda_j^{\alpha})}}{1+e^{-(\lambda_i^{\alpha}+\lambda_j^{\alpha})}}.
\label{pl}
\end{align}
The Lagrangian multipliers  $\lambda_{i}^{\alpha}$ are fixed by the $N M$ conditions
\begin{align}
\bar{k}_i^{\alpha}=\sum_{j\neq i}p_{ij}^{\alpha},
\end{align}
where $\bar{k}_i^{\alpha}$ is the expected degree of node $i$ in layer $\alpha$.
This last expression, in the case in which there is a structural cutoff $\bar{k}_i^{\alpha}<\sqrt{\avg{\bar{k}^{\alpha}}N}$, has a very simple solution given by
\begin{align}
p_{ij}^{\alpha}=\frac{\bar{k}_i^{\alpha}\bar{k}_j^{\alpha}}{\avg{\bar{k}^{\alpha}}N}.
\label{puc}
\end{align}
Note that here $\avg{\,}$ indicates average over the nodes. This is an example of canonical multiplex networks. The  entropy of this multiplex ensemble is called {\it Shannon entropy} and is given by
   \begin{align}
 S&=-\sum_{\vec{G}}P_C(\vec{G})\ln P_C(\vec{G})\nonumber \\
 &=-\sum_{\alpha=1}^M\sum_{i<j}[p_{ij}^{\alpha}\log p_{ij}^{\alpha}+(1-p_{ij}^{\alpha})\log(1-p_{ij}^{\alpha})].
 \label{Sh}
 \end{align}
These ensembles can display  degree correlations in the different layers, but they are uncorrelated ensembles in the sense that they  satisfy Eq.~(\ref{uc0}) and Eqs.~(\ref{ucorr}).\\

The multiplex networks in this ensemble can be generated by putting a link between node $i$ and node $j$ with probability $p_{ij}^{\alpha}$ given by Eq.~(\ref{pl}) or, in presence of the structural cutoff, by Eq.~(\ref{puc}).
\item {\it ii) Fixed  degree sequence }\\
Here we also consider the microcanonical multiplex network ensembles  in which we  fix the exact degree sequence $\{k_i^{\alpha}\}$ in every layer.
In this case the probability of a multiplex network is given by
\begin{align}
P_M(\vec{G})=\frac{1}{Z} \prod_{\alpha=1}^M \prod_{i<j}\delta\left(k_i^{\alpha},\sum_j a_{ij}^{\alpha}\right),
\end{align}
where $Z$ is a normalization constant.
We call the entropy of this ensemble {\it Gibbs entropy } $N\Sigma$ to distinguish it from the {\it Shannon entropy} $S$ of the canonical ensembles.
We have
\begin{align}
N\Sigma&=\sum_{\vec{G}}P_M(\vec{G})\ln P_M(\vec{G}),\nonumber\\
&=S-N\Omega.
\end{align}
where $S$ is the Shannon entropy of the conjugated canonical model given by Eq.~(\ref{Sh}) calculated assuming that the expected degree of each node $i$ is given exactly by its  degree $k_i$,  and  where $\Omega$ is given by
\begin{equation}
N\Omega=-\sum_{\alpha=1}^M \sum_{i=1}^N\log \pi_{k_i^{\alpha}}(k_i^{\alpha})
\end{equation}
where $\pi_{y}(x)=\frac{1}{x!}y^x\exp[-y]$ is the Poisson distribution with average $y$ calculated at $x$.
Therefore, the Gibbs entropy $N\Sigma$ is lower than $S$,  and the microcanonical ensemble is not equivalent in the thermodynamic limit to the conjugated canonical ensemble.
The multiplex networks in this ensemble can be generated by generating the different networks in the different layers with  the configuration model without allowing for multiple edges and  tadpoles in the networks.
\end{itemize}
These types of networks are the most simple examples of multiplex, and many dynamical processes have been first studied on these ensembles, by analyzing also the effect of correlation between the degrees in different layers.
In Ref.~\cite{Bianconi13} it has been shown that if a uncorrelated multiplex network ensemble is formed by sparse networks
(e.g. Poisson networks, exponential networks with finite average degree, scale-free networks with power-law exponent $\gamma>2$),  the expected overlap of the links between any two layers of the multiplex is negligible. Therefore, these ensembles cannot be used to describe multiplex networks where the overlap of the links is significant.

\paragraph{Correlated multiplex networks ensembles}
The correlated multiplex ensembles are ensembles of networks in which the links in different layers are correlated.
In particular, in the {\it correlated } multiplex network ensembles, Eq.~(\ref{uc0}) does not hold, i.e.
\begin{equation}
P(\vec{G})\neq \prod_{\alpha=1}^M P_{\alpha}(G^{\alpha}).
\end{equation}
 In this correlated ensembles, the entropy  $S$ cannot be expressed as the sum of the entropies $S^{\alpha}$, i.e.
 \begin{align}
 S\neq \sum_{\alpha}S^{\alpha}=-\sum_{\alpha=1}^M P_{\alpha}(G^{\alpha})\log P_{\alpha}(G^{\alpha}).
 \end{align}
Furthermore, one has that it exists at least a pair of layers  $\alpha$ and $\alpha'$  and a pair of nodes $i$ and $j$
for which
\begin{equation}
\avg{a_{ij}^{\alpha}a_{ij}^{\alpha'}}\neq\avg{a_{ij}^{\alpha}}\avg{a_{ij}^{\alpha'}}.
\end{equation}

Examples of correlated multiplex ensembles are multiplex layers with
given multidegree sequence,   directed networks with given directed
multidegree sequence, weighted networks with given multidegree and
multistrength of the links, etc. \cite{Bianconi13,Menichetti}.

In particular, correlated multiplex networks can display a significant overlap of the links.
Let us consider here a simple case of a duplex  $M=2$ in which we either fix {\em(i)} the expected multidegree sequence (canonical multiplex network ensemble) or {\em (ii)} the exact multidegree sequence (microcanonical multiplex network ensemble).
\begin{itemize}
\item{\it (i) Fixed expected multidegree sequence }\\
The probability  $P_C(\vec{G})$ of a multiplex network $\vec{G}$ in which we fix  the expected multidegree sequence $\{\bar{k}_i^{\vec{m}}\}$  is given by
\begin{align}
P_C(\vec{G})=\prod_{i<j} \sum_{\vec{m}}\left[p_{ij}^{\vec{m}} A_{ij}^{\vec{m}}\right]
\end{align}
where $p_{ij}^{\vec{m}}$ is the probability of the multilink $\vec{m}=(m_1,m_2,\ldots, m_{\alpha})$  between node $i$ and node $j$, and $ {\bf A}^{\vec{m}}$ is the multiadjacency matrix with elements  $A_{ij}^{\vec{m}}=\prod_{\alpha=1}^M \left[a_{ij}^{\alpha}m_{\alpha}+(1-a_{ij}^{\alpha})(1-m_{\alpha})\right]$.
For a sparse network, the probabilities $p_{ij}^{\vec{m}}$ of a multilink $\vec{m}\neq\vec{0}$ can be expressed in terms of the Lagrangian multipliers $\lambda_i^{\vec{m}}$ as
\begin{align}
p_{ij}^{\vec{m}}=\frac{e^{-(\lambda_i^{\vec{m}}+\lambda_j^{\vec{m}})}}{\sum_{\vec{m}}e^{-(\lambda_i^{\vec{m}}+\lambda_j^{\vec{m}})}},
\label{plm}
\end{align}
while $p_{ij}^{\vec{0}}$ is fixed by the normalization condition $\sum_{\vec{m}}p_{ij}^{\vec{m}}=1$.
The Lagrangian multipliers  $\lambda_{i}^{\vec{m}}$ are fixed by the conditions
\begin{align}
\bar{k}_i^{\vec{m}}=\sum_{j\neq i}p_{ij}^{\vec{m}},
\end{align}
where $\bar{k}_i^{\vec{m}}$ is the expected multidegree $\vec{m}\neq\vec{0}$ of node $i$.
This last expression, in the case in which there is a structural cutoff  on the multidegree $\bar{k}_i^{\vec{m}}<\sqrt{\avg{\bar{k}^{\vec{m}}}N}$ has a very simple solution given by
\begin{align}
p_{ij}^{\vec{m}}=\frac{\bar{k}_i^{\vec{m}}\bar{k}_j^{\vec{m}}}{\avg{\bar{k}^{\vec{m}}}N},
\label{pucv}
\end{align}
for every multilink $\vec{m}\neq\vec{0}$.
This is an example of canonical multiplex networks. The  entropy of this multiplex ensemble is called again {\it Shannon entropy} and is given by
\begin{align}
 S&=-\sum_{\vec{G}}P_C(\vec{G})\ln P_C(\vec{G})\nonumber \\
 &=-\sum_{\vec{m}}\sum_{i<j}[p_{ij}^{\vec{m}}\log p_{ij}^{\vec{m}}].
 \label{Shm}
 \end{align}
In order  to construct a correlated multiplex ensemble with given multidegree sequence it is sufficient to follow the following scheme.
\begin{itemize}
\item
Calculate the probability $p_{ij}^{\vec{m}}$ to have a multilink $\vec{m}$ between node $i$ and $j$ using Eq.~(\ref{plm}) or directly Eq.~(\ref{pucv}) in the presence of a structural cutoff on the multidegrees.
\item
For every pair of node $i$ and $j$, draw a multilink $\vec{m}$ with probability $p_{ij}^{\vec{m}}$ and consequently  put a  link in every  layer $\alpha$ where $m_{\alpha}=1$, and put no link in every layer $\alpha$ where $m_{\alpha}=0$.
\end{itemize}
\item {\it (ii) Fixed  multidegree sequence }\\
The conjugated multiplex network ensemble of the previous example is the ensemble in which we  fix the exact multidegree sequence $\{k_i^{\alpha}\}$ in every layer.
In this case the probability of a multiplex network is given by
\begin{align}
P_M(\vec{G})=\frac{1}{Z} \prod_{\vec{m}} \prod_{i<j}\delta\left(k_i^{\vec{m}},\sum_j a_{ij}^{\vec{m}}\right),
\end{align}
with $Z$ being a normalization constant.
We call again the entropy of this ensemble {\it Gibbs entropy } $N\Sigma$ to distinguish it from the {\it Shannon entropy} $S$ of the canonical ensemble.
We have
\begin{align}
N\Sigma&=\sum_{\vec{G}}P_M(\vec{G})\ln P_M(\vec{G}),\nonumber\\
&=S-N\Omega.
\end{align}
where $S$ is the Shannon entropy of  Eq.~(\ref{Sh}) calculated assuming that the expected multidegrees of a node are given by the exact multidegree of the nodes in the microcanonical ensemble, and where  $\Omega$ is given by
\begin{equation}
N\Omega=-\sum_{\vec{m}|\sum_{\alpha=1}^M m_{\alpha}>0} \sum_{i=1}^N\log \pi_{k_i^{\vec{m}}}(k_i^{\vec{m}}).
\end{equation}

\end{itemize}
\paragraph{Spatial multiplex network ensembles}
When the multiplex networks are embedded in a real or in a hidden space (as in the case of the airport
multiplex network in Ref.~\cite{CardilloSR13}), the overlap of the links can emerge naturally from the correlations induced by the distance in the embedding space.
Intuitively, if in every layer links are more likely between nodes that are closer in the embedding space,
then one can observe non-negligible overlap of the links because two nodes that are close in the embedding space are more likely to be connected to each other in every layer of the multiplex. We have that, if the  probability
$P(\vec{G}|\{\vec{r}_i\})$ of a network $\vec{G}$ with $N$ nodes on positions $\{\vec{r}_i\}$ is given by
\begin{align}
P(\vec{G}|\{\vec{r}_i\})=\prod_{\alpha=1}^MP_{\alpha}({G}^{\alpha}|\{\vec{r}_i\}),
\end{align}
where $P(G^{\alpha}|\{\vec{r}_i\})$ is the probability of a network in layer $\alpha$, then  we can observe a significant overlap of the links \cite{halu2013}.

\subsection{Ensembles of networks of networks}

In this case, there are multiple degrees of freedom that may  characterize these complex structures, and
therefore several types of networks of networks can be considered:

\begin{itemize}
\item
{\it Networks of networks with fixed supernetwork}, i.e., networks of networks where if a layer $\alpha$ is connected with layer $\beta$, every node $(i,\alpha)$ is connected to its replica node $(i,\beta)$ in layer $\beta$ {\cite{BD,Gao:gbhs2011}}.
\item {\it Networks of networks with given superdegree distribution}, where every node $(i,\alpha)$ is connected to
$q=q_{\alpha}$ replica nodes $(i,\beta)$ of randomly chosen layers $\beta$ \cite{BD2}.
\item{\it Networks of networks with fixed supernetwork  and random permutations of the labels of the nodes}, i.e. network of networks  where if a layer $\alpha$ is connected with layer $\beta$, any node $(i,\alpha)$ is connected to a single node $(j,\beta)$, where $j$ is taken from a random permutation $\pi_{\alpha,\beta}$ of the indices $\{i\}$ \cite{Gao2012GinestraNatPhys,Gao:gbhs2012,Gao:gbsxh2013}.
\item{\it  Networks of networks with multiple interconnections}, where every node $(i,\alpha)$ has $k_i^{\alpha,\beta}$ connections with nodes in layer $\beta$ \cite{LeiSo09}.
\end{itemize}
In order to distinguish the links between nodes of the same layers and
the links between nodes of different layers, we will call the latter
interlinks.

\subsubsection{Case I: Networks of networks with fixed supernetwork and links allowed only between replica nodes}
\label{par:netonet1}
\begin{figure}[!t]
 \centering
   \includegraphics[width=0.8\textwidth]{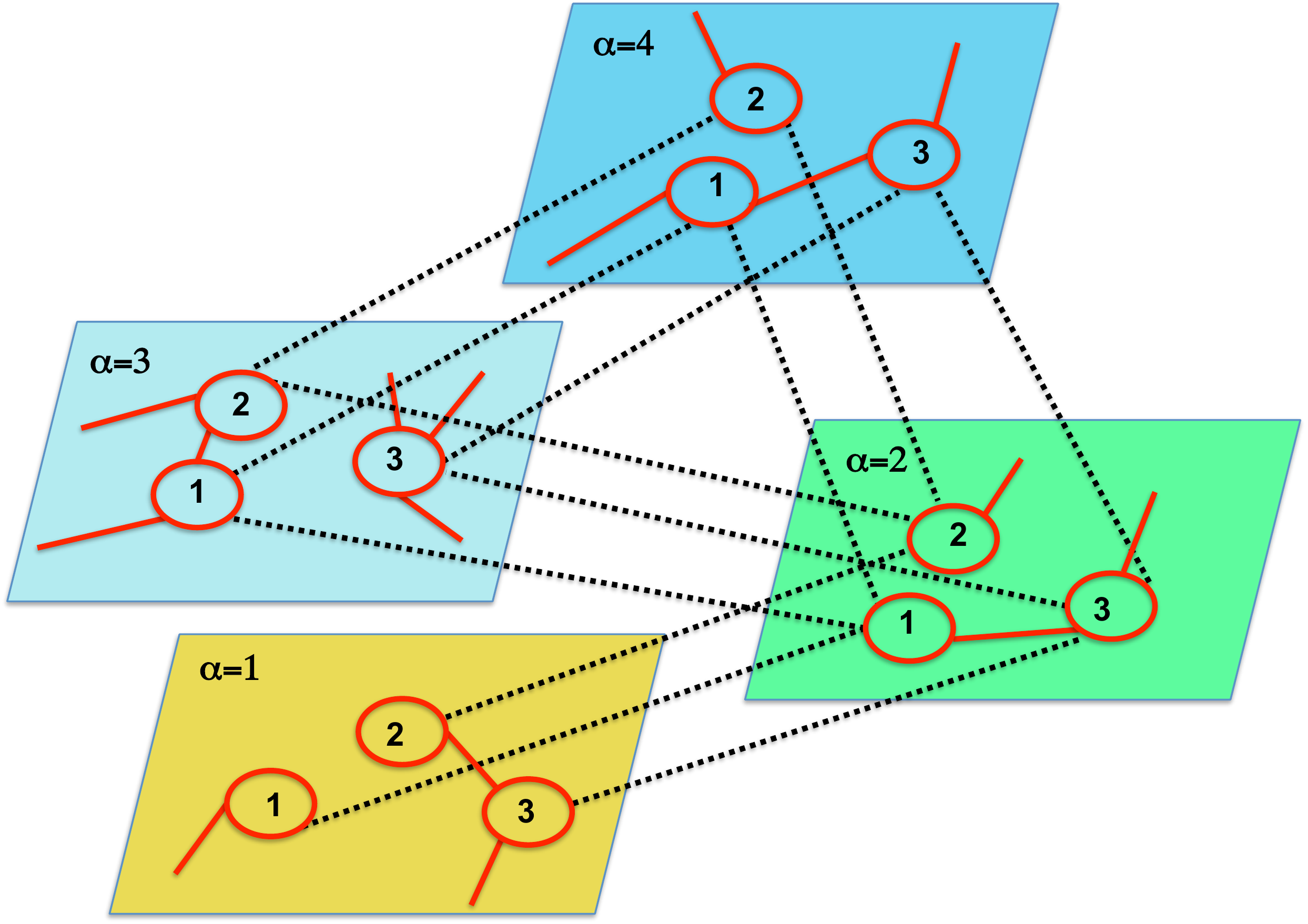}
 \caption{(Color online). Schematic view of a typical network of networks with given supernetwork. Interdependencies (interlinks
between nodes from different levels) are shown by the black dashed
lines. Intralinks between nodes within layers are shown as solid red
lines. Reprinted figure from Ref.~\cite{BD}.}
 \label{fig:netonet1}
\end{figure}

In this subsection we will consider a network of networks in which if network $\alpha$ is connected with
network $\beta$, each node $(i,\alpha)$ of network $\alpha$ is connected to the  node $(i,\beta)$ of network $\beta$
(see Fig.~\ref{fig:netonet1}). The  supra-adjacency matrix of the
network of networks has  elements $a_{i\alpha,j\beta}=1$
if there is a link between node $(i,\alpha)$ and node $(j,\beta)$, and zero otherwise. In these
specific
networks,
$a_{i\alpha,j\beta}=0$
if both $i\neq j$ and $\alpha\neq \beta$.
The network of networks is characterized by  an adjacency matrix $A_{\alpha\beta}$, such that for every node $i=1,2,\ldots, N$, $A_{\alpha\beta}=a_{i\alpha,i\beta}$. The adjacency matrix $A_{\alpha\beta}$ is the adjacency matrix of  the supernetwork, ${\cal G}$, of the network of networks.
We can consider ensemble of networks where the interlinks between different layers are fixed by the adjacency matrix $A_{\alpha\beta}$, but the network in a given layer is random.
In particular, we can choose a canonical  ensemble of networks of networks
in which each layer is a uncorrelated network and each node has a given expected
degree $\bar{k}^{\alpha}_i$ in each layer $\alpha$.
In this case, the probability $P(\{\bf a\})$  of the supra-adjacency matrix ${\bf a}$ is given by
\begin{align}
P(\{\mathbf a\})&=\prod_{\alpha=1}^M\prod_{i=1}^N\left\{\delta(a_{i\alpha,i\beta},A_{\alpha\beta})\prod_{\beta\neq \alpha}\prod_{j\neq i}\delta(a_{i\alpha,j\beta},0)\right\}\times\\
& \times\prod_{\alpha=1}^M\prod_{i<j}\left[\left(\frac{\bar{k}_i^{\alpha}\bar{k}_j^{\alpha}}{\avg{\bar{k}^{\alpha}}N}\right)a_{i\alpha,j\alpha}+\left(1-\frac{\bar{k}_i^{\alpha}\bar{k}_j^{\alpha}}{\avg{\bar{k}^{\alpha}}N}\right)(1-a_{i\alpha,j\alpha}) \right].
\end{align}
An alternative is to consider  an ensemble of networks of networks in which the network in each layer is built
using the configuration model, i.e. each node has given  degree $k_i^{\alpha}$ in layer $\alpha$.
In this case, the probability of the supra-adjacency matrix $\mathbf a$ is
\begin{align}
P(\{{\bf a}\})= \frac{1}{Z} \prod_{\alpha=1}^M\prod_{i=1}^N\left\{\delta(a_{i\alpha,i\beta},A_{\alpha\beta})\prod_{\beta\neq \alpha}\prod_{j\neq i}\delta(a_{i\alpha,j\beta},0)\right\}\prod_{\alpha=1}^M\prod_{i=1}^N\delta(k_i^{\alpha},\sum_{j=1}^N a_{i\alpha,j\alpha}).
\end{align}

\subsubsection{Case II: Networks of networks with given superdegree distribution} \label{par:netonet2}

\begin{figure}[!t]
 \centering
   \includegraphics[width=0.8\textwidth]{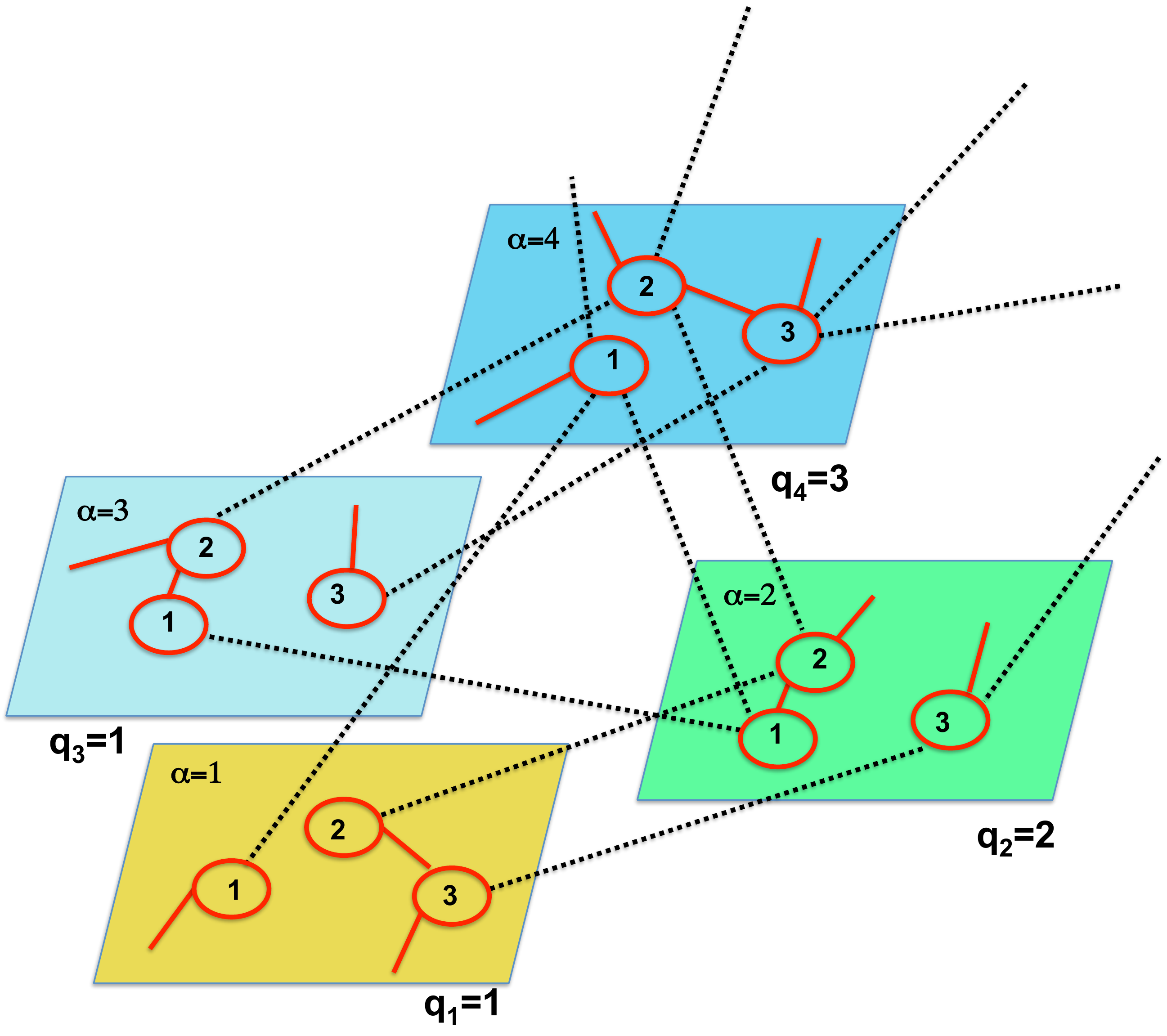}
 \caption{(Color online). Schematic representation of a network of
   networks with given superdegree distribution. Each layer $\alpha$
   has assigned given superdegree $q_{\alpha}$ but each node
   $(i,\alpha)$ can be linked to any $q_{\alpha}$ replica
   nodes. Reprinted figure from Ref.~\cite{BD2}.}
\label{fig:netonet2}
\end{figure}

In this case,  each node $(i,\alpha)$ can be linked only to  its replica nodes $(i,\beta)$,
but there is no fixed supernetwork between the layers. Nevertheless, we assume that each layer $\alpha$ has a given
superdegree $q_{\alpha}$, i.e. in these networks of networks every node $(i,\alpha)$ is linked to  $q=q_{\alpha}$ replica nodes $(i,\beta)$ of randomly chosen layers $\beta$ \cite{BD2} (see Fig.~\ref{fig:netonet2}).

Moreover, each network (layer) $\alpha$ is generated randomly according to a canonical or a microcanonical network ensemble.
If fact, we can consider a  microcanonical version of this model in which the
degree sequence in  each layer  $\{k_i^{\alpha}\}$ is fixed, or a canonical version of this model in which the expected degree sequence $\{\bar{k}_i^{\alpha}\}$ is fixed.
Therefore, for  the  canonical  network of networks ensemble,
 every network of networks with a supra-adjacency matrix ${\bf a}$ has a probability $P({\bf a})$ given by
\begin{align}
P({\bf a})=\frac{1}{Z}\prod_{\alpha=1}^M\prod_{i=1}^N\left\{ \delta\left(q_{\alpha},\sum_{\beta\neq \alpha} a_{i\alpha,i\beta}\right)\left[\left(\frac{\bar{k}_i^{\alpha}\bar{k}_j^{\alpha}}{\avg{\bar{k}^{\alpha}}N}\right)a_{i\alpha,j\alpha}+\left(1-\frac{\bar{k}_i^{\alpha}\bar{k}_j^{\alpha}}{\avg{\bar{k}^{\alpha}}N}\right)(1-a_{i\alpha,j\alpha}) \right]\prod_{\beta\neq \alpha}\prod_{j\neq i}\delta\left(a_{i\alpha,j\beta},0\right)\right\}.
\label{Pa2c}
\end{align}
Alternatively, one can consider the microcanonical network of networks ensemble in which every network of networks
with a supra-adjacency matrix ${\bf a}$ has a probability $P({\bf a})$ given by
\begin{align}
P({\bf a})=\frac{1}{Z}\prod_{\alpha=1}^M\prod_{i=1}^N\left\{\delta\left(k^{\alpha}_i,\sum_{j\neq i}a_{i\alpha,j\alpha}\right)\right.\left.\ \delta\left(q_{\alpha},\sum_{\beta\neq \alpha} a_{i\alpha,i\beta}\right)\right\}\prod_{\alpha=1}^M\prod_{i=1}^N\prod_{\beta\neq \alpha}\prod_{j\neq i}\delta\left(a_{i\alpha,j\beta},0\right).
\label{Pa2}
\end{align}
\subsubsection{Case III: Networks of networks with fixed supernetwork  and random permutations of the labels of the nodes}
\label{par:netonet1bis}
\begin{figure}[!t]
 \centering
   \includegraphics[width=0.8\textwidth]{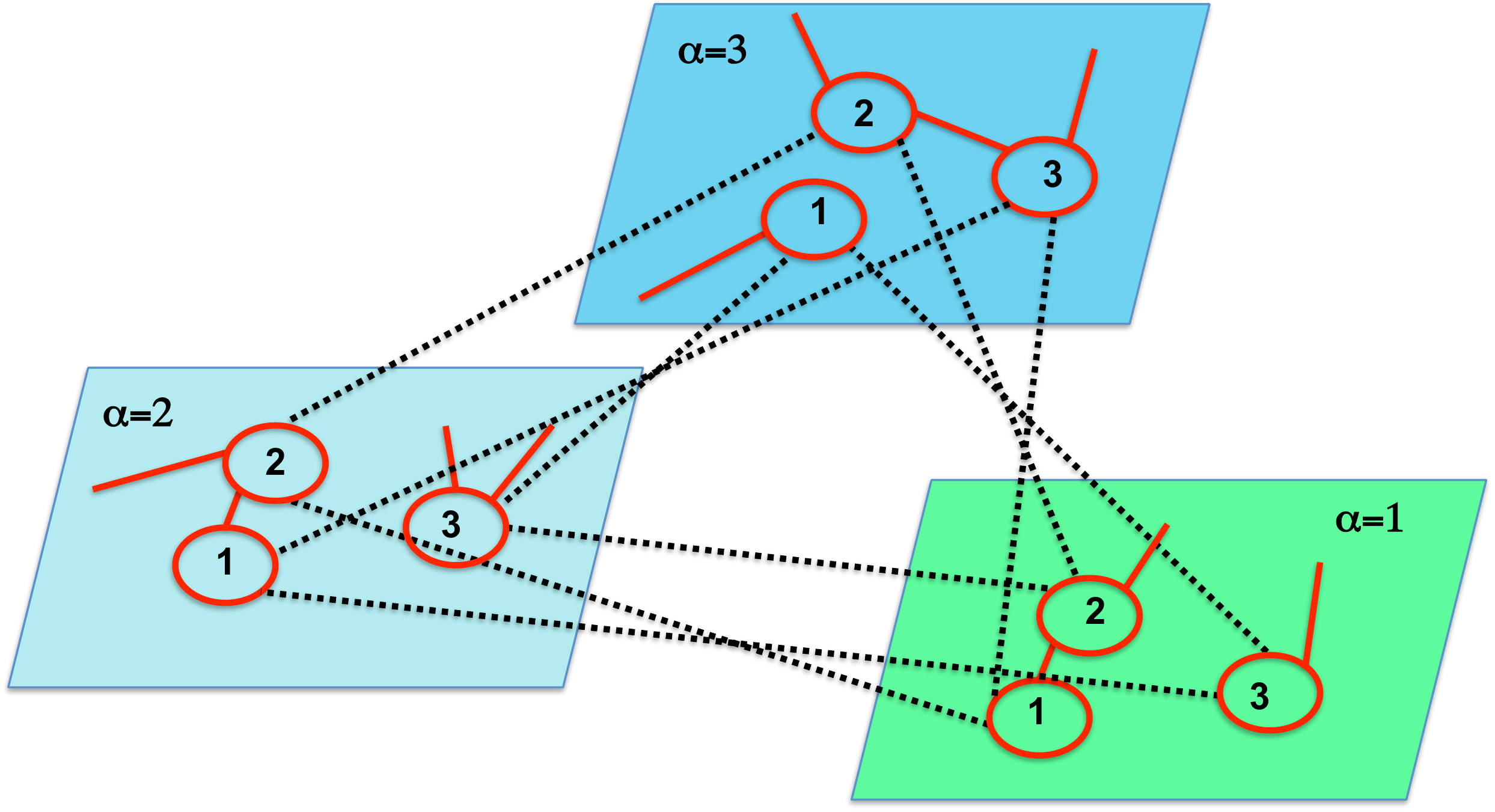}
 \caption{(Color online). Schematic representation of a network of
   networks with fixed supernetwork and random permutations of the
   labels of the nodes (case III). The supernetwork can be arbitrary but in the drawing we show a specific realization in which  the supernetwork is a loop formed by $M=3$ layers.
 } \label{fig:netonet1bis}
\end{figure}

We here consider networks of networks in which if network $\alpha$ is connected with
network $\beta$, each node $(i,\alpha)$ of network $\alpha$ is
connected to a single  node $(j,\beta)$ of network $\beta$  randomly  chosen between the nodes of layer $\beta$  (see Fig.~\ref{fig:netonet1bis}). The one to one mapping between the nodes is performed by defining a permutation $\pi_{\alpha,\beta}$ of the indices $\{i\}$ such that $j=\pi_{\alpha,\beta}(i)$ if and only if node $(i,\alpha)$ is linked to node $(j,\beta)$. In order to define a undirected network of networks, we need to  impose that the permutation $\pi_{\beta,\alpha}$ is the inverse permutation of $\pi_{\alpha,\beta}$, enforcing that if $j=\pi_{\alpha,\beta}(i)$ then $i=\pi_{\beta,\alpha}(j)$. The network of networks  supra-adjacency matrix has  elements $a_{i\alpha,j\beta}=1$
if there is a link between node $(i,\alpha)$ and node $(j,\beta)$,
i.e. if and only if $j=\pi_{\alpha,\beta}(i)$, and zero otherwise. In
these specific networks,  for $\beta\neq \alpha$,
$a_{i\alpha,j\beta}=0$
if  $j\neq \pi_{\alpha,\beta}(i)$.
The network of networks is characterized by  an adjacency
matrix $A_{\alpha\beta}$, such that for every node $i=1,2,\ldots, N$,
$A_{\alpha\beta}=a_{i\alpha,\pi_{\alpha,\beta}(i)\beta}$ is the adjacency matrix of the supernetwork.\\
The present case differs substantially from case I. For instance, consider the case of a supernetwork forming a
loop of size $M$ between the different layers.
In case I, starting from  each node $(i,\alpha)$ and following only
interlinks between different layers, we can reach only  $M$ other
nodes, while in case III (as each permutation
$\pi_{\alpha,\beta}(i)$ is random) the number of different nodes that
can be reached following interlinks
can be significantly higher than $M$ (see Fig.~\ref{fig:netonet1bis}).
If the supernetwork is a fully connected network of $M$ nodes, case I
 describes a multiplex with all the replica
nodes in the different layers connected to each other,
while case III  does not reduce to the multiplex case.\\

One can consider an ensemble of networks where the interlinks between
different layers are fixed by the adjacency matrix $A_{\alpha\beta}$
and the permutations $\pi_{\alpha,\beta}(i)$, but the network in a given layer is random.
In particular, we can choose a canonical  ensemble of networks of networks
in which each layer is a uncorrelated network, and each node has a given expected
degree $\bar{k}^{\alpha}_i$ in each layer $\alpha$.
In this case, the probability $P(\{\bf a\})$ of the supra-adjacency matrix ${\bf a}$ is given by
\begin{align}
P(\{\mathbf a\})&=\prod_{\alpha=1}^M\prod_{i=1}^N\left\{\delta(a_{i\alpha,\pi_{\alpha,\beta}(i)\beta},A_{\alpha\beta})\prod_{\beta\neq \alpha}\prod_{j\neq \pi_{\alpha,\beta}(i)}\delta(a_{i\alpha,j\beta},0)\right\}\times\\
& \times\prod_{\alpha=1}^M\prod_{i<j}\left[\left(\frac{\bar{k}_i^{\alpha}\bar{k}_j^{\alpha}}{\avg{\bar{k}^{\alpha}}N}\right)a_{i\alpha,j\alpha}+\left(1-\frac{\bar{k}_i^{\alpha}\bar{k}_j^{\alpha}}{\avg{\bar{k}^{\alpha}}N}\right)(1-a_{i\alpha,j\alpha}) \right].
\end{align}
An alternative is to consider  an ensemble of networks of networks in which the network in each layer is built
using the configuration model, i.e. each node has given  degree $k_i^{\alpha}$ in layer $\alpha$.
In this case, the probability of the supra-adjacency matrix $\mathbf a$ is
\begin{align}
P(\{{\bf a}\})=\frac{1}{Z} \prod_{\alpha=1}^M\prod_{i=1}^N\left\{\delta(a_{i\alpha,\pi_{\alpha,\beta}(i)\beta},A_{\alpha\beta})\prod_{\beta\neq \alpha}\prod_{j\neq \pi_{\alpha,\beta}(i)}\delta(a_{i\alpha,j\beta},0)\right\}\prod_{\alpha=1}^M\prod_{i=1}^N\delta(k_i^{\alpha},\sum_{j=1}^N a_{i\alpha,j\alpha}).
\end{align}

\subsubsection{Case IV: Networks of networks with multiple interconnections}
\label{par:netonet3}
\begin{figure}[!t]
 \centering
   \includegraphics[width=0.8\textwidth]{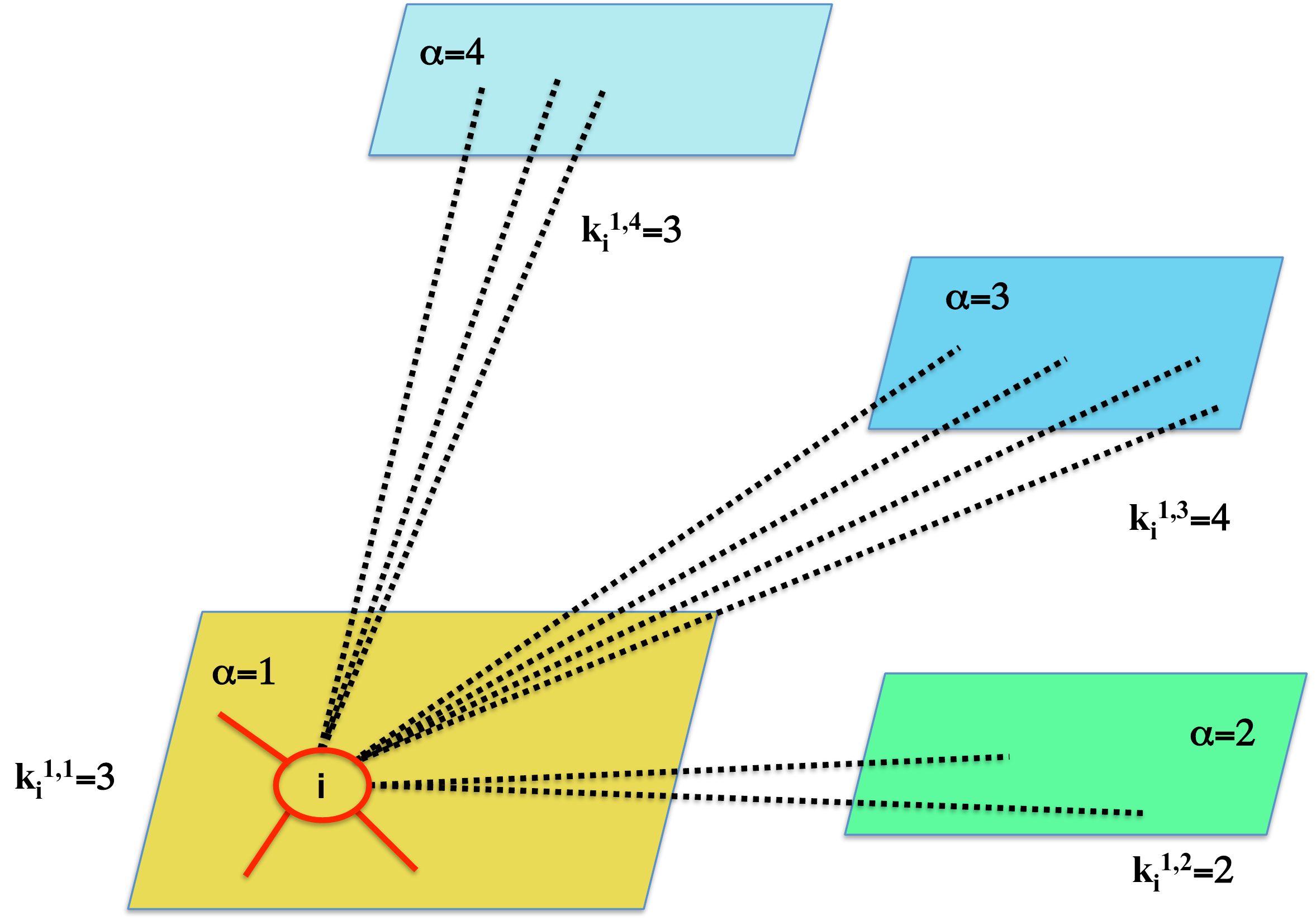}
 \caption{(Color online). Schematic representation of a network of networks with multiple interconnections. Each node $(i,\alpha)$ has $k_i^{\alpha\beta}$ links with nodes in layer $\beta$.
 } \label{fig:netonet3}
\end{figure}

The last example assumes that the nodes of a layer can be linked to any other node in other layers,
and that also multiple interlinks between one node and the node of other layers are allowed.
This model has been used, for instance, in Ref.~\cite{LeiSo09}.
The model considers every node $(i,\alpha)$ as having $k_i^{\alpha\beta}$ connections with nodes in layer $\beta=1,2,\ldots, M$,
and therefore it requires a number $M N$ of parameters (the degrees $k_i^{\alpha,\beta}$) which is much larger than the number of parameters needed for the other models proposed before, as long as  $M\gg 1$.
The typical structure of a network of networks in this ensemble is shown in Fig.~\ref{fig:netonet3}.
Once again, one can have a  microcanonical version of this model (in which the
degree sequences  $\{k_i^{\alpha\beta}\}$ are fixed), or a canonical version (in which the expected degree sequences $\{\bar{k}_i^{\alpha\beta}\}$ are fixed).
For the canonical ensemble the supra-adjacency matrix ${\bf a}$ has a probability $P({\bf a})$ given by
\begin{align}
P({\bf a})=\prod_{\alpha=1}^M\prod_{\beta\leq \alpha}\prod_{i=1}^N\prod_{j<i}\left[\left(\frac{\bar{k}_i^{\alpha\beta}\bar{k}_j^{\beta,\alpha}}{\avg{\bar{k}^{\alpha\beta}}N}\right)a_{i\alpha,j\beta}+\left(1-\frac{\bar{k}_i^{\alpha\beta}\bar{k}_j^{\beta,\alpha}}{\avg{\bar{k}^{\alpha\beta}}N}\right)(1-a_{i\alpha,j\beta}) \right],
\label{Pa3c}
\end{align}
while for the microcanonical ensemble the probability $P({\bf a})$ is
\begin{align}
P({\bf a})=\frac{1}{Z} \prod_{\alpha=1}^M\prod_{\beta=1}^M\prod_{i=1}^N\left\{\delta\left(k^{\alpha\beta}_i,\sum_{j\neq i}a_{i\alpha,j\beta}\right)\right\}.
\label{Pa3}
\end{align}


\section{Resilience and percolation}\label{sec:percolation}

The present section is devoted to review the rather huge literature regarding
two very relevant, and somehow correlated, processes: resilience of the multilayer networks' structure under
random failures (and/or cascades of failures), and percolation.

Resilience is the ability of a network to maintain its main topological structure under the action of
some kind of damages. It is well known that single-layer heterogeneous networks, such as scale-free (SF) networks, are particularly resilient to random damages, whereas they generally are extremely fragile to attacks targeting the nodes with higher degrees.
This is, actually, one important motivation that the scientific community has given
to the observed SF degree distribution: both technological (e.g. the
Internet) and biological networks (e.g. the protein-protein interaction and other biological networks in the cell) display a SF nature because they must be robust against random failures \cite{cnsw00,ceah00,ceah01,cohen-2002,vm02,clmr03,clmr04}.
In a multilayer scenario, nevertheless, conclusions and predictions are far different, and indeed the available literature
allows one to conclude that, if a network is interdependent
with other networks, its robustness properties can be strongly affected.

On the other side, percolation (from the Latin {\it percolare}, whose meaning is ``to filter'' or ``to trickle through'') refers, in classical physics, to the movement and filtering of fluids through porous materials. Within the last fifteen years, the main formalism and predictions of percolation theory have been applied also to complex networks, for the study of the emergence (and robustness) of a giant connected component in the graph having a size of the order of the size of the entire network. In this framework, statistical physics concepts (such as scaling theory and phase transitions) have been used to characterize network percolation properties, and study percolation thresholds.
Along this section we will extensively discuss how the traditional percolation theory has been recently generalized to the context of multilayer networks.

\subsection{The fragility of multilayer interdependent networks}
It is a matter of fact that in  many  multilayer networks some nodes of a layer
are interdependent on nodes in other layers. A node is interdependent on another node
in another layer if it needs the other node to function in order to function itself properly.
Examples of interdependent networks are found in complex infrastructures,
which are increasingly interconnected, in economy where networks of banks, insurance companies,
and firms interact with each other, in transportation networks, where railway networks, airline networks
and other transportation networks are interdependent, or in biology where the cell is alive only if all its
biological networks are functioning at the same time.

When two or more networks are interdependent, a fraction of node failures in one layer can trigger
a  cascade of failures that propagate in the multilayer network (as it was the case, for instance of the 2003
blackout in Italy \cite{Buldyrev10}),  and this is indicative of an increasing fragility
of these structures with respect to the single-layer cases \cite{Buldyrev10,Vespignani}.
Therefore, new aspects of the  robustness of networks emerge when interdependencies are taken into account.
Moreover, the impact of the topology of the interdependent networks on their robustness properties
is also unexpected: in fact, in absence of correlations, interdependent SF networks are more fragile
than Poisson networks when a constant average degree is kept \cite{Buldyrev10}.
When we consider multiplex networks or networks of networks with given supernetwork and links allowed only between replica nodes as described
in Sec.~\ref{par:netonet1}, the system responds globally to the damage inflicted in a single layer.
At any given time, either all the layers have a percolating cluster, or none of them has one \cite{BD}.
When instead one considers a network of networks with given superdegree sequence as described in
Sec.~\ref{par:netonet2} , layers with different number of interdependencies respond differently
to the inflicted damage. In this latter case, the layers with higher number of interdependencies,
i.e. with higher superdegree $q_{\alpha}$, are more fragile than those with smaller superdegree
\cite{BD2}. Indeed, in these layers the percolation cluster disappears for  a fraction of initial damaged nodes
$1-p$ smaller than the one that is necessary to disrupt layers with smaller number of interdependencies.
 Finally, if we consider a  network of networks with fixed supernetwork and random permutations of the label of the nodes, as described in Sec.~\ref{par:netonet1bis} one has to distinguish between the case in which the supernetwork is a tree, and that in which the supernetwork contains loops \cite{Gao2012GinestraNatPhys,Gao:gbsxh2013}. If the supernetwork is a tree, all the layers start to percolate when  the fraction of initially damaged nodes is less than $1-p_c$, i.e. either all the layers percolate or none of them does \cite{Gao:gbhs2011}. In the case in which the supernetwork contains loops and the supernetwork  is  a random network with given superdegree distribution, the layers $\alpha$ with higher superdegree $q_{\alpha}$ are more fragile than those with smaller superdegree. In this case the mutually connected component is described by the same equations derived  for the network of networks described in  Sec.~\ref{par:netonet2} \cite{BD2}.

The increased fragility of interdependent networks opens new important scientific questions.
A major one is the  design of optimal strategies to mitigate the dramatic effect of interdependencies
and allow for correlated topologies  with increased resilience under random damage.
This might have consequence for the infrastructure design of the new generation,
for reducing risk in financial networks, and for assessing the robustness of the biological networks in the cell.
Notice that, while for random interdependent networks percolation is clearly  sharpened by the presence of interdependencies,  the question weather interdependencies always sharpen the percolation transition has been matter of debate \cite{Son2011GinestraPRL,Havlin_debate, Grassberger_reply}.

\begin{figure}
 \centering    
\includegraphics[width=\textwidth]{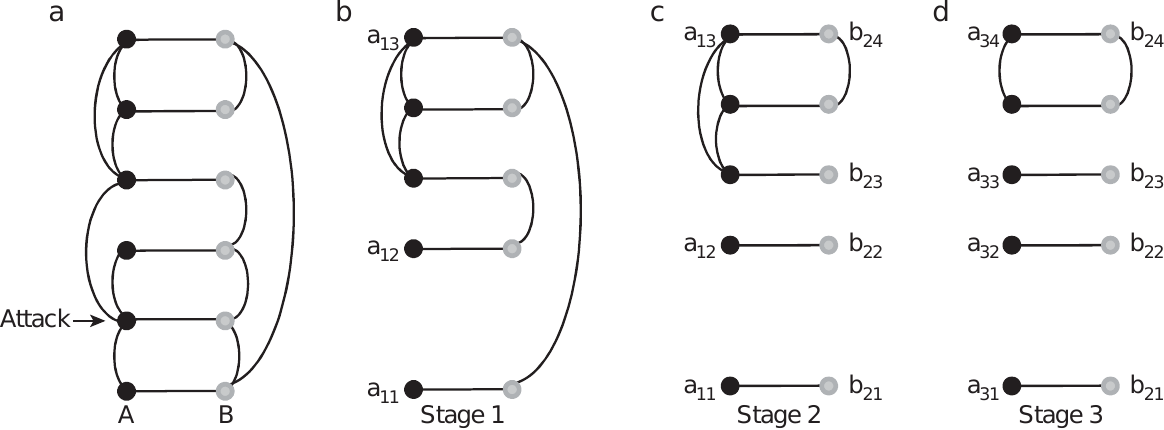}
 \caption{Schematic description on how to find the mutually connected component in a multilayer
interdependent network (duplex) considering the cascading failures propagating from one layer to the other one as described in Sec.~\ref{par_per}. The set of active nodes that remains at the end of the cascade is the mutually connected component. Reprinted figure from Ref.~\cite{Buldyrev10}. Courtesy of S. Havlin.}
    \label{fig:avalanches}
\end{figure}

\subsection{Percolation in interdependent networks}
\label{par_per}
After the {pioneering and} seminal work of Ref.~\cite{Buldyrev10}, it became clear that, in presence of interdependencies,
the robustness of   multilayer networks can be evaluated by calculating the size of their
{\em  mutually connected giant component} (MCGC) when a random damage affects a fraction $1-p$ of the nodes in the system.
The MCGC of a multilayer network is the largest component that remains
after the random damage propagates back and forth in the different layers.

In Fig.~\ref{fig:avalanches}, it is shown how to construct the MCGC
for the case of a multiplex formed by two layers (layer A and layer B) where the replica nodes are interdependent.
Assume that a group of nodes in layer A is damaged, all the nodes in the giant component of layer A
are active while the nodes that are not in the giant component are not active.
Then consider the layer B. All the nodes of layer B that are interdependent with nodes that are not
active in layer A, are damaged. Next, set as active all the nodes that remain in the giant
component of layer B,  and set as not active all the nodes that are not in the giant component.
By repeating the algorithm alternating the analysis of layer A and layer B, it is possible
to characterize an avalanche of failure events  that propagates from one layer to the other one,
 until the avalanche does not propagate any more.
The set of active nodes that remains at the end of the iteration is the MCGC.
Between any two nodes in this  mutually connected component there are at least two paths,
one in one layer A and one in layer B, that connect the two nodes and that pass
only through nodes that belong to the mutually connected component. In this case,
the MCGC is also called the {\em viable cluster} of the network\cite{Baxter}.

The size of the MCGC of a multilayer network
 emerges discontinuously at a critical value $p=p_c$,
at least in the case of random networks in which there is no overlap of the links
in the different layers and every node is interdependent at least on another node in another layer.
This is in contrast with the theory of percolation in single-layer networks,
where the giant component emerges continuously at a second order phase transition.
Moreover, if one approaches the value $p=p_c$, the interdependent networks
are affected by cascading failures that propagate throughout the structure.
Other works have extended the original formalism to  network of networks \cite{Gao:gbhs2011,Gao:gbsxh2013,Gao2012GinestraNatPhys}, network with partial interdependencies \cite{Parshani10}  with multiple dependencies links \cite{Shao2011GinestraPRE} and exploring the effect of targeted attack \cite{Havlin_target_attack}.
{ Subsequent studies on the subject have been carried out
  by Son et al. \cite{Son}.}
The Authors call this algorithm an `` epidemic spreading'' process,
but the algorithm can be more suitably termed as the ``message passing algorithm''
for this generalized percolation problem ~\cite{mezard2009,Parisi,hartman2005}.
Here we will try to present all this material in a pedagogical way, starting
from a generalization of the algorithm of Ref.~\cite{Son} that can be used both for
multiplex and for networks of networks.

\subsubsection{The mutually connected giant component of a multilayer network}
In this subsection, we will consider a special type of a  multilayer network
formed  by $M$ layers  $\alpha=1,2,\ldots, M$, each one formed by interactions between $N$ nodes $i=1,2,\ldots,N$.
Every node  will be indicated by the pair $(i,\alpha)$,
which denotes the label $i$ of the nodes in the specific layer $\alpha$.
We furthermore call all the nodes $(i,\alpha)$ characterized by the same label $i$
but belonging to different layers  $\alpha$ the ``replica nodes''.
Every node $(i,\alpha)$ can be connected to nodes $(j,\alpha)$  within the same layer
with ``connectivity links'', or with its ``replica nodes'' $(i,\beta)$ in other layers with ``interdependency links''.
In Ref.~\cite{Shao2011GinestraPRE}, it is further allowed to one node in layer
$\alpha$ to be interdependent to more than one node in a given layer $\beta$,
but here (if not explicitly otherwise stated) we make the simplifying assumption
that every node in layer $\alpha$ can be interdependent at most to a node in layer $\beta$, i.e.
on its ``replica node''. This framework allows to treat at the same
time, multiplex networks and networks of networks  as described in Sec.~\ref{par:netonet1} and Sec.~\ref{par:netonet2}.
We define the network of networks with a supra-adjacency matrix of elements $a_{i\alpha,j\beta}=1$
if there is a link between node $(i,\alpha)$ and node $(j,\beta)$ and zero otherwise.
Therefore, here we will treat multilayer networks in which $a_{i\alpha,j\beta}=0$
if both $i\neq j$ and $\alpha \neq \beta$.

The MCGC is defined in Ref.~\cite{Son}
as  the set of nodes $(i,\alpha)$ that satisfy the following recursive set of equations:
 \begin{enumerate}[{\em a)}]
\item  at least one neighbor $(j,\alpha)$ of node $(i,\alpha)$ in layer $\alpha$ is in the MCGC;
\item  all the interdependent nodes $(i,\beta)$ of node $(i,\alpha)$ are in the mutually
connected giant component.
\end{enumerate}
For a given multilayer  network which is locally tree-like,  it is easy to construct a ``message passing''
algorithm \cite{Parisi, mezard2009, hartman2005} that allows us to determine if node
$(i,\alpha)$ is in the mutually connected component.
This approach has been proposed in Ref.~\cite{Son} and  further investigated in Ref.~\cite{Watanabe2014GinestraPRE}.
We denote by $\sigma_{i\alpha \to j\alpha}=1,0$
the  message within a layer, from node $(i,\alpha)$ to node $(j,\alpha)$,
$\sigma_{i\alpha \to j\alpha}=1$ indicating that node $(i,\alpha)$ is in the mutually connected component
when we consider the cavity graph constructed by removing the link $(i,j)$ in layer $\alpha$.
Furthermore, let us denote by  $S'_{i\alpha \to i\beta}=0,1$
the  message between the ``replicas'' $(i,\alpha)$ and $(i,\beta)$ of node $i$
in layers $\alpha$ and $\beta$. $S'_{i\alpha \to i\beta}=1$
indicates that the node $(i,\alpha)$ is in the mutually connected component
when we consider the cavity graph by removing the link between node $(i,\alpha)$ and node $(i,\beta)$.
 In addition, we indicate with  $s_{i\alpha}=0$ a node that is removed
as an effect of the damage inflicted to the network, otherwise $s_{i\alpha}=1$.

If the node $(i,\alpha)$ is not damaged, i.e. $s_{i\alpha}=1$, the message $\sigma_{i\alpha\to j\alpha}$
will be equal to one if all the messages coming to node $(i,\alpha)$ from the interdependent nodes
$(i,\beta)$ are equal to one, and if at least one message $\sigma_{\ell\alpha\to i\alpha}$ coming from a node $(\ell,\alpha)$ neighbor of node $(i,\alpha)$ in layer $\alpha$ and different from node $(j,\alpha)$, is equal to one.
If the node $(i,\alpha)$ is instead damaged, the message $\sigma_{i\alpha\to j\alpha}$ will be equal to zero.

If the node $(i,\alpha)$ is not damaged, i.e. $s_{i\alpha}=1$, the message $S'_{i\alpha\to i\beta}$
will be equal to one if all the messages coming to node $(i,\alpha)$ from the interdependent
nodes $(i,\gamma)$ different from $(i,\beta)$ are equal to one,
and if at least one message $\sigma_{\ell\alpha\to i\alpha}$ coming from a node $(\ell,\alpha)$ neighbor of node $(i,\alpha)$ in layer $\alpha$, is equal to one.
If the node $(i,\alpha)$ is instead damaged, the message  $S'_{i\alpha\to i\beta}$ will be equal to zero.

Therefore, the message passing equations are of the following form:
\begin{align}
\sigma_{i\alpha\to j\alpha} &= s_{i\alpha}\  \prod_{\beta\in {\cal N}_i(\alpha)} \ \ \ S'_{i\beta \to i\alpha}
\left[1-\prod_{\ell\in N_{\alpha}(i)\setminus j}\ (1-\sigma_{\ell\alpha \to i\alpha})\right],
\nonumber
\\
S'_{i\alpha \to i\beta} &= s_{i\alpha}\ \prod_{\gamma\in {\cal N}(\alpha)\setminus \beta}\ S'_{i\gamma \to i\alpha}\left[1-\prod_{\ell\in N_{\alpha}(i)}\ \ (1-\sigma_{\ell\alpha \to i\alpha})\right],
\label{mp1}
\end{align}
where $N_{\alpha}(i)$ indicates the set of nodes $(\ell,\alpha)$ which are neighbors of node $i$
in layer $\alpha$, and ${\cal N}_i(\alpha)$ indicates the layers $\beta$ of the nodes $(i,\beta)$
interdependent on the node  $(i,\alpha)$.

Finally, $S_{i\alpha}=1$ ($S_{i\alpha}=0$) indicates that  the node $(i,\alpha)$ is (is not)
in the mutually connected component, namely
\be
S_{i\alpha} = s_{i\alpha}\!\!\!\!\prod_{\beta\in {\cal N}_i(\alpha)}\!\!\!\!S'_{i\beta \to i\alpha}
\left[1-\!\!\!\!\prod_{\ell\in N_{\alpha}(i)}\!\!\!\!(1-\sigma_{\ell\alpha \to i\alpha})\right].
\label{S}
\ee
In other words, a node $(i,\alpha)$ is in the mutually connected component if it is not damaged
(i.e. $s_{i\alpha}=1$), if all the messages $S'_{i\beta\to i\alpha}$ coming from the interdependent nodes are equal to one,
and if at least one message coming from a neighbor node $(\ell,\alpha)$ in layer $\alpha$ is equal to one.
In Ref.~\cite{BD}, it has been shown that these message passing equations can be solved
explicitly for every network of networks of this type including both cases described in Sec.~\ref{par:netonet1} and Sec.~\ref{par:netonet2}  giving, for the messages $\sigma_{i\alpha \to j\alpha}$  within each layer
\begin{align}
\sigma_{i\alpha\to j\alpha}&=s_{i\alpha}\left[1-\prod_{\ell\in N_{\alpha}(i)\setminus j}(1-\sigma_{\ell\alpha \to i\alpha})\right]\!\!\!\ \ \prod_{\beta\in {\cal C}_i({\alpha})\setminus \alpha}\!\!\left\{s_{i\beta}\left[1-\!\!\!\prod_{\ell\in N_{\beta}(i)}(1-\sigma_{\ell\beta\to i\beta})\right]\right\},
\label{mes2}
\end{align}
where ${\cal C}_i(\alpha)$ is the set of layers $\beta$ such that the  ``replica nodes'' $(i,\beta)$
are connected to the the node $(i,\alpha)$ by paths going through the interdependency links.
Finally, the Eqs.~(\ref{S}) for $S_{i\alpha}$ can be written exclusively as functions of the messages
$\sigma_{i\alpha\to j\alpha}$ as
\begin{align}
S_{i\alpha}&=s_{i\alpha}\prod_{\beta\in {\cal C}_i({\alpha})}\!\!\left\{s_{i\beta}\left[1-\!\!\!\prod_{\ell\in N_{\beta}(i)}(1-\sigma_{\ell\beta\to i\beta})\right]\right\}.
\label{Smes2}
\end{align}

For the network of networks described in Sec.~\ref{par:netonet1bis} or
Sec.~\ref{par:netonet3} a similar calculation can be carried out
showing that in networks of networks of any type, the mutually
connected component can be defined as the following:
A node $(i,\alpha)$ is in the mutually connected giant component if and only if:
\begin{enumerate}[{\it a)}]
\item it has at least one neighbor node $(j,\alpha)$ in layer $\alpha$ that belongs to the mutually connected component;
\item all the nodes $(\ell,\beta)$ that can be reached from node $(i,\alpha)$ by interdependent links have at least one neighbor node that belongs to the mutually connected component.
\end{enumerate}

\subsection{Interdependent multiplex networks}
A multiplex network can be considered as a network of networks of the type described in Sec.~\ref{par:netonet1},  in which each layer is connected with all the other layers (the supernetwork formed by the layer is a fully connected network) and each node is connected only to its replica nodes in the other layers.
Let us consider a  multiplex network in which every layer $\alpha$ has a degree distribution $P_{\alpha}(k)$
and in which every node $(i,\alpha)$ is interdependent on every  one of its replica nodes.
In order to evaluate  the expected size  of the mutually connected component,
we average the messages over this ensemble of the network of networks.
We indicate with $\sigma_{\alpha}$ the average message within a layer
$\avg{\sigma_{i\alpha \to j\alpha}}=\sigma_{\alpha}$.

The  equations for the average messages $\sigma_{\alpha}$  are given in terms of the parameter
$p$ and  the generating functions $G_0^{\alpha}(z)$ and $G_1^{\alpha}(z)$ defined as
\begin{align}
G_0^{\alpha} &= \sum_{k}P_{\alpha}(k)z^k\nonumber \\
G_1^{\alpha} &= \sum_k \frac{kP_{\alpha}(k)}{\avg{k}_{\alpha}}z^{k-1},
\label{G0G1}
\end{align}
where $\avg{k}_{\alpha}$ indicates the average degree in layer $\alpha$, i.e. $\avg{k}_{\alpha}=\sum_k kP_{\alpha}(k)$.
We assume here that the damage of the nodes in the different layers is correlated,
and that therefore a node $(i,\alpha)$ is damaged with probability $p$ together with all its replica nodes. We have
\begin{align}
P(\{s_{i,a}\})=\prod_{i=1}^M \left[p\prod_{\alpha=1}^M\delta(1,s_{i,\alpha})+(1-p)\prod_{\alpha=1}^M\delta(0,s_{i,\alpha})\right],
\label{Psia}
\end{align}
where $\delta(x,y)$ indicates the Kronecker delta.
This assumption is needed if we interpret the set of different replica nodes $(i,\alpha)$ in a multiplex like a single node having different types of interactions.
Therefore, using Eq.~(\ref{mes2}), in Ref.~\cite{Son} it was found that
\be
\sigma_{\alpha} = p\left\{ \prod_{\beta\neq \alpha} \left[1-G_0^{\beta}(1-\sigma_{\beta})\right]\right\}\left[1-G_1^{\alpha}(1-\sigma_{\alpha})\right].
\label{e110a}
\ee

Moreover, using Eq.~(\ref{Smes2}), the probability for $S_{\alpha}=\Avg{S_{i\alpha}}=S$
that a randomly chosen node  belongs to the mutually connected component is
\be
S=p\prod_{\beta=1}^M\left[1-G_0^{\beta}(1-\sigma_{\beta})\right].
\label{e120a}
\ee
Equivalent equations have been previously found in Refs.~\cite{Buldyrev10,Gao:gbhs2011} using an alternative derivation.
In particular, if all layers $\beta$ have the same topology, then the average messages
 within different layers $\beta$ are all equal $\sigma_{\beta}=\sigma$, and are given by
\be
\sigma = p\left[1-G_0(1-\sigma)\right]^{M-1}[1-G_1(1-\sigma)].
\label{e130}
\ee
Furthermore, the probability $S=\avg{S_{i\alpha}}$ that a node in a given layer is in the mutually connected component is given by
\be
S=p\left[1-G_0(1-\sigma)\right]^{M}.
\label{e140}
\ee
In Ref.~\cite{Havlin_target_attack} the case of targeted attack toward high or low degree nodes have been considered. It has been found in this context that in contrast with the single-layer network scenario, SF interdependent networks are difficult to defend  using the strategy of protecting high degree nodes.

\subsubsection{Case of a multiplex  formed by $M$  Poisson networks  with the same average degree}
\begin{figure}
 \centering
   \includegraphics[width=0.75\textwidth]{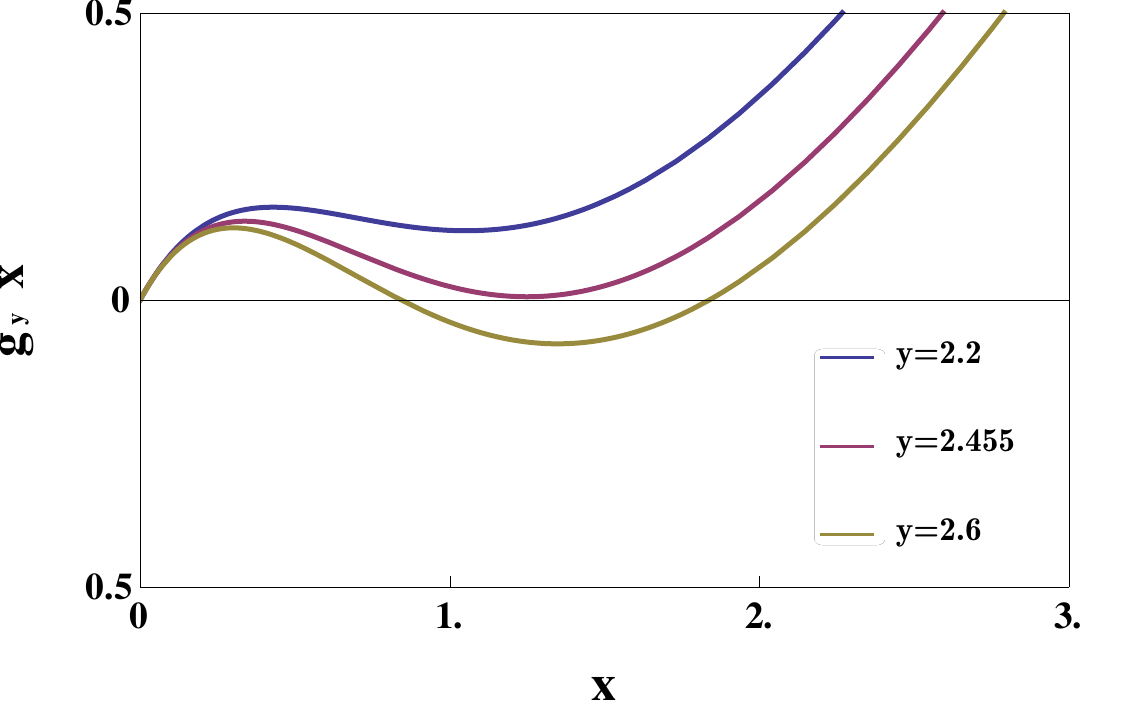}
 \caption{(Color online). The function $g_{cp=y}(x=Sc)=x-y(1-e^{-x})^2$ plotted for different values of $y=cp$.
The solution of the equation $g_{cp}(Sc)=0$ determines the fraction $S$ of nodes in the mutually connected component of  a duplex formed by two Poisson networks with average degree $c$, when a fraction $1-p$ of nodes has been damaged.
As $cp<2.45541\ldots$, only the trivial solution $S=0$ exists, while at $cp=2.45541\ldots$ a non trivial solution $S>0$ emerges discontinuously. }
    \label{fig:gx}
\end{figure}

In the case of a multiplex formed by $M$ Poisson networks with  the same average degree
$\avg{k}_{\alpha}=c$, we have that the equation for the order parameter becomes
\begin{align}
\sigma=S=p\left(1-e^{-cS}\right)^M.
\end{align}
For every finite  $M\ge 2$ it is possible to show that the MCGC
emerges discontinuously at $p=p_c(M)$. The equation for the order parameter $\sigma$ can be written as
\begin{align}
g_{y=cp}(Sc=x)=x-y \left(1-e^{-x}\right)^M=0.
\end{align}
The equation $g_y(x)=0$ has always a solution $\sigma=0$, but at $y=cp=y_c$
another solution appears discontinuously when the function $g_y(x)$ is tangential to
the $x$ axis at the point $x=x_c,y=y_c$. Therefore, in order to find the critical values $x=x_c$ and $y=y_c$,
we can impose the set of equations
\begin{align}
g_{y}(x)&=0\nonumber \\
\left.\frac{dg_y(x)}{dx}\right.&=0.
\end{align}
In Fig.~\ref{fig:gx} we plot the function $g_y(x)$ for $M=2$,
showing that at $y=y_c=2.45541\ldots$ the equation $g_y(x)$ develops a non trivial solution at $Sc=x_c=1.25643\ldots$.

\begin{figure}
 \centering
 \includegraphics[width=0.5\textwidth]{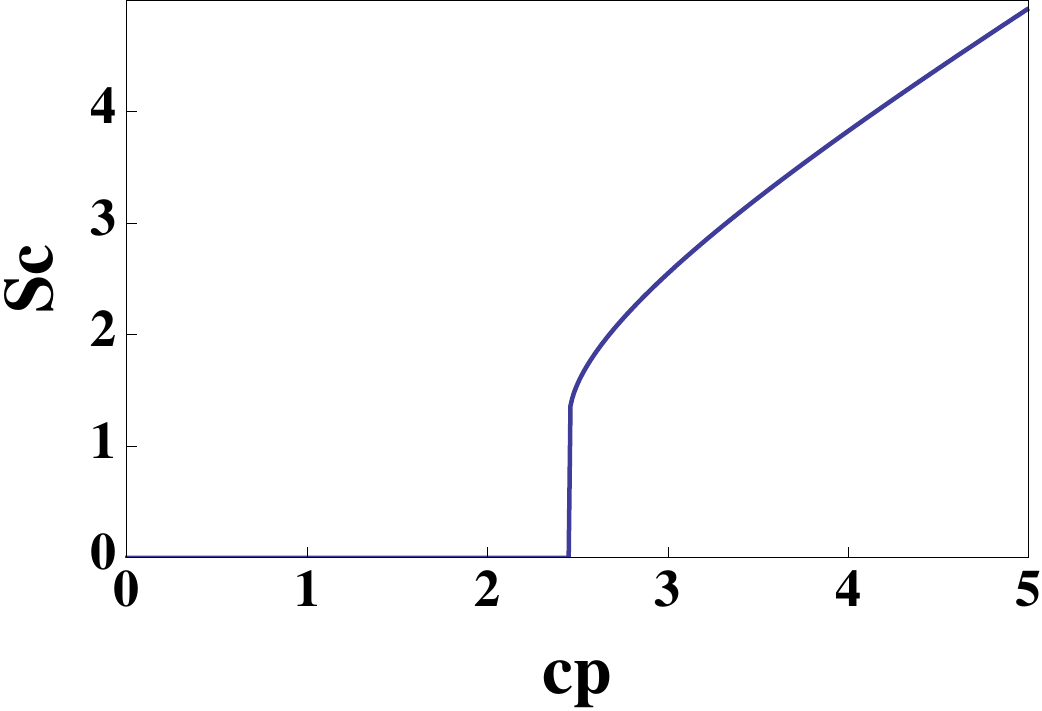}
 \caption{The quantity  $Sc$ as a function of the product $cp$,  where $S$ is the fraction of nodes in the mutually connected component of a duplex network formed by two Poisson networks with average degree $c$, and $1-p$ is the fraction of nodes initially damaged in the multiplex  network. }
\label{fig:S}
\end{figure}
In Fig.~\ref{fig:S} we plot the order parameter $S$ as a function of $y=cp$, for the case of a duplex,
i.e.  $M=2$. It is clear that the order parameter has a discontinuity at $y=y_c=2.45541\ldots$.
Moreover, it can be shown that the order parameter has a singularity for $y\to y_c^{+}$.
In fact, { it has been shown \cite{Baxter,Parshani10}} that the transition is a discontinuous
{\em hybrid transition} characterized by a square-root singularity at the phase transition,
in other words the order parameter, close to the transition, scales as
\begin{align}
S-S_c\propto (cp-y_c)^{1/2},
\end{align}
for $cp-y_c=\epsilon$ and $0<\epsilon\ll 1$.

 \subsubsection{Case of a duplex network formed by two Poisson networks with different average degree}
\begin{figure}
 \centering
   \includegraphics[width=\textwidth]{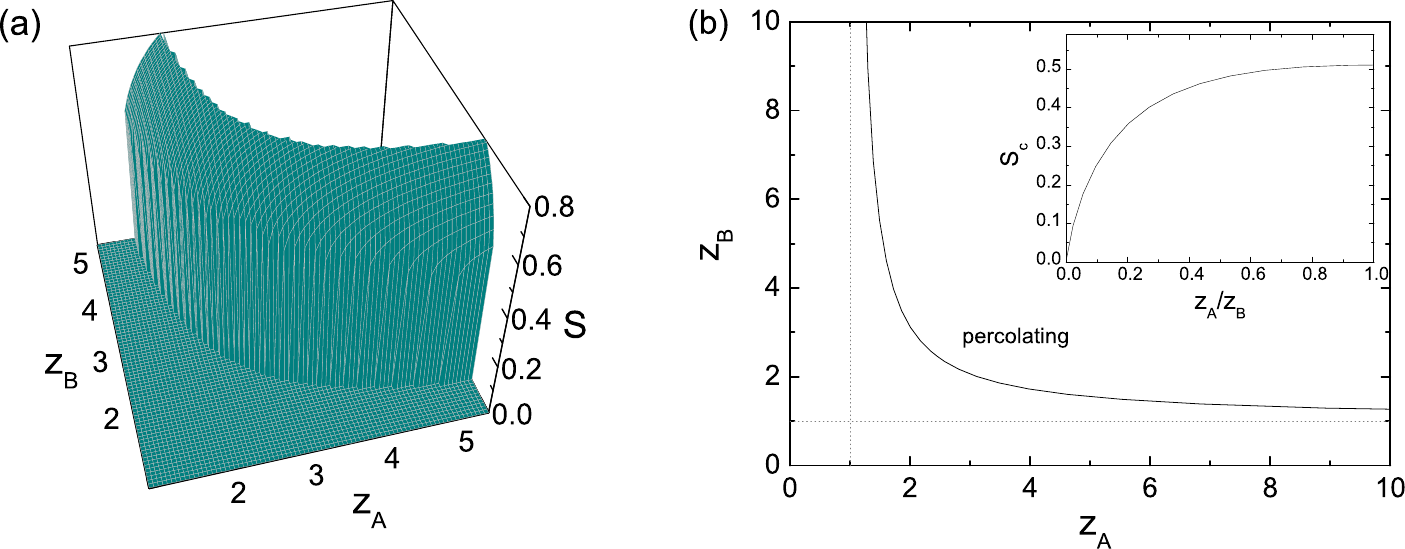}
 \caption{Phase diagram (for $p=1$) of a multiplex formed by two Poisson networks (network A and network B) with average degrees $\avg{k}_A=z_A$ and $\avg{k}_B=z_B$ respectively. Reprinted figure from Ref.~\cite{Son}.
\copyright \,\,IOP Publishing Ltd and Sissa Medialab srl. Reproduced by permission of IOP Publishing. All
      rights reserved.}
    \label{fig:Poisson2}
\end{figure}

 In Ref.~\cite{Son}, the case of a duplex network formed by two Poisson networks (network A and network B)
with different degree distribution (with average degree respectively $\avg{k}_A=z_A$ and $\avg{k}_B=z_B$)
has been studied. The equations determining the emergence of the MCGC
reduce to a single equation for $S=\sigma_A=\sigma_B$ given by
 \begin{align}
S=p(1-e^{-z_A S})(1-e^{-z_BS}).
 \end{align}
 Therefore, the phase diagram only depends on the parameters $(pz_A,p z_B)$.
 Also in this case, the emergence of the mutually connected component is described
by a discontinuous hybrid transition, at the points of the phase diagram satisfying simultaneously
the following set of equations
 \begin{align}
 h(S)&=S-p(1-e^{-z_A S})(1-e^{-z_BS})=0\nonumber \\
 h^{\prime}(S)&=0.
 \end{align}
 The phase diagram of the model is shown in Fig.~(\ref{fig:Poisson2}) for $p=1$.

\subsubsection{Multiplex networks formed by layers of scale-free networks}
\begin{figure}
 \centering
   \includegraphics[width=0.7\textwidth]{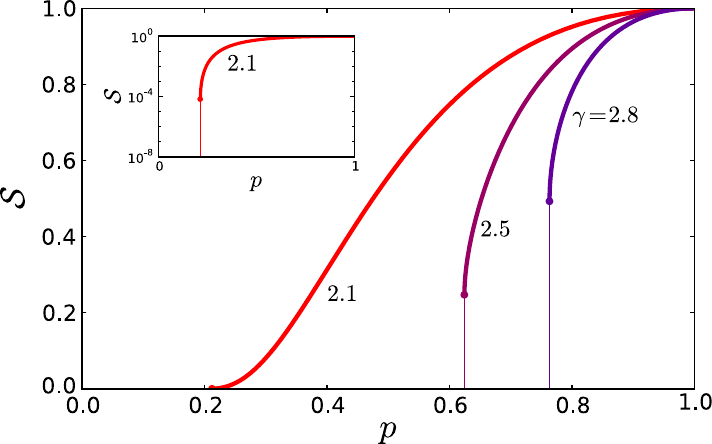}
 \caption{(Color online). Fraction of nodes $S$ in the MCGC  as a function of  $p$  for two symmetric
power-law distributed networks with, from right to left, power-law
exponent $\gamma=2.8 , 2.5$, and $\gamma=2.1$ and same minimal
degree. The height of the jump becomes very small as the power-law
exponent $\gamma$ approaches 2, but is not zero, as seen in the inset,
which show $S$ vs. $p$ on a logarithmic vertical scale for
$\gamma=2.1$.    Reprinted figure with permission from
   Ref.~\cite{Baxter}. \copyright\, 2012 by the American Physical Society.}
    \label{fig:sf}
\end{figure}

\begin{figure}[t!]
 \centering
   \includegraphics[width=0.8\textwidth]{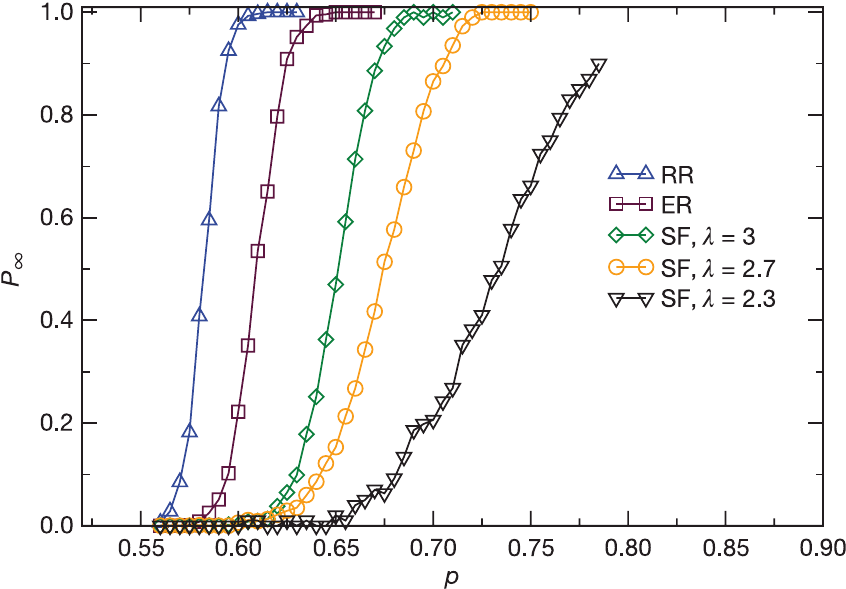}
 \caption{(Color online). Simulation results for the fraction of nodes $p_{\infty}=S$  in the mutually connected component as a function of $p$ for SF networks  with power-law exponent $\lambda= 3, 2.7, 2.3$, an ER network and a Random Regular (RR) network, all with an average degree $\avg{k}=4$. The simulation results are performed at finite number of nodes $N$, and therefore are not clearly discontinuous due to finite size effects.
Nevertheless, it can be seen already that $p_c$ is higher for broader distributions. Reprinted figure from Ref.~\cite{Buldyrev10}. Courtesy of S. Havlin.}
    \label{fig:sf2}
\end{figure}

In the case of single-layer networks, SF networks (with power-law degree distribution
$P(k)=Ck^{-\gamma}$ and $\gamma\in (2,3]$) are known to be much more robust to random failures than
networks with a finite second moment $\avg{k^2}$, e.g. of Poisson networks \cite{cnsw00,ceah00,ceah01}.
Indeed, for SF networks of size $N$, the giant component emerges if the fraction of nodes
that are not initially damaged $p>p_c$ with $p_c\to 0$ as $N\to \infty$.
Therefore, these networks continue to have the giant component also if most of the nodes are
initially damaged. A finite percolation threshold in these networks is a finite size effect, i.e.
the percolation threshold disappears in the limit $N\to \infty$.
The situation is completely different for multiplex networks formed by interdependent SF networks \cite{Buldyrev10}.
The percolation threshold $p=p_c$ for the  mutually connected component is finite also
for multiplex networks formed by layers of SF networks \cite{Buldyrev10}.
In the case in which the layers of the multiplex have the same power-law degree distribution
$p(k)=Ck^{-\gamma}$, the mutually connected component emerges discontinuously for $p=p_c$ as a hybrid phase
transition with a square root singularity \cite{Buldyrev10,Baxter}. In these networks, $p_c$
is not vanishing but remains finite as in the case of a multiplex
in which the layers are formed by Poisson networks \cite{Buldyrev10,Baxter}.
In Fig.~\ref{fig:sf} we show the size of the mutually connected component for two SF networks
with the same power-law exponent $\gamma$. As $\gamma\to 2$, and keeping the minimal degree constant,
the discontinuity of $S$ becomes smaller and smaller but is not vanishing as long as $\gamma>2$.
In Ref.~\cite{Buldyrev10} it was shown that actually another major difference exists
between SF single-layer and multiplex networks. In fact, if we consider a multiplex
formed  by two SF networks with the same degree distribution,
and compare their percolation threshold $p_c$ keeping the average degree $\avg{k}$ constant,
changing only the power-law exponent we obtain that the percolation  threshold $p=p_c(\gamma)$
increases as $\gamma$ is decreased. This implies that multiplex formed by broader SF are more fragile than
multiplex with a steeper degree distribution (see Fig.~\ref{fig:sf2}). This surprising result shows
that the robustness of multiplex networks has very new and unexpected features.

\subsubsection{Cascading failures}

In the case of a  multiplex network, one can follow the propagation of
the failures back and forth from one network to the others { \cite{Buldyrev10,Gao:gbhs2012,Gligor}}. In the specific case of a duplex,
it is possible to characterize how the failure in one network propagates in the other layer in the following way.
We start form the layer A and we consider the equations determining the size $S_A^{(0)}$
of the giant component when the random damage to a fraction $1-p$ of the nodes has been inflicted to the network.
In particular, $S_A^{(0)}$ will depend on the average message $\sigma_A^{(0)}$ indicating
what is the probability that a node at an end of a link in layer A is in the giant component of the same layer.
Then, we consider layer B, where we assume that all the nodes whose replica nodes in layer A
are not in the giant component of layer A are damaged.
Therefore, we consider all the nodes in layer B that remain in the giant component of layer B
after this damage propagating from layer A has been inflicted.
The probability that the nodes are not damaged is $p[1-G_0(1-\sigma_A^{(0)})]$ and, using this probability
one can calculate the probability $\sigma_B^{(1)}$ that a random link in layer B reaches a node in the giant component of layer B and the probability $S_B^{(1)}$ that a random node in layer B
is in the remaining giant component of layer B.
Iterating this process back and forth from one layer to the other,
it is possible to describe the propagation of cascading events in the interconnected network.
At the end of this iterative procedure, the remaining nodes are the nodes in the mutually connected component.
We will call  $\sigma_A^{(n)}$ and $\sigma_B^{(n)}$  the probability that,
in following a link either in layer A or in layer B, one reaches a node in the giant component of layer A or in the giant component of layer B at the $n^{th}$ step of this cascading process.
Similarly, we will call by $S_A^{(n)}$ and $S_B^{(n)}$ the probability that a node in layer A
or in layer B is in the giant component of layer A or of layer B.
The propagation of failures from one network to the other can be described by the following equations
 for the probabilities $\sigma_A^{(n)}$, $\sigma_B^{(n)}$
\begin{align}
\sigma_{A}^{(0)}&=p[1-G_1^{A}(1-\sigma_A^{(0)})]\nonumber \\
\sigma_{B}^{(1)}&=p[1-G_{0}^{A}(1-\sigma_A^{(0)})][1-G_1^B(1-\sigma_B^{(1)})]\nonumber \\
\ldots& \nonumber \\
\sigma_{A}^{(2m+2)}&=p[1-G_0^{B}(1-\sigma_B^{(2m+1)})][1-G_1^A(1-\sigma_A^{(2m+2)})]\nonumber \\
\sigma_{B}^{(2m+3)}&=p[1-G_{0}^{A}(1-\sigma_A^{(2m+2)})][1-G_{1}^B(1-\sigma_B^{(2m+3)})],
\end{align}
and by the following equations for the probabilities $S_A^{(n)}$, $S_B^{(n)}$
\begin{align}
S_A^{(0)}&=p[1-G_0^{A}(1-\sigma_A^{(0)})]\nonumber \\
S_{B}^{(1)}&=p[1-G_{0}^{A}(1-\sigma_A^{(0)})][1-G_0^B(1-\sigma_B^{(1)})]\nonumber \\
\ldots &\nonumber \\
S_{A}^{(2m+2)}&=p[1-G_0^{B}(1-\sigma_B^{(2m+1)})][1-G_0^A(1-\sigma_A^{(2m+2)})]\nonumber \\
S_{B}^{(2m+3)}&=p[1-G_{0}^{A}(1-\sigma_A^{(2m+2)})][1-G_{0}^B(1-\sigma_B^{(2m+3)})].
\end{align}
{See Refs.~\cite{Gao:gbhs2012,Gligor}} for the general equations describing the cascades of failures in multiplex networks with a generic value of $M$.
Eventually, these cascading events stop when the damage does not propagate further in the duplex network.
Close to  the phase transition $p\simeq p_c$, the cascades become  systemic events and
propagate until the mutually connected component completely disappears.
This phenomenon reveals that the interdependent networks
can be much more fragile than single-layer networks and are prone to abrupt cascading failures.

\begin{figure}
 \centering
  \includegraphics[width=0.7\textwidth]{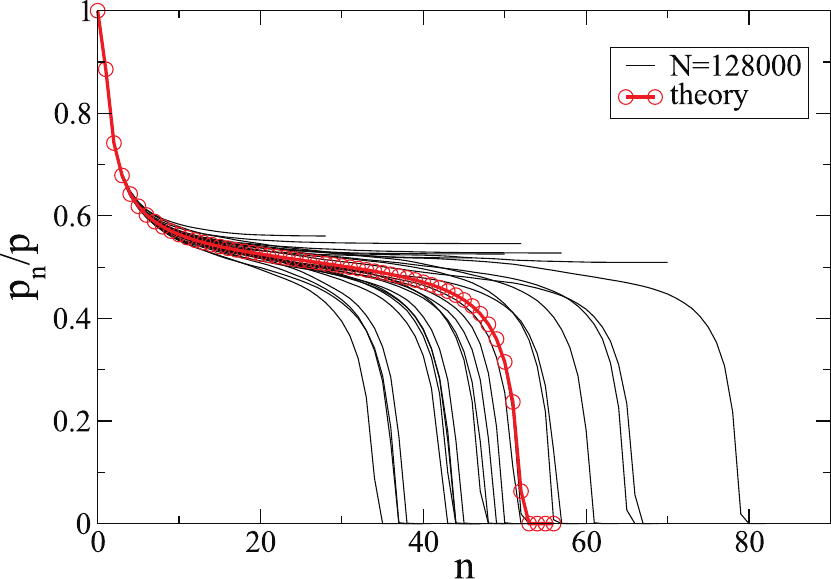}
 \caption{(Color online). The fraction of nodes $p_n$ in the mutually connected component
remaining after $n$ steps of the propagation of the damage back and forth between the two layers of a duplex network.
In this case, the  duplex has two layers with Poisson degree distribution and the same average $\avg{k}$.
Moreover, $p$ is set at the value  $p=2.455/\avg{k}<p_c$, and the total number of nodes $N$ is given by $N=12800$.
The red circles indicate the theoretical predictions valid in the limit $N\to \infty$. Reprinted figure from Ref.~\cite{Buldyrev10}. Courtesy of S. Havlin. }
    \label{fig:pn}
\end{figure}

In Ref.~\cite{Buldyrev10} the size of the network after $n$ steps of propagation of the failures
from one network to the other was first studied and characterized. In particular,
$p_n$ indicates the fraction of non-damaged nodes in the network after $n$ steps.
In other words
\begin{align}
p_n=\left\{\begin{array}{ccc} S_A^{(n)} &\mbox{for} & n=2m,\,\, m=1,2,\ldots\\
S_B^{(n)}&\mbox{for} & n=2m+1,\,\,  m=1,2,\ldots.\end{array}\right.
\end{align}
In Fig.~\ref{fig:pn} the subsequent values of $p_n$ are plotted for different realizations
of a duplex Poisson network with $p=2.45/\avg{k}<p_c$.
The solid red line indicates the theoretical predictions valid in the limit $N\to \infty$.
These predictions are very good for  small $n$, but for large values of $n$
the simulations start to show finite size fluctuations. In particular,
the simulations result  either in a multiplex network with a non zero mutually connected
giant component or in a multiplex network without it,
while the deterministic theory predicts that in the $N\to \infty$
the mutually giant component disappears in the network.
In Ref.~\cite{Buldyrev10} an analysis of the finite size effects was first performed,
yielding the following scenario for the average number of steps $\avg{n}$
of a cascade of failures as a function of $p$ and $N$:
\begin{itemize}
\item {\em Case $p<p_c$}
The average number of steps $\avg{n}$ of the cascade of failures can be approximated by
\begin{align}
\avg{n}\sim \frac{1}{\sqrt{p_c-p}}.
\end{align}
\item {\em Case $p=p_c$}
The probability $P(n)$ that a critical avalanche of failures has $n$ steps scales like
\begin{align}
P(n)=C\frac{e^{-\alpha \frac{n}{N^{1/4}}}}{N^{1/4}},
\end{align}
where $C>0$ and $\alpha>0$ are constants, independently on the number of nodes $N$ in the multiplex.
Therefore, the average number of steps of a critical avalanche of failures  $\avg{n}$ is diverging
with the multiplex size $N$ as
\begin{align}
\avg{n}\sim N^{1/4}.
\end{align}
\item{\em Case $p>p_c$}
The average number of steps $\avg{n}$ of the cascade of failures can be approximated by
\begin{align}
\avg{n}\sim \frac{\ln N}{\sqrt{p-p_c}}.
\end{align}
\end{itemize}

\subsubsection{Partial interdependence}
If the networks are only partially interdependent, the transition becomes continuous\cite{Parshani10,Son}.
Different versions of this model have been proposed modulating the degree distribution,
and the correlations between the layers \cite{parshani2011,Havlin_dep2,Havlin_dep3,Havlin_SF_dep}.
Here we show how a passage between a discontinuous and a continuous transition can occur in the
simplest setting of a multiplex, and in particular in a duplex, formed by ER networks.
Let us consider a multiplex of $M$ layers where each node $(i,\alpha)$
is interdependent on each of its replica nodes $(i,\beta)$ with probability $r$.
In this case, the message passing Eqs.~$(\ref{mes2})$ still remain valid, where
${\cal C}_i(\alpha)$ defines  all the set of layers $\beta$ for which the node $(i,\alpha)$
is interdependent on its replica node $(i,\beta)$.  We assume that every layer $\alpha$ has
degree distribution $P_{\alpha}(k)$, and that every node $(i,\alpha)$
is interdependent on every  one of its replica nodes.
We will describe the emergence of the mutually connected component by averaging  the messages
over links in each layer.
We indicate with $\sigma_{\alpha}$ the average message within a layer
$\avg{\sigma_{i\alpha \to j\alpha}}=\sigma_{\alpha}$.
Similarly to the case of a fully interdependent multiplex, we will  assume that the damage of the nodes
in the different layers is correlated and that, therefore, we have $P(\{s_{i\alpha}\})$
given by Eq.~(\ref{Psia}) that we rewrite here for convenience, i.e.
\begin{align}
P(\{s_{i,a}\}=\prod_{i=1}^M \left[p\prod_{\alpha=1}^M\delta(1,s_{i,\alpha})+(1-p)\prod_{\alpha=1}^M\delta(0,s_{i,\alpha})\right].
\end{align}

The equations for the average messages within a layer are given in terms of the parameter
$p$ and the generating functions $G_0^{\alpha}(z)$ and $G_1^{\alpha}(z)$ defined in Eq.~(\ref{G0G1}).
Using Eq.~(\ref{mes2}) we find
\be
\sigma_{\alpha} = p\left\{ \prod_{\beta\neq \alpha} \left[1-rG_0^{\beta}(1-\sigma_{\beta})\right]\right\}\left[1-G_1^{\alpha}(1-\sigma_{\alpha})\right].
\label{e110p}
\ee
Moreover, using Eq.~(\ref{Smes2}) we can derive the probability for $S_{\alpha}=\Avg{S_{i\alpha}}=S$
that a randomly chosen node  belongs to the mutually connected component
\be
S=p\left\{ \prod_{\beta\neq \alpha}\left[1-rG_0^{\beta}(1-\sigma_{\beta})\right] \right\}\left[1-G_0^{\alpha}(1-\sigma_{\alpha})\right].
\label{e120p}
\ee
In particular, if all layers $\beta$
have the same topology, then the average messages within different layers
are all equal $\sigma_{\beta}=\sigma$, and are given by
\be
\sigma = p\left[1-rG_0(1-\sigma)\right]^{M-1}[1-G_1(1-\sigma)].
\label{e130p}
\ee
Furthermore, the probability $S=\avg{S_{i\alpha}}$
that a node in a given layer is in the mutually connected component is given by
\be
S=p[1-rG_0(1-\sigma)]^{M-1}\left[1-G_0(1-\sigma)\right].
\label{e140p}
\ee

\begin{figure}[t!]
 \centering
 \includegraphics[width=0.75\textwidth]{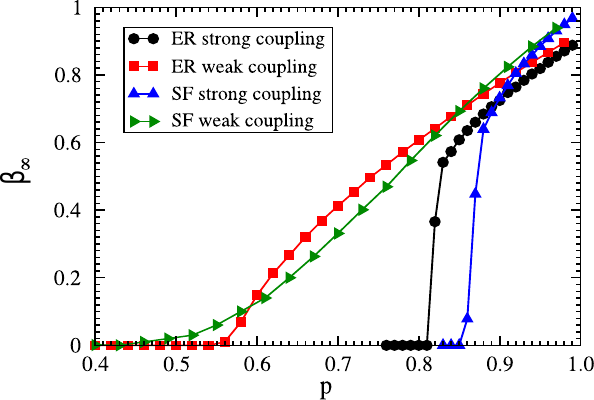}
 \caption{(Color online). Simulation results for the fraction of nodes $\beta_{\infty}=S$
in the MCGC of  a duplex formed by  $N = 50,000$ nodes, as a function of $p$. For strong coupling (high value of $r$) between the networks, there is a discontinuity in $\beta_{\infty}$
(Poisson networks-circle-, and SF networks-up rectangle). For weak coupling (low value of $r$) between the networks, the change in $\beta_{\infty}$ is described by a continuous  transition
(Poisson networks-square, and SF networks-right rectangle). Reprinted figure with permission from
   Ref.~\cite{Parshani10}. \copyright\, 2010 by the American Physical Society.}
    \label{fig:partial}
\end{figure}

For $r=0$ the nodes of each layer have no interdependencies with the nodes of the other layers,
therefore each layer will percolate independently, and the percolation transition will be a
second order continuous phase transition.
On the contrary, when $r=1$ we have complete interdependency of each node of a given layer
on all its replica nodes in the other $M-1$ layers, and the transition will be a discontinuous hybrid transition.
Figure~\ref{fig:partial} reports simulations results of these two limits, in duplex networks
of different degree distributions.
In particular, there is a tricritical point $(r=r_c,p=p_c)$
for which the transition changes its nature, from discontinuous (for $r>r_c$) to continuous (for $r<r_c$).

Let consider a duplex network formed by two Poisson networks with the same average degree $\avg{k}=c$.
The order parameter $\sigma=S$ is the solution of the equation
\begin{align}
S=p\left(1-re^{-cS}\right)\left(1-e^{-cS}\right).
\end{align}

The points of the hybrid  phase transition can be found by imposing
\begin{align}
h(S)&=S-p\left(1-re^{-cS}\right)\left(1-e^{-cS}\right)=0,\nonumber \\
\frac{dh(S)}{dS}&=0,\nonumber\\
\end{align}
and by looking for a non trivial solution $S>0$.
This solution can be found for any $r>r_c$ characterizing the level of partial interdependence of the tricritical point of the model.
For $r<r_c$, the emergence of the mutually connected component is dictated by a
second order phase transition at a critical value of $p$ that can be obtained by imposing the conditions
$h(0)=0, h^{\prime}(0)=0$.
The tricritical point separates these two regimes, and results \cite{Son} from imposing
$h(0)=0,h^{\prime}(0)=0,$ and $h^{\prime\prime}(0)=0$.
This system of equations yields the parameters of the tricritical point:
\begin{align}
r_c&=\frac{1}{3},\nonumber \\
cp_c&=\frac{3}{2}.
\end{align}
The avalanches of failures in the partially interdependent networks have been characterized in
Refs.~\cite{Parshani10,parshani2011,Havlin_dep2,Havlin_dep3}, where the Authors observe a different
scenario depending on the nature of the phase transition in the network.
In particular, it has been observed,  that if the system is close to the hybrid discontinuous phase transition,
the avalanches of failures are very long. In fact, the order parameter $p_n$ decays very slowly
with the number of steps $n$ of the propagation of the damage back and forth in the two layers,
and finally decays to a zero value.
On the contrary, the critical avalanches of failures close to the second order phase transition
are characterized by an order parameter $p_n$ that decays very rapidly with the number of
steps $n$ of the propagation of the damage.


\subsubsection{Percolation in multiplex networks with multiple support-dependence relations}
In Ref.~\cite{Shao2011GinestraPRE}, the case of a duplex is considered
 in which each node  $(i,\alpha)$ can be at the same time connected to several nodes
$(j,\beta)$ in layer $\beta\neq \alpha$ and can have multiple connections within the same layer $\alpha$.
It is assumed that a node $(i,\alpha)$ is in the mutually connected component
if the following two conditions are met:
\begin{enumerate}[{\it a)}]
\item
there is at least one neighbor $(j,\alpha)$ of node $(i,\alpha)$ in layer $\alpha$ that is in the mutually connected component;
\item
for every layer $\beta\neq \alpha$, there is at least one node $(j,\beta)$ connected to node $(i,\alpha)$ which is in the mutually connected component.
\end{enumerate}
Assuming that the layers are two (layer A and layer B) and that the degrees
$k_i^{\alpha\beta}$ are drawn from a Poisson degree distribution $P_{\alpha\beta}(k^{\alpha\beta})$
with $\avg{k^{A,A}}=a$, $\avg{k^{BB}}=b$, $\avg{k^{AB}}=\tilde{a}$ and $\avg{k^{BA}}=\tilde{b}$,
 it has been shown in Ref.~\cite{Shao2011GinestraPRE} that the equations determining the probability
$\sigma_A$ ($\sigma_B$) that a node in layer A (B) is in the  mutually connected component are given by
\begin{align}
\sigma_{A}&=p^A\left[1-e^{-\tilde{a}\sigma_B}\right] \left[1-e^{-{a}\sigma_A}\right],\nonumber \\
\sigma_{B}&=p^B\left[1-e^{-\tilde{b}\sigma_A}\right] \left[1-e^{-{b}\sigma_B}\right],
\end{align}
where $1-p^A$ ($1-p^B$) is the initial fraction of damaged node in layer A (B). For these networks,
the percolation is hybrid and discontinuous, but in the limit $\tilde{a},\tilde{b}\to \infty$ it becomes continuous.

\subsection{Interdependent networks of networks}
Following
Refs.~\cite{BD,BD2,Gao2012GinestraNatPhys,Gao:gbhs2011,Gao:gbhs2012,Gao:gbsxh2013},
we will consider { three} major types  of networks of networks.
\begin{itemize}
\item
{\it Case I:Networks of networks with fixed supernetwork and interdependent links allowed only between  replica nodes},  where every node $(i,\alpha)$ of layer $\alpha$ is interdependent on its replica node $(i,\beta)$ in the same set of layers $\beta$ \cite{BD}. This ensemble of network of networks has been described in Sec.~\ref{par:netonet1}.
\item
{\it  Case II: Networks of networks with fixed supradegree distribution}, where every node $(i,\alpha)$
is interdependent on $q=q_{\alpha}$ replica nodes $(i,\beta)$ of randomly chosen layers $\beta$ \cite{BD2}.
This ensemble has been described in Sec.~\ref{par:netonet2}.
\item
{\it Case III: Networks of networks with fixed supernetwork and random permutation of the labels of the nodes,} where every node $(i,\alpha)$ is interdependent on a node $(\pi_{\alpha,\beta}(i),\beta)$ if layer $\alpha$ is interdependent on layer $\beta$, where $\pi_{\alpha,\beta}$ is a random permutation of the labels $\{i\}$ of the nodes \cite{Gao2012GinestraNatPhys,Gao:gbhs2011,Gao:gbhs2012,Gao:gbsxh2013}.

\end{itemize}
\subsubsection{Case I: Networks of networks with fixed supernetwork and interdependent links allowed only between  replica nodes}

In this subsection, we will consider networks of networks in which, if network $\alpha$ is interdependent with
network $\beta$, each node $(i,\alpha)$ of network $\alpha$ is interdependent on node $(i,\beta)$
of network $\beta$, and vice versa.
This ensemble has been introduced in Sec.~\ref{par:netonet1}, and it has been graphically represented in  Fig.~\ref{fig:netonet1}. The corresponding supra-adjacency matrix has elements $a_{i\alpha,j\beta}=1$
if there is a link between node $(i,\alpha)$ and node $(j,\beta)$, and zero otherwise. In these
specific networks, $a_{i\alpha,j\beta}=0$ if both $i\neq j$ and $\alpha\neq \beta$.
The graph of interdependencies is characterized by an adjacency matrix $A_{\alpha\beta}$
such that, for every $i$, $A_{\alpha\beta}=a_{i\alpha,i\beta}$.
The  matrix $A_{\alpha\beta}$ is the adjacency matrix of the supernetwork ${\cal G}$.

In Ref.~\cite{BD} it has been shown  that for this type of networks of
networks,
where each node can be interdependent  only on its replica nodes in  the other layers,
the supernetwork can be a tree or can contain loops, and in both cases, as long as the supernetwork is connected,
the size of the mutually connected component is determined by the same equations
determining the MCGC in a  multiplex network formed by the same
layers.
This result is strongly dependent on the type of considered networks of networks.
For example,  for  networks of networks of case III the presence of
loops in the supernetwork change significantly the mutually connected
component
\cite{Gao2012GinestraNatPhys,Gao:gbhs2011,Gao:gbhs2012,Gao:gbsxh2013}.

In Ref.~\cite{BD} it has been shown  that the problem of a wide range of networks of networks can be
 actually reduced to the well studied problem of multiplex networks.
The main technical results of Ref.~\cite{BD} has been to show  that the messages $\sigma_{i\alpha \to j\alpha}$
within each layer are given by  Eq.~(\ref{mes2}), that we rewrite here for convenience
\begin{align}
\sigma_{i\alpha\to j\alpha}&= s_{i\alpha}\left[1-\prod_{\ell\in N_{\alpha}(i)\setminus j}(1-\sigma_{\ell\alpha \to i\alpha})\right]
\ \ \!\!\!\prod_{\beta\in {\cal C}_i({\alpha})\setminus \alpha}\!\!\left\{s_{i\beta}\left[1-\!\!\!\prod_{\ell\in N_{\beta}(i)}(1-\sigma_{\ell\beta\to i\beta})\right]\right\},
\label{mes2b}
\end{align}
where ${\cal C}_i(\alpha)$ is the set of layers $\beta$ such that the  ``replica nodes'' $(i,\beta)$
are connected to the node $(i,\alpha)$ by paths going through the interdependency links, and
where $N_{\alpha}(i)=\ell$ indicates that set of  label $\ell$ of nodes  $(\ell,\alpha)$
that are neighbors of node $(i,\alpha)$ in layer $\alpha$.
Ref.~\cite{BD} considered the case  in which  the supernetwork is given
(fixed by its adjacency matrix $A_{\alpha\beta}$),
and where each layer $\alpha$ is generated independently from a configuration model with a degree distribution
$P_{\alpha}(k)$.
The supernetwork is, therefore, not random, while the layers are infinite uncorrelated random networks.
It is assumed that each  layer $\alpha$ has a degree sequence $\{k_i^{\alpha}\}$,
and that the degrees of the replica nodes in the different layers are uncorrelated.
Furthermore, it is assumed that the nodes  $(i,\alpha)$ are removed from layers
with probability $1-p_{\alpha}$ (i.e. they are instigators of cascading failures).
Consequently, the probability $P(\{s_{i\alpha}\})$ is given by
\be
P(\{s_{i\alpha}\})=\prod_{\alpha=1}^M\prod_{i=1}^N p_{\alpha}^{s_{i\alpha}}(1-p_{\alpha})^{1-s_{i\alpha}}.
\ee
In order to evaluate  the expected size of the mutually connected component,
the messages  are averaged over this ensemble.
If $\sigma_{\alpha}$ indicates the average message within a layer
$\avg{\sigma_{i\alpha \to j\alpha}}=\sigma_{\alpha}$, the  equations for the average messages
within a layer are given in terms of the parameters
$p_{\beta}=\avg{s_{i\beta}}$
and the generating functions $G_0^{\beta}(z)$ and $G_1^{\beta}(z)$ are defined as
\begin{align}
G_0^{\beta} &= \sum_{k}P_{\beta}(k)z^k, \nonumber \\
G_1^{\beta} &= \sum_k \frac{kP_{\beta}(k)}{\avg{k}_{\beta}}z^{k-1}.
\end{align}
Without loss of generality, we can consider a connected supernetwork ${\cal G}$ formed by $M$ layers.
In fact, if the supernetwork is not connected, we can consider each connected component of the
supernetwork separately. Using Eq.~(\ref{mes2b}), the expression for $\sigma_{\alpha}$ is given by
\be
\sigma_{\alpha} =
p_{\alpha}\!\!\!\!\prod_{\beta\in
{\cal G} \setminus \alpha}\!\!\!\!\left\{p_{\beta}[1-G_0^{\beta}(1{-}\sigma_{\beta})]\right\}\![1-G_1^{\alpha}(1{-}\sigma_{\alpha})].
\label{e110}
\ee
Moreover, using Eq.~(\ref{Smes2}), the probability $S_{\alpha}=\avg{S_{i\alpha}}$
that a randomly chosen node in layer $\alpha$
belongs to the mutually connected component is given by
\be
S_{\alpha}=\prod_{\beta\in
{\cal G}}
\left\{p_{\beta}[1-G_0^{\beta}(1-\sigma_{\beta})]\right\}.
\label{e120}
\ee
Thus, as long as the supernetwork is connected,  the resulting mutual component is determined
by the same equations governing the case of a  mutually connected component between the same layers,
where the only difference between Eq.~(\ref{e110}) and Eq.~(\ref{e110a}) and between Eq.~(\ref{e120}) and Eq.~(\ref{e120a}) depends on the different choice of the probability $P(\{s_{i\alpha}\})$ for the initial damage inflicted in the network.
In the case in which all layers $\beta$ have the same topology and
equal probabilities $p_{\beta}=p$, the average messages  within different layers
are all equal $\sigma_{\beta}=\sigma$, and are given by
\be
\sigma = p^{M}\left[1-G_0(1-\sigma)\right]^{M-1}[1-G_1(1-\sigma)].
\label{e130pp}
\ee
Furthermore, the probability $S=\avg{S_{i\alpha}}$ that a node in a given layer is in the
mutually connected component is given by
\be
S=\left\{p[1-G_0(1-\sigma)]\right\}^{M}.
\label{e140pp}
\ee
Therefore, the size of the mutually connected component always depends on the number of layers $M$
in the connected supernetwork.

When we include partial interdependence between the nodes, by introducing the probability $r$ that a node is interdependent on a linked node in another layer, the Eqs. (\ref{e110})-(\ref{e120}) become respectively
\be
\sigma_{\alpha} =
p_{\alpha}\!\!\!\!\prod_{\beta\in
{\cal G} \setminus \alpha}\!\!\!\!\left\{p_{\beta}[1-rG_0^{\beta}(1{-}\sigma_{\beta})]\right\}\![1-G_1^{\alpha}(1{-}\sigma_{\alpha})],
\label{e110b}
\ee
and
 \be
S_{\alpha}=\prod_{\beta\in
{\cal G}\setminus \alpha}
\left\{p_{\beta}[1-rG_0^{\beta}(1-\sigma_{\beta})]\right\}[1-G_0^{\alpha}(1{-}\sigma_{\alpha})].
\label{e120b}
\ee

From these  results, four main conclusions can be drawn:
\begin{enumerate}[{\it a)}]
\item
the transition is hybrid and discontinuous as soon as $M\ge 2$,
\item
the critical value $p_c$ always depends on the number of layers $M$,
\item
when the mutually connected component emerges at $p=p_c$, every layer of the network of networks
contains a finite fraction of nodes that is in the mutually connected component, i.e.
in each layer the percolation cluster emerges at the same time at $p=p_c$,
\item
if we include a partial interdependence of the nodes, the transition  changes from discontinuous to continuous at a tricritical point, but also in this case either all layers  percolate, or any of them does.
\end{enumerate}

\subsubsection{Case II: Networks of networks with given superdegree of the layers}

Networks of networks can have a structure more random than the one considered in the above subsection.
In Ref.~\cite{BD2}, the attention has been addressed to the case in which every node of a network (layer) $\alpha$
is connected   with  $q_{\alpha}>0$ randomly chosen replicas  in some other networks,
and it is interdependent of these nodes with probability $r$.
Each network (layer) $\alpha$ is generated from a configuration model with  the same degree distribution
$P_{\alpha}(k)=P(k)$, and each node $(i,\alpha)$ is connected to $q_{\alpha}$ other ``replica'' nodes
$(i,\beta)$ chosen uniformly at random. We further assume that the degree sequence in  each layer
is $\{k_i^{\alpha}\}$ and that the degrees of the replicas of node $i$ are uncorrelated.
This is the ensemble described in Sec.~\ref{par:netonet2}, and graphically represented in Fig.~\ref{fig:netonet2}.
Every network of networks in this ensemble has a supra-adjacency matrix ${\bf a}$ with probability $P({\bf a})$ given by
\begin{align}
P({\bf a})&=\prod_{\alpha=1}^M\prod_{i=1}^N\left\{\delta\left(k^{\alpha}_i,\sum_{j\neq i}a_{i\alpha,j\alpha}\right)\right.\left.\ \delta\left(q_{\alpha},\sum_{\beta\neq \alpha} a_{i\alpha,i\beta}\right)\right\}\prod_{\alpha=1}^M\prod_{i=1}^N\prod_{\beta\neq \alpha}\prod_{j\neq i}\delta\left(a_{i\alpha,j\beta},0\right).
\label{Pa0}
\end{align}
The initially inflicted random damage is defined by the variables $s_{i\alpha}$.
In particular, if $s_{i\alpha}=0$, the node $(i,\alpha)$ is initially damaged, otherwise $s_{i\alpha}=1$.
The following probability $P(\{s_{i\alpha}\})$ is considered:
\begin{align}
P(\{s_{i\alpha}\})=\prod_{\alpha=1}^M\prod_{i=1}^N p^{s_{i\alpha}}(1-p)^{1-s_{i\alpha}}.
\label{Ps2}
\end{align}
In order to quantify the expected  size of the mutually connected component in this ensemble,
Ref.~\cite{BD2} considers the average of the messages over this ensemble.
The message passage equations for this problem are given by Eqs.~(\ref{mes2}).
Let us first consider the case $r=1$, where all the links between the different layers indicate
an interdependency between the two linked replica nodes.
The equations for the average message within a layer are given in terms of the parameter
$p=\avg{s_{i\alpha}}$, and the generating functions $G_0^{k}(z),G_1^{k}(z),G_0^q(z), G_1^q(z)$ are given by
\begin{align}
G_0^k(z)=\sum_k P(k) z^k
,\  &
G_1^k(z)=\displaystyle\sum_k \displaystyle\frac{kP(k)}{\Avg{k}}z^{k-1}
,
\nonumber
\\[5pt]
G_0^q(z)=\sum_q P(q) z^q
,\  &
G_1^q(z)=\displaystyle\sum_q \displaystyle\frac{qP(q)}{\Avg{q}}z^{q-1}.
\end{align}
In particular, if  $\sigma_q$ indicates the average messages within a layer $\alpha$ of degree $q_{\alpha}=q$, we obtain
\begin{align}
\sigma_{q}&=p\sum_{s}P(s|q)\left[\sum_{q'}\frac{q' P(q')}{\Avg{q}}p[1-G_0(1-\sigma_{q'})]\right]^{s-1}[1-G_1(1-\sigma_{q})],
\label{sq}
\end{align}
where $P(s|q)$ indicates the probability that a node $i$ in layer $\alpha$ with $q_{\alpha}=q$
is in a connected component ${\cal C}(i,\alpha)$ of the local supernetwork of cardinality
(number of nodes) $|{\cal C}(i,\alpha)|=s$.
\begin{figure}
 \centering
   \includegraphics[width=\textwidth]{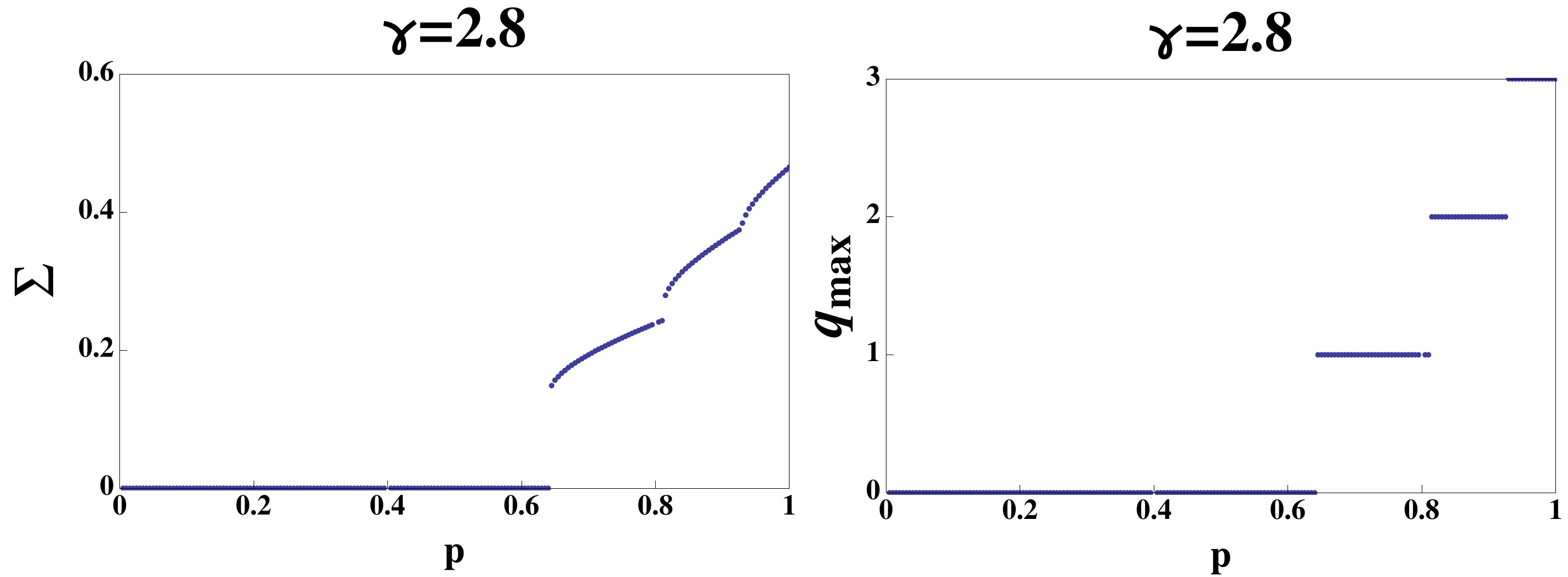}
 \caption{Plots of the order parameter $\Sigma$ (left panel) and of the maximal superdegree
of  percolating layers $q_{max}$ (right panel) vs. $p$ for a configuration model with a Poisson $P(k)$
distribution  with average $\avg{k}=c=20$ and a SF $P(q)$ distribution with $\gamma=2.8$,
and with  minimal degree $m=1$ and maximal superdegree
$Q=10^3$. Reprinted figure from Ref.~\cite{BD2}. }
    \label{fig:netonet2_per}
\end{figure}

Similarly, the probability that a node $i$ in a layer $\alpha$ with superdegree $q_{\alpha}=q$
is in the mutually connected component $S_q=\Avg{S_{i\alpha}}$ is given by
\begin{align}
S_q&=p\sum_{s}P(s|q)\left[\sum_{q'}\frac{q' P(q')}{\Avg{q}}p[1-G_0(1-\sigma_{q'})]\right]^{s-1}[1-G_0(1-\sigma_{q})].
\label{Sq}
\end{align}
In Ref.~\cite{BD2} it has been shown that Eqs.~(\ref{sq}) and Eqs.~(\ref{Sq}) can be expressed in the
following simplified way:
\begin{align}
\sigma_q&=p(\Sigma)^q[1-G_1^k(1-\sigma_q)]
,
\nonumber
\\[5pt]
S_q&=p(\Sigma)^q[1-G_0^k(1-\sigma_q)]
,
\nonumber
\\[5pt]
\Sigma &=\left[\sum_{q'}\frac{q' P(q')}{\Avg{q}}p[1-G_0^k(1-\sigma_{q'})]\right] \sum_{q'}\frac{q' P(q')}{\Avg{q}}(\Sigma)^{q'-1}.
\label{e160}
\end{align}
Therefore, in this ensemble the parameter $\Sigma$ determines both $\sigma_q$ and $S_q$ for any value of the superdegree $q$. For this reason,  $\Sigma$ can be considered the order parameter.

\begin{figure}
 \centering
  \includegraphics[width=0.6\textwidth]{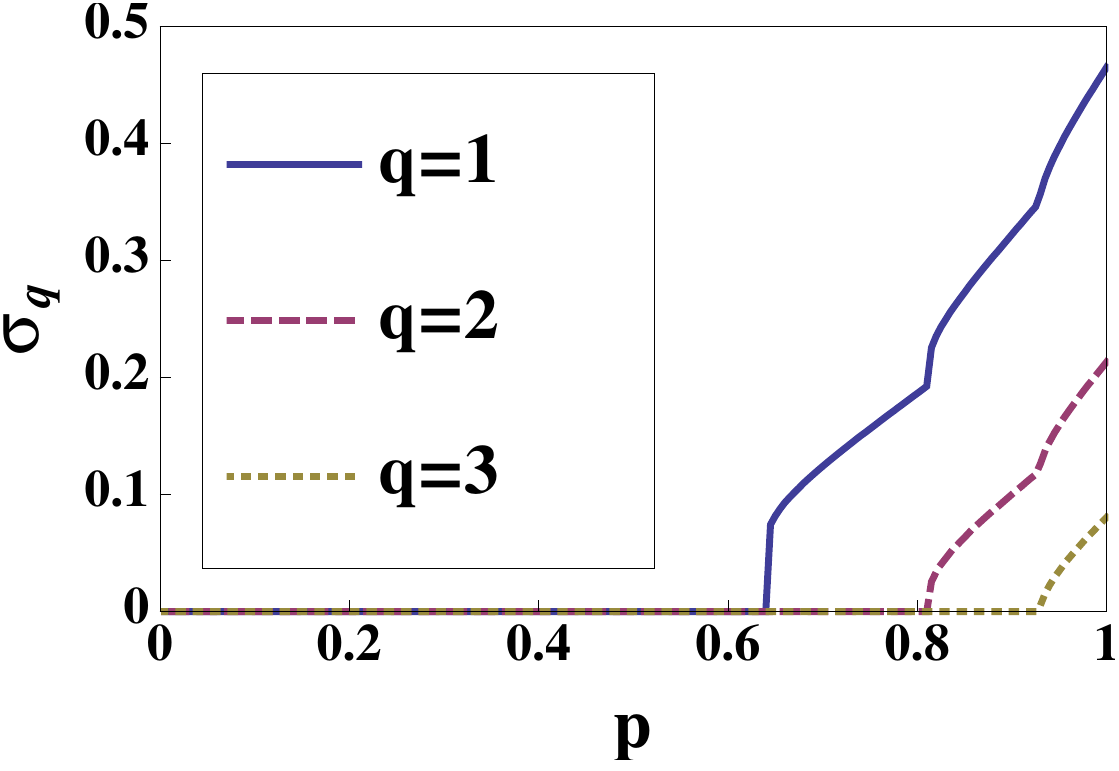}
 \caption{(Color online). Plot of $\sigma_q=S_q$ (fraction of nodes in
   a layer of superdegree $q$, belonging to the mutual component)
   vs. $p$ for different values $q=1,2,3$ in the configuration model
   of the network of networks having a Poisson $P(k)$ distribution
   with average $\Avg{k}=c=20$ and a SF $P(q)$ distribution with
   $\gamma=2.8$, minimal degree $m=1$ and maximal superdegree
   $Q=10^3$. For each value of $q=1,2,3$, $\sigma_q$ emerges
   discontinuously, with a jump, which becomes smaller and smaller
   with increasing $q$. The emergence of $\sigma_2$ is accompanied by
   a discontinuity of $\sigma_1(p)$. The emergence of $\sigma_3$ is
   accompanied by discontinuities of $\sigma_1(p)$ and  $\sigma_2(p)$.
   Reprinted figure from Ref.~\cite{BD2}.  }
    \label{fig:netonet2b_per}
\end{figure}


From the study of these equations, the following scenario can be drawn in the limit $M\to \infty$:
\begin{enumerate}[{\it a)}]
\item
 if the degrees $q_{\alpha}$ are heterogeneous, the percolation transitions are multiple,
each one corresponding to the emergence of a percolation cluster in each layer with a different value of $q$;
\item
 each of these transitions is hybrid and discontinuous, and
\item
the layers with higher number of interdependencies are more fragile than those with a smaller number of interdependencies.
\end{enumerate}
The increased fragility of those layers with higher superdegree is in sharp contrast 
with what happens in the percolation of single layers,
in which nodes of high degree are the more robust.
In Fig.~\ref{fig:netonet2_per}, the order parameter $\Sigma$ is shown for a power-law distribution
$P(q)\sim q^{-\gamma}$, with $\gamma=2.8$ and a Poisson distribution $P(k)$ with average $\avg{k}=c=20$.
The multiple discontinuous phase transitions are apparent. Moreover, the degree $q_{max}$
corresponding to the degree of interdependencies of the percolating layers with maximal value of $q_{\alpha}$ is shown.
Each transition, characterized by a discontinuity of the order parameter $\Sigma$, corresponds to
the activation of the layers with increasing superdegree $q_{\alpha}$.
Figure~\ref{fig:netonet2b_per} shows that the size of the percolation cluster $\sigma_q$
in the  layers with  $q$ number of interdependencies have multiple discontinuities
in correspondence with the discontinuities of the order parameter $\Sigma$.
Moreover, it is shown that the percolating cluster of layers with higher superdegree $q$
is disrupted before the percolating cluster of layers with smaller superdegree.

For $r<1$, Eqs.~(\ref{e160}) are modified as follows:
\begin{align}
\sigma_q&=p(r\Sigma+1-r)^q[1-G_1^k(1-\sigma_q)],
\nonumber
\\[5pt]
S_q&=p(r\Sigma+1-r)^q[1-G_0^k(1-\sigma_q)],
\nonumber
\\[5pt]
\Sigma &=\left[\sum_{q'}\frac{q' P(q')}{\Avg{q}}p[1-G_0^k(1-\sigma_{q'})]\right] \sum_{q'}\frac{q' P(q')}{\Avg{q}}(r\Sigma+1-r)^{q'-1}.
\end{align}
We note here that, in the case in which the local supernetwork is regular,
i.e. $P(q)=\delta(q,m)$, and each layer is formed by a Poisson network with $\avg{k}=c$,
one finds $\Sigma[r\Sigma+1-r]=\sigma_m=\sigma$ satisfying the following equation
\begin{align}
\!\!\!\sigma=p\left[\frac{1}{2}\left(1-r+\sqrt{(1-r)^2+4r\sigma}\right)\right]^{m}(1-e^{-c\sigma}).
\label{sigmam0}
\end{align}

The emergence of the mutually connected component in the configuration model of a network of networks
with $r<1$ can always display multiple percolation transitions corresponding to the activation of
layers in increasing value of $q_{\alpha}$. Nevertheless, these transitions can be either continuous or discontinuous,
depending on the value of $r$.

\subsubsection{Case III: Networks of networks with fixed supernetwork and random permutation of the labels of the nodes}

Here we consider the robustness of networks of networks described in Sec.~\ref{par:netonet1bis} where every node $(i,\alpha)$ is interdependent on a node $(\pi_{\alpha,\beta}(i),\beta)$ if layer $\alpha$ is interdependent on layer $\beta$, and $\pi_{\alpha,\beta}$ is a random permutation of the labels $\{i\}$ of the nodes.

In Refs.~\cite{Gao2012GinestraNatPhys,Gao:gbhs2011,Gao:gbhs2012,Gao:gbsxh2013}
this structure for the network of networks is implicitly assumed and
they demonstrate that the structure of the supernetwork matters. In fact, if the supernetwork ${\cal G}$ is a  tree, then the size of its mutual component is
determined by the same equations that determine the size of the MCGC
in a network of networks of case I. On the other hand, it was found that, for a supernetwork  with loops, this is not any more true \cite{Gao2012GinestraNatPhys,Gao:gbhs2012,Gao:gbsxh2013}.
In particular, in
Refs.~\cite{Gao2012GinestraNatPhys,Gao:gbhs2012,Gao:gbsxh2013} the
size of the MCGC for a network of networks  was explicitly derived for the case of  a regular random  supernetwork.
The size of the mutually connected component when each layer is a Poisson network of average degree  $c$ and  the  supernetwork is regular (i.e. the superdegree $q$ of each layer has distribution $P(q)=\delta(q,m)$) is given by
\begin{align}
\!\!\!\sigma=p\left[\frac{1}{2}\left(1-r+\sqrt{(1-r)^2+4r\sigma}\right)\right]^{m}(1-e^{-c\sigma}),
\label{sigmam1}
\end{align}
where here  partial interdependency  of the links is taken into
account and $r$ indicates the probability that a node is
interdependent on a linked node in another layer
\cite{Gao2012GinestraNatPhys,Gao:gbsxh2013}.
We observe here that the Eq.~(\ref{sigmam1}) is equivalent to Eq.~(\ref{sigmam0}) derived for networks of networks belonging to case II.
This result can be generalized. In fact if  the supernetwork is a
random network with given superdegree distribution the mutually
connected component of a network of network of case III
follows the same equations determining the MCGC in  the network of network of case II.
Therefore the following general  scenario holds for networks of networks in case III:
\begin{itemize}
\item
If the supernetwork is a tree,  the MCGC is determined by the same equations determining the MCGC in a network of networks of case I.
Therefore, in this case either all the layers are percolating or none does.
\item
If the supernetwork is a random network with given superdegree
distribution, the MCGC is determined by the same equations determining
the MCGC in a network of networks of case II. 
Therefore, in this case, there are multiple phase transitions and layers with higher superdegree are more fragile than layers with smaller superdegree.
\end{itemize}

\subsection{Effects of correlations and embedding space on percolation properties of multilayer networks}
The correlations on multiplex networks have important consequences on their percolation properties.
Here we will consider the effects of overlap of the links, degree correlations, and those of the embedding space.

\subsubsection{Percolation in networks with overlap of the links}

The overlap of the links can change significantly the properties of the percolation transition.
In fact, in the limit in which we have a multiplex of $M$ totally overlapping layers, the mutually connected giant component
of the multiplex is identical to the giant component of a single layer, and the percolation transition is of the second order.

For these reasons, it would be natural to expect that, when the overlap between the layers is over a threshold value, the emergence of the mutually connected component is continuous, while when the overlap is below such a critical value, the emergence of the mutually connected component is described by a hybrid transition.

This problem was tackled by several groups \cite{CellaiLo13,Hu,Li,MinGo2013}, finding different results.
In Ref.~\cite{CellaiLo13}, the Authors have studied the phase diagram corresponding to the message passing algorithm finding a
tricritical point for a non vanishing value of the overlap of the links. In \cite{Hu} the mutually connected component has been studied with different analytical techniques. In this latter case, the Authors do not find any tricritical point, and the percolation transition remains discontinuous for any finite fraction of links without overlap.

 \begin{figure}
 \centering
 \includegraphics[width=0.7\textwidth]{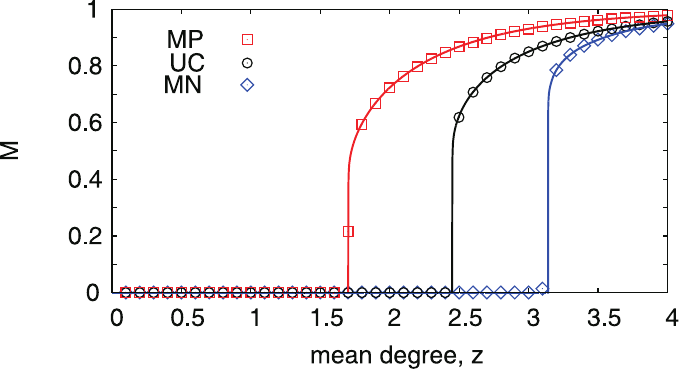}
 \caption{(Color online). The mutually connected component for a
   duplex of Poisson networks with different level of degree
   correlations. The symbol MP indicates maximally-positive degree
   correlations, the symbol MN indicates maximally-negative degree
   correlations, the symbol UC indicates no degree correlations.
   Reprinted figure with permission from
   Ref.~\cite{MinGo2013}. \copyright\, 2014 by the American Physical Society.}
    \label{fig:per_Goh_corr}
\end{figure}

\subsubsection{Percolation in networks with degree correlations}

The presence of degree correlations in multiplex networks can modify drastically
the percolation threshold $p_c$ \cite{Same_Buldyrev,parshani2010}.
The general equations for the percolation transition in correlated multiplex have been first derived in Ref.~\cite{Baxter},
 while Refs.~\cite{Same_Buldyrev,parshani2010, MinGo2013} considered the case of single-layer networks with the help
of the message passing algorithm of Ref.~\cite{Watanabe2014GinestraPRE}.
Ref.~\cite{Same_Buldyrev} considered multiplex networks in which the replica nodes have the same degree.
Refs.~\cite{Scala_assortativity,parshani2010} studied the robustness of the interdependent world-wide
airport network and the world-wide port network with positive degree correlations and clustering.
Finally, in Ref.~\cite{MinGo2013} different types of correlation between the degree of the nodes in the different layers have been considered.
The Authors investigated the maximal changes induced by the degree correlations
for duplex with the same degree sequences, but either uncorrelated by the degree or maximally-positive
correlated or maximally-negative correlated.
For maximally-positive correlated multiplex networks, the percolation threshold $p_c$ is  smaller than that
for the maximally-negative correlated case. The networks with uncorrelated degrees
in the different layers have a percolation threshold $p_c$ that is in between the previous two.
In Fig.~\ref{fig:per_Goh_corr} the dependence of the fraction of nodes in the MCGC of correlated Poisson networks
is shown as a function of the average degree of the nodes.
In Ref.~\cite{MinGo2013}, the role of targeted attack of multiplex networks is also investigated and discussed.

Other papers have investigated the effect of the degree correlations within a layer on the percolation properties of the network, and in particular assortativity \cite{Scala_assortativity}, or the effect of clustering \cite{Havlin_int_Clust}.
Finally, in Ref.~\cite{Braustein_corr}, the effect of degree correlations has been studied on a duplex
network with partial interdependence.


\subsubsection{Percolation in spatial multiplex networks}
Many networks are not only interdependent, but also embedded in a physical space.
For instance, infrastructures are mostly spatial and therefore it is essential
to investigate the role of space in determining their robustness.
Ref.~\cite{Son2011GinestraPRL} has shown that when the multiplex networks are embedded in real space,
the interdependencies do not always sharpen the percolation transition.
The topic is the subject of debate in the literature \cite{Havlin_debate,Grassberger_reply}.
Moreover, the nature of the percolation transition can depend on the typical distance in space of
the interdependent nodes. In fact, by modulating the typical distance from two interdependent nodes,
the transition can change from discontinuous (for large typical distances), to continuous
(for typical distances lower than a threshold value) \cite{Li2012GinestraPRL}.
This model has been extended in several directions \cite{Bashan,heal,attack_spatial}.
For example, in Ref.~\cite{Bashan} partial interdependence was
considered {\parbox[t]{\linegoal}{ and the Authors found that, in contrast to
  unembedded networks, for any fraction of dependency links the system collapses in an abrupt transition.}}
 {Reference}~\cite{heal} tackled
the problem of a network that recovers after each cascade. Furthermore, Ref.~\cite{attack_spatial}
studied the case of  { localized} attack percolation in spatial networks.

\subsection{Other percolation problems}
\subsubsection{Classical percolation}
The percolation of multilayer networks can also be studied in the absence of interdependencies.
This is the case in which all the connections, either those between nodes in the same layer and those
between nodes in different layers, have the same role and all the links are ``dependency links''.
In particular, one can consider the giant component as a function of the fraction of damaged nodes
in the multilayer structure. In the context of multilayer transportation systems,
involving different means of communication such as buses, undergrounds, trains, etc.,
one station will be in the giant component if by following at least a path using any number of
transportation systems, it is possible to reach a finite fraction of other stations.
 In this case, a node of the structure belongs to the giant component of the multiplex
if at least one of its links connects the node to the giant component.
 This problem has been studied in Ref.~\cite{LeiSo09}, considering the  model of networks of networks
described in Sec.~\ref{par:netonet3}, and graphically shown in Fig.~\ref{fig:netonet3}.
 Previously similar problems have been already addressed in the context of networks with different mixing patterns
\cite{Newman_mixing}, or epidemic spreading \cite{AllNo09,Vazquez06}.
As in the interdependent network case, the percolation threshold  depends on the different types of degree correlations between the layers. Ref.~\cite{MinGo2013} investigated and discussed also the role of
targeted attacks of multiplex networks. Finally, in Ref.~\cite{Basu} the percolation transition has been characterized for a network in which each layer is a  subgraph of some underlying network.

\subsubsection{Percolation of  antagonistic networks}
In Ref.~\cite{ZhaoGB2013}, the Authors considered the case in which there is antagonism between
the different layers of a multiplex. In particular, they studied the case of a duplex, and they assumed
that  the function or activity of a node is incompatible with the function or
activity of the replica node in the other layer.
The nodes in layer $\alpha$ are indicated in the following by $(i,\alpha)$,
while the nodes in layer $\beta$ are indicated by $(i,\beta)$, with $i=1,2,\ldots,N$.
The two nodes $(i,\alpha)$ and $(i,\beta)$ are replica nodes with an antagonistic interaction.
In order to determine if a node $(i,\alpha)$  is in the percolation cluster of the $\alpha$,
or if a node $(i,\beta)$ is in the percolation cluster of the layer $\beta$, the following algorithm has been proposed.

A  node $(i,\alpha)$ is part of the percolation cluster in layer $\alpha$ if  the following recursive set of
conditions is satisfied:
\begin{enumerate}[{\it a)}]
\item  at least one neighbor $(j,\alpha)$ of node $(i,\alpha)$ in layer $\alpha$ is in
the percolation cluster of layer $\alpha$;
\item  none of the nodes  $(j,\beta)$, neighbors of the replica node $(i,\beta)$, are in the percolation cluster
of layer $\beta$.
\end{enumerate}
Similarly, a  node $(i,\beta)$ is part of the percolation cluster in layer $\beta$
if the following recursive set of conditions is satisfied:
\begin{enumerate}[{\it a)}]
\item  at least one neighbor $(j,\beta)$ of node $(i,\beta)$ in layer $\beta$ is in the percolation cluster of layer $\beta$;
\item  none of the nodes  $(j,\alpha)$, neighbors of the replica  node $(i,\alpha)$, are in the percolation cluster
of layer $\alpha$.
\end{enumerate}
For a given multilayer network which is locally tree-like it is possible to construct a ``message passing''
algorithm that determines if node $(i,\alpha)$ belongs to the percolation cluster of layer $\alpha$.

We denote by $\sigma_{i\alpha \to j\alpha}=1,0$
the  message within a layer, from node $(i,\alpha)$ to node $(j,\alpha)$.
$\sigma_{i\alpha \to j\alpha}=1$ indicates that node $(i,\alpha)$ is in the percolation cluster of layer $\alpha$
when we consider the cavity graph by removing the link $(i,j)$ in network $\alpha$.
 In addition, we indicate with  $s_{i\alpha}=0$
a node that is removed from the network as an effect of the damage inflicted to the network, otherwise
$s_{i\alpha}=1$.

\begin{figure}[t!]
 \centering
  \includegraphics[width=0.7\textwidth]{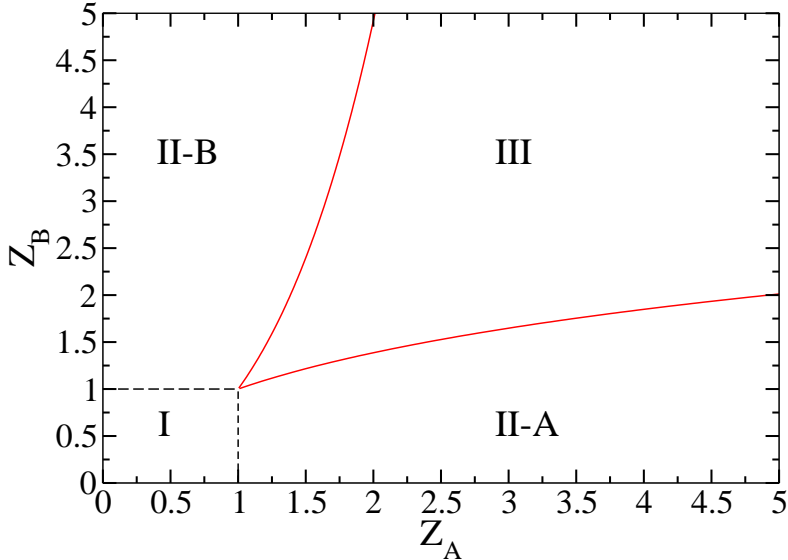}
\caption{Phase diagram of the percolation of two antagonistic Poisson
  networks with average degree $\avg{k}_A=z_A$ and $\avg{k}_B=z_B$ for
  $p=1$. In region I none of the two networks is percolating. In
  Region II-A  network A is percolating but not network
  B. Symmetrically, in region II-B network B is percolating but not
  network A. In region III there are two bistable solutions to the percolation problem and, depending on the initial conditions,
either network A or network B is percolating. Reprinted figure with permission from Ref.~\cite{ZhaoGB2013}. \copyright
      IOP\, Publishing Ltd and Sissa Medialab srl. Reproduced by
      permission of IOP Publishing. All rights reserved.}
 \label{fig:ant}
\end{figure}

The message passing equations take the following form:
 \begin{align}
\sigma_{i \alpha\to j \alpha}&=s_{i \alpha}\left[1-\prod_{\ell\in N_{\alpha}(i)\setminus j}(1-\sigma_{\ell \alpha \to i \alpha})\right] \prod_{\ell\in N_{\beta}(i)}(1-\sigma_{\ell \beta\to i\beta}),\nonumber \\
\sigma_{i \beta\to j \beta}&=s_{i \beta}\left[1-\prod_{\ell\in N_{\beta}(i)\setminus j}(1-\sigma_{\ell \beta \to i \beta})\right] \prod_{\ell\in N_{\alpha}(i)}(1-\sigma_{\ell \alpha\to i\alpha}).\nonumber \\
\label{mes2a}
\end{align}
$S_{i\alpha}=1$ ($S_{i\alpha}=0$) indicates that the node $(i,\alpha)$ is (is not) in the percolation cluster of layer $\alpha$.
The variables $S_{i\alpha}$ can be expressed in terms of the messages, i.e.
\begin{align}
S_{i \alpha}&=s_{i \alpha}\left[1-\prod_{\ell\in N_{\alpha}(i)}(1-\sigma_{\ell \alpha \to i \alpha})\right] \prod_{\ell\in N_{\beta}(i)}(1-\sigma_{\ell \beta\to i\beta}),\nonumber \\
S_{i \beta}&=s_{i \beta}\left[1-\prod_{\ell\in N_{\beta}(i)}(1-\sigma_{\ell \beta \to i \beta})\right] \prod_{\ell\in N_{\alpha}(i)}(1-\sigma_{\ell \alpha\to i\alpha}).\nonumber \\
\label{Sa}
\end{align}
If we consider a duplex formed by two random networks with degree distribution $P_A(k)$ and $P_B(k)$ respectively,
we can average the messages in each layer getting the equations for $\sigma_{A}=\avg{\sigma_{i\alpha\to j\alpha}}$ and $\sigma_{B}=\avg{\sigma_{i\beta\to j\beta}}$, that read as
\begin{align}
\sigma_{A}&=p[1-G_1^{A}(1-\sigma_A)]G_{0}^{B}(1-\sigma_B)\nonumber \\
\sigma_B&=p[1-G_1^{B}(1-\sigma_B)]G_0^{A}(1-\sigma_A).
\end{align}
The probabilities $S_A$, (or $S_B$) that a random node in layer A (or layer B) is in the percolation cluster of layer A (or layer B) are given by
\begin{align}
S_{A}=p[1-G_0^{A}(1-\sigma_A)]G_{0}^{B}(1-\sigma_B)\nonumber \\
S_B=p[1-G_0^{B}(1-\sigma_B)]G_0^{A}(1-\sigma_A).
\end{align}
This novel percolation problem has surprising features \cite{ZhaoGB2013}. For example, for a duplex network
formed by two Poisson networks with average degree respectively $\avg{k}_A=z_A$ and $\avg{k}_B=z_B$,
the phase diagram is shown in Fig.~\ref{fig:ant}, and displays a bistability of the solutions for some parameter values.
In region I, for $z_A<1$ and $z_B<1$ none of the two Poisson networks are percolating.
In Region II-A only network A contains a percolation cluster. Symmetrically, in region II-B only network B
contains a percolation cluster. Nevertheless, in region III, the solution of the model is bistable,
and depending on the initial condition of the message passing algorithm, either network A or network B is percolating.
Therefore, this percolation problem can display an hysteresis loop, as shown in Ref.~\cite{ZhaoGB2013}.

\subsubsection{K-core percolation}
In Ref.~\cite{k_core}, the ${\bf k}$-core of multiplex networks was defined as the set of nodes between
which there are at least $k_{\alpha}\leq 2$ paths connecting them in each layer $\alpha=1,2,\ldots, M$.
It has been shown that the ${\bf k}$-core on a duplex emerges always discontinuously and for every degree distribution
as a function of the fraction $1-p$ of removed nodes, following a hybrid transition, with the only exception
of the trivial $(1,1)$-core. The characterization of the  ${\bf k}$-core composition of a multiplex networks is a significantly richer problem than the characterization of the $k$-core structure of single-layer networks.
Moreover, it has been shown that multiplex networks are also less robust than their counterpart single-layer ones,
if one considers the destruction of the  ${\bf k}$-cores induced by random removal of the nodes.

\subsubsection{Viability in multiplex networks and  weak percolation}

Different models have been recently proposed that generalize the concept of percolation to interdependent networks.
Ref.~\cite{Min_Goh} assumed that each layer has a certain number of resource nodes.
Each node of the multiplex will then belong to the viable cluster of the resource network,
if it is connected by at least one path to a resource node in each layer.
The model displays  discontinuity, bistability, and hysteresis in the fraction of viable nodes with respect to the
 density of networks and the fraction of resource nodes. The model can be used in the context of infrastructures,
to show that the recovery from damage can also be extremely slow and difficult, due to hysteresis.
In Ref.~\cite{Weak_per_Davide} the weak bootstrap percolation and the weak pruning percolation have been proposed.
The weak bootstrap percolation describes the recovery of multiplex networks, while the weak pruning percolation
 describes the deactivation/damage  propagating processes in a multiplex structure.
While bootstrap percolation and pruning percolation on single-layers are equivalent, Ref.~\cite{Weak_per_Davide}
shows that they do differ in the multiplex case. When layers are formed by Poisson networks,
the weak bootstrap percolation includes both continuous and discontinuous transitions,
while the weak pruning percolation has only a continuous transition when defined on a duplex,
and a discontinuous hybrid transition when defined on a multiplex with 3 or more layers.

\subsection{Cascades on multilayer networks}
\label{par_cascades}
In multilayer networks, cascades can be caused not only by interdependent percolation but also
by other processes including cascades of loads
\cite{BruSo12,Goh_sandpile} described, for instance, by the Bak-Tang-Wiesenfeld sandpile model \cite{Bak},
 by the threshold cascade model \cite{BruLee12} proposed by Watts in Ref.~\cite{Watts_threshold},
or by novel cascade processes \cite{Lee_Brummitt_Goh}.
In Ref.~\cite{BruSo12} cascades of loads have been proposed to study the robustness of multilayer
infrastructures and power-grids. In that work, it has been shown that moderately increasing
the interlayer connectivity can reduce the size of the largest cascade in each layer, and it is therefore advantageous.
Nevertheless, it has been shown that too much interconnections can become detrimental for the stability
of the multilayer system because it allows single layers to inflict lager cascades to neighbor layers, and furthermore
new interconnections increase the capacity and the total possible load, eventually generating larger cascades.

In Ref.~\cite{Goh_sandpile} the role of multiplexity in the sandpile dynamics was further explored,
showing that the avalanche size distribution is not affected by the multiplexity of the network.
However, the detailed dynamics is affected by the multiplexity.
For example, high-degree nodes in  multiplex networks fail more often than in single-layer networks.
In Ref.~\cite{BruLee12} cascades of adoption of a behavior, such as ideas or financial strategies,
or movement to join,  are studied with the Watts model \cite{Watts_threshold}.
It is shown that multiplexity  can actually increase the network vulnerability to global cascades.
This has relevance for advertising strategies that can become more effective  with an  increasing
multiplexity, or for banks that might become more vulnerable with every new layer in the multiplex.
In Ref.~\cite{Lee_Brummitt_Goh} a novel cascade model, inspired by the Watts threshold model,
has been considered in which some nodes adopt a behavior once enough neighbors in all the layers have also adopted it,
and some other nodes adopt the behavior only if enough neighbors in some layers do.
This model can either inhibit or facilitate the cascade of events. Moreover, the model in some region of the phase space,
allows for abrupt emergence of cascading events, revealing some novel mechanisms for the onset of
avalanches  that can be useful to characterize civil movements, or product adoption.

\section{Spreading processes and games}
\label{sec:spreading}

As in the case of single-layer networks, the characterization of multilayer networks has to be completed with the knowledge about how the structural features of these networks affect their functioning. The ubiquity of multilayer networks in natural, social and technological systems demands to revisit most of the dynamical settings that were previously successfully addressed.
This is the goal of this section. The most relevant processes that have been studied include stochastic processes such as random walks, epidemic spreading and evolutionary game dynamics. In addition, each layer can sustain different coevolving processes, that are influencing each other. Thus,
the multilayer formalism allows to study the important scenario in which different dynamical
processes interplay, such as the competition between different virus strains in a population.

The section is structured as follows. The first part is devoted to
diffusion processes starting from the simple cases of linear diffusion and
random walks. As a generalization of these
processes, we also address the onset of congestion in communication
systems. In this latter framework, the diffusion in the network takes
into account the limited capacity of the nodes for sending and storing
incoming packets. There,
the interest lies on uncovering how the multiple routes between network layers affects the onset of jamming in the system.

The second part of the section focuses on spreading processes. In this
case, the main goal is to revisit the methods developed for epidemic
spreading on complex networks and generalize them to the framework of
multilayer. As usual in this context, the main concern is the prediction of the
outbreak threshold. However, other measures such as the effect of
potential control measures,
the coevolution of the disease spreading, and the transmission of information about its impact 
can be addressed by coupling the dynamical processes in each of the network layers. 
In addition to disease spreading, another type of contact process, i.e. opinion dynamics, will be reviewed. In this setting,
the goal is identifying the conditions for reaching agreement or consensus that, in the context of multilayer networks,
can be extended to a problem for the coevolution of different opinions in a system of agents.

Finally, the third part is concerned with the emergence of cooperative traits, and their survival under the framework of natural selection. 
This topic is quite challenging since defective behaviors are seen to provide more benefits at an individual level, 
leading to the extinction of cooperation. In the last decade,
spatial evolutionary games have been extensively investigated showing that cooperation is promoted thanks
to the possibility of forming compact cooperative clusters. This mechanism coined {\em network reciprocity} turned out to
be enhanced in single heterogeneous networks. One interesting question
naturally arises: could network reciprocity be further enhanced? Many
different ways of coupling the evolutionary dynamics in each network
layer have been proposed, and we will here systematically review
the main achievements in this direction.

\subsection{Diffusion on multilayer networks}
One of the most general issues on networks is the study of transport processes across network topologies \cite{blanchard,tadic}. These studies, indeed, are good proxies for many real dynamics involving navigation and flows of many kinds,
data in computer networks, resources in technological ones or goods and humans in transportation systems.
The variety of such processes ranges from simple diffusive dynamics and random walks
\cite{noh,yangpre,fontoura,condamin,gogarlat} to sophisticated protocols aimed at routing information \cite{sole,guimera,echenique,demartino}. These efforts allowed to characterize many central features of networks that cannot be captured
from intrinsically local properties of nodes, and to design optimal navigation schemes suited to real network architectures.

\subsubsection{Linear Diffusion}

The study of transport processes in multilayer networks has allowed to add an intrinsic ingredient of real transportation and communication systems, which is the possibility of driving flows by using multiple channels, or modes.
The first attempt to address this issue was done by G\'omez et al. in Ref.~\cite{GoDi13}, in which simple linear diffusion is generalized to the case of multiplex networks. The time-continuous equations for the linear diffusion in a multiplex
composed of $M$ networked layers (each of them having one representation of each of the $N$ nodes) are:
\begin{equation}
\frac{dx_i^{\alpha}}{dt}= D_\alpha \sum_{j=1}^{N} w_{ij}^{\alpha}(x_j^{\alpha}-x_i^{\alpha}) + \sum_{\beta=1}^M D_{\alpha\beta} (x_i^{\beta}-x_i^{\alpha})\;,
\label{diffusioneqs}
\end{equation}
where  $x^{\alpha}_i(t)$ ($i=1,\ldots,N$ and $\alpha=1,\ldots,M$) denotes the state of node $i$ in layer $\alpha$. The diffusion coefficients are, in general, different for each of the intralayer diffusive processes ($D_{\alpha}$ with $\alpha=1,\ldots,M$) and for each of the interlayer diffusion dynamics ($D_{\alpha\beta}$ with $\alpha$, $\beta=1,\ldots,M$).
In this way, the first term in the right-hand side of Eq.~(\ref{diffusioneqs}) accounts for the intralayer
diffusion, while the second one represents those interlayer processes.
Being a multiplex, the latter ones take place between the replicas of a given node in each of the layers.

The above dynamical equations can be casted in the usual form by defining the Laplacian ${\cal L}$ (usually called supra-Laplacian) of the multiplex  as an $M N\times M N$ matrix of the form:
\begin{equation}
{\cal L}=\left(
  \begin{array}{cccc}
    D_1 {\bf L^{1}} & 0 & \dots & 0\\
    0 & D_2{\bf L^{2}} & \dots & 0\\
    \vdots & \vdots & \ddots & \vdots\\
    0 & 0 & \dots & D_M {\bf L^{M}}\\
  \end{array}
  \right) +\left(
\begin{array}{cccc}
   \sum_{\beta\neq 1} D_{1\beta} {\bf I} & -D_{12}{\bf I} & \dots & -D_{1M} {\bf I}\\
    -D_{21}{\bf I} & \sum_{\beta\neq 2} D_{2\beta} {\bf I}  & \dots & -D_{2M} {\bf I}\\
    \vdots & \vdots & \ddots & \vdots\\
    -D_{M1}{\bf I}  & -D_{M2}{\bf I}  & \dots & \sum_{\beta\neq M}  D_{M\beta} {\bf I} \\
  \end{array}\right) \;,
  \label{supraLaplacianrepe}
\end{equation}
where {\bf I} is the $N\times N$  identity matrix and ${\bf L^{\mathbf \alpha}}$ is the usual $N\times N$ Laplacian matrix of the network layer $\alpha$ whose elements are $L^{\alpha}_{ij}=s^{\alpha}_i\delta_{ij}-w_{ij}^{\alpha}$ , and $s^{\alpha}_i$
is here the strength of the representation of node $i$ in layer $\alpha$,
$s^{\alpha}_i=\sum_j w_{ij}^{\alpha}$. With this definition of ${\cal L}$, the diffusion equation (\ref{diffusioneqs}) can be formulated as $\dot{{\bf x}}={\cal L}{\bf x}$.

\smallskip

For the sake of simplicity, one considers that the intralayer diffusion coefficients are set to $D_{\alpha}=1$ $\forall \alpha$ (without any loss of generality, as each coefficient $D_{\alpha}$ can be absorbed  by the terms of the corresponding Laplacians
${\bf L^{\alpha}}$) and that the interlayer coefficients are equal to $D_{\alpha\beta}=D_x$ $\forall$ $\alpha,\beta$.  The simplified Laplacian thus reads:
\begin{equation}
{\cal L}=\left(
  \begin{array}{cccc}
    {\bf L^{1}} & 0 & \dots & 0\\
    0 & {\bf L^{2}} & \dots & 0\\
    \vdots & \vdots & \ddots & \vdots\\
    0 & 0 & \dots & {\bf L^{M}}\\
  \end{array}
  \right) +D_{x}\cdot\left(
\begin{array}{cccc}
   (M-1) {\bf I} & -{\bf I} & \dots & -{\bf I}\\
    -{\bf I} & (M-1){\bf I}  & \dots & - {\bf I}\\
    \vdots & \vdots & \ddots & \vdots\\
    -{\bf I}  & -{\bf I}  & \dots & (M-1)  {\bf I} \\
  \end{array}\right)
\;.
\label{supraLaplacian2}
\end{equation}
The above formulation allows to study the role of the interlayer coupling $D_x$ in the diffusion properties of the multiplex,
 and relating them with those of the set of isolated network  layers ($D_x=0$). In fact, given the symmetry of the
Laplacian (\ref{supraLaplacian2}), the solution of the linear system can be expressed in terms of the eigenvectors and eigenvalues of ${\cal L}$ whose time evolution reads $\phi_i(t)=\phi_i(0){\mbox e}^{-\lambda_i t}$. Thus, the diffusion time-scale, i.e. the time $\tau$ that the multiplex needs to relax to the stationary solution, is given by the inverse of the
smallest non-zero eigenvalue of ${\cal L}$ that, for a connected multiplex, is the second eigenvalue (from the smallest $\lambda_1=0$, to the largest one), that is $\tau=1/\lambda_2$.

In Ref.~\cite{GoDi13} the Authors focused on the case of two layers, $M=2$, and through perturbation limit they were able to
relate the value of $\lambda_2$ for the weak ($D_x\ll 1$) and strong ($D_x \gg 1$) interlayer coupling limits.
In Ref.~\cite{SoDo13}, Sol\'e  et al. extended the perturbative analysis to the full spectrum of
${\cal L}$, $\{\lambda_1=0,\lambda_2,\ldots,\lambda_{M\cdot N}\}$, and for an arbitrary number $M$ of layers.
The main results of Ref.~\cite{SoDo13} concerning the smallest non-trivial eigenvalues of ${\cal L}$ are the following:
\begin{itemize}
\item For $D_x\ll 1$ the set of the $M-1$ smallest non-trivial eigenvalues of ${\cal L}$ increases linearly with $D_x$.
\item For $D_x\gg 1$  the set of the $N-1$ smallest non-trivial eigenvalues  of ${\cal L}$ are approximately the $N-1$ non-zero eigenvalues of a Laplacian corresponding to a simple network of $N$ nodes resulting from taking the average of the layers composing the multiplex, so that the weights of each link in this latter network read as $w^{av}_{ij}=\sum_{\alpha}w_{ij}^{\alpha}/M$.
\end{itemize}

\begin{figure}[t!]
 \centering
   \includegraphics[width=\textwidth]{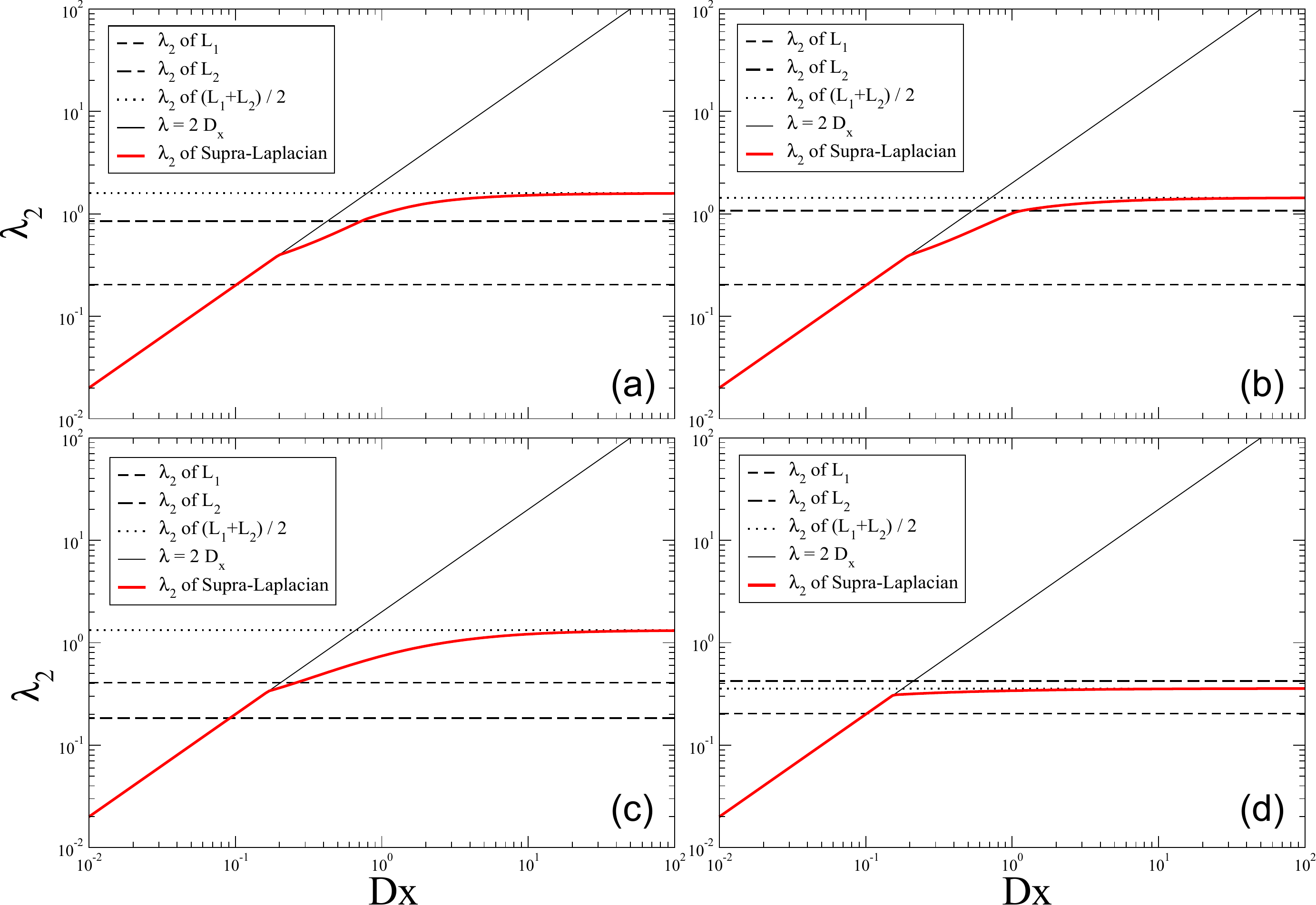}
 \caption{(Color online). Evolution of $\lambda_2(D_x)$ for
   multiplexes composed of two layers and $N=10^3$. The layers are of
   the form: (a) scale-free (SF) with $P(k)\sim k^{-2.5}$ and
   Erd\H{o}s-R\'enyi  network with average degree $\langle
   k\rangle=8$, (b) same SF network as in (a) and a Watts-Strogatz
   small-world network with $\langle k\rangle=8$ and replacement
   probability of 0.3, (c) same SF network as in (a) and another SF
   network with  $P(k)\sim k^{-3}$ and (d) same SF network as in (a)
   and the same SF network plus $400$ additional links randomly
   placed. It is clear that in the first three examples superdiffusion
   shows up for large values of $D_x$, whereas for the multiplex in
   (d) this is not the case. Reprinted figure with permission from
   Ref.~\cite{GoDi13}. \copyright\, 2013 by the American Physical Society.
}
    \label{fig:diffusion}
\end{figure}

The main result of Refs.~\cite{GoDi13,SoDo13} appears evident when looking at the strong coupling
regime, $D_x\gg 1$. While for $D_x\ll 1$ the time scale associated to diffusion goes as $\tau\propto (M \cdot D_x)^{-1}$,
the case $D_x\gg 1$ shows that the multiplex is always faster than the
slowest one of the composing layers.
Moreover, although not guaranteed, {\em superdiffusion} (i.e. the fact that the multiplex relaxes faster than any of its
composing layers) is also possible. In Fig.~\ref{fig:diffusion} we
show the evolution of $\lambda_2(D_x)$
for four multiplexes (with and without superdiffusion) composed of two layers.

\smallskip

\begin{figure}[t!]
\begin{center}
\includegraphics[width=0.95\textwidth]{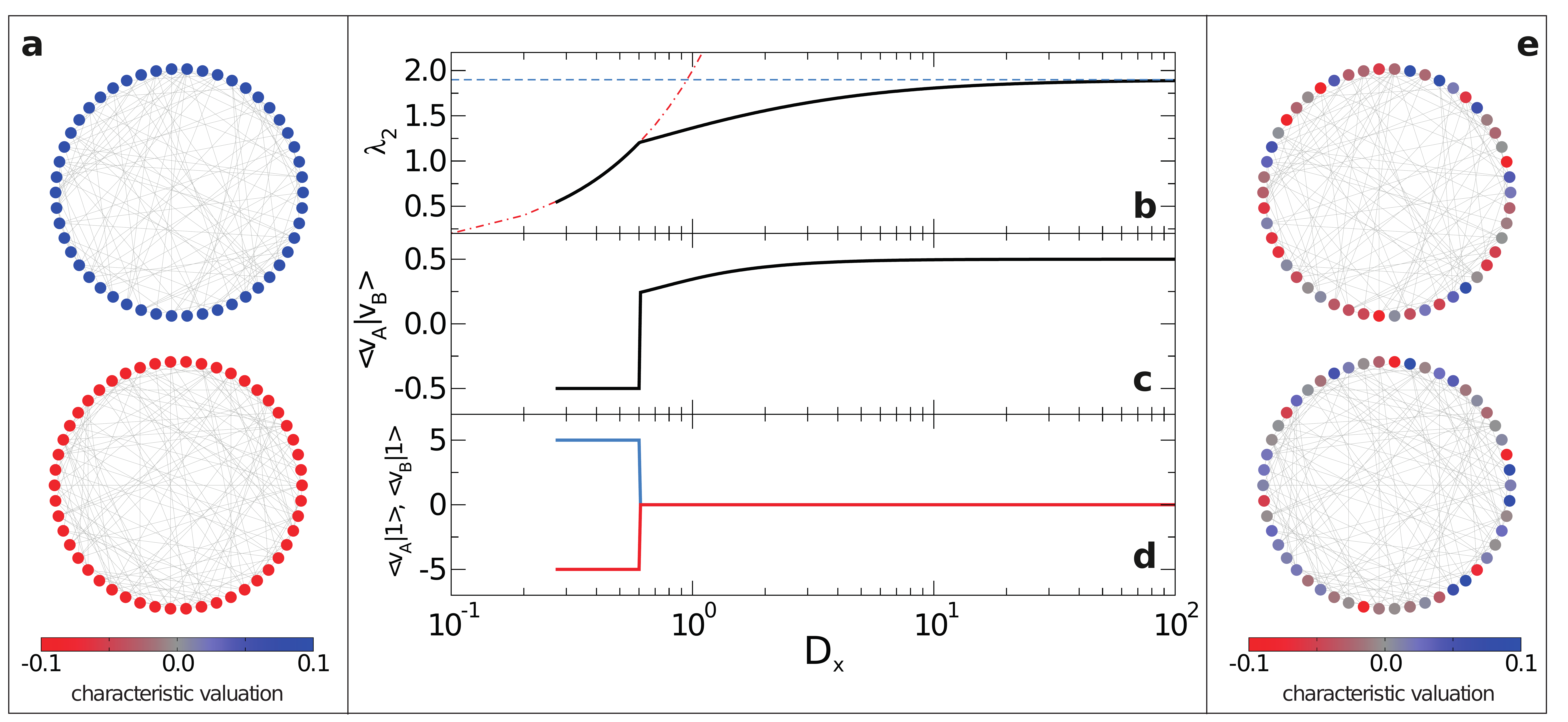}
\end{center}
\caption{(Color online). Evolution of $\lambda_2$ and its associated eigenvector, ${\bf v_2}$, as a function of $D_x$
for a multiplex composed of two Erd\H{o}s-R\'enyi networks of $N=50$ nodes and average degree $\bar{k}=5$. In this example, the critical point is
$(D_x)_c=0.602(1)$.  {\bf a)} Values of the components in  ${\bf v_2}$ (characteristic valuations) for the nodes in the two layers for $D_x=0.602$ (just before the onset of the transition). {\bf b)} $\lambda_2$ as a function of $D_x$ (black line).
The discontinuity of the first derivative of $\lambda_2$ is very clear. The transition between the two known different
regimes ($2\cdot D_x$, blue dashed line, and $\frac{\lambda_2\left({\bf L^A}+{\bf L^B}\right)}{2}$, red dash-dotted line) is evident.
{\bf c)} Projection of ${\bf v_2^B}$ into ${\bf v_2^A}$ as function of $D_x$. {\bf d)} Projection of the unit vector into ${\bf v_2^A}$ and ${\bf v_2^B}$ as functions of
$D_x$. These two projections indicate the sum of all components of ${\bf v_2^A}$ and ${\bf v_2^B}$ respectively. {\bf e)} Values of the components in  ${\bf v_2}$ (characteristic valuations) for the nodes in the two layers for  $D_x=0.603$ (just after the onset of the transition). Reprinted figure from Ref.~\cite{radicchi}. Courtesy of A. Arenas.
}
\label{fig:structuraltransition}
\end{figure}

From the evolution of $\lambda_2(D_x)$ shown in Fig.~\ref{fig:diffusion} it is clear that the diffusive properties (as described by $\tau$) of the multiplex do not evolve homogeneously with the interlayer coupling, and there is a value $(D_x)_c$ (that depends on the composition of the multiplex) in which the linear increase of $\lambda_2(D_x)$ changes abruptly.
This phenomenon is explained by Radicchi and Arenas in Ref.~\cite{radicchi} as a structural transition
from a decoupled regime (in which layers behave as independent) to a systemic regime
(in which layers are indistinguishable).
This change is also explained in terms of the eigenvector ${\bf v_2}$
associated to the smallest non-zero eigenvalue of the Laplacian of the multiplex, $\lambda_2$,
since it provides useful information about the modular structure of a network, as it is the key ingredient of the spectral partitioning method developed by Fiedler \cite{Fiedler,Newman2010,havlinbook}.
In Ref.~\cite{radicchi}, the Authors focused on multiplexes composed
of 2 layers A and B and recovered, by solving a minimization problem, the same result as in Ref.~\cite{GoDi13} for the behavior of $\lambda_2$ for small and large values of $D_x$.
Then, they analyzed the evolution with $D_x$ of the components of the vector ${\bf v_2}$ focusing on the difference between its first $N$ components (those corresponding to the first layer $A$), ${\bf v_2^A}$, and the remaining $N$
(those corresponding to the second layer $B$), ${\bf v_2^B}$. They observed, as in Ref.~\cite{GoDi13},
that for $D_x\leq (D_x)_c$ the components of ${\bf v_2}$ are of the
form ${\bf v_2^A}=-{\bf v_2^B}$ (see panel a) of
Fig.~\ref{fig:structuraltransition}), each one of them having all the $N$ entries identical.
On the other hand, when $D_x\gg 1$, the components of each subvector have the same sign (see panel e) of Fig.~\ref{fig:structuraltransition}). Thus, in this limit,
the partition of the multiplex into two layers is no longer possible. Furthermore, they  found an upper bound for the value
$(D_x)_c$ as:
\begin{equation}
(D_x)_c \leq \frac{1}{4}\lambda_2({\bf L^A}+{\bf L^B})\;,
\end{equation}
where ${\bf L^A}$ and ${\bf L^B}$ are the Laplacian matrices of layers $A$ and $B$, respectively.
The structural phase transition showing the exact value of $(D_x)_c$ is apparent in panel c) of Fig.~\ref{fig:structuraltransition} when measuring the evolution of the projection of ${\bf v_2^B}$ into ${\bf v_2^A}$ as function of $D_x$.

\subsubsection{Random walks}

\begin{figure}[t!]
 \centering
   \includegraphics[width=0.8\textwidth]{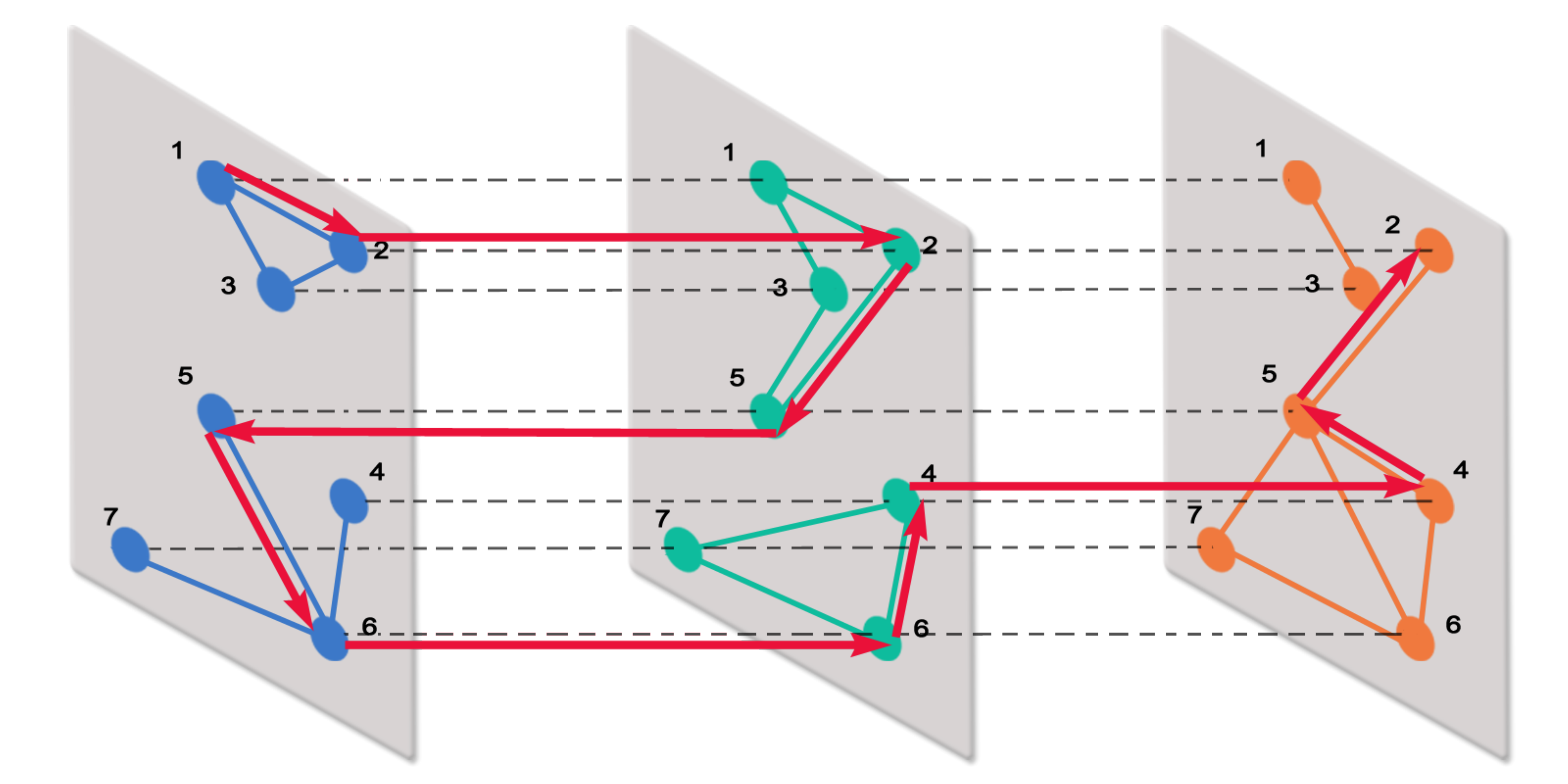}
 \caption{(Color online). Path (sequence of arrows) of a random walker navigating a multiplex network composed
of $N=7$ nodes and $M=3$ layers. In this example, the walker is neither allowed to switch between layer 1 and layer 3
in one time step nor to change node and layer
simultaneously. Reprinted figure from Ref.~\cite{Domenico13}. Courtesy
of A. Arenas.}
    \label{fig:rwpath}
\end{figure}

Although linear diffusion provides already a deep understanding of the differences between the behavior of single and multilayer
networks, the connection with diffusion phenomena on real systems is not so straightforward due to the discrete nature of flows in
real communication and transport systems. In Ref.~\cite{Domenico13} De Domenico et al. generalized random walks in monolayer
graphs to the case of multiplex networks. Figure~\ref{fig:rwpath} is an illustration of a path of a random walker (RW) exploring a multiplex network composed of $N=7$ nodes and $M=3$ layers, in which intralayer and interlayer transitions are shown.

In Ref.~\cite{Domenico13} a general form for the evolution of the occupation probabilities of a RW in a multiplex of $M$ layers and $N$ nodes is written as
\begin{equation}
\label{eq:rwstep}
p_{i,\alpha}(t+1)=\Pi_{ii}^{\alpha\alpha}p_{i,\alpha}(t)
+\sum_{\substack{\beta=1\\\beta\neq\alpha}}^{M} \Pi_{ii}^{\beta\alpha}p_{i,\beta}(t)
+\sum_{\substack{j=1\\ j\neq i}}^{N} \Pi_{ji}^{\alpha\alpha}p_{j,\alpha}(t)
+\sum_{\substack{\beta=1\\\beta\neq\alpha}}^{M}\sum_{\substack{j=1\\ j\neq i}}^{N}\Pi_{ji}^{\beta\alpha}p_{j,\beta}(t).
\end{equation}
where $p_{i,\alpha}(t)$ is the probability of finding the walker at time $t$ in the representation of node $i$ in layer $\alpha$. The transition probabilities are encoded in ${\bf \Pi}$ so that $\Pi_{ii}^{\alpha\alpha}$ is the probability that the walker does not move,
$\Pi_{ii}^{\alpha\beta}$ is the probability that a walker changes from layer $\alpha$ to $\beta$ staying in the same node $i$, $\Pi_{ij}^{\alpha\alpha}$ is the probability that the walker hops from node $i$ to $j$ within the same layer $\alpha$,
and $\Pi_{ij}^{\alpha\beta}$ is the probability that the walker changes from node $i$ to node $j$, and also from layer $\alpha$ to $\beta$ (this last movement being not actually represented in Fig.~\ref{fig:rwpath}).

By setting the values of each transition probability, the Authors of Ref.~\cite{Domenico13} studied different kinds of RWs.
As in the case of single-layer networks, the mathematical formulation of RWs makes use of the adjacency matrix,
${\cal A}$, of the multiplex (also called supra-adjacency matrix) in order to write the expressions for the transition
probabilities between nodes and layers. The matrix is of the form
\begin{equation}
\mathcal{A}=\left(
\begin{array}{cccc}
  \mathbf{W^{1}} & D_{12}\mathbf{I} & \dots & D_{1M}\mathbf{I} \\
  D_{21}\mathbf{I} & \mathbf{W^{2}} & \dots & D_{2M}\mathbf{I} \\
  \vdots & &\ddots &  \vdots\\
  D_{M1}\mathbf{I} &   D_{M2}\mathbf{I} & \dots & \mathbf{W^{M}} \\
\end{array}
\right)\;,
\label{eq:supradj}
\end{equation}
where ${\bf W^{\alpha}}$ is the adjacency matrix of layer $\alpha$ with elements $\{w_{ij}^{\alpha}\}$. Then, the generalization of the classical RW to a multiplex is obtained by setting
\begin{align}
\Pi_{ii}^{\alpha\alpha}&=(1-q)\;,\\
\Pi_{ii}^{\alpha\beta}&=q\frac{D_{\alpha\beta}}{s_{i}^{\alpha}+S_{i}^{\alpha}}\;,\\
\Pi_{ii}^{\alpha\beta}&=q\frac{w_{ij}^{\alpha}}{s_{i}^{\alpha}+S_{i}^{\alpha}}\;,\\
\Pi_{ii}^{\alpha\beta}&=0\;.
\end{align}
In the above expressions $S_{i}^{\alpha}=\sum_{\beta=1}{M}
D_{\alpha\beta}$ is the strength of the interlayer coupling of the
representation of node $i$ in layer $\alpha$, and $q$ is the hopping
rate such that $1-q$ accounts for the
possibility of the random walker to wait in a given vertex and in a
certain layer.
In fact, considering the typical assumption in classical RWs, one can set $q=1$ so that
$\Pi_{ii}^{\alpha\alpha}=0$ (i.e. the walkers never remain in the same position) and, following again the simplification done for linear diffusion, one can further assume that $D_{\alpha\beta}=D_x$ $\forall \alpha$, $\beta$.

With the help of these assumptions, one can study the diffusion
properties of RWs.
In Ref.~\cite{Domenico13},
the coverage of the walks (measuring the number of nodes visited at least once in a given time window $\tau$)
is studied in detail for multiplexes composed of two layers.
The main result is that for large enough values of $D_x$, the time to
cover the multiplex is smaller than the whole required to explore each
one of the layers separately.

\smallskip

Another important measure (when dealing with RWs) is the concept of {\em communicability} \cite{comm1,comm2},
that quantifies the number of possible paths communicating two given nodes of a network.
In the same framework of classical RWs in multiplex, communicability has also been extended in Ref.~\cite{communicability}
to account for the all the walks between any pair of nodes in a multiplex.
Indeed, considering the supra-adjacency matrix of Eq.~(\ref{eq:supradj}), the number of walks of length $L$ between the representation
of a node $i$ in layer $\alpha$ and the representation of node $j$ in layer $\beta$ is given by different entries of the $L^{\mathrm{th}}$
power of the supra-adjacency matrix, ${\cal A}^L$. Considering the definition of communicability in single-layer networks,
one is interested in assigning more importance to shorter walks than to long ones.
In this way, one defines the communicability matrix ${\bf G}$ as
\begin{equation}
{\bf G} = {\bf I} + {\cal A} + \frac{{\cal A}^2}{2!} + \cdots = \sum^\infty_{L=0} \, \frac{{\cal A}^L}{L!}=\exp({\cal A})\,.
\end{equation}
One can further express the communicability matrix as
{\begin{equation}
{\bf G} =\exp({\cal A}) =
\left(
\begin{array}{cccc}
  \mathbf{G^{1}} & {\bf G^{12}} & \dots & {\bf G^{1M}} \\
  {\bf G^{21}} & \mathbf{G^{2}} & \dots & {\bf G^{2M}} \\
  \vdots & &\ddots &  \vdots\\
  {\bf G^{M1}} &   {\bf G^{M2}} & \dots & {\bf G^{{M}}} \\
\end{array}
\right)\;,
\end{equation}}
where ${\bf G^{\alpha}}$  is a $N\times N$ matrix containing the
communicability between the representations of every pair of nodes
within layer $\alpha$  of the multiplex. It is important to note that
${\bf G}^{\alpha} \ne \exp ({\bf W^\alpha})$, since ${\bf G^{\alpha}}$
takes into account those paths that,  while connecting the
representations of two nodes within the same layer, can include hops
to any other layer different from $\alpha$ due to the coupling
$D_x$. Obviously, if $D_x=0$, then one has
\begin{equation}
{\bf G} =
\left(
\begin{array}{cccc}
   \exp ({\bf W^1}) & 0 & \dots & 0 \\
   0 & \exp ({\bf W^2}) & \dots & 0 \\
  \vdots & &\ddots &  \vdots\\
  0 & 0 & \dots & \exp ({\bf W^M}) \\
\end{array}
\right)\;.
\end{equation}
This means that communicability naturally accounts for the fact that, in a multiplex, information between the representation of two nodes within the same layer flows actually in parallel at the different layers.

\subsubsection{Diffusion-based centrality measures}

Another relevant aspect of diffusive processes is the evaluation of the importance that the nodes of a network have on its global organization. To this aim, RWs have intensively been used to assess the importance of nodes based on their occupation number, i.e. the probability of finding a walker in each of the nodes in the steady state of the dynamics.
In Ref.~\cite{Domenico13}, the Authors study the set of occupation numbers in the steady state of the dynamics,
$\{ p_{i,\alpha}^*\}$, for the general family of walks described in Eq.~(\ref{eq:rwstep}),
including the classical one, by assigning different values for the transition probabilities included in ${\bf \Pi}$.
In Ref.~\cite{manliocent}, the Authors focused on the generalization of the classical RWs (with $q=1$).
Since the occupation probabilities in a multiplex refer to each representation of a given node $i$ on a given layer
$\alpha$, $p_{i,\alpha}^*$, in order to evaluate the centrality of a given node $i$ one must aggregate the occupation numbers
for each of its representations in all the layers
\begin{equation}
p_{i}^*=\sum_{\alpha=1}^M p_{i,\alpha}^*\;,
\end{equation}
so that $p_{i}^*$ accounts for the probability of finding the walker in node $i$, regardless of the layer.

\smallskip

One of the most successful applications of RWs to derive centrality measures is the well-known PageRank centrality method
\cite{marchiori,BrinPage1998}, which is at the core of the most widely-used search engine of the World-Wide-Web, {\em Google}.
PageRank is based on a diffusion process that mimics the navigation through webpages as the motion of a random walker
following hyperlink pathways. Whereas these link-by-link hops are followed most of the times, there is a small probability $r$
that at a given time step the walker performs a long distance hop to an arbitrary node of the system.
In this case, the transitions encoded in ${\bf \Pi}$ should be reformulated in the following way
${\bf \hat{\Pi}}=r{\bf{\Pi}}+(1-r){\bf H}\;,$
with $r\in[0,1]$, where $(1-r)$ is the probability of performing a long distance hop and the elements of ${\bf H}$ are defined as $H_{ij}^{\alpha\beta}=1/(M N)$.

In Ref.~\cite{Halu13}, Halu et al. proposed a different version of the PageRank dynamics that is not as general as the above
in the sense that there is a privileged layer that influence the remaining layers of the multiplex.
In the case of a $2$-layer multiplex, the nodes in the first one are characterized by means of the usual PageRank on single-layer
networks, whereas the transitions between adjacent nodes of the second layer are biased by the centrality of their counterparts
in the first layer.

Finally, in Ref.~\cite{manliocent} the Authors use the formalism provided by RWs in a multiplex to compute other centrality
measures not directly computed from the occupation numbers of the nodes, but with the number of paths that cross them.
Examples are betweenness and closeness centralities, which in principle are designed to account for shortest paths, but that can be extended to RW paths.

\subsubsection{Routing and congestion phenomena}

The processes described in this section are  proxies of real
diffusion dynamics in transportation and information interconnected networks. In
fact, most of the data and human flows in such systems are far from random, and
most of the times they follow paths that minimize a distance or certain other cost functions.
This is the case of shortest-path diffusion through which the agents (such as data
packets or humans) go from an origin node to a destination one following the
path that minimizes the number of nodes visited along their trip (in the case of
unweighted networks), or the sum of the inverse of the  weights associated to each link contained
in it (for weighted ones). In this context, the problem of congestion arises when
the so-called capacity of nodes is limited, i.e. the number of packets
that a node can send to its neighbors per unit time is bounded. However, in a
time step, it may occur that a node receives more packets than its capacity.
When this event happens for a macroscopic part of the system, congestion shows
up.

The usual set up for studying congestion is a situation  in which a number of
packets, $pN$, is injected in the network per unit time, being the origin
and destination nodes of these packets randomly assigned, and $p$ the
probability of an information packet is created. Packets are delivered
from the origin to their destination (where they disappear) by hopping between
adjacent nodes. As congestion results in the unbalance between the number of
packets introduced in the network and that of delivered ones, the following
order parameter \cite{alexcong1,alexcong2}

\begin{equation}
\rho=\lim_{t\rightarrow\infty}\frac{A(t+T)-A(t)}{p N T}
\end{equation}
is used to monitor the transition to congestion. In the latter equation,
$A(t)$ is the number of packets that the network is handling at some time $t$, and thus
$\rho$ measures the normalized ($0\geq \rho\geq 1$) increase of packets per
unit time. Many works in single-layer networks (see for instance
Refs.~\cite{alexcong1,alexcong2,tadic,toroczkai2004,echenique1,echenique}) have
studied the transition from the free-flow regime ($\rho=0$) to the congested one
($\rho>0$) at some critical load of the network $p_c$. Many different ways of
delivering the packets on top of a variety of network topologies have been
introduced in order to delay the onset of congestion.

\begin{figure}
\begin{center}
\includegraphics[width=0.8\textwidth]{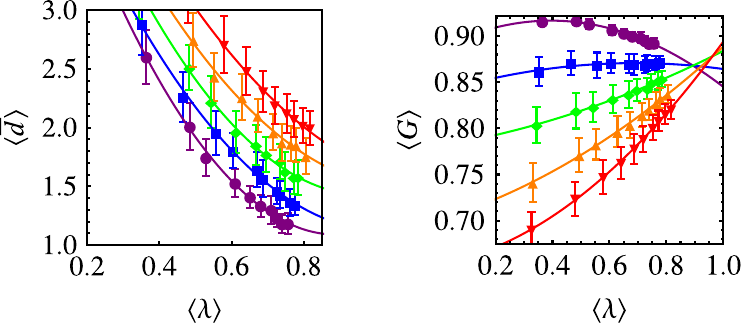}
\end{center}
\caption{(Color online). Evolution of the the average Gene coefficient $\avg{G}$ of
the distribution of paths across the edges (left panel) and the average distance $\avg{\bar{d}}$
between nodes (right panel) as functions of the average value of the interlayer coupling
$\avg{\lambda}$ in the model introduced in Ref.~\cite{MorBar12}.
The different curves correspond to different organization of the
origin-destination matrix $T_{ij}$. This organization is initially centralized
(one node is the unique destination of all the packets) and then randomized with
some probability $p$. Reported data correspond to: $p=0$ (purple dots), $0.2$ (blue squares),
$0.4$ (green diamonds), $0.6$ (orange triangles), and $0.8$ (red
inverted triangles).  Reprinted figure with permission from
   Ref.~\cite{MorBar12}. \copyright\, 2012 by the American Physical Society.}\label{fig:barthelemy}
\end{figure}

Recently, several works have addressed the problem of congestion
\cite{MorBar12,routing,zhuopa2011,tan2014} on top of interconnected
networks. In Ref.~\cite{MorBar12}, Morris and Barth\'elemy analyzed an interconnected
system formed by two layers which partially resembles a multiplex. In fact, the
second layer is composed of a subset of nodes of the first one so that each node
in the second layer is directly connected with its representation in the first
network. Origin and destinations can be only chosen in the first (large) layer
so that, given that the agents travel following the shortest path, when networks
are uncoupled the agents do not make use of the second platform. However, coupling
enters into play when the diffusion of agents between adjacent nodes is faster
in the second of the layers. In this case, agents may find short paths between
their origin and destination making use of the second layer. To account for this
coupling, the Authors measure the coupling between layers as

\begin{equation}
\lambda=\sum_{i\neq j} T_{ij}\frac{\sigma_{ij}^{coupled}}{\sigma_{ij}}\;,
\end{equation}
where $T_{ij}$ is the fraction of agents traveling from node $i$ to node $j$
(of the first layer) whereas $\sigma_{ij}$ and $\sigma^{coupled}_{ij}$ are,
respectively, the number of shortest paths between the former two nodes $i$ and
$j$, and those paths that contain edges from both network layers. Obviously
$\lambda\in[0,1]$, being $\lambda$ small (large) in the weakly (strongly) coupled
regime. In fact, as $\lambda$ increases, the average distance between two nodes
in the first layer decreases as an effect of the use of the second layer.
However, this effect comes together with the accumulation of the used paths on a
very small subset of edges which can lead to a saturation of the system due to
congestion. This latter effect is shown by measuring the Gene coefficient
$G\in[0,1]$ that accounts for the distribution of paths across the edges, so that
small (large) values of $G$ correspond to a spread (concentrated) distribution of the
flows across the edges (see Fig.~\ref{fig:barthelemy}). The Authors show that it
is possible to obtain an optimal coupling $\lambda^{\star}$ that balances the
aforementioned effect, so to obtain an optimal communication multiplex.

The congestion diagram $\rho(p)$ has been studied for usual routing protocols
implemented on top of interconnected networks in Refs.~\cite{routing,zhuopa2011,tan2014}.
As described above, congestion can be
directly connected to the limited capacity of nodes and links for handling the
flows. In these latter works, the Authors consider interconnected networks in
which some pairs of nodes of different networks are linked, thus providing the
communication between the two networks and the mix of their flows. The main
conclusion ~\cite{routing,tan2014} is that the more is the interconnection between
networks, the better is the performance of the system in terms of the critical value
$p_c$ from which the system gets congested.
Moreover, in Ref.~\cite{tan2014} the Authors show that when nodes of similar degrees
from different networks are
linked, the critical value $p_c$ raises considerably, pointing out that the
connection of the hubs of the different networks helps to alleviate the
saturation of both networks.

\subsection{Spreading processes}

Among the possible dynamical scenarios of relevance, the analysis of
the spreading of diseases and the design of contention policies have
capitalized most of the literature in the field of network science
\cite{vespbook08,Newman_SIAM,Boccaletti2006,dgm08,Dorogovtsev_epidemics,
  Boguna_epidemics}. This interest has
been rooted in the successful extension of the classical compartmental
models used in mathematical epidemiology
\cite{Maybook,hethcote,daley,Murray} beyond the mean-field equations,
so to incorporate networked architectures that capture realistic
interaction patterns. From the physical point of view, the study of
epidemic models on top of networks has allowed to extend a paradigm of
non-equilibrium phase transitions beyond lattice models \cite{Marro}.
In addition, the study of the spreading of infectious diseases on top of networks
gives also insight on other
dynamical processes of interest, such as the information and rumor spreading in social networks.
Thus, the recent interest on multilayer networks has provided a natural way to extend and improve our
understanding of the basics of spreading dynamics in real-world complex systems.

\smallskip

Let us first briefly review two typical compartmental models:
the {\em Susceptible-Infected-Susceptible}  (SIS) and the {\em Susceptible-Infected-Recovered} (SIR) models.
In both cases susceptible individuals can be infected when they are in contact with an already infected agent.
In this case, and considering the time discrete version of SIS and SIR dynamics,
there is some probability $\beta$ that the individual takes the infection.
Once infected, agents recover with some probability $\mu$ per unit time.
The difference between SIS and SIR models is the following: in the SIS model recovered individuals
become susceptible of contracting the disease, i.e. there is no acquired immunity after suffering the infection
(this setting suits with most of the typical sexually transmitted diseases).
At the opposite, in the SIR case, recovered individuals are immune to the disease, and they cannot be infected again.
From the dynamical point of view, SIS and SIR models differ in their stationary regime:
while for the SIR model the dynamics dies out once there is no infected individual in the population
(so that only susceptible and recovered agents coexist), for the SIS model sustained epidemic activity
may persist over time as a consequence of the reinfections of recovered agents.

\smallskip

In any case, the key question is whether the spread of the disease causes an epidemic scenario in which a
macroscopic part of the population is affected.
The typical control parameter in both SIR and SIS models is the ratio between the infection and recovery probabilities,
$\beta/\mu$, while the order parameter that quantifies the impact of the disease depends on the type of dynamics used.
In the case of the SIR model, the impact of the disease is given by the number of recovered individuals in the final state
of the dynamics. For the SIS model, instead, the order parameter is the number of infected agents in the sustained endemic
state, that allows to track a non-equilibrium transition between the healthy and the epidemic phases.

\subsubsection{Disease spreading on single-layer networks}

We start by briefly summarizing the basic equations that capture a disease spreading process on top of a single-layer network.
To this aim, we will focus on the SIS model, although other compartmental dynamics such as the SIR model
can be casted under the same framework, and we will review the microscopic formalism developed in Ref.~\cite{EPLnostre}.

As usual, we consider a network of $N$ nodes that, in the most general case, is weighted.
We represent the connections between the nodes by the entries $\{w_{ij}\}$ of a $N\times N$ weighted adjacency
matrix ${\bf W}$. Under the SIS framework, at each time step, an infected node makes a number $\lambda$
of trials to transmit the disease to its neighbors with probability $\beta$ per unit time.
Since this process has no memory, the probability $p_i(t)$ that node $i$ is infected at time $t$
follows a Markovian dynamics defined as
\begin{equation}
p_i(t+1) =  (1-q_i(t))(1-p_i(t)) + (1-\mu) p_i(t)\;,
\label{markov}
\end{equation}
where $q_i(t)$ is the probability that node $i$ is not being infected by any neighbor at time $t$,
\begin{equation}
q_i(t) = \prod_{j=1}^{N} (1-\beta r_{ji} p_j(t))\;,
\label{capin}
\end{equation}
and  $r_{ij}$ are the elements of a $N\times N$ matrix ${\bf R}$ representing the probability that node $i$ is
in contact with node $j$, that is
\begin{equation}
r_{ij} = 1- \left( 1- \frac{w_{ij}}{s_i} \right)^{\lambda}\;.
\label{mcon}
\end{equation}
The first term on the right hand side of Eq.~(\ref{markov}) is the probability that node $i$ is
susceptible ($1-p_i(t)$) and is infected ($1-q_i(t)$) by at least one of its neighbors, whereas the second one
is the probability that node $i$ is infected at time $t$ and does not recover.
The form of matrix ${\bf R}$ in Eq.~(\ref{mcon}) allows to capture different kinds of interactions under
the same mathematical formulation. For instance, the case $\lambda=1$ (one contact between $i$ and $j$ per unit time)
corresponds to a {\em Contact Process} (CP), so that $r_{ij} =
w_{ij}/s_i$, being $s_i$ the total strength of node $i$. Another general situation is that corresponding
to $\lambda\rightarrow\infty$,  which yields $r_{ij} = a_{ij}$, leading to a {\em Reactive Process} (RP),
regardless of whether the network is weighted or not.
The latter limit is what is typically adopted for the SIS dynamics.

\smallskip

The phase diagram of the SIS model is then obtained by searching for the stationary solution
$\{p_i^*\}$ of Eq.~(\ref{markov}), that results in the following equation
\begin{equation}
p^*_{i} = (1-q^*_{i}) + (1-\mu) p^*_{i} q^*_{i}\;,
\label{pista}
\end{equation}
and computing the usual order parameter of the SIS model as
\begin{equation}
\rho = \frac{1}{N}   \sum_{i=1}^{N} p^*_{i}\;.
\end{equation}
Equation~(\ref{pista}) has always the trivial solution $p_i=0$, $\forall i=1,\ldots,N$
(corresponding to $\rho=0$). However, other solutions corresponding to $\rho>0$ are possible, and
they represent the existence of an epidemic state. The relevant question is thus what is the
critical value of the control parameter $\beta/\mu$ for which $\rho>0$
is a solution of the SIS dynamics.

\smallskip

The calculation of this critical point is performed by considering that, when the value $\beta/\mu$
is close to the critical one the probabilities $p_i \approx \epsilon_i$ ($0<\epsilon_i \ll 1$).
The substitution in Eq.~(\ref{capin}) and Eq.~(\ref{pista}) yields
\begin{align}
q_{i} &\approx 1 - \beta \sum_{j=1}^{N} r_{ji} \epsilon_{j}\;,
\label{qiap}
\\
0&=\sum_{j=1}^{N} \left( r_{ji} - \frac{\mu}{\beta} \delta_{ji} \right ) \epsilon_j \;,
\label{final}
\end{align}
where $\delta_{ij}=1$ for $i=j$, and $\delta_{ij}=0$ otherwise. The system of Eqs.~(\ref{final}) shows that
a non-trivial solution is possible when $\mu/\beta$ is an eigenvalue of the matrix ${\bf R}$.
Thus, the problem of finding the epidemic threshold (the minimum value of $\beta/\mu$ giving rise to an epidemic)
is tantamount to computing the spectra of the contact matrix ${\bf R}$. In particular, since we are interested in
finding the smallest as possible value of $\beta/\mu$, we are interested in the largest eigenvalue,
$\Lambda=\max(\lambda_1,\lambda_2,\ldots,\lambda_N)$, of the matrix ${\bf R}$.
For a given value of $\mu$, the critical value for the probability of infection, $\beta_c$,
that represents the onset of the epidemic phase, is
\begin{equation}
\beta_c=\frac{\mu}{\Lambda({\bf R})}\;.
\label{betac}
\end{equation}

\smallskip

It is worth analyzing the two limiting cases of CP and RP. In the first case, one obtains the trivial
result that the only non-zero solution corresponds to $\beta_c=\mu$, because the matrix ${\bf R}$ is a
transition matrix whose maximum eigenvalue is always $\Lambda=1$. For the RP corresponding to the
SIS spreading process, instead,  the largest eigenvalue must be calculated  \cite{chungPNAS}.
 We observe here that the determination of the epidemic threshold for
 single-layer networks is a subject of active debate 
\cite{Dorogovtsev_epidemics, Boguna_epidemics}. Indeed, there is one
potential problem in the derivation of Eq.~(\ref{betac})  coming from
the fact that  the maximal eigenvector of the matrix $r_{ij}$ might be
localized on an infinitesimal fraction of nodes in the network. 
In this case, the order parameter would be non vanishing only on an
infinitesimal fraction of the network and, therefore, it is not
possible to speak of a true phase transition at $\beta_c$ 
determined by Eq.~(\ref{betac}).  Finally, for annealed networks which rewire their connections on the same time scale of the dynamical
process and keep their degree sequence constant, one can approximate
$r_{ij}$ as  $r_{ij}=\avg{a_{ij}}=\frac{k_ik_j}{\avg{k}}$. 
This matrix has maximal eigenvalue $\Lambda^{\prime}=\langle k^2 \rangle /
\langle k \rangle$, recovering for these networks \cite{Guerra2010,NPHMF} the classical result reported in
Ref.~\cite{vespiromu}.

\smallskip

The above result was originally obtained by means of the so-called heterogeneous mean-field (HMF) approximation
developed in Refs.~\cite{vespiromu,vrpre,bogupas}. The HMF assumes that the contact matrix $r_{ij}$ 
can be approximated as $r_{ij}=\avg{a_{ij}}$, and considers that all
nodes having the same degree behave equally which, in terms of the microscopic formulation, means that $p_i(t)=p_j(t)$
whenever $k_i=k_j$. Under this assumption, valid if the network is
dynamically rewiring its connections keeping the degree distribution
fixed, one ends up with a system of equations for each degree class
$k$ present in the network. 

\smallskip

By assuming $\rho_k(t)$ to be the probability that a node of degree $k$ is infected and assuming  $r_{ij}=\avg{a_{ij}}=\frac{k_ik_j}{\avg{k}}$, the HMF equations read
\begin{equation}
\rho_k(t+1)=\rho_k(t)-\mu\rho_k(t)+k\beta(1-\rho_k(t))\sum_{k^{\prime}}P(k^{\prime}|k)\rho_{k^{\prime}}(t)\;,
\label{HMF}
\end{equation}
where $P(k^{\prime}|k)$ is the conditional probability that, given a node of degree $k$, this is connected
with another one with degree $k^{\prime}$. The stationary solution, $\{\rho_k^*\}$, of the above set of equations
gives the value of the order parameter $\rho=N\sum_k P(k)\rho_k^*$.
In analogy with the microscopic formulation, one is interested in the non-trivial solutions (yielding $\rho>0$) of
\begin{equation}
\mu\rho_k^*=k\beta(1-\rho_k^*)\sum_{k^{\prime}}P(k^{\prime}|k)\rho_{k^{\prime}}^*\;.
\end{equation}
In Refs.~\cite{vespiromu,vrpre,bogupas} it is shown that the above equation has non-trivial solutions when
$\beta/\mu>\langle k\rangle/\langle k^2\rangle$ (in agreement with the result obtained via the microscopic formulation).
It is important to note that this result points to the fact that for degree-heterogeneous networks (such as SF networks)
the epidemic threshold can be extremely small, since the second moment of the degree distribution $P(k)$ becomes very large.
On its turn, the evidence of an almost absent epidemic threshold in real systems
defied the classical results on epidemiology, and demanded for a change in the usual contention
policies applied in systems in which an homogeneous structure was assumed.

\subsubsection{Disease spreading on interconnected networks}

In this context, two main kinds of multilayer networks have capitalized the attention in the recent years:
interconnected, and multiplex graphs. As far as disease spreading dynamics is concerned, both  substrates have their
own relevance, and the generalization of compartmental models such as the SIS and SIR to these
substrates has stimulated a great interest. The case of interconnected networks considers the spreading
of a disease in several network layers, each node of each network representing a different agent.
One example is the case of sexually transmitted diseases. Just think of three networks of sexual contacts:
one composed of agents sharing heterosexual relationships, and two other graphs of homosexual (gay and lesbian) contacts.
These three networks are connected through edges in which one (or both) of the connected nodes is a bisexual agent.
Obviously, the values of the parameters describing the spreading dynamics, such as the infection rates,
depend in principle on the nature of the link through which the contagion is produced.
In this setting, the key question is to relate the epidemic onset of the whole system with those corresponding
to the isolated networks.

\smallskip

Different works have addressed the above issue in both the SIS \cite{Saumell,Sahneh,VanMieghem,CoBa13} and SIR
\cite{Dickinson,HinSin13} frameworks. For the SIS case, different approaches point out
to a same central result: {\em An epidemic may appear in the interconnected system even for infection rates for which a disease would be unable to propagate in each isolated network}.
This result is obtained both via the HMF approximation \cite{Saumell}, and via the analysis of the microscopic equations and the spectral properties of the associated connectivity matrix \cite{Sahneh,VanMieghem, CoBa13}.

As pointed above, the SIS dynamics considers different parameters depending on the nature of the contacts and the agents.
Focusing on the case of two populations $A$ and $B$, $\beta^{aa}$ ($\beta^{bb}$) is the infectious rate between
individuals belonging to the same network A (B) whereas $\beta^{ab}$ ($\beta^{ba}$) is the infectious rate from a node
in network A (B) to another one in B (A). The recovery rates are also fixed to $\mu^a$ ($\mu^b$) for individuals
in population A (B). In this way, one can write the microscopic equations based on the adjacency matrix of the
whole interconnected system, which in the case of two coupled networks read as
\begin{equation}
\hat{\bf A}={\bf A}+{\bf C}=%
\begin{bmatrix}
{\bf A_{1}} & 0\\
0 & {\bf A_{2}}%
\end{bmatrix}+
\begin{bmatrix}
0 & {\bf C_{12}} \\
{\bf C_{21}} & 0%
\end{bmatrix}
\label{eq:dsic}
\end{equation}
where submatrices ${\bf A_{1}}$ (${\bf A_2}$) define the connections between nodes of network A (B), while ${\bf C_{12}}$ (${\bf C_{21}}$) define, on their turn, the connections from nodes of network A (B) to those in B (A). In Ref.~\cite{CoBa13},
the exact microscopic equations for generic SIS dynamics are written, extending Eq.~(\ref{markov}),
which in the case of the usual {\em Reactive} scenario (and for two populations) yields
\begin{equation}
\vec{p}(t+1)=(\vec{1}-\vec{p}(t))*(\vec{1}-\vec{q}(t))+(\vec{1}-\vec{\mu})*\vec{p}(t)\;.
\label{reaceq}
\end{equation}
Note that $\vec{\mu}$ is a vector whose components can take values $\mu^a$ and $\mu^b$ depending on the network in
which the corresponding agent is placed. In addition, the expression for the probability that a node $i$ is not
infected by any neighbor (in network A or B), $q_{i}(t)$, reads as
\begin{equation}
q_i(t)=\prod_{j}(1- {R}_{ij}p_{j}(t))\;,
\label{eq:q}
\end{equation}
where the reaction matrix ${\bf R}$ of the system incorporates the infection rates into the adjacency matrix of the interconnected networks:
\begin{equation}
{\bf R}=%
\begin{bmatrix}
\beta^{aa}{\bf A_{1}} & \beta^{ab}{\bf C_{12}}\\
\beta^{ba}{\bf C_{21}} & \beta^{bb}{\bf A_{2}}%
\end{bmatrix}\;.
\end{equation}
As in the case of single-layer networks, the spectral properties of this reaction matrix yield
the critical behavior of the system,  i.e. the epidemic onset of the interconnected system. As a consequence,
one can study different combinations of the transmission parameters
to assess the conditions for the emergence of an epidemic as in Ref.~\cite{Sahneh}.

\begin{figure}[t!]
\begin{center}
\includegraphics[width=0.6\textwidth]{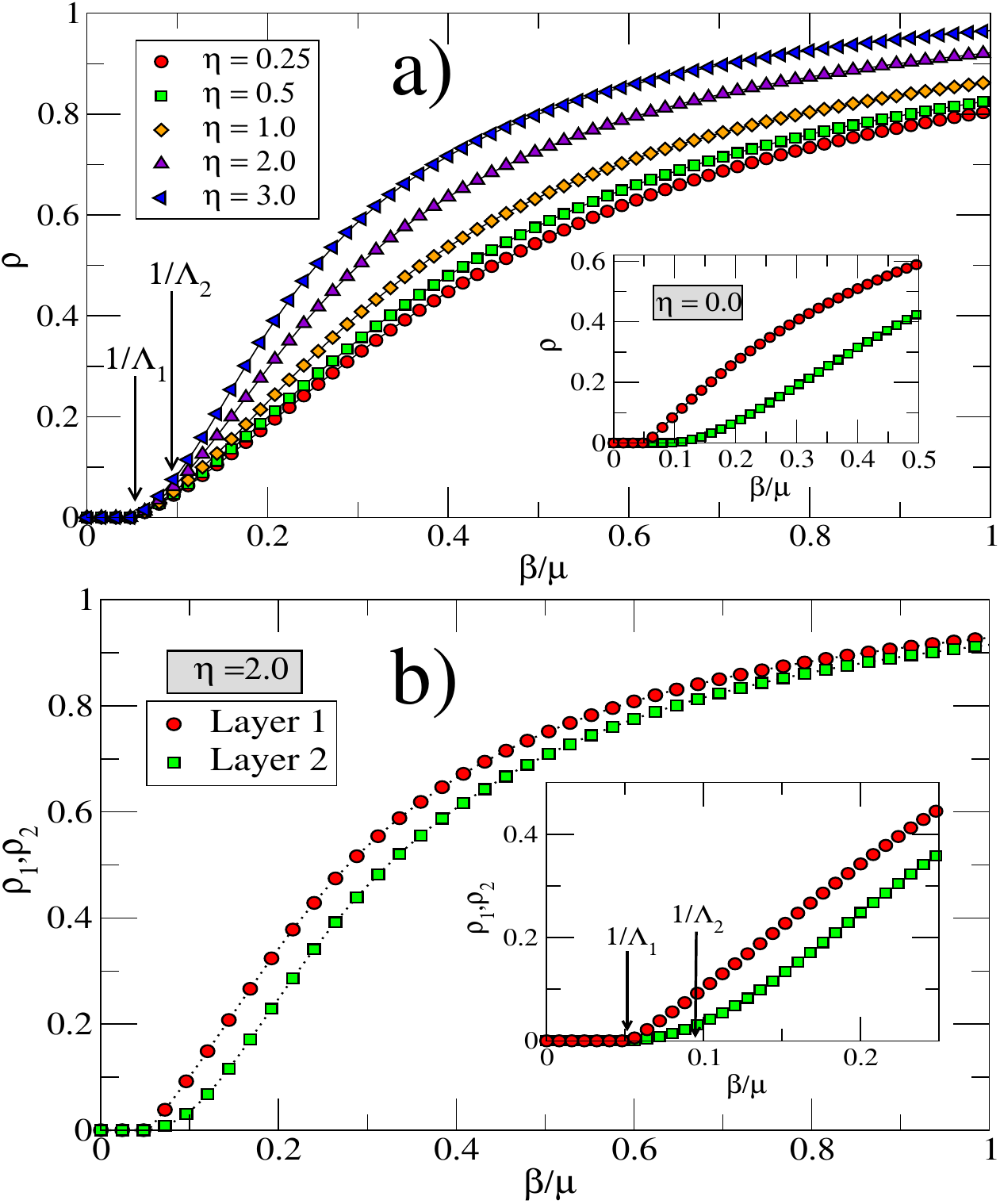}
\end{center}
\caption{(Color online). Panel (a): Fraction of infected individuals
  ($\rho$) at the steady state as a function of the control parameter
  $\frac{\beta}{\mu}$ for two interconnected networks of $N=10^4$
  nodes each for different values of the ratio
  $\eta=\frac{\delta}{\beta}$. The arrows show the inverse of the
  largest eigenvalues of the two coupled networks. The inset shows the
  case when the networks are uncoupled.  In panel (b) the fraction of
  infected individuals is shown for each of the networks and for
  $\eta=2.0$. The inset is a zoom around the critical point showing
  how the epidemic in the second network is triggered by the onset in
  the first one. Reprinted figure with permission from
   Ref.~\cite{CoBa13}. \copyright\, 2013 by the American Physical Society.}
\label{fig:SISinter}
\end{figure}

A useful insight about the behavior of the epidemic onset is achieved by treating equally the propagation
properties of the disease within each layer ($\beta^{aa}=\beta^{bb}=\beta$ and $\mu^a=\mu^b=\mu$),
and by symmetrizing the propagation between them $\beta^{ab}=\beta^{ba}=\delta$.
Consequently, one can derive interesting results based on perturbation theory \cite{CoBa13} and spectral graph theory
\cite{Sahneh,VanMieghem}. For instance, the works reported in Refs.~\cite{VanMieghem,CoBa13} show that the epidemic
threshold $\beta_c$ is given by the largest eigenvalue of a  matrix defined as
${\cal A}={\bf A}+\frac{\delta}{\beta}{\bf C}$, so that
\begin{equation}
\beta_c=\frac{\mu}{\Lambda({\cal A})}.
\end{equation}
In fact, since the spectral radius of a submatrix is always smaller or equal than that of the whole matrix,
one obtains that
\begin{equation}
\Lambda({\cal A})\geq \max{[\Lambda({\bf A_{1}}),\Lambda({\bf A_{2}})]}\;,
\end{equation}
i.e. the epidemic threshold of the whole system is always smaller than those of the isolated networks
(see Fig.~\ref{fig:SISinter}).
Moreover, in Ref.~\cite{VanMieghem} the Authors analyzed the behavior of the largest eigenvalue
$\Lambda({\cal A})$ as a function of the composing networks, and concluded that whenever two nodes from A and B
with large eigenvector components (corresponding to  $\Lambda({\bf A_{1}})$
and $\Lambda({\bf A_{2}})$, respectively) are linked,
the epidemic threshold of the entire system reduces dramatically.

The fact that the interconnected system is always more prone to the epidemic than the set of isolated networks
can be explained in terms of a perturbative analysis that assumes that $\delta\ll\beta$.
In Ref.~\cite{CoBa13}, the Authors used the former limit to explore the behavior when either
{\em (i)} there is one dominant layer (e.g. $\Lambda({\bf A_{1}})\gg\Lambda({\bf A_{2}})$) and  {\em (ii)}
both layers have roughly the same epidemic onset ($\Lambda({\bf A_{1}})\sim\Lambda({\bf A_{2}})$).
In the first case, they show that the maximum eigenvalue of the whole system is equal to that corresponding
to the dominant layer $\Lambda({\cal A})=\Lambda({\bf A_{1}})$, whereas the nontrivial stationary
solution at $\beta_c$ yields
\begin{equation}
\bigtriangleup \vec{v}= \left(
\begin{array}{c}
\vec{p}^*_1  \\
\frac{\delta}{\beta\Lambda} \vec{p}^*_{1}
\end{array}\right),
\end{equation}
where $\vec{p}^*_1 $ is the stationary solution in the dominant network at its epidemic onset.
It is clear that the onset of an epidemic in the dominant layer automatically induces the spreading in the other one.

The second limit ($\Lambda({\bf A_{1}})\sim\Lambda({\bf A_{2}})$) reveals the strong synergetic effect of the
coupling between layers. If isolated, the networks would have a roughly simultaneous epidemic onset located
at their corresponding $\beta_c$. However, the interconnection of layers makes a correction of the maximum eigenvalue as
\begin{equation}
\Lambda({\cal A})=\max[\Lambda({\bf A_{1}}),\Lambda({\bf A_{2}})]+\frac{\delta}{\beta}
\frac{\vec{p}^*_1{\bf C_{12}}\vec{p}^*_2+\vec{p}^*_2{\bf C_{21}}\vec{p}^*_1}{(\vec{p}^*_1)^T\vec{p}^*_1+(\vec{p}^*_2)^T\vec{p}^*_2}
\end{equation}
thus pointing out that the critical point of the system is smaller than any of the isolated layers, and that the
correction to its value depends on the product of eigenvector centralities between connected nodes belonging to different networks, as pointed out in Ref.~\cite{VanMieghem}.

The HMF approximation, although being a coarse-grained picture, also recovers the above result.
In this case, one can easily generalize Eq.~(\ref{HMF}) to the case of interconnected graphs.
Once again, in the case of two interconnected graphs A and B, the HMF equations in their
time-continuous version read \cite{Saumell},
\begin{equation}
\frac{d \rho^A_{\mathbf{k}_a}}{dt}=-\mu^A\rho^A_{\mathbf{k}_a} +
\beta^{aa} (1- \rho^A_{\mathbf{k}_a}) k_{aa}  \sum _{\mathbf{k}'_a} \rho^A_{\mathbf{k}'_a} P_{AA} (\mathbf{k}'_a | \mathbf{k}_a ) +\beta^{ba}(1- \rho^A_{\mathbf{k}_a})  k_{ab} \sum _{\mathbf{k}'_b} \rho^B_{\mathbf{k}'_b} P_{AB} (\mathbf{k}'_b | \mathbf{k}_a ) \;,
\label{HMFinter}
\end{equation}
for the evolution of the fraction of infected individuals in network A having $k_{aa}$ neighbors in network A
and $k_{ab}$ neighbors in network B [the evolution for $\rho^B_{\mathbf{k}_b}$ is analogous to Eq.~(\ref{HMFinter})].
Here the degree classes in each network, say A, is characterized by a vector degree ${\bf k}_a\equiv(k_{aa},k_{ab}$),
and the conditional probabilities used in Eq.~(\ref{HMFinter}) make use of this vectorial classification of nodes.
As in the case of the HMF equations in single-layer networks, in order
to derive the epidemic onset, one is interested in the limit where the densities of infected individuals are small, and therefore one
works with the linear equation
\begin{equation}
\frac{d \vec{\rho} }{dt}=-\vec{\rho}+\mathbf{C} \vec{\rho},
\end{equation}
where $\vec{\rho}\equiv (\rho^A_{\mathbf{k}_a},\rho^B_{\mathbf{k}_b})$, and
\begin{equation}
\mathbf{C}=
\left(
\begin{array}{cc}
\beta^{aa} k_{aa} P_{AA} (\mathbf{k}'_a | \mathbf{k}_a ) &
\beta^{ba} k_{ab} P_{AB} (\mathbf{k}'_b | \mathbf{k}_a ) \\[0.5cm]
\beta^{ab} k_{ba} P_{BA} (\mathbf{k}'_a | \mathbf{k}_b ) &
\beta^{bb} k_{bb} P_{BB} (\mathbf{k}'_b | \mathbf{k}_b )
\end{array}
\right).
\end{equation}
Obviously, whenever the maximum eigenvalue of matrix $\mathbf{C}$ satisfies $\Lambda(\mathbf{C})<1$,
any initial fraction of infected individuals decays to the trivial solution $\vec{\rho}=\vec{0}$.
As a result, the critical epidemic point is defined by $\Lambda(\mathbf{C})=1$.
The analysis of this condition in uncorrelated networks was carried out in Ref.~\cite{Saumell},
yielding that $\Lambda(\mathbf{C})\geq\max[\Lambda(A),\Lambda(B),\Lambda(AB)]$,
 being $\lambda(A)$ and $\Lambda(B)$ the maximum eigenvalues of the isolated networks, and $\Lambda(AB)$
the one corresponding to that of the bipartite network obtained by considering only the links between nodes
belonging to different networks. Once again, the HMF approximation allows to conclude that the coupled system can
sustain an epidemic even when the isolated networks would be free of disease.

\smallskip

For the SIR model  the treatment is quite similar \cite{Dickinson,HinSin13}. However, in Ref.~\cite{Dickinson},
based on the analogy between bond percolation and the SIR model, the Authors unveiled one important difference with
respect to the case of the SIS dynamics. Whereas for strongly coupled networks the value of the epidemic onset
of the whole system is smaller than that of the isolated networks (as in the SIS model),
when the coupling is weak (i.e. the internetwork links are much less than those connecting individuals
within the same network), the system can sustain an epidemic state in one network while not affecting the remaining ones.
These mixed states point out that for the SIR dynamics there is a critical internetwork coupling,
below which the network behaves quite similarly to a set of unconnected networks.

\subsubsection{Disease spreading on multiplex networks}

As pointed out many times already, multiplex networks are,
from the structural point of view, just particular cases of multilayer networks in which each node
is represented in each of the layers (so that the layers have the same number $N$ of nodes).
In addition, each layer is connected to another one via $N$ connections resulting from the
links that each node establishes with its representations in the remaining layers.
Given that the results shown for the disease spreading on top of an interconnected network are rooted in the spectral
properties of its contact matrix, the case of a single disease spreading on top of a multiplex network can be analyzed
in the same way by considering that the ($M N\times M N$) adjacency matrix is of the particular form
\begin{equation}
\label{eq:multiplexA}
\hat{\bf A}={\bf A}+{\bf C}=%
\begin{bmatrix}
{\bf A_{1}} & 0 & \cdots & \cdots & 0\\
0 & {\bf A_{2}} & 0 & \cdots & 0 \\%
\cdots & \cdots & \cdots & \cdots & \cdots \\%
0 & \cdots & \cdots & \cdots & {\bf A_M}
\end{bmatrix}+
\begin{bmatrix}
0 & {\bf I} & \cdots & \cdots & {\bf I}\\
{\bf I} & 0 & {\bf I} & \cdots & {\bf I}\\
\cdots & \cdots & \cdots & \cdots & \cdots \\
{\bf I} & \cdots & \cdots & {\bf I} & 0
\end{bmatrix}\;.
\end{equation}
However, a multiplex automatically sets the constraint that when a node gets infected by the disease
in one layer, and therefore it  becomes a spreader of the disease in any of the other layers.
This ingredient can only be approached (in the interdependent framework)
by setting $\beta^{ab}=\beta^{ba}=1$ and $\beta^{aa},\beta^{bb},\mu^a,\mu^b\ll 1$,
which is a far different case with respect to the weakly coupled regime usually studied in interdependent networks.

\smallskip

In fact, in Ref.~\cite{Cuenda} the Authors analyzed the SIS model in a multiplex in a similar way to
Ref.~\cite{CoBa13},  i.e. by writing the microscopic dynamics.
In this case, the expression for the probability that a node $i$ is not infected by any neighbor (in network A or B),
$q_{i}(t)$, reads
\begin{equation}
q_i(t)=\prod_{\alpha=1}^M\prod_{j=1}^N(1- \beta^{\alpha}(A_{\alpha})_{ij}\cdot p_{j}(t))\;,
\label{eq:q2}
\end{equation}
where ${\bf A_{\alpha}}$ is the adjacency matrix in layer $\alpha$. Note that, at variance with Eq.~(\ref{eq:q}),
the vector containing the probability that any node is infected, $\vec{p}$, has $N$ components since now
the state of a node must be the same in any of the layers composing the multiplex.
One can note that the only difference with an epidemic spreading on the projected network (the network resulting from the sum of the layers) is that the transmission rates $\beta^{\alpha}$ are different for each of the layers.
The analysis of the microscopic dynamics relates the epidemic onset with the maximum eigenvalue of the matrix, defined as
\begin{equation}
R_{ij}=1-\prod_{\alpha=1}^M(1-\beta^{\alpha}(A_{\alpha})_{ij})\;.
\end{equation}
Again, it is shown that the epidemic onset is always smaller than those of the isolated networks.

\begin{figure}[t!]
\begin{center}
\includegraphics[width=0.7\textwidth]{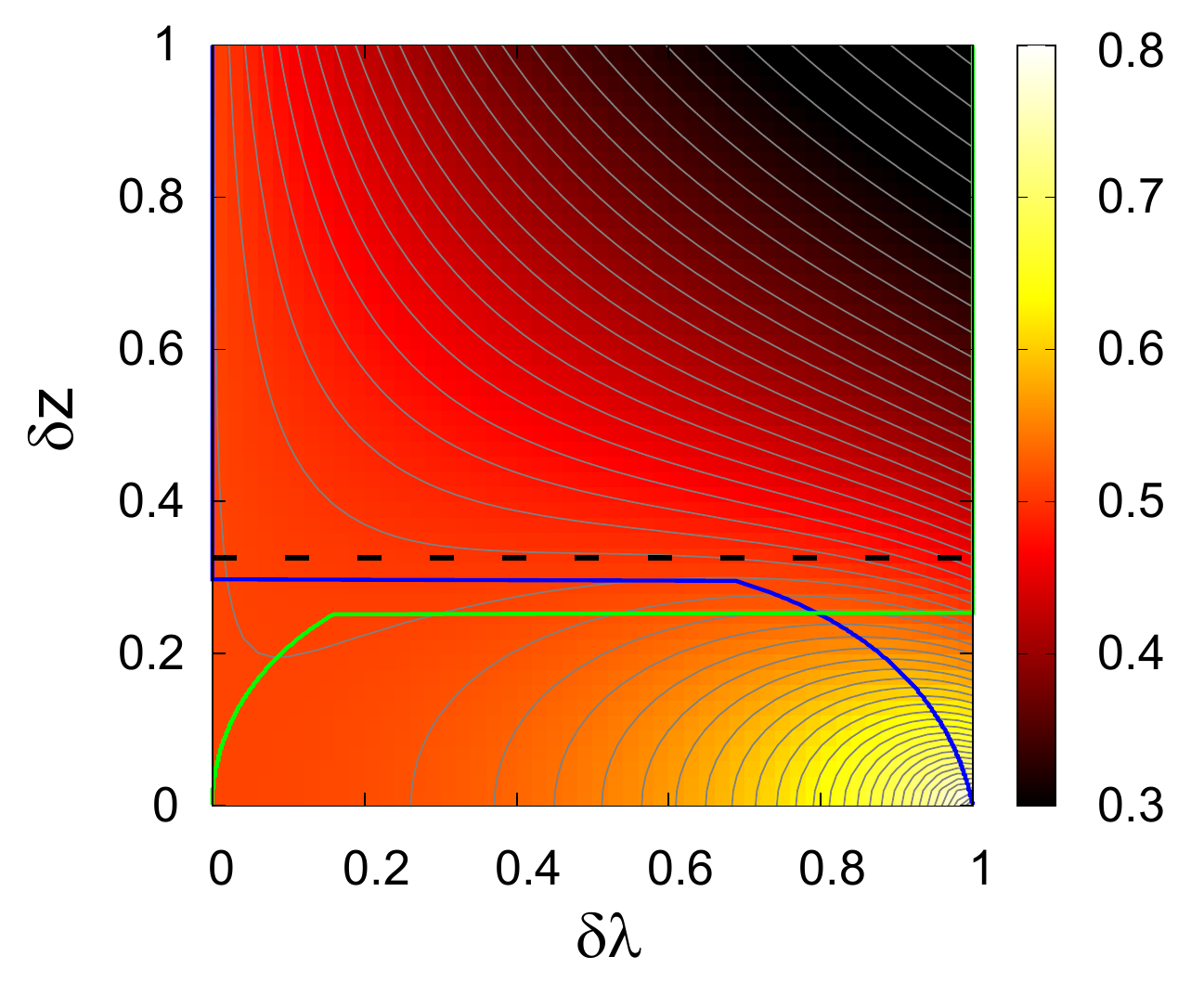}
\end{center}
\caption{(Color online). The contour plot shows the epidemic threshold $\beta_c$ as a function of
$\delta z$ and $\delta\lambda$ for a two layer multiplex composed of two ER networks.
The dashed line (black) is a boundary of $\partial\beta_c/\partial\delta\lambda>0$ for all
$\delta\lambda$, and the green (blue) line indicates the lowest
(highest) value of $\beta_c$ at a given $\delta z$.
$\beta_c$ shows non-monotonic dependence on $\delta
\lambda$. Reprinted figure from Ref.~\cite{MinGoh}. Courtesy of B. Min.}
\label{fig:SIRmultiplex}
\end{figure}

The SIR model in multiplex networks has been analyzed in Refs.~\cite{MinGoh,Yagan}.
In Ref.~\cite{MinGoh}, the Authors made use of a two-layers network in which a SIR disease
switches layer at the expense of a cost.
They derived the epidemic threshold, obtaining a similar result to that of SIR dynamics in interconnected networks
\cite{Dickinson} when the coupling between networks is strong: the epidemic threshold cannot be larger than those
of the isolated networks. Obviously, multiplexity incorporates naturally the constraint that the network layers
are far from the weak coupling regime in which mixed states appeared.
In this work, the Authors derived an interesting result concerning the behavior of $\beta_c$
in a multiplex composed of two ER graphs as a function of the difference, $\delta\lambda$,
between the intralayer and interlayer transmission rates and the difference, $\delta z$,
between the average degree of both layers. It was shown (see Fig.~\ref{fig:SIRmultiplex})
that the multiplex epidemic threshold $\beta_c$ grows with $\delta\lambda$
(when the intralayer transmission rate increases with respect to the interlayer one),
for low values of $\delta z$. However, this trend is reversed when $\delta z$ grows enough, so that
$\beta_c$ decreases with $\delta\lambda$.

In Ref.~\cite{Yagan}, the SIR dynamics is addressed on top of a multiplex containing one physical layer
and a second one in which only a subset of the nodes in the first layer takes part.
Importantly, the size of this subset is tunable so that they are able to derive
the conditions for the spreading of the disease as a function of the size of the second layer.

\paragraph{Competing epidemics}

The particular setting of multiplex networks provides the best suited structural backbone
for the study of the dynamical interplay between different dynamical processes taking place within the same set of nodes.
In this way, multiplexity offers the possibility of incorporating different network layers
for each dynamical process at work. This possibility leads to approach a well studied problem in single-layer networks:
the competition between the spreading of two different diseases. In Ref.~\cite{newmancomp}, Newman proposed
a model in which two consecutive SIR diseases spread within a single-layer network.
For the first disease at work, the critical properties and the diffusion of the infections remain the same, as usual.
However, the second disease finds it more difficult to spread since those agents recovered from the first of the
diseases gain immunity also to the second one. The main result is that, even for SF networks,
the epidemic threshold for the coexistence of both diseases is larger than zero.
Subsequently, other works extended the framework to the case of simultaneous spreading of diseases
\cite{Karrercomp}, or to the SIS model \cite{Wangcomp}.

\smallskip

The further extension of those works to multiplex networks is natural, and adds the possibility of
exploring the effects of having two different transmission networks
for each one of the diseases,
and thus incorporating the fact that different diseases may have different transmission channels.
Funk and Jansen were the first to address this issue in Ref.~\cite{funkcomp} by considering,
as in Ref.~\cite{newmancomp}, that the disease spreads in a consecutive way.
To this end, they considered a multiplex composed of two networks and a SIR model so that the first
spreading takes place in one of the networks leaving a set of recovered nodes.
These nodes are also set as recovered in the second layer
(as if they were fully immunized) prior to the second spreading that takes place in this second layer.
The main result is that a positive correlation between the degrees of the nodes in both layers enhances the
immunization of the network to the second spreading
(increasing the effective epidemic threshold for the coexistence of two epidemics).
This approach has been generalized in Ref.~\cite{marceaucomp} to the scenario of two SIR diseases
whose spreading takes place simultaneously in the two layers of the multiplex, allowing to study the influence of delay between two spreading processes and the effect of full and partial immunity.

\smallskip

\begin{figure}
\begin{center}
\includegraphics[width=\textwidth]{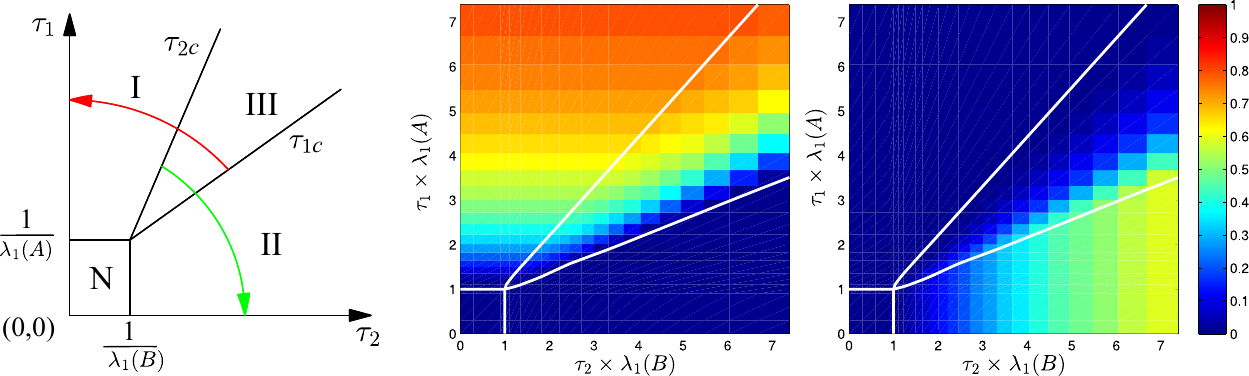}
\end{center}
\caption{(Color online). (Left panel) The survival regions diagram in the SI$_1$I$_2$S  model in two-layer multiplex. Four possible scenarios appear:
extinction region (N) where both viruses die-out, mutual extinction region I, where virus 1 survives and virus 2 dies out,
mutual extinction region II, where only virus 2 survives and virus 1 dies out, and finally the coexistence region III,
 where both viruses survive and persist in the population.
The red arrow shows the survival region of virus 1 (regions I and III) and the green
arrow shows the survival region of virus 2 (regions II and III).
(Middle and right panels) The two contour plots show the results from numerical simulations of the
SI$_1$I$_2$S model for a multiplex composed of two SF networks of $N=1000$.
The fraction of infected individuals in the steady state of the dynamics in network 1 (middle panel)
and network 2 (right panel) are shown. The white lines are theoretical threshold curves accurately separating the survival
regions as in the left panel. Reprinted figure from
Ref.~\cite{sahnehcomp}. Courtesy of C. Scoglio and F.D. Sahneh.}
\label{fig:SI1I2Scomp}
\end{figure}

The study of interacting epidemics has been also applied to the SIS model in Refs.~\cite{Wei1,Wei2,sahnehcomp}.
In fact, the extension there to multiplex networks leads to a modification of the standard SIS model
into a SI$_1$I$_2$S model with the so-called {\em mutual exclusivity}, so that when an agent is affected
by one of the diseases, ${\mbox I}_{\alpha}$, spreading in layer $\alpha$, this automatically confers immunity to the
disease running in the other network layer. Obviously, once this agent becomes susceptible again, it may be infected
by any of the two viruses. The analysis can be tackled by writing the microscopic equations of the SI$_1$I$_2$S
 defined by two adjacency matrices ${\bf A_1}$ and ${\bf A_2}$
and considering different disease parameters for each of the two spreading viruses:
$\beta_1$ ($\beta_2$) for the rate of transmission of disease $1$ ($2$) and $\mu_1$ ($\mu_2$)
for the rate of the transition from I$_1$ (I$_2$) to the susceptible state.
In Ref.~\cite{sahnehcomp}, the Authors made use of a coarse-grained microscopic approach
shown in Ref.~\cite{NINFA} for SIS, and that represents an intermediate step between the accurate microscopic
equations \cite{EPLnostre} and the HMF approximation.
The set of equations for the probabilities that agents are infected by the two viruses read as
\begin{align}
p^1_i(t+1)&=p^1_i(t)-\mu_1p^1_i(t)+\beta_1(1-p^1_i(t)-p^2_i(t))\sum_{j=1}^N (A_1)_{ij}p^1_j(t),\\
p^2_i(t+1)&=p^2_i(t)-\mu_2p^2_i(t)+\beta_2(1-p^1_i(t)-p^2_i(t))\sum_{j=1}^N (A_2)_{ij}p^2_j(t).
\label{SIScomp}
\end{align}

\smallskip

As usual, this system of equations can be linearized around the trivial solution,
$p_i^{\alpha}=0$ $\forall$ $i$,$\alpha$, and the conditions under which an epidemic in either population $1$,
population $2$ or both will hold, can be derived.
The conditions will depend on both control parameters $\tau_1=\beta_1/\mu1$ and $\tau_2=\beta_2/\mu_2$.
A trivial statement is that whenever $\tau_1<1/\Lambda({\bf A_1})$ and $\tau_2<1/\Lambda({\bf A_2})$
there is no epidemic in any of the network layers. However, a relevant scenario appears when any of
the control parameters (or both) is (are) above this threshold, since competition between the spread of the
two viruses appears. In Ref.~\cite{sahnehcomp}, the Authors performed an analytical derivation based
on perturbation theory and numerical simulations to derive the phase diagram as a function of
$\tau_1$ and $\tau_2$. Analytical and numerical results are shown in Fig.~\ref{fig:SI1I2Scomp}, highlighting
the existence of four regimes:
the free phase (N), epidemic in population $1$ and free phase in $2$ (I),
epidemic in population $2$ and free phase in $1$ (II),
and finally the coexistence of both epidemic states in layers $1$ and $2$ (III).
Interestingly, from the diagrams, it is clear that when the infection rate is large enough in one population,
say $1$, it hinders the epidemic onset in $2$, even when $\tau_2>1/\Lambda({\bf A_2})$.
A further relevant result of Ref.~\cite{sahnehcomp} is that whenever the two network layers are
identical, the coexistence region disappears.

\paragraph{Interaction between social dynamics and epidemic spreading}

As mentioned above, the multiplex structure allows to couple different dynamical processes involving
the same set of agents being, in principle, different the structural backbone on top of which each process operates.
In this way, disease spreading has been recently coupled to social dynamics \cite{granellprl},
in order to take into account how the social awareness about the spreading of a disease can influence
the impact of the latter. To this aim, the Authors coupled the SIS spreading taking place on a physical
network with another compartmental stochastic model in which the state of the agents can be of two kinds:
Aware ($A$) and Unaware ($U$). The first dynamics is completely characterized by the usual two parameters:
the transmission, $\beta$, and recovery, $\mu$, rates.
The second dynamics takes place on the second layer (a kind of virtual network in which information about
the epidemic flows) and the transition between states $A$ and $U$ is analogous to that in the SIS model:
the direct contact of an $A$ agent with another one in $U$ produces two $A$ agents with probability
$\lambda$, whereas an aware individual can become unaware with probability $\delta$.
The coupling between the two dynamical models is done by reducing the transmission rates
from infected individuals to aware and susceptible ones by a factor $\gamma$ ($\gamma=0$ means that aware
individuals become fully immunized). In addition, the state unaware-infected
is not admitted and, in a time discrete-version, it becomes aware-infected with probability $1$ (if not recovered).

\smallskip

\begin{figure}[t!]
\begin{center}
\includegraphics[width=0.7\textwidth]{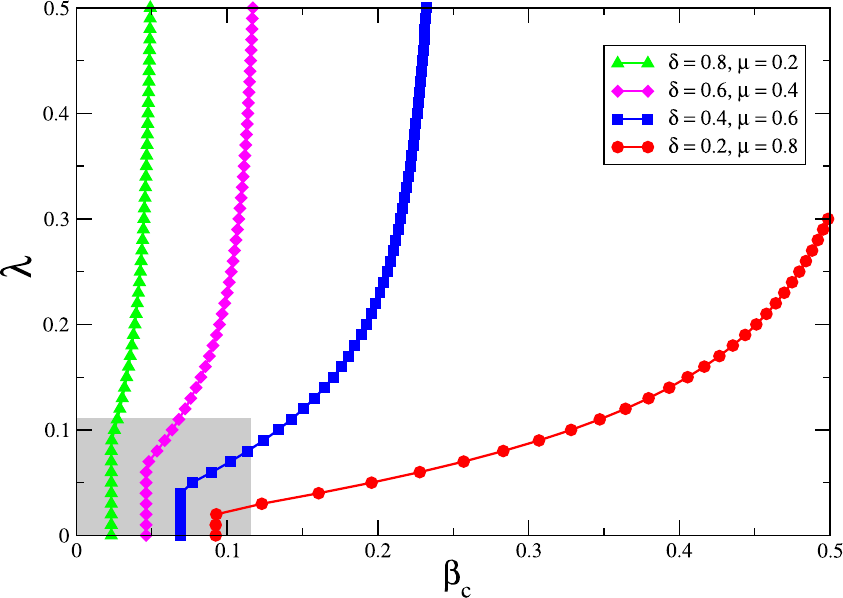}
\end{center}
\caption{(Color online). Dependence of the onset of the epidemics $\beta_c$ as a function of $\lambda$
for different values of the recovery rates $\delta$ and $\mu$ for a two-layer multiplex composed of:
{\em (i)}  a SF network of $N=1000$ and exponent $2.5$ in the physical layer, where SIS dynamics takes place, and
{\em (ii)} the same SF network of the physical layer plus 400 extra random links (non-overlapping with previous) in the virtual layer, where Aware-Unaware dynamics works.
The shaded rectangle shows the free phase (in which all individuals are Unaware and Healthy)
when the layers are uncoupled. The boundaries are defined by the bound of the structural characteristics of
each layer: $1/\Lambda({\bf A_1})$ and $1/\Lambda({\bf A_2})$. Reprinted figure with permission from
   Ref.~\cite{granellprl}. \copyright\, 2013 by the American Physical Society.}
\label{fig:granellprl}
\end{figure}

As in Ref.~\cite{sahnehcomp}, when isolated, the two dynamical processes display a critical point at some
values of the respective control parameters, $\beta/\mu$ and $\lambda/\delta$.
In particular: $\beta_c=\mu/\Lambda({\bf A_1})$ and $\lambda_c=\delta/\Lambda({\bf A_2})$,
where ${\bf A_1}$ and ${\bf A_2}$ are the adjacency matrices of the first layer where SIS and Aware-Unaware
dynamics take place respectively.
Once again, the coexistence of these processes changes the onset of the epidemic state.
In particular, when the infection rate of the SIS dynamics is large enough, the infected agents trigger
the mechanism of awareness in the virtual layer, thus lowering the incidence of the disease and increasing
the epidemic onset. This result is shown in Fig.~\ref{fig:granellprl} for several values of
$\mu$ and $\delta$. Interestingly, as the ratio $\delta/\mu$ decreases,
the value of the epidemic onset is more influenced by the Aware-Unaware dynamics.

\paragraph{{Impact of interlayer topological properties}}
Except for the influence of dynamical processes located in other layers,
the network properties between different layers also play a significant role in the diffusion of disease.
The first viewpoint comes from the investigation of coupling density, which tries to reproduce the empirical
fact that not all the nodes are present on each layer of social networks. In Ref.~\cite{Buono},
a fraction $q$ of nodes are shared by two layers enabling to form partially overlapped networks.
Since there exists the same disease in both layers, the equal transmission probability is assumed,
whereas the spreading of infection starts from one random node. The
main aim becomes here to explore the
effect of the overlapping fraction $q$ on the disease propagation with reference to the SIR dynamics model.
Based on the theoretical analysis and simulations, the Authors found that the epidemic threshold of
multiplex networks depends on both topology and the coupling density.
In the limit of $q \to 1$, the epidemic threshold is minimum, and both layers have the identical
ratio of recovered nodes at any transmission probability.
This is because nearly each node belongs to two layers at the same time.
However, with $q$ decreasing, the epidemic threshold monotonously enhances, so as to reach the
maximal value (at $q \to 0$) dominated by the isolated threshold of the layer with larger propagation ability.
This point means that, even though there exists negligible differences between the system composed
of two isolated networks and the system connected by just a few interlayer connections,
these links can make the epidemic threshold of the isolated network with lower spreading ability
discontinuously change the threshold of the other layer. This finding is shedding light
into the non-medical intervention for the control of the disease propagation.

Of particular interest is distinguishing how the disease spreads if
the nodes are coupled by a certain rule from the case of simple overlapping.
 In a recent investigation \cite{Zhao},
a two-route transmitted disease employing SIR model is introduced into the multiplex framework,
where two transmission routes take place on each layer, respectively.
Moreover, the Authors defined two
new measures to evaluate the influence of interlayer links.
The first is the degree-degree correlation (DDC) between
layers: positive (negative) values indicate that high (low)
degree nodes are inclined to have more (less)
connections within another layer.
The other quantity is the average similarity of neighbors
(ASN) from different
layers of nodes, which measures how many joint neighbors one node possesses in both layers.
The Authors found that, even if both layers are below their respective thresholds,
the disease can still propagate across the multiplex system.
If the value of DDC is larger, a lower disease threshold and a smaller outbreak size are supported.
The epidemic threshold and disease size keep both robust against the change of ASN.

\subsubsection{{Voluntary prevention measures}}

{Except for the outstanding achievements of the prediction of the outbreak threshold,
another task of utmost urgency, during the study of epidemiology, is searching for the effective
policy or strategy to control the epidemic spreading, and to warrant public health
\cite{gilligan2002ZhenABR,world2010ZhenWHO}. This point seems particularly remarkable with worldwide pandemics
of severe acute respiratory syndrome (SARS), bird flu and new H1N1 flu
\cite{fouchier2003ZhenNature,small2007ZhenPRL,fraser2009ZhenScience}. A great number of immunization scenarios,
such as targeted immunization \cite{pastor2002ZhenPRE},
ring immunization \cite{muller2000ZhenJMB}, and acquaintance immunization \cite{cohen2003ZhenPRL},
have been proposed, but the majority of them, based on the premise of compulsory requirements,
completely neglects the individuals' willingness and desires,
which instead follow  many social factors including religious beliefs,
the human rights and social total expenses as well
\cite{eguiluz2000ZhenPRL,nowak1990ZhenPsR,funk2009ZhenPNAS,fine1986ZhenOUP}.
In this sense, voluntary prevention measures, incorporating both studies of infectious disease and spontaneous human
behavior traits, become popular programs \cite{bauch2005ZhenPRSB,cneter2008ZhenAtlanta},
namely, the so-called ``disease-behavior feedback dynamics study''.

\paragraph{Voluntary vaccination on single-layer networks}

Among several existing control measures, vaccination attracts the hugest attention from theoretical
and experimental perspectives \cite{geoffard1997ZhenPrinceton,ibuka2014ZhenPlosONE,francis1997ZhenJPE,bauch2003ZhenPNAS}.
 During the seasons of disease spreading, the high risk of infection usually stimulates individual vaccination enthusiasm,
while high expenses or side effects of a vaccine may lead to the opposite case.
Thus, individuals face a dilemma: to vaccinate or not, both of which ways can be managed as two basic strategies
with respective cost functions. To resolve this dilemma, game theory has been integrated into individual
decision-making processes, where nodes hold the prevention strategy according to risk and expenses
in the epidemic campaign \cite{bauch2004ZhenPNAS}. In a recent research \cite{zhang2010ZhenNJP},
where nodes are only allowed to make rational selection (i.e. taking vaccination only if its expense is low),
the Authors found that disease outbreak can be more effectively inhibited on SF networks than on random networks.
Compared with early theoretical predictions of disease thresholds \cite{vespiromu},
this interesting outcome indicates that voluntary vaccination can effectively control the spreading of
diseases in real-world networks. Similarly, when the imitation dynamics with noise replaces the deterministic
rule in the disease-behavior study, multiple effects spontaneously arise:
for perfect vaccine, vaccination behavior of SF networks beats that of random networks,
while imperfect vaccine enables the promotion of vaccination and eradication of disease to act better on
homogeneous connectivity patterns \cite{cardillo2013ZhenPRE}. In Ref.~\cite{fu2011ZhenPRSB} it is found that the
spatial population structure shows a ``double-edged sword'' effect, which fosters vaccine uptake at the low cost
of vaccination, but causes vaccination to plummet when the cost goes beyond a certain threshold.
More specifically, if a small fraction of individuals consistently holds vaccinating in flu-like diseases,
these committed vaccinators can efficiently avoid the clustering of susceptible individuals and stimulate
other imitators to take vaccination, thus contributing to the promotion of vaccine uptake \cite{liu2012ZhenPRE}.

\paragraph{Voluntary prevention behavior on multilayer networks}

\begin{figure}
\begin{center}
\includegraphics[width=0.7\textwidth]{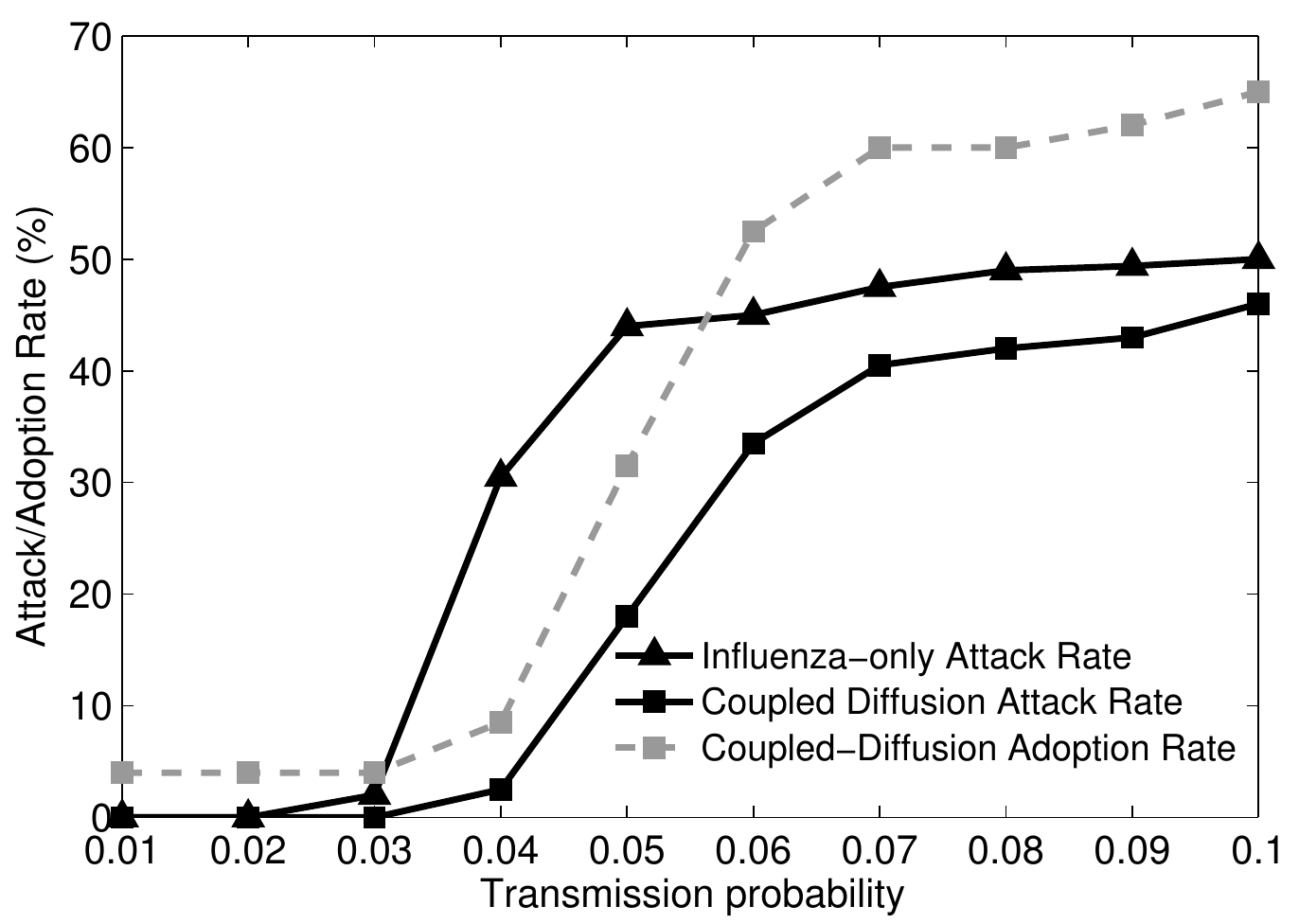}
\end{center}
\caption{Attack or adoption rate vs.  the transmission probability in disease-only diffusion processes
and coupled diffusion processes, where influenza is selected as the potential disease.
Here, the adoption rate denotes the percentage of nodes adopting prevention behavior.
Reprinted figure from Ref.~\cite{mao2012ZhenSSM}. \copyright\, 2012 with
permission from Elsevier.
}
\label{fig:vaccina2}
\end{figure}

Although both infection diseases and prevention measures diffuse simultaneously through networks and interact with each
other, few existing studies have coupled them together. Recently,
the research of disease-behavior interactions on top of multilayer structures has also received attention.
In the first conceptual work \cite{mao2012ZhenSSM}, two diffusion processes are implemented in the multiplex viewpoint,
where two networks, which are fully or partially coupled via the interlayer links,
involve the same individuals yet different intralayer connections.
Under such a framework, one network is regarded as the infection network to transmit diseases,
and the remaining communication network channels individual influence concerning preventive behavior.
Moreover, at variance from the aforementioned case where vaccination is
the principal measure on single networks,
other prevention behaviors, such as wearing facemasks, washing hands frequently, taking pharmaceutical drugs,
and avoiding contact with sick people, are considered as the main strategies, which seem particularly
reasonable before attaining effective vaccine \cite{stohr2004ZhenScience}.

Based on the online surveys, the threshold model is selected, whereas only once the proportion of
adopters among intralayer neighbors of communication networks reaches the perceived threshold of adoption
pressure (or the proportion of infected individuals exceeds the perceived risk of infection on the infection network),
the node can adopt the preventive behavior.
Surprisingly, the Authors found that, as compared to the widely used disease-only model,
the coupled diffusion model leads to a lower frequency of infection. Figure~\ref{fig:vaccina2} shows that,
as preventive behavior protects a certain people from being infected,
the threshold and peak of disease outbreak are effectively suppressed in the new embedded scenario.
Furthermore, the sensitivity analysis identifies that the structure of infection network has dramatic
influence on both epidemic and adoption percentages, while the variant of communication network produces less effects.

After such a seminal research, the role of multiplex networks in the coupled disease-behavior
system has been further investigated. Recently, a triple diffusion process,
comprising disease transmission, information flow regarding disease and preventive behaviors against disease,
was integrated into multilayer social networks \cite{mao2014ZhenAG}.
 Among the interaction of the three diffusion dynamics, the negative and positive feedback loops are formed.
Here it is worth mentioning that, as compared with Ref.\cite{granellprl} involving the information diffusion as well,
the effect of prevention behavior is of particular evidence. Under the help of agent-based method,
this new conceptual framework can reasonably replicate the observed trends of influenza infection
and online query frequency in the empirical population.

Finally, a similar scenario was proposed in Ref.~\cite{bauch2013ZhenScience},
where the social and biological dynamics processes are embedded into a multiplex viewpoint.
Under such a disease-behavior system, biological contagion spreads on one layer,
while the main control measure, vaccination, is carried out in the social network.
At variance with the case of an isolated network, the Authors suggested that the resulting
coupling can bring positive or negative impact on the vaccination coverage during a vaccine scare, i.e.
 accelerating or decelerating the elimination of disease, which can
 get support from the empirical examples.
Due to the variety of factors on the social layer,
the epidemic trajectories and the uptake of control measures can also change to a large extent.
Moreover, inspired by these findings, the Authors suggested that the coupled disease-behavior model,
which gets empirical validation, can be used as a predictive tool to explore the optimal strategies for public health.

}

\subsubsection{{Opinion formation dynamics}}

{Understanding the emergence of a macroscopically ordered state is one of the based questions in natural and social sciences.
To elucidate this issue, dozens of experimental and theoretical frameworks have been proposed
from viewpoints of statistical physics, economics and sociology
\cite{dgm08,redner2001ZhenCUP,castellano2009ZhenRMP,hegselmann2002ZhenJASSS,fowler2008ZhenJP}.
Opinion formation dynamics, as one simple and paradigmatic dynamical process,
has attracted particular attention. Up to now, various spin-like setups have been put forward to study the
evolution of opinion dynamics, which include the Sznajd model \cite{sznajd2000ZhenIJMPC},
majority rule model \cite{krapivsky2003ZhenPRL,wu2010ZhenPRE},
bounded confidence model \cite{hegselmann2002ZhenJASSS,kozma2008ZhenPRE},
and voter model \cite{holley1975ZhenAP,suchecki2005ZhenPRE}, to name but a few.
Among the existing scenarios, the voter model is attracting the
most remarkable attention \cite{redner2001ZhenCUP,sood2005ZhenPRL}, and therefore it is also
the main candidate in this subsection.

The basic definition of the voter model is rather simple: each agent (voter) is endowed with a binary variable $s=\pm1$.
 At each time step, an agent $i$ is chosen along with one of its
 neighbors $j$ and $s_{i}=s_{j}$, i.e., the agent adopts the opinion
 of the neighbor. In this sense, the transition rate of individual
 opinion is proportional to the fraction of neighboring nodes in the opposite state.
The dynamical process is repeated until the states with all sites equal (consensus) are absorbing.
In spite of being extremely simple, the voter model produces rich dynamical behaviors
\cite{shao2009ZhenPRL,suchecki2005ZhenPRE}, especially from the angles of spatial interaction.
In one-dimensional lattices, the dynamics is exactly the same as the zero-temperature Glauber dynamics
\cite{glauber2004ZhenJMP}. Turning to two-dimensional lattices, the growth of domains can makes the interfaces
quite rough, at odds with usual coarsening systems \cite{dornic2001ZhenPRL}.
 Moreover, a series of extended versions, such as the vacillating voter model \cite{lambiotte2007ZhenJSM},
nonlinear voter model \cite{molofsky1999ZhenTPB}, heterogeneous voter model \cite{masuda2010ZhenPRE}, and constrained voter model \cite{vazquez2003ZhenJPA} have also been suggested. In the following
we will present more details on the voter model.

\paragraph{Voter model on single-layer networks}
As it is well-known, the property of spatial topology usually has a
nontrivial influence on
the dynamical models.
In this sense, the ordering dynamics of voter model should not be exceptional.
If employing one complete graph, the mean consensus time $\langle T \rangle$ can be easily obtained by using the
Fokker-Planck equation \cite{slanina2003ZhenBPJB}. The consideration
of a heterogeneous connectivity can make the
mean consensus time $\avg{T}$ to become closely related with the
structural network details.
On SF networks with exponent $\gamma$, direct voter model makes $\langle T \rangle$ to scale as the size
$N$ of the network for $\gamma>3$ and sublinearly for $\gamma \le 3$
\cite{sood2005ZhenPRL,castellano2005ZhenPRE,suchecki2005ZhenEPL}. Non
trivial relationships between $\avg{T}$ and $N$ are also found in
other versions of the voter model as in Ref.~\cite{sood2008ZhenPRE}.
For small-world networks, the plateau formed by active
interfaces hinders the diffusion of opinions, so that the consensus eventually relies on the linear system size \cite{castellano2003ZhenEPL,vilone2004ZhenPRE}. Along this research line, Zschaler et al. recently examined
the effect of directed networks on the order-disorder transition \cite{zschaler2012ZhenPRE}.
The impact of other spatial mechanisms, such as social influence and heterogeneous beliefs,
are also of particular evidence \cite{latane1981ZhenAP,galam2005ZhenPRE}.

\begin{figure}[t!]
\begin{center}
\includegraphics[width=0.8\textwidth]{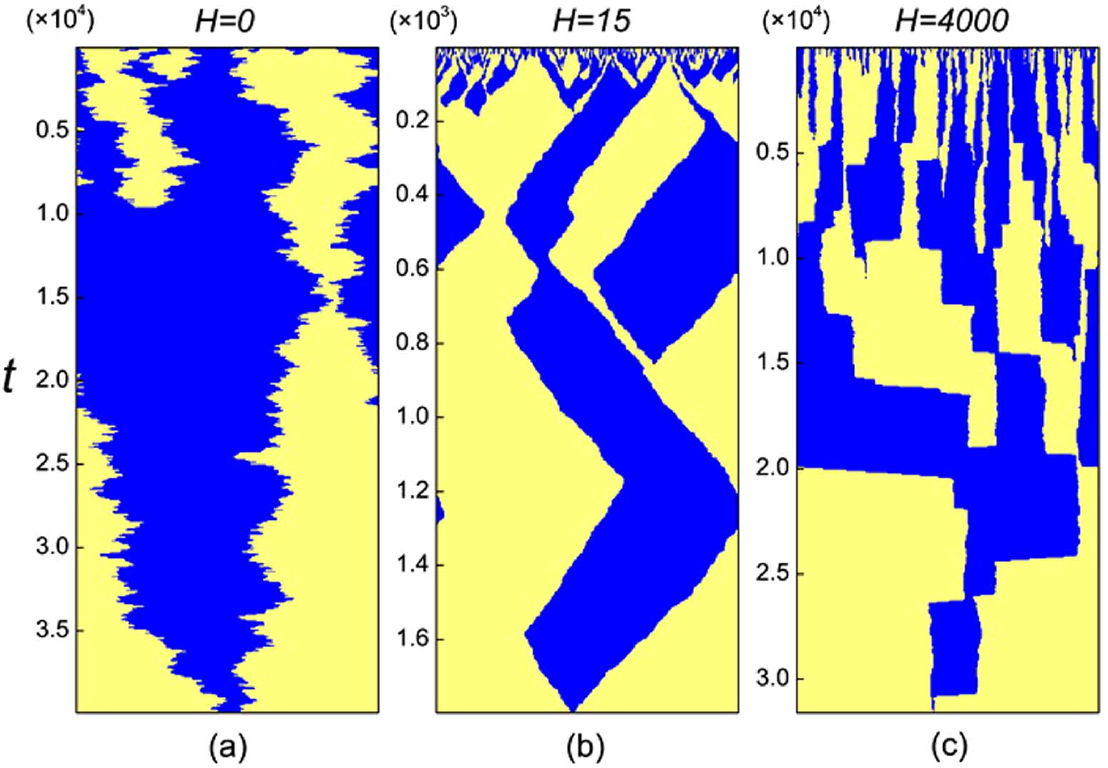}
\end{center}
\caption{(Color online). Snapshots of the evolution of opinion dynamics
with three typical freezing periods $H$. From left to right, the values of $H$ are 0, 15 and 4000.
In each panel, the vertical axis corresponds to the elapsed time, while the yellow and blue areas
on the horizontal lines represent the opinion clusters.
The total time spent to reach a global consensus, i.e. the consensus time $\langle T \rangle$,
under the scenario of intermediate value ($H=15$) is far less than two other cases.
The size of networks is $N=500$, $q=0.01$ is the switching ratio during the freezing period.
Reprinted figure from Ref.~\cite{wang2014ZhenSR}.}
\label{fig:opinion1}
\end{figure}

At variance with the aforementioned processes on static networks or in a Markovian memoryless fashion,
coevolution voter model has also been duly studied.
We first focus on non-Markovian patterns. In recent researches \cite{stark2008ZhenPRL,stark2008ZhenACS},
Stark et al. explored the effect of a memory-dependent transition rate,
which is determined by the persistence time of individual current opinion.
The longer a voter keeps its opinion, the less probable is that it switches the status in the long run.
The Authors found that the process towards reaching a macroscopic consensus was accelerated by slowing
down the microscopic dynamics. Looking at another recent report \cite{wang2014ZhenSR},
where individual ability $q$ of opinion switching is weakened during a freezing period $H$,
the Authors revealed that an intermediate freezing period can lead to the fastest consensus,  i.e. to the smallest
$\langle T \rangle$. The essence of this accelerated consensus is attributed to the biased random walk
of the interface between adjacent opinion clusters (as shown in Fig.~\ref{fig:opinion1}).
Moreover, it is worth checking how the dynamical topology affects the diffusion of opinions.
Under such a framework, an agent can adjust its connections with a probability (otherwise it updates its opinion).
The state of the system on coevolving networks can be characterized by the interface density
$\rho$ quantifying the fraction of edges linking nodes with different states (active links).
When $\rho=0$ the system is in the active phase, while for $\rho \ne 0$ it keeps the frozen phase.
Based on the mean-field approximation approach, Ref.~\cite{vazquez2008ZhenPRL} revealed that
the competition between opinion imitation and the rewiring links drives a fragmentation transition
from the active to the frozen phase.

To mimic some social phenomena and check voting phenomenology,
many modifications have also been suggested based on the original voter model.
The most renowned extension is the presence of the ``zealot'', who never changes its opinion \cite{mobilia2003ZhenPRL}.
For low-dimension networks, this type of agent can influence the
selection of all the agents, and lead to the general
consensus in the form of its own opinion. However, once higher-dimension networks are considered
it becomes rather difficult to inspire the consensus in the whole system, but the ``zealot'' just induces a
partial bias in the local neighborhood \cite{mobilia2007ZhenJSM,masuda2010ZhenPRE,mobilia2005ZhenPRE}.
 In addition, the voter model has also been applied to biological and ecological problems,
where people try to protect the diversity of species and fixation in the evolution of competing species \cite{antal2006ZhenPRL,zillio2005ZhenXiv}.

\begin{figure}
\begin{center}
\includegraphics[width=0.7\textwidth]{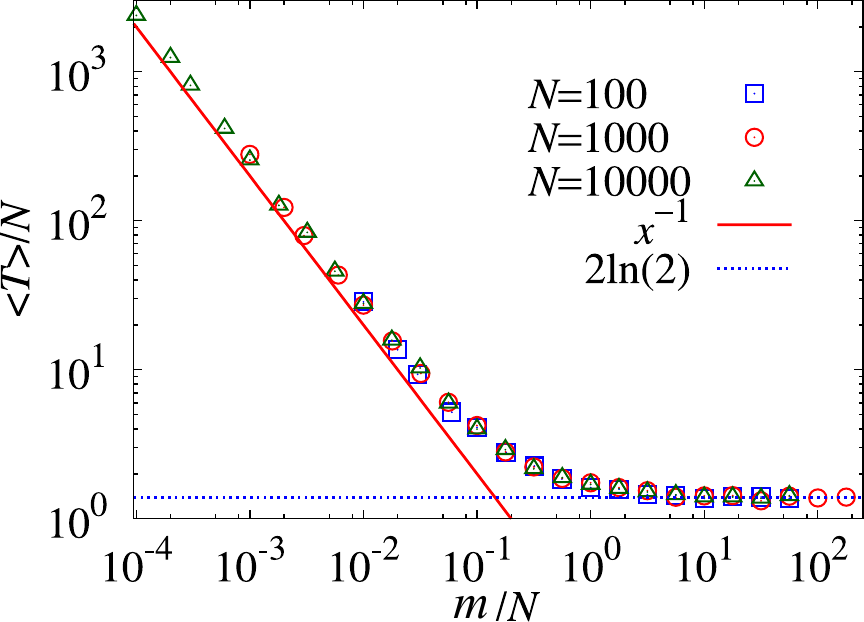}
\end{center}
\caption{(Color online) Relationship between the normalized mean consensus time, $\langle T \rangle /N$,
and the number of interlayer links per node, $m/N$, for different network sizes. It is clear that there exist
two consensus regimes. The solid line represents the relationship  $\langle T \rangle /N \propto (m/N)^{-1}$
as guides to the eye. Reprinted figure from
Ref.~\cite{masuda2014ZhenArx}. Courtesy of N. Masuda.}
\label{fig:opinion2}
\end{figure}

\paragraph{Voter model on multilayer networks}

With the fast development of network science, the research of opinion dynamics has also been extended
to multilayer networks. One potential approach is to create interlayer links,
through which opinions can be exchanged between different networks.
In a recent research \cite{masuda2014ZhenArx}, the Author considered
the case of a two-clique graph,
which is composed of 2 cliques interconnected by $m$ interlayer links.
If the size of both cliques is $N$, each node has $m/N$ interlayer links on average.
These links are either regularly placed, or distributed at full random.
At each time step, one link is randomly selected with equal probability $1/[N(N-1)+m]$,
and then the opinion imitation takes place between two endpoints of this link.
In this sense, the impact of the number of interlayer links per node becomes of particular interest.
Through strictly mathematical analysis and computational simulations, the Authors unveiled the scaling
relationship between the time to consensus and the number of interlayer links per node.
If there exist many interlayer links per node, the mean consensus time is $\langle T \rangle=O(N)$, i.e.
the system reaches the fast consensus regime, which agrees with the expectation of a complete graph
\cite{sood2008ZhenPRE}. However, once the number of interlayer links per node is much smaller than unity,
the mean consensus time is $\langle T \rangle=O(N^{2})$, i.e. the system attains the slow consensus regime.
The crossover between these two regimes takes places at approximately one interlayer link per node,
as shown in Fig.~\ref{fig:opinion2}.
Thus, these results suggest that the sparse interlayer connections between two networks extremely decelerate
the ordering state for the voter model dynamics.

In Ref.~\cite{diakonova2014ZhenWiv}, a coevolution voter model is incorporated into a multilayer framework,
where a fraction $q$ of nodes has the interlayer links across two networks.
On each layer $l$ ($l=1, 2$), individual opinion and intralayer link are allowed to evolve in accordance
with their own topological temporal scales.
Node $i$ selects one random neighbor $j$ in the same layer.
If both have different opinions, node $i$ copies the state of $j$ with probability $p_{l}$,
otherwise (with probability $1-p_{l}$) it severs the connection with $j$ and draws a new link
to one randomly chosen node in the same layer possessing a state identical to itself.
The case $p_{1}=p_{2}$ preserves the symmetric multilayer setup, namely, the time scales are the same in both layers.
At variance, for $p_{1}\ne p_{2}$, an asymmetric multilayer case takes place, namely,
different topological time scales characterize the two layers. In the
symmetric case, the Authors
found that interlayer connectivity plays a critical role in determining the state transition of systems.
Once again, the absorbing fragmentation transition appears at a certain threshold value of $p_{l}$.
Moving to the asymmetric case, the Authors unveiled an anomalous shattered fragmentation transition,
where the layer dynamics splits into many isolated components possessing different opinions, namely
an explosive growth of the number of disconnected components.
Moreover, Ref.~\cite{diakonova2014ZhenWiv} also identified the critical degree of interlayer
connectivity to stop the fragmentation of one layer by coupling it to another layer that has no fragments.

Finally, it is interesting to consider the application of opinion dynamics in the multiplex perspective.
In a recent research \cite{halu2013ZhenEPL} a simple model of opinion formation dynamics
is proposed to describe the parties competing for votes within a political election.
At variance with previous setups of voter model \cite{redner2001ZhenCUP,sood2005ZhenPRL},
each opinion or party is modeled here as one social network where the contagion dynamics takes place.
Every agent, at the election day, can choose either to be active in
only one of the two networks (vote for one party)
or to be inactive, namely, not to vote on both networks. Based on the simulated annealing algorithm,
the Authors showed a rich phase diagram.
In a wide region of the phase diagram, the competition between the two parties allows
for the emergence of pluralism, namely two antagonistic parties gather a finite share of the total votes.
The densely connected network usually wins the election campaign, regardless of the initial distribution.
Nevertheless, small perturbations could break the existing
equilibrium, and a small minority of committed agents can alter the election outcome,
which is consistent with recent claims \cite{xie2011ZhenPRE}.
}

\subsection{Evolutionary games on multilayer networks}

The study of evolutionary games on top of single-layer networks has attracted
a lot of interest as a connector between statistical physics, evolutionary dynamics and social sciences  \cite{szabo:2007,roca:2009a,jrsiPGG}.
This is a very appealing research topic since the dynamics
implemented in the network arises from an evolutionary approach \cite{nowak:2006a}.
In addition, the study of evolutionary games on networks has been one of the most studied
avenues to understand the emergence of cooperation in different contexts \cite{nowak:2006b}.
The emergence of such a collective phenomenon, cooperation, is a key issue that arises when studying seemingly diverse evolutionary puzzles, such as the origin of multicellular organisms \cite{maynard-smith:1995},
the altruistic behavior of humans and primates  \cite{kappeler:2006}, or the way
advanced animal societies, such as ant colonies, work \cite{wilson:2000,henrich:2004}, among others.

\smallskip

Previous research on evolutionary games on networks has mainly focused on social dilemmas, in which agents
can take one of two strategies: {\em Cooperation} (C) and {\em Defection} (D).
Among these dilemmas, the {\em Prisoner's Dilemma} game (PDG) represents the most addressed paradigm
when studying the emergence of cooperation \cite{rapoport:1966,axelrod:1984}.
This game is fully described by the following payoff matrix
\begin{equation}\label{PD}
{\bf \Pi}=
\bordermatrix{
  & {\mbox C} & {\mbox D} \cr
{\mbox C} & {R} & {S} \cr
{\mbox D} & {T} & {P} \cr} =
\bordermatrix{
  &  {\mbox C} & {\mbox D} \cr
{\mbox C} & 1 & 0 \cr
{\mbox D} & b>1 & 1>\epsilon\geq 0 \cr}\;,
\end{equation}
showing that when two cooperators meet, each one obtains a ``{\em Reward}''
($R=1$) whereas the interaction between two defectors causes a ``{\em Punishment}'' ($P=\epsilon$) to both.
However, cooperation is hampered by the players' ``{\em Temptation}'' to defect,
since defecting against a cooperator yields more payoff than cooperating ($T=b>1$),
and by the risk arising from cooperating with a defector, as it yields the lowest payoff,
usually called ``{\em Suckers}'', ($S=0$) \cite{macy:2002}.
These payoffs satisfy the ranking $T>R>P\ge S$.
The social dilemma is thus established, since mutual cooperation yields both an individual
and total benefit higher than that of mutual defection.

\smallskip

The pairwise nature of the game is translated to a population scale by making the agents interacting
(playing) with each other, and accumulating the payoff obtained from each interaction.
After each round of the game, the strategies of the agents are updated so that those agents with less
payoff are tempted to copy (replicate) the strategy of those fittest individuals.
In unstructured populations, in which players are well-mixed, evolutionary dynamics leads all the
individuals to defection \cite{hofbauer:1998,sigmund:2010}. However, the existence of a network of
interactions, so that each agent can only play with those  directly connected to it,
the population can sometimes promote the emergence of cooperation.
This mechanism promoting cooperation was coined as {\em network reciprocity} \cite{nowak:1992},
and it was observed to be substantially enhanced when the network substrate is SF \cite{santosprl, gogardprl}.

\smallskip

Very recently \cite{jrsiPGG} the attention of the community  has been extended to the $n$-player ($n>2$)
generalization of the PDG, also called Public Goods Game (PGG) \cite{kagel:1995}.
Under this setting, each cooperator contributes a fixed investment $v$ to the public good whereas defectors do not
contribute. The total contribution of the $n$ agents is then  multiplied by an enhancement
factor $1<r<n$ and finally the result is equally distributed between all $n$ members of the group.
The payoff of cooperators and defectors in a group with a fraction $x$ of cooperators is, respectively
\begin{align}
f_C(x)&=(rx-1) v\;,
\\
f_D(x)&=rxv.
\end{align}
Thus, defectors obtain the same benefit of cooperators at zero cost, i.e. they free-ride
on the effort of cooperators. As in the PDG, the evolutionary outcome of the PGG differs
if played on a well-mixed population, in which the dynamics ends up in a full defector population,
or on a structured one. In particular, Brandt et al. \cite{brandt:2003,hauert:2003} showed that,
when games are restricted to local interactions, cooperation is promoted so that full cooperation
is observed for values of the enhancement factor $r<n$.
This result was later generalized to complex networks \cite{santos:2008} and bipartite graphs
\cite{chaosgogard,eplgogard}.

\subsubsection{Pairwise interaction games}

\paragraph{Games on interconnected networks}
The study about emergence of cooperation in the PDG in networked populations has been also extended
to the case of multilayer systems. One of the approaches is to
consider that each layer is composed of a different set of agents, so that each of them plays both with those neighbors
within the same network layer and with those placed in the other ones \cite{floriapre,jiangsrep}.
In Ref.~\cite{floriapre} the Authors considered the case of 2 interconnected networks in which each
agent plays a different game depending on whether the neighbor belongs to its same network, or to the other one.
In particular, they considered two possible payoff matrices, one for those interactions between
nodes of the same network layer, ${\bf \Pi}_{intra}$, that takes the same form of Eq.~(\ref{PD}),
and the following one for interlayer interactions
\begin{equation}\label{inter}
{\bf \Pi}_{inter}=\bordermatrix{
  &  {\mbox C} & {\mbox D} \cr
{\mbox C} & 1 &  0 \cr
{\mbox D} & b>1 & \epsilon<0 \cr}.
\end{equation}
Thus, if ${\bf A_{1}}$ and ${\bf A_2}$ are the adjacency matrices of the two networks and
${\bf C_{12}}={\bf C_{21}}$ is the matrix containing the interlayer connections
(so that the total adjacency matrix reads as in Eq.~(\ref{eq:dsic})),
the payoff of a node $i$ belonging to network layer $\alpha$ is
\begin{equation}
f_{i}^{\alpha}(t)=\sum_{j=1}^{N_1}({\bf A_{\alpha}})_{ij}\cdot \vec{s_i}^{\alpha}(t)^T{\bf \Pi}_{intra}\vec{s_j}^{\alpha}(t)+\sum_{j=1}^{N_2}({{\bf C_{12}}})_{ij}\cdot \vec{s_i}^{\alpha}(t)^T{\bf \Pi}_{inter}\vec{s_j}^{\beta}(t)\;,\;\; {\mbox{with}}\;\; \alpha\neq\beta\;,
\end{equation}
where $\vec{s_{i}}^{\alpha}$ is a two-components vector denoting the strategy of the node $i$
in layer $\alpha$ at time $t$: $\vec{s_{i}}^{\alpha}$ reads $(1,0)^T$ for Cooperation and $(0,1)^T$ for Defection.

\begin{figure}
\begin{center}
\includegraphics[width=0.70\textwidth]{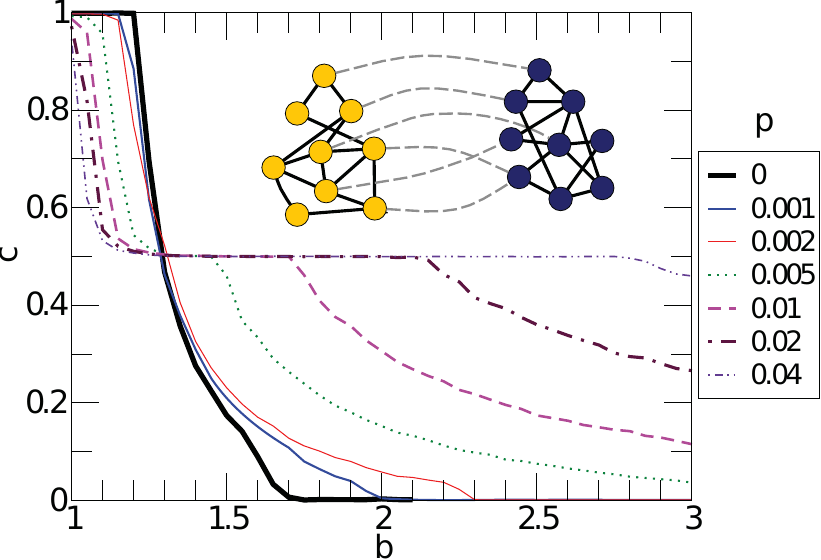}
\end{center}
\caption{(Color online) Fraction of cooperators $c$ in the population 1 as a function of $b$,
for different values of the fraction $p$ of interpopulation contacts.
The population 1 (of size $N_1=10^3$) has been coupled to a smaller population 2 ($N_2=10^2$).
While initial strategies in population 1 are equally probable (random initial conditions),
the population 2 starts from the absorbent state of fully defection. Other parameters are $r=0$, $\epsilon= -0.4$.
Both populations have a random (Erd\H{o}s-R\'enyi) network of contacts
with average degree $\langle k \rangle=6$. Reprinted figure with permission from
   Ref.~\cite{floriapre}. \copyright, 2012 by the American Physical Society.}
\label{fig:floriapre}
\end{figure}

Thus, nodes gain payoff according to a PDG when playing with neighbors in the same layer,
whereas for interlayer connections the payoff remains the same except for those interactions between two
defectors, as the {\em Punishment} becomes negative being the smallest possible payoff in this matrix, i.e.
the payoff ranking evolves into $T>R>S>P$. This latter change turns the PDG into a {\em Snowdrift} game (SDG).
In this way, the exploitation that defectors make of cooperators,
regardless of the layer where they are located,
is hampered by the payoff loss when meeting a defector belonging to the other network layer.
The results in Ref.~\cite{floriapre} reveal that, both for well-mixed layers and networked ones,
it is possible to find polarized states in which cooperation dominates in one of the layers
whereas defection does in the other. Figure~\ref{fig:floriapre} shows the fraction of cooperators, $c$,
as a function of the {\em Temptation} $b$.
When there are no connections between network layers ($p=0$), so that the two populations play a PDG,
the diagram follows the usual trend in single-layer networks:
cooperation resists for small values of $b>1$ and then suddenly decreases so that $c=0$
for $b\gg 1$. However, polarized states (here revealed by the plateau at $c=0.5$)
appear as soon as the two layers are connected, i.e. even for a small number of interlayer links,
thus showing the importance that the interconnection between layers has on the evolutionary outcome.

\smallskip

In Ref.~\cite{jiangsrep} the Authors addressed a similar interconnected system of $2$ networks,
but considering that both intra and interlayer interactions are governed also by the same Eq.~(\ref{PD}), i.e.
the payoff matrix of the PDG. In this case, the Authors focus on the emergence of cooperation as a function
of the density of interlayer links and the topology of each layer at work.
In general, for low density of interlayer connections, the overall cooperation of the system is enhanced.
In this regime, it was found that those agents connecting the two network layers are more likely
to cooperate than those sharing only intralayer links. Furthermore, although in general
the interconnection of layers promotes cooperation in the whole system, when looking at the level of single layers,
the effect of interconnection can produce a reduction of cooperation in one of the layers that
is compensated by the increase in the other one.

\smallskip

Aside from involving payoff via playing games with nodes of other networks,
another potential approach is to allow direct information transmission or exchange between layers.
In a recent research \cite{santos2014ZhenSR}, where PDG and SDG are respectively implemented on different layers,
a biased imitation is proposed: a node with probability $p$ can mimic the strategy of intralayer neighbor
in the same network, otherwise the internetwork neighbor is chosen as the strategy donor.
Under such a framework, the Authors try to explore how the cooperation trait varies as a function
of a biased probability $p$. The biased probability unveils a multiple effect,
which is robust against the change of population structures. A slight decline of the biased probability
starting from $p=1$ promotes the cooperation behavior in the PDG layer. On the contrary, this decrease
impairs the evolution of cooperation in the SDG subpopulation. Ref.~\cite{santos2014ZhenSR} also verifies
that qualitatively valid observations can be extended to a much broader parameter space in virtue of the mean-field method.

\paragraph{Games on multiplex networks}

\begin{figure}
\begin{center}
\includegraphics[width=0.90\textwidth]{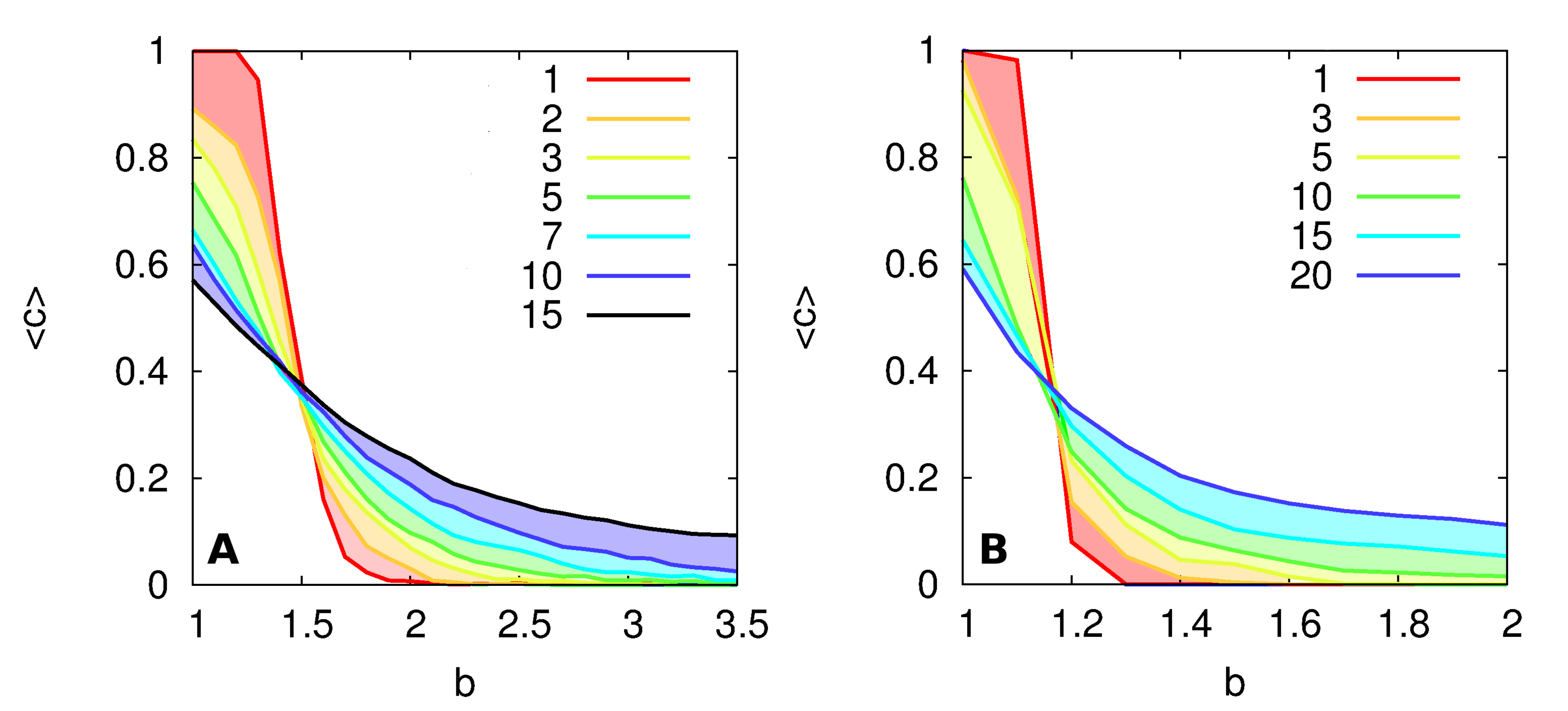}
\end{center}
\caption{(Color online). Cooperation diagrams of multinetworks.
Average level of cooperation $\langle c\rangle$  as a function of the temptation $b$ to defect for several multinetworks with different number of layers $M$ (the number of layers in indicated in the legend).
In panel {\bf A} the network layers are ER graphs with $\langle k\rangle=3$ (sparse graphs) while in panel
{\bf B} we have $\langle k\rangle=20$. In both cases $N=250$ nodes. As it can be observed,
the resilience of cooperation increases remarkably as the number of
layers $M$ grows. Reprinted figure from Ref.~\cite{gogardsrep}.}
\label{fig:gogardsrep}
\end{figure}

A second stream of works addressed the study of evolutionary dynamics in multilayer networks \cite{gogardsrep}.
In this way, all the $M$  layers have the same number of nodes, say $N$,
since each node is the representation of a particular agent,
and the only links between layers are those connecting the nodes representing the same agent.
The assumption of a multiplex imposes dynamical correlations between the states
(or strategies) of the replica nodes.
In Ref.~\cite{gogardsrep} this correlation is implemented by considering that an agent $i$ can,
in principle, take different strategies in each of the layers.
For instance, in the PDG an agent $i$ can be a defector in $n$ layers, while it cooperates in the remaining
$M-n$ layers. The total payoff of a node $i$ in layer $\alpha$ at time $t$ reads
\begin{equation}
f_{i}^{\alpha}(t)=\sum_{j=1}^{N}A_{ij}^{\alpha}\vec{s_i}^{\alpha}(t){\bf \Pi}\vec{s_j}^{\alpha}(t)\;,
\end{equation}
where the payoff matrix is defined as in Eq.~(\ref{PD}).
In addition, the total payoff of agent $i$ is the sum of all the payoffs accumulated
by the different representations across the $M$ layers
\begin{equation}
f_i(t)=\sum_{\alpha=1}^{M}f_{i}^{\alpha}(t).
\end{equation}

In Ref.~\cite{gogardsrep} the update of the strategy used by an agent $i$ at layer $\alpha$
depends on $f_i(t)$ (instead of taking into account only the payoff accumulated
at this layer $f_i^{\alpha}(t)$), thus coupling the evolution of the states of the nodes
in a given layer with the dynamical states of the rest of the layers.
Figure~\ref{fig:gogardsrep} shows that the correlation between layers influences the cooperation levels
of the system in a way that, by increasing the number $M$ of layers,
the resilience of cooperation is enhanced at the expense of decreasing the level of
cooperation observed for small values of the temptation $b$ to defect.

Moreover,  motivated by the fact that each agent is usually member of different networks in human societies,
a completely distinct setup is suggested in Ref.~\cite{wang2014ZhenPRE}.
The system is there composed of 2-layer SF networks.
One network, named interaction layer, is used for the accumulation of payoffs,
and the other serves as the updating network for strategies or states.
The Authors show that, with all the possible combinations of degree mixing,
breaking the symmetry through assortative mixing in one layer and/or disassortative mixing
in the other as well as preserving the symmetry by means of assortative mixing in both layers,
impedes the evolution of cooperation.
The essence of this scenario for the cooperation inhibition can be attributed
to the collapse of giant cooperation clusters, especially as compared to the case of
single-layer SF networks \cite{wang2014ZhenPRE}.

\paragraph{Games on interdependent networks}

The third realm of the evolutionary game in multilayer networks turns to that of interdependent networks.
It is worth mentioning that this form possesses evident differences with the aforegoing situations.
As compared to the case of interconnected networks, the direct interaction of game and strategy
exchanged between network layers are strictly forbidden.
On the other hand, the corresponding nodes connected via the
internetwork links represent different agents, at variance with the multiplex scenario, thus both the accumulation of payoff and strategy updating
are just limited to the same network  layer. In this sense, the internetwork links are mainly used to create
mutual influence between different networks.
Since utility and strategy are prime elements in evolutionary game theory,
creating interdependence through both factors becomes a natural choice.
The framework of interdependent network games is then the following: agents of each layer do not engage
in any game interaction beyond their own layer, but just refer utility or strategy information of other
systems during the process of strategy updating with their intranetwork neighbors.

A recent research suggests the case of 2 interdependent networks,
network $\alpha$ and network $\beta$, in which each node obtains the payoff by playing the games
with its intralayer neighbors \cite{wang2014ZhenSR2}.
Then the interdependent correlation between networks (the node-to-point interdependence) is realized by defining an utility function as follows:
\begin{equation}
U_{i}^{\alpha}(t)=(1-\gamma)\cdot f_{i}^{\alpha}(t)+\gamma\cdot f_{i'}^{\beta}(t),
\end{equation}
where $f_{i}^{\alpha}(t)$, $f_{i'}^{\beta}(t)$ represent the payoffs of a given node $i$ in network $\alpha$ and
that of corresponding partner $i'$ from network $\beta$, and $\gamma\in[0,1]$ is the parameter
accounting for the coupling strength between layers.
The function $U_{i}^{\alpha}(t)$ is then used for the update of the strategy of agent $i$
at layer $\alpha$. Such an expression is a symmetric function,  i.e. the payoff of the node
in network $\beta$ also accounts for the same part ($1-\gamma$) in its utility.
It is worth noting that setting $\gamma=0$ decouples both networks,
and the model gets back to the traditional single-layer network case \cite{szabo:2007}.
With the increment of the coupling correlation, the utility of one network depends more and more on the other system. In particular, for $\gamma=1$, the utility of one node becomes fully determined by the other
network, and the  updating strategy becomes the process of tossing a coin.

\begin{figure}
\begin{center}
\includegraphics[width=\textwidth]{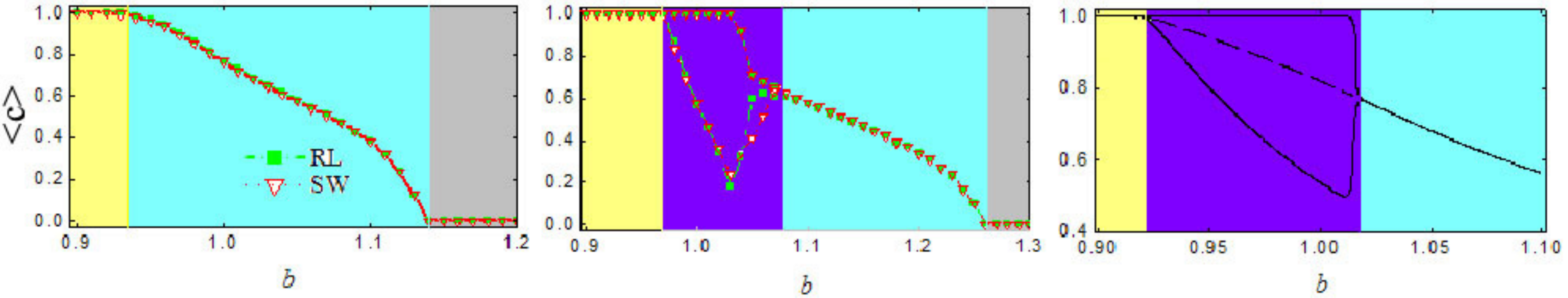}
\end{center}
\caption{(Color online). Cross sections of phase diagrams for $\gamma=0.4$ (left) and $\gamma=0.9$ (middle)
on the interdependent square lattices (green squares) and small-world networks (red triangles)
with a fraction of rewired links equal to $0.05$. The colored regions correspond to different sections:
yellow is pure cooperators section, cyan is mixed strategies section,
gray is pure defectors section and purple is symmetry breaking section.
The lines in right panel denote the results of strategy-couple pair approximation for $\gamma=0.9$,
which is qualitatively similar to the case of the middle panel.
The dashed line represents an unstable solution of the analytical
approach. Note that different ranges for the horizontal axis are used
in the three panels, just for the guide of the eye.
Reprinted figure from Ref.~\cite{wang2014ZhenSR2}.}
\label{fig:jqsrep}
\end{figure}
\smallskip

The Authors selected the PDG as the paradigmatic model to explore the
effect of the coupling strength.
They found that there exists a threshold value $\gamma_t$
($\gamma_t\simeq 0.5$) in the system, such that, when the coupling strength $\gamma$ is smaller than this threshold, the symmetry of cooperation level in both networks is maintained. However, as the value of $\gamma$ exceeds $\gamma_t$, there is a spontaneous symmetry breaking of the fraction of cooperators between different networks.
Figure~\ref{fig:jqsrep} shows the typical cross sections of phase diagrams for different
coupling degree $\gamma$. In the case of $\gamma=0.4$,
where cooperation level is well promoted, one can observe the existence of three sections:
pure cooperators section, mixed strategies section and pure defectors section.
If $\gamma$ is set as $0.9$, a novel section emerges: the symmetry breaking section.
Moreover, this observation is universally effective on different networks and can be qualitatively
predicted by the strategy-couple pair approximation approach,
in which an accurate prediction of the phase transition is difficult, due to the fact that
long-range interactions are neglected among nodes in the network \cite{szabo:2007}.

\subsubsection{The public goods game}

The first work on this game in multilayered networks \cite{wang_z_epl12}
addressed a simple system composed of two regular lattices that are coupled in the usual
coupling way (each node in a layer has a link with its corresponding partner in the other one).
The PGG is played according to the usual rules, so that in each of the layers,
an individual constitutes a group together with its $4$ nearest {intralayer} neighbors,
thus taking part  simultaneously in $5$ different PGGs.
As described for PDG, the coupling between layers comes from an utility function that couples
the payoff obtained by the agent $i$ in layer $\alpha$, $f_i^\alpha$,
with that obtained by its partner $i'$ in the other layer $\beta$, $f_{i'}^\beta$.
In this case, the coupling takes the form
\begin{equation}
U_i=\gamma \cdot f_i^{\alpha}+(1-\gamma) \cdot f_{i'}^{\beta}.
\label{coupling}
\end{equation}
Here the utility function, at variance with Ref.~\cite{wang2014ZhenSR2},
is the same for the two layers, and it incorporates a new ingredient since there is one dominating
layer depending on the value of $\gamma$ (namely, a biased master-slave interdependence fashion):
when $\gamma< 0.5$ ($\gamma>0.5$) layer $\alpha$ ($\beta$) is  dominated by layer $\beta$ ($\alpha$).
As usual, the utility function is used for updating the strategies of the nodes in each layer.
The most interesting result obtained in Ref.~\cite{wang_z_epl12} is that,
when one layer almost dominates the other, cooperation is extraordinarily favored in the slave layer.
For $\gamma \rightarrow 0$ and $\gamma \rightarrow 1$, the corresponding master layer
behaves almost equally to an isolated square lattice showing more weakness to defection than the slave
one, in which the relation between the strategy and the utility function is almost random.
These results are explained in terms of time-scale separation for the spread of strategies
that favors the development of cooperator clusters in the slave layer.

Many other kinds of interlayer couplings can be introduced in the utility function.
In Ref.~\cite{wangsrep2013-2}, using a two-layer {topology} composed of two identical 2-dimensional lattices,
the Authors followed a similar approach to that used in
Ref.~\cite{wang2014ZhenSR2} for the PDG, since each one of the layers has its own utility
function, which in its turn depends on what is the state of the other layer.
In this case, the utility function for the agent $i$ in layer $\alpha$ is defined as
\begin{equation}
U_{i}^{\alpha}=(1-\gamma-\gamma^{\prime}) \cdot f_{i}^{\alpha}+\gamma \cdot \left(\sum_{j=1}^{N}A_{ij}^{\alpha}f_{j}^{\alpha}\right)+\gamma^{\prime} \cdot \left(\sum_{{j'}=1}^{N}A_{i'j'}^{\beta}f_{j'}^{\beta}\right),
\label{eq:pggsrep}
\end{equation}
where the terms multiplied by $\gamma$ and $\gamma^{\prime}$ account for the average payoff of the neighbors of $i$ in the same layer $\alpha$ and that of those neighbors of the {corresponding partner}
$i^{\prime}$ in the other layer $\beta$, respectively.
The effect of the coupling parameters $\gamma$ and $\gamma^{\prime}$ is the promotion of cooperation although
this effect is more pronounced for $\gamma$ than for the interlayer coupling $\gamma^{\prime}$
{(see the left panel of Fig.~\ref{fig:gogardsrep2}).
This observation was given the name of {\it interdependent network reciprocity}.
The essence of this mechanism requires the simultaneous formation of correlated cooperator
clusters on both networks, which can get further support from the synchronized evolution of strategy pairs
between networks, as shown in the right panel of Fig.~\ref{fig:gogardsrep2}.
If synchronization does not emerge or if the coordination process is disturbed,
interdependent network reciprocity fails, resulting in the total collapse of cooperation.
Moreover, this promoting effect is proven to take place also in systems involving more than 2 layers,
as long as the coordination between the networks is not disturbed.

Another way to model  the interlayer coupling  via the utility function
is the probabilistic interconnection introduced in Ref.~\cite{wangjstat2012},
in which two lattices are coupled in an interconnected way.
However, two {nodes of both layers} are effectively coupled with probability $p$,
so that not all the possible $N$ pairs of interlayer links are active.
 When an interlayer link is active,  the utility function of player $i$ in layer $\alpha$ is influenced,
so that apart from the $5$ PGG in which it is enrolled within the corresponding layer
$\alpha$, it also participates in that centered on its {partner $i^{\prime}$} in the other layer $\beta$.
In this way, the Authors showed that there is an optimal interlayer connection probability $p$
that maximizes the cooperation level.

The information of strategy also plays a crucial role in affecting the trait of other systems.
In Ref.~\cite{szolnokinjp2013}, the Authors proposed an alternative coupling between two lattices. The coupling is done via the update rule by which the agents decide their strategies
(to cooperate or to defect) for the next round of the game instead that via the utility function.
The probability that agent $i$ in layer $\alpha$ adopts the strategy of a neighboring node $j$
in the same layer depends here on the difference between their {payoff} functions and on the fact that the corresponding {partner} of $i$ in layer $\beta$ has the same or a different strategy.
If the strategies of both {the node $i$ and its partner $i^{\prime}$} are the same, the probability
of changing is decreased with respect to the usual one when the network layers are isolated.
This coupling is seen not only to enhance cooperation in both PDG and PGG, but also to sustain
the spontaneous emergence of correlated behaviors between the two networks.
If the range of information sharing is extended to groups, this promotion effect is even more evident.

 \subsubsection{Partial overlap and spatial coevolution}
\begin{figure}
\centering
\includegraphics[width=0.9\textwidth]{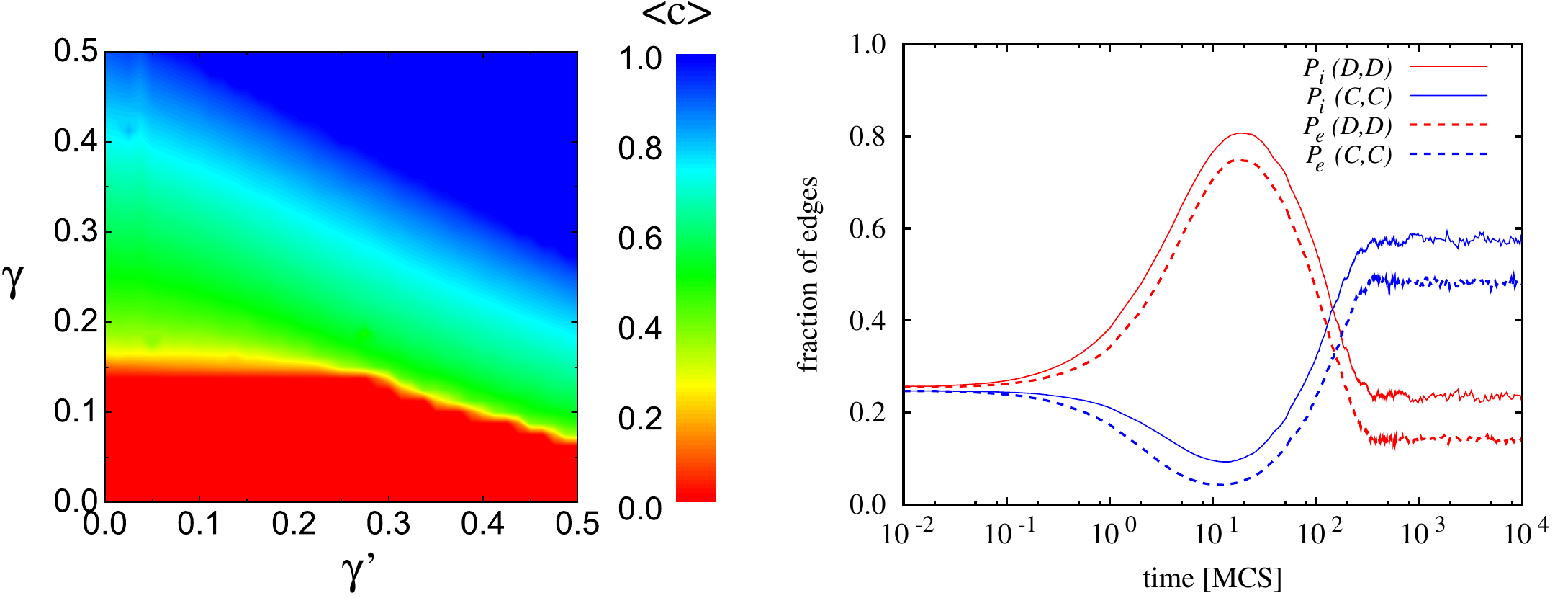}
\caption{(Color online) The contour plots show the stationary fraction of cooperators $\avg{c}$
as a function of the coupling parameters $\gamma$ and $\gamma^{\prime}$ entering in Eq.~(\ref{eq:pggsrep}) for $r=3.5$ in PGG (left panel). Time courses of strategy pairs within and between networks (right panel),
where $P_i$ and $P_e$ denote the corresponding pair configuration probabilities within (internal)
and between (external) networks, respectively. The parameters are $r=3.5$, $\gamma=0.15$
and $\gamma^{\prime}=0.5$ in right panel. Reprinted figure from Ref.~\cite{wangsrep2013-2}.}
\label{fig:gogardsrep2}
\end{figure}

While the effect of interlayer nodes has been extensively identified, the question of what would be
the best overlapping level for the resolution of social dilemmas remains unclear.
To answer this issue, one scenario of partial overlapping is suggested in Ref.~\cite{wangsrep2013},
where two networks possessing the same number $N$ of nodes compose the interdependent system.
Before any interaction, a fraction $\rho$ of the nodes of one layer, $\rho N$ nodes,
which are chosen uniformly at random, are linked to their corresponding partners in the other one,
and the residual nodes have no internetwork partners. Due to the existence of interdependence,
the utilities of these overlapping nodes are not simply payoffs obtained from the interactions
with the intralayer neighbors in the same network,
but they rather involve the performance of their partners in the other network.
Here the correlation between the payoffs obtained by a given node $i$ and its partner $i^{\prime}$
is done by defining the following utility function:
\begin{equation}
U_{i}^{\alpha}(t)=f_{i}^{\alpha}(t)+\gamma \cdot f_{i^{\prime}}^{\beta}(t),
\label{eq:wzsrep}
\end{equation}
with $\gamma\in[0,1]$ determining the strength of coupling between layers.
In particular, if the node $i$ has no external links,
its utility simply equals to its own payoff ($U_{i}^{\alpha}(t)=f_{i}^{\alpha}(t)$).
The utility function $U_{i}^{\alpha}(t)$ is then used for the update of the strategy of agent $i$
at layer $\alpha$. As in the setup of Ref.~\cite{wang2014ZhenSR2},
the utility function has here a symmetric form.

Under such a scenario, cooperation is studied as a function of $\rho$ and $\gamma$.
The Authors found that an intermediate density of sufficiently strong coupling between networks
(more accurately, $\gamma \rightarrow 1$ and $\rho\simeq 0.5$) warrants an optimal regime
for cooperation to survive.
This observation
is robust to the variations of the strategy updating dynamics, the interaction topology, and the governing
social dilemma, thus suggesting a high degree of universality. Similarly to Ref.~\cite{jiangsrep}
it is observed that,
within a layer, cooperation tends to fixate preferentially in those nodes that are connected with
their counterpart in the other layer.

Furthermore, it is instructive to examine how the cooperative behavior survives on the coevolution framework,
in which the heterogeneity is usually ubiquitous \cite{perc2010ZhenPlos,perc2010ZhenBS}.
In a recent research \cite{wang2014ZhenJTB}, where no interlayer links exist between two independent
structured populations, a utility threshold is designed to manage the emergence of temporary interlayer links.
If the current utility of node $i$ exceeds this threshold, it is rewarded by one  interlayer link to the
corresponding partner belonging to the other network in the next time step;
otherwise the external link is not formed. As for the evaluation of utility, Eq.~(\ref{eq:wzsrep}) is retained.
It is easy to estimate that, as long as the utility of all nodes drops below the threshold,
the external link is terminated, which returns to decoupled single-layer networks.
The Authors showed that an intermediate threshold gives rise to distinguished players
that act as strong catalysts of cooperative behavior in both PDG and PGG. However,
there also exist critical utility thresholds beyond which distinguished players are no longer able to percolate.
The interdependence between the two populations then vanishes, and the survival of cooperation becomes a hard task.
Finally, this demonstration of the spontaneous emergence of
optimal interdependence by means of a simple coevolutionary rule based on the self-organization
of reward and fitness is proven to be universally effective on different interaction networks.

\begin{figure}[t!]
\begin{center}
\includegraphics[width=0.7\textwidth]{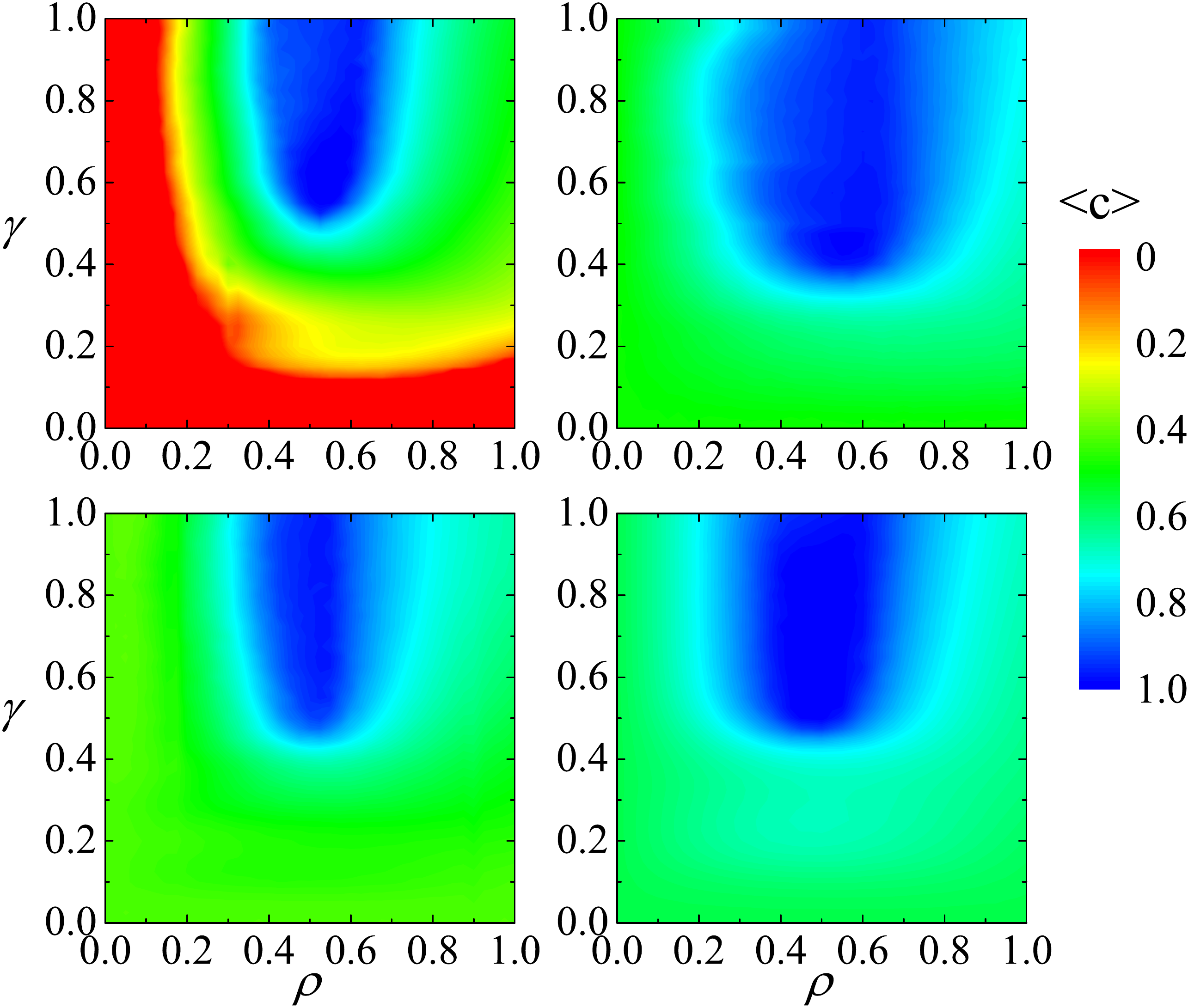}
\end{center}
\caption{(Color online) The contour plots show the stationary fraction of cooperators
  $\langle c\rangle$ as a function of the coupling ratio $\rho$ and coupling degree $\gamma$
for  different cases.
Both top panels and the left bottom one refer to a prisoner dilemma underlying game,
whereas the snowdrift game  is used in the bottom right panel. Square lattices are used except
for the bottom left panel, which uses triangle lattices.
Similarly, the top right panel uses proportion imitation rule \cite{santosprl,gogardprl}, while Fermi update rule \cite{SzaboPRE1998,szabo:2007} is held in the other ones. It is obvious that the proposed scenario of partial overlap is universally effective.
Reprinted figure from Ref.~\cite{wangsrep2013}.}
\label{fig:wzsrep}
\end{figure}

At variance with the scenario of utility threshold, another setup referring to  coevolution between
strategy and network structures considers the strategy transmission ability \cite{wang2014NJP}.
Initially, each node $i$ of two independent networks has a low and identical transmission ability $w_i$.
If node $i$ succeeds in passing its strategy to one of its intranetwork neighbors,
its ability enhances by $\Delta$, $w_i=w_i+\Delta$; otherwise $w_i$ becomes $w_i-\Delta$.
A threshold $w_{th}$ is selected as the criterion of creating an external link:
 only when the transmission ability satisfies $w_i \ge w_{th}$, player $i$ is allowed
to have an interlayer link to its corresponding partner in the other network.
If $w_i$ drops below the threshold $w_{th}$, the external link is terminated.
The making and breaking of the interlayer links between the two networks thus serves
as individual rewarding and punishment, that reflect the evolutionary success within a network.
Here, the utility of a node with (without) interlayer link is in accordance with Ref.~\cite{wangsrep2013},
 yet using the fixed strength of coupling. In this sense, the question becomes what are
the optimal conditions for resolving social dilemma.

The Authors showed that, due to coevolutionary rule, the interdependence between networks
self-organizes so as to yield optimal conditions for the evolution of cooperation even under
extremely adverse conditions. In this case, approximately half of the nodes form an interlayer link,
while the other half is denied the participation in the activities beyond their host network.
 Moreover, due to the spontaneous emergence of a two-class society, with only the upper class being
allowed to control and take advantage
of the external information, Authors verified that cooperative nodes having the interlayer links
are much more competent than defectors in sustaining compact clusters of followers.
Such an asymmetric exploitation confers them a strong evolutionary advantage that
may resolve even the toughest of social dilemmas.


\section{Synchronization}
\label{sec:synchro}
Networked systems, i.e. ensembles of distributed dynamical
systems that interact over complex wirings of
connections, are prominent candidates to study the emergence
of collective organization in many areas of science~\cite{AlbertBarabasi2001,Boccaletti2006}.

In particular, a special interest has been devoted to the analysis
of synchronized states (from the Greek \greek{σύγχρονος},
meaning
``sharing the same time''), as such states play a crucial role in
many natural phenomena, like as the emergence of coherent global behaviors
in both normal and abnormal brain functions~\cite{epilepsis}, or the
food web dynamics in ecological systems~\cite{maccan98}.

In the past years, the emergence of synchronized
states~\cite{booksPhaseSynchr} has been extensively
reported and studied for the case of mono-layer
networks, with the emphasis focusing on how the
complexity of the layer topology influences the
propensity of the coupled units to synchronize~\cite{complNetSync1,complNetSync2,barahona02,barahona02b}.
While the literature on this subject is indeed extremely
large, we point the reader to a few monographic
reviews and books~\cite{Boccaletti2006,ArenasPR2008,potsdam,potsdam2,potsdam3},
whose consultation may provide guidance and help
in understanding the major accomplishments, results
and related concepts.

As soon as the multilayer nature of connections is brought into the game,
the subject must be divided into two different scenarios, which have to be
individually treated.

In the first case, the different layers correspond to different
connectivity configurations that alternate to define a time-dependent
structure of coupling amongst a given set of dynamical units. In
this scenario, the basic problem is to describe and assess how
the existence and stability conditions for the state in which the
network's dynamical units are synchronous are modified with respect
to the classic (monolayer) situation, in which the coupling between
the networked units is time-independent.

The second case, instead, is the scenario where the different
layers are \emph{simultaneously} responsible for the coupling
of the network's units, with explicit additional layer-layer
interactions to be taken into account. It is worth stressing
that here an \emph{inter}layer synchronized state is a different
and much more general concept, as it does not necessarily require
\emph{intra}layer synchronization. In other words, the dynamical
units within each layer can be out of synchrony, but synchronized
with the corresponding units in all other layers.

As a consequence, this section is divided into two main parts,
each reporting on one of the two described scenarios,
and is intended to summarize for each case the extension of the
main analytical treatments and approaches that are traditionally used
to deal with synchronization of monolayer networked
systems.

\subsection{Synchronization in multilayer networks with alternating layers}
We start with the case in which the network under consideration
is composed of a number $N$ of identical dynamical systems, which
interact via a time-dependent structure of connections, each of
which corresponding, at a given time, to one of the different layers.

Very recent investigations on synchronization of time-varying networks
include the treatment of complete synchronization in temporal Boolean
networks~\cite{Li2013IreneNN}, the analysis of stability of synchronized
states in switching networks in the presence of time delays~\cite{Yao2009IreneAuto},
and the investigation of the effects of temporal networking interactions
when compared with the corresponding aggregate dynamics where all interactions
are present permanently~\cite{Masuda2013IrenePRL}.

We focus here on the situation of an ensemble of identical
units where the wiring of connections is switching between
the different layers following a periodic or aperiodic sequence.
It is important to remark that detailed studies of the latter
case actually started much before the relatively recent introduction
of the concept of multilayer networks. In particular, a widely
studied case is the limit of the so-called \emph{blinking networks}~\cite{blinking,blinking2},
where the switching between different configurations happens
rapidly, i.e. with a characteristic time scale much shorter
than that of the networked system's dynamics. Under these
conditions, it was found that synchronous motion can be established
for sufficiently rapid switching times even in the case where
each visited wiring configuration (each network layer) would
prevent synchronization under static conditions.

Yet, the limit of blinking networks is not a suitable framework
for an adequate description of several natural phenomena. For instance,
modeling processes such as \emph{mutations} in biological systems~\cite{mutation},
or \emph{adaptations} in social systems or financial market dynamics~\cite{adaptation}
would demand the very opposite limit of time-varying networks whose
structure evolution takes place over characteristic time scales
that are commensurate with, or even secular with respect to, those
of the nodes' dynamics.

Thus, in the following, we will concentrate on the case in which synchronized states
appear in dynamical networks, \emph{without making any explicit a priori assumption}
on the time scale of the variation of the coupling wiring.

\subsubsection{The Master Stability Function for time dependent networks}\label{master-temporal}
The starting point is the generic equation describing the evolution
of an ensemble of $N$ identical systems, networking via a time dependent
structure of connections:
\begin{equation}\label{eq1}
\dot{\mathbf x}_i = \mathbf{f}(\mathbf x_i) - \sigma \sum_{j=1}^N G_{ij}(t)\mathbf h[\mathbf x_j],\quad i = 1, \ldots, N\:.
\end{equation}

Here, dots denote temporal derivative, ${\mathbf x}_i \in \mathbb{R}^m$ is the $m$-dimensional vector
describing the state of the $i^\mathrm{th}$ network unit, and $\mathbf{f}(\mathbf x):\mathbb{R}^m\to\mathbb{R}^m$
and $\mathbf h[\mathbf x]:\mathbb{R}^m\to\mathbb{R}^m $ are two vectorial functions describing the
local dynamics of the nodes and the output from a node to another, respectively.

Furthermore, $G_{ij}(t)\in\mathbb{R}$ are the time varying elements
of an $N\times N$ symmetric matrix specifying the evolution, in strength
and topology, of the network connections. The matrix $G(t)$ satisfies
the defining properties of a Lagrangian matrix: it has zero row-sums
($\sum_{j}G_{ij}(t)=0\ \forall i$ and $\forall t$), strictly positive
diagonal terms ($G_{ii}(t)>0\ \forall i$ and $\forall t$) and strictly
non-positive off-diagonal terms ($G_{ij}(t)\leq 0\ \forall i\neq j$
and $\forall t$).

Our goal is to assess the stability of the synchronized solution
$\mathbf x_1(t)=\mathbf x_2(t)=\dotsc =\mathbf x_N(t)\equiv\mathbf x_s(t)$,
whose existence is indeed warranted by the zero row-sum property of $G(t)$.

To this purpose, we first notice that $G(t)$ is symmetric,
and therefore it is always diagonalizable and, at all times,
admits a set $\lambda_i(t)$ of real eigenvalues and a set
$\mathbf v_i(t)$ of associated orthonormal eigenvectors, such
that $G(t)\mathbf v_i(t)=\lambda_i(t)\mathbf v_i(t)$ and $\mathbf v_j^{\mathrm T}\cdot\mathbf v_i=\delta_{ij}$.

The zero row-sum condition and the Gershgorin's circle theorem~\cite{gershgorin} further
ensure that:
\begin{enumerate}
\item the spectrum of eigenvalues is such that $\lambda_i(t)\geq 0\ \forall i$
and $\forall t$;
\item the smallest eigenvalue $\lambda_1(t)\equiv 0$, with associated
eigenvector $\mathbf v_1(t)=\frac{1}{\sqrt{N}}\lbrace 1,1,\dotsc,1\rbrace^{\mathrm T}$,
entirely defines the synchronization manifold ($\mathbf x_i(t)=\mathbf x_s(t)\ \forall i$);
\item all the other eigenvalues $\lambda_i(t)$ ($i=2, \dotsc, N$) have associated
eigenvectors $\mathbf v_i(t)$ that form a basis of the space transverse to the manifold
of $\mathbf x_s(t)$ in the $m \times N$-dimensional phase space of Eq.~(\ref{eq1}).
\end{enumerate}
 Notice
that all the eigenvalues except the first are strictly positive for connected graphs
($\lambda_i(t)>0$ for $i=2, \dotsc, N$).

Consider then $\delta\mathbf x_i(t)=\mathbf x_i(t)-\mathbf x_s(t)=(\delta x_{i,1}(t),\dotsc,\delta x_{i,m}(t))$,
i.e. the deviation of the $i^\mathrm{th}$ vector state from the synchronized state, and construct the $N \times m$
column vectors $\mathbf{X}=(\mathbf{x}_1^{\mathrm T},\mathbf x_{2}^{\mathrm T},\dotsc,\mathbf x_N^{\mathrm T})$
and $\delta\mathbf X=(\delta\mathbf x_1^{\mathrm T},\dotsc,\delta\mathbf x_N^{\mathrm T})$. It is straightforward
to show that the linearization of Eq.~(\ref{eq1}) around the synchronized solution leads, in linear order of $\delta\mathbf X$,
to
\begin{equation}\label{eq:var}
 \delta\dot{\mathbf X} = \left[I_N\otimes\mathrm J\mathbf f(\mathbf x_s) - \sigma G(t)\otimes\mathrm J\mathbf h(\mathbf x_s)\right]\delta\mathbf X\:,
\end{equation}
where $\otimes$ denotes the direct product, and $\mathrm J$ is the Jacobian operator.

The arbitrary state $\delta\mathbf X$, at each time, can be written
as a sum of direct products of the eigenvectors of $G(t)$ and a time-dependent
set of row-vectors $\eta_i\left(t\right)$, i.e.
\begin{equation*}
 \delta\mathbf X=\sum_{i=1}^N\mathbf v_i(t)\otimes\eta_i(t)\:,
\end{equation*}
with $\eta_i(t)=(\eta_{i,1},\dotsc,\eta_{i,m})$. Therefore, applying
$\mathbf v_j^{\mathrm T}$ to the left side of each term in Eq.~(\ref{eq:var}),
one finally obtains
\begin{equation}\label{variation}
  \ddt\eta_j=\mathbf K_j\eta_j\ \ \boxed{-\sum_{i=1}^N\mathbf v_j^{\mathrm T}(t)\cdot\ddt\mathbf v_i(t)\eta_i}\:,
\end{equation}
where $j=1,\dotsc,N$ and
$\mathbf K_j=\left[\mathrm J\mathbf f(\mathbf x_s) - \sigma\lambda_j(t)\mathrm J\mathbf h(\mathbf x_s)\right]$.

The key point here is noticing that the boxed term in Eqs.~(\ref{variation})
is what differentiates them from the classical equations used to calculate
the Master Stability Function (MSF) ~\cite{barahona02}, as it explicitly accounts
for the time variation of the eigenvector basis. In the following, we will
first discuss the case where such an extra term vanishes, and later the general
case of constructing a quasi-static evolution among different network layers,
deriving a proper analytical expression for it.

\subsubsection{Commutative layers}
Eqs.~(\ref{variation}) transform into a set of $N$ variational equations of the form
\begin{equation}\label{vari1}
 \ddt\eta_j = \mathbf K_j\eta_j\:
\end{equation}
if $\sum_{i=1}^N\mathbf v_j^{\mathrm T}(t)\cdot\ddt\mathbf v_i(t)\eta_i=0$,
i.e. if all eigenvectors are constant in time.

This condition can be satisfied in two different ways, namely
either the coupling matrix $G(t)$ is constant, as in the case
of a monolayer network originally considered in Ref.~~\cite{barahona02},
or the temporal evolution occurs through commutative layers.
The latter case means that starting from an initial wiring $G_0\equiv G(t=0)$,
\emph{the coupling matrix $G(t)$ must commute with $G_0$ at all
times $t$} ($G_0G(t)=G(t)G_0\ \forall t$). This was explicitly
discussed in Ref.~\cite{Boccaletti2006StefanoPRE}, where it was
shown that:
\begin{enumerate}
\item an evolution along commutative graphs
(i.e., different commutative layers of a multilayer network)
can be constructed by starting from any initial wiring $G_0$,
and
\item such an evolution provides a much better condition
for the stability of the synchronous state than that given by
a monolayer network.
\end{enumerate}

To show this, the first step is constructing explicitly
the set of commutative layers through which the evolution
takes place. To this purpose, one can notice that any initial
symmetric coupling matrix can be written as $G_0=V\Lambda_0V^{\mathrm T}$,
where $V=\lbrace\mathbf v_1,\dotsc,\mathbf v_N\rbrace$
is an orthogonal matrix whose columns are the eigenvectors
of $G_0$, and $\Lambda_0=\mathrm{diag}(0,\lambda_2(0),\dotsc,\lambda_N(0))$
is the diagonal matrix formed by the eigenvalues of $G_0$.

At any other time $t$, a zero row-sum symmetric commuting
matrix $G(t)$, defining the layer visited by the structure
evolution at that time, can be constructed as
\begin{equation} \label{pippo}
 G(t)=V\Lambda(t)V^{\mathrm T}\:,
\end{equation}
where $\Lambda(t)=\mathrm{diag}(0,\lambda_2(t),\dotsc,\lambda_N(t))$
with $\lambda_i(t)>0\ \forall i>1$. It is easy to show that $G(t)$ is
positive semidefinite, and since the canonical basis vectors are not
collinear with the eigenvector $\mathbf v_1$ we have $G_{ii}>0\ \forall i$.
Thus, Eq.~(\ref{pippo}) provides a way to generate all possible layers
that commute with $G(0)$, are positive semidefinite with zero row-sum,
and dissipatively couple each network unit. Henceforth, we refer to
this set of matrices as the dissipative commuting set (DCS) of $G(0)$.
It is important to notice that while not all members of the DCS are
true Lagrangian matrices (it is in fact possible for their off-diagonal
components to be positive), nevertheless all commuting Lagrangian matrices
are contained in the DCS.

In Ref.~\cite{Boccaletti2006StefanoPRE}, the Authors proposed a general method to select
a set of commutative matrices within the DCS. The method is based on starting from a given
layer graph $G_0$, and producing a large set of different realizations of the same graph.
This allows the estimation of the initial distribution $p(\lambda_0)$ of the
non-vanishing eigenvalues of the set. Then, one can directly construct $\Lambda(t)$ to
use in Eq.~(\ref{pippo}) by randomly drawing a set of eigenvalues either according to $p(\lambda_0)$,
or uniformly between $\lambda_2(0)$ and $\lambda_N(0)$. The former approach can be implemented
either by using the spectrum of a different realization of $G_0$ (the so-called eigenvalue
surrogate method~\cite{Boccaletti2006StefanoPRE}), or by extracting directly the eigenvalues
from the distribution $p(\lambda_0)$. The latter method, instead, simply corresponds to the
random selection of the eigenvalues from a uniform distribution. In either case, the random
extractions can be performed in an ordered ($0\leq \lambda_2(t) \leq \dotsc \leq \lambda_N(t)$) or unordered
way, and one only needs to extract $N-1$ eigenvalues, since $\lambda_1(t)$ must be always~0.

In addition, Ref.~\cite{Boccaletti2006StefanoPRE} also discusses how the different
choices for the construction of commutative layers lead to modifications of the main
topological structures. In general, the resulting matrix $G(t)$ is a dense matrix
that can be associated to a undirected weighted network, whose weight matrix $W(t)$
has elements $W_{ii}(t)=0$, and $W_{ij}(t)=|G_{ij}(t)|$ for $i \neq j$. The main
properties that were studied in Ref.~\cite{Boccaletti2006StefanoPRE} are the average
value of the strength distribution, the average local clustering coefficient,
and the average shortest path length, defined as follows~\cite{AlbertBarabasi2001,Boccaletti2006,NohPRE66,lm01}.

For weighted networks, the distance between two adjacent
nodes $i$ and $j$ is $l_{ij}=\frac{1}{W_{ij}}$, and the distance along
a path $\lbrace n(1),n(2),\dotsc,n(m)\rbrace$ can be expressed as $L_{n(1)\rightarrow n(m)}=\sum_{k=1}^{m-1}l_{n(k)n(k+1)}$.
Then, the distance between two non-adjacent nodes $i$ and
$j$ is the length of one of the shortest paths connecting them:
\begin{equation*}
 \ell_{ij}=\min_{\mathrm{all paths}\ i\rightarrow j}L_{i\rightarrow j}\:.
\end{equation*}
With these definitions, a network's average shortest path
length is~\cite{AlbertBarabasi2001,Boccaletti2006,NohPRE66,lm01}
\begin{equation*}
\langle\ell\rangle = \frac{2}{N(N-1)}\sum_{\substack{i,j\\i\neq j}}\ell_{ij}\:.
\end{equation*}

Similarly, the clustering coefficient of node $i$ can be defined as
\begin{equation*}
 c_i=\frac{2}{s_i(N-2)}\sum_{j,m}W_{ij}W_{im}W_{jm}\:,
\end{equation*}
where $s_i=\sum_jW_{ij}$ is the strength of node $i$.
The average local clustering coefficient
$\langle C\rangle$ is then given by $\langle C\rangle =\frac{1}{N}\sum_i c_i$,
while the strength distribution $P(s)$ characterizes
the heterogeneity of the network~\cite{NohPRE66,lm01}.
\begin{table}
\begin{center}
\begin{tabular}{|c|c|c|c|}
\hline & $\langle s\rangle$ & $\langle C\rangle$ &
$\langle\ell\rangle$ \tabularnewline \hline \hline Initial
condition $G_0$ (scale-free)& $9.94$ & $5.5773\times 10^{-4}$ & $5.5144$
\tabularnewline \hline Eigenvalue surrogate method & $11.6881$ &
$8.4730\times 10^{-4}$& $5.2758$ \tabularnewline \hline Choosing from
$p(\lambda_0)$ (ordered)& $12.7042$ & $9.1235\times 10^{-4}$& $5.2325$
\tabularnewline \hline Choosing from $p(\lambda_0)$ (unordered)&
$142.8746$ & $0.0543$ & $3.7144$ \tabularnewline \hline Uniform
distribution (ordered)& $123.6484$ & $0.0634$ & $0.9570$ \tabularnewline
\hline Uniform distribution (unordered)& $521.7568$& $0.6125$& $1.1098$
\tabularnewline \hline
\end{tabular}\end{center}
\caption{\label{tab:stat}Statistical properties of the commuting
layers. The table reports the average strength $\langle s\rangle$,
the average local clustering coefficient $\langle C\rangle$ and
the average shortest path length $\langle\ell\rangle$ for the different
methods of selecting the set of commuting layers. In all cases the
results are ensemble averages over 100 different realizations of
networks of size $N=500$. Table reproduced figure with permission from
   Ref.~\cite{Boccaletti2006StefanoPRE}. \copyright\, 2006 by the American Physical Society.}\label{table}
\end{table}

Table~\ref{table} summarizes the behavior of these quantities
for the different eigenvalue selection strategies described above. The initial
layer $G_0$ is the Laplacian matrix of a scale-free (SF) network grown using the
preferential attachment algorithm~\cite{BA}, and whose strength distribution
follows a power-law $P(s)\sim s^{-\gamma}$. Figure~\ref{fig:degree} illustrates
the strength distributions of the network layers constructed with different
methods. It is evident that the eigenvalue surrogate method, as well as the
ordered selection of eigenvalues from the initial distribution $p(\lambda_0)$,
\emph{do not} change the SF character of the strength distribution
(thick red and dashed blue lines in panel~a). At the same time, they yield
values for $\langle s\rangle=\sum_{\lbrace s\rbrace}sP(s)$,
$\langle C\rangle$ and $\langle\ell\rangle$ that deviate only slightly from
those of the original graph (see Table~\ref{table}). This indicates that both
methods are convenient strategies to construct a commuting set of layers
through which the evolution of the graph substantially preserves all the main
topological features of the initial condition. Conversely, choosing the eigenvalues
from a uniform distribution, either in an ordered or in a unordered way, produces
layers that completely destroy the original SF strength distribution
(thick red solid line and dashed blue line in panel~b). These correspond to
structural configurations for which each measured quantity wildly differs
from its value in $G_0$. Finally, the intermediate solution of choosing the
eigenvalue spectrum from the initial $p(\lambda_0)$ in a unordered way, while
still preserving the power-law tail of the strength distribution (solid black
line in panel~b), still changes substantially the topological structure of
the underlying network.

\begin{figure}[t!]
\centering
\includegraphics[width=0.7\textwidth]{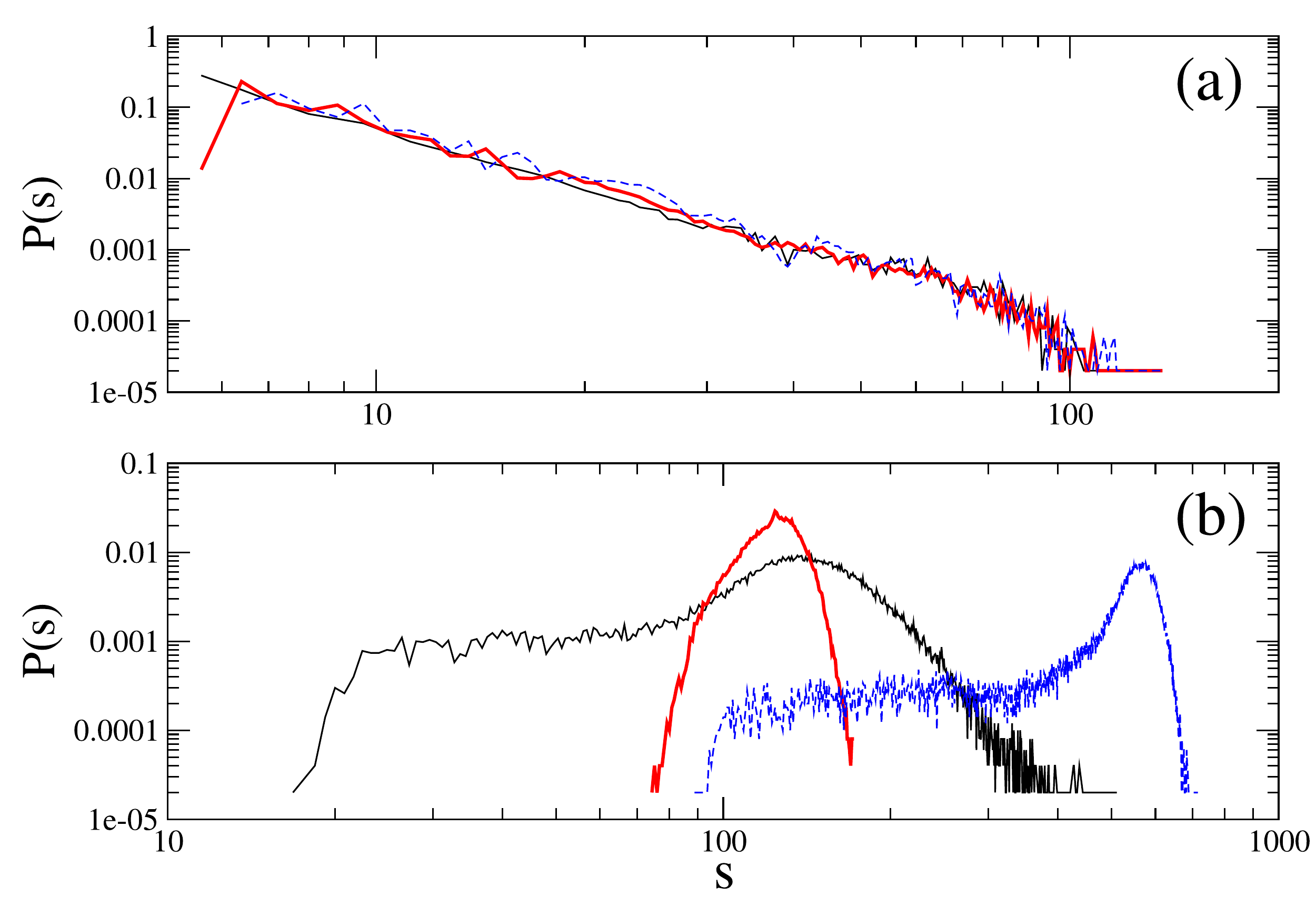}
\caption{\label{fig:degree}(Color online). Strength distributions $P(s)$ of the commuting layers. (a) Initial condition
$G_0$ (solid black line), $G(t)$ constructed by the eigenvalue surrogate
method (thick solid red line), and $\Lambda(t)$ obtained by randomly
choosing the eigenvalues from $p(\lambda_0)$ in an ordered way (dashed
blue line). (b) Eigenvalues randomly chosen from $p(\lambda_0)$ in a
unordered way (solid black line), eigenvalues chosen from a uniform distribution
in an ordered way (thick solid red line), and in a unordered way (dashed
blue line). The details of the statistical ensembles are the same as
for Table~\ref{tab:stat}. Reprinted figure with permission from
   Ref.~\cite{Boccaletti2006StefanoPRE}. \copyright\, 2006 by the American Physical Society.
}
\end{figure}

\subsubsection{Enhancing synchronization via evolution through commutative layers}
Besides the analysis of changes in the structural properties of the layers belonging
to the DCS and generated via different methods, one also needs to study the consequences
of the evolution through commutative layers on the stability of the synchronization
solution.

Because of the commutativity of the layers, Eq.~(\ref{variation})
takes the form given by Eq.~(\ref{vari1}).

Replacing $\sigma\lambda_j(t)$ by $\nu$ in the kernel
$\mathbf K_j$, the problem of stability of the synchronization
manifold becomes equivalent to studying the $m$-dimensional
parametric variational equation
\begin{equation*}
 \dot\eta = \mathbf K_\nu\eta\:,
\end{equation*}
where $\mathbf K_\nu =\left[\mathrm J\mathbf f(\mathbf x_s)-\nu\mathrm J\mathbf h(\mathbf x_s)\right]$.
This allows to plot the curve $\Lambda_{\max}$\textit{vs.}\ $\nu$, where $\Lambda_{\max}$ is the largest
of the $m$ associated conditional Lyapunov exponents. This curve is classically called the Master Stability
Function ~\cite{barahona02}.

For $G(t)=\mbox{constant}$, the synchronized state is transversely
stable if all $\lambda_i$ ($i=2,\dotsc, N$), multiplied by the
same coupling strength $\sigma$, fall in the range where $\Lambda_{\max}(\nu)<0$.
When $G(t)$ evolves inside the DCS, the eigenvectors are fixed
in time, and one can make the hypothesis that at time $T=k\mathrm{dt}$,
the modulus of each eigenmode $\eta_i$ ($i\neq 1$) is bounded
by the corresponding maximum conditional Lyapunov exponent
\begin{equation*}
 \frac{|\eta_i(T)|}{|\eta_i(0)|}\leq\prod_{n=0}^{k-1}\exp\left(\Lambda_{\max}\left(\sigma\lambda_i\left(n \mathrm{dt}\right)\right)\mathrm{dt}\right)\:.
\end{equation*}

It immediately follows that the transverse stability
condition for  the synchronization manifold is now that
$\forall i\neq 1$
\begin{equation}\label{cond2}
\boxed{S_i=\lim_{T\to\infty}\frac{1}{T}\int_0^T\Lambda_{\max}\left(\sigma\lambda_i\left(t'\right)\right)\mathrm{dt'}<0}\:.
\end{equation}

It is extremely important to notice that condition~(\ref{cond2})
\emph{does not} require $\Lambda_{\max}\left(\sigma\lambda_i(t)\right)$
to be negative at all times.

In fact, one can construct a commutative evolution
such that \emph{at each time} there exists at least
one eigenvalue $\lambda_i(t)$ for which
$\Lambda_{\max}\left(\sigma\lambda_i(t)\right)>0$,
and yet obtain a transversely stable synchronization
manifold.

The very relevant conclusion is that a multilayer network
where different layers are visited at different times may
produce a stable synchronized motion even when the emergence
of a synchronous behavior of the units would be prevented
by using any of the layers as a constant wiring configuration.
Similar conclusions were independently reached in Ref.~\cite{Amritkar2006StefanoChaos}
for fast switching times between commutative layers.

A simple example is a periodic evolution between
network layers, with period $T_p=(N-1)\tau$, during
which $G(t)$ is given by
\begin{equation}\label{periodic}
G(t)=\sum_{l=1}^{N-1}G_l\chi_{\left[\left(l-1\right)\tau,l\tau\right)}\:,
\end{equation}
where $\chi_{\left[\left(l-1\right)\tau,l\tau\right)}$ denotes
the characteristic function of the interval $\left[\left(l-1\right)\tau,l\tau\right)$.
{Here, the matrices $G_l$ defining the layers are constructed,
as before, as $G_l=V_1\Lambda_lV_1^{\mathrm{T}}$. Given
$\Lambda_1=\mathrm{diag}(0,\lambda_2,\dotsc,\lambda_N)$, the set of the other $N-2$ matrices $\Lambda_l$
 is constructed by subsequently shifting the sequence of the eigenvalues of
$G_1$, i.e. $\Lambda_2=\mathrm{diag}(0,\lambda_3,\lambda_4,\dotsc,\lambda_N,\lambda_2)$,
$\Lambda_3=\mathrm{diag}(0,\lambda_4,\lambda_5,\dotsc,\lambda_2,\lambda_3)$, etc.

If, for instance, $\Lambda_{\max}\left(\sigma\lambda_2\right)>0$,
then all the layers will have a direction in the phase space along which
the synchronization manifold is transversely unstable. However, if $\sum_{j=2}^{N-1}\Lambda_{\max}\left(\sigma\lambda_j\right)<0$,
then condition~(\ref{cond2}) will be satisfied \emph{in all directions} transverse
to the synchronization manifold, and therefore the synchronized solution
will be transversely stable.

\begin{figure}[t!]
\centering
\includegraphics[width=0.70\textwidth]{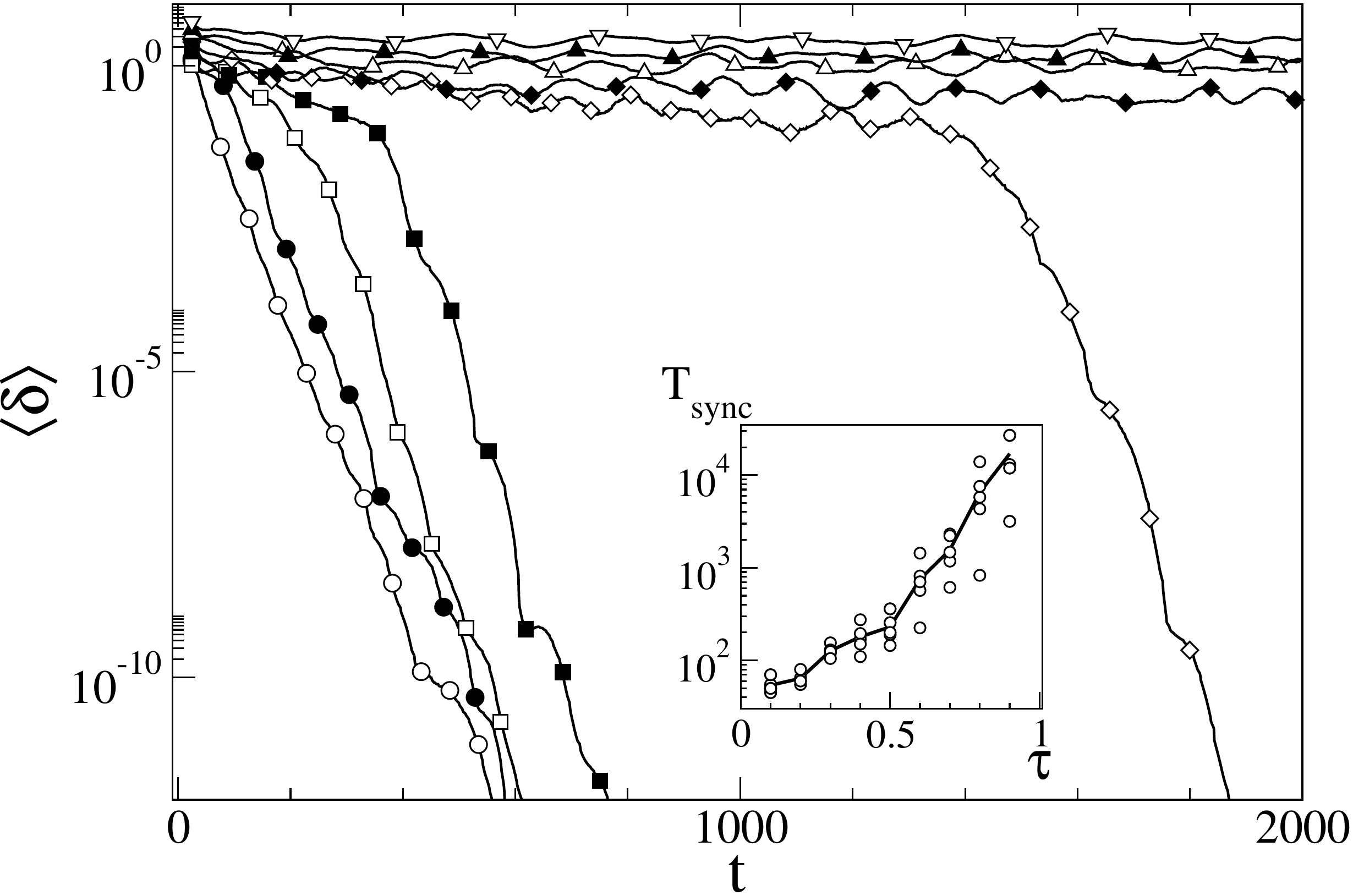}
\caption{\label{fig:degree1} Time evolution of synchronization error $\langle\delta\rangle$ for $\tau=0.2$
(empty circles), $0.3$ (filled circles), $0.4$ (empty squares), $0.5$
(filled squares), $0.6$ (empty diamonds), $0.7$ (filled diamonds), $0.9$
(empty up triangles), $1.1$ (filled up triangles), and $1.5$ (empty down
triangles). In all cases, the results are ensemble averages over~5 different
random initial conditions for a dynamical network of $N=200$ identical
chaotic R\"ossler oscillators. Inset: Synchronization time $T_{sync}$ \textit{vs}.\ switching
time $\tau$ (open circles refer to different initial conditions, solid
line is the ensemble average). Reprinted figure with permission from
 Ref.~\cite{Boccaletti2006StefanoPRE}. \copyright\, 2006 by the American Physical Society.}
\end{figure}

As an example of such an extreme situation, Fig.~\ref{fig:degree1}
shows the results of an evolution of a network where the initial
layer is a SF graph of $N=200$ R\"ossler chaotic oscillators,
each obeying Eq.~(\ref{eq1}) with
\begin{equation*}
\mathbf x\equiv(x,y,z),\ \mathbf f(\mathbf x)=\left(-y-z, x+0.165y, 0.2+z\left(x-10\right)\right),\ \mathbf h[\mathbf x]=y\:.
\end{equation*}
The evolution of the wiring follows Eq.~(\ref{periodic}).
For $\sigma=0.03$ one finds that $\sum_{j=2}^{N-1}\Lambda_{\max}\left(\sigma\lambda_j\right)\approx-9.158$,
but $\Lambda_{\max}\left(\sigma\lambda_j\right)$ is positive
for the first~80 eigenvalues, i.e., the synchronization
manifold would be unstable in \emph{at least~80 different
transverse directions in each of the layers}, if taken
as fixed. However, defining the synchronization error as
\begin{equation*}
\langle\delta\rangle(t)=\sum_{j=2}^N\frac{|x_i-x_1|+|y_i-y_1|+|z_i-z_1|}{3(N-1)}\:,
\end{equation*}
and the synchronization transient $T_{sync}$ as the time
needed for $\langle\delta\rangle(t)$ to become smaller
than $0.1$, we see from the data of Fig.~\ref{fig:degree1}
that the multilayer network is always able to synchronize,
with a $T_{sync}$ that scales almost exponentially with $\tau$.

It is important to remark that, while this numerical
example is used here just to illustrate the enhancement
of synchronization in multilayer commuting graphs under
extreme conditions, the validity of condition~(\ref{cond2})
is general, and by no means limited to periodic evolutions
of the connectivity matrices.

A further point that should be stressed is that the multilayer
commuting case is substantially different from the fast switching
procedure described in Ref.~\cite{blinking}. Indeed, the stability
condition \emph{does not} impose in principle any limitation
on the switching time $\tau$, making it possible to enhance
the collective functioning of the network even for wirings evolving
over secular (yet finite) time scales.

\subsubsection{Non-commutative layers}\label{NCL}
So far, we considered the case of an evolution through commutative layers, clearly demonstrating the importance
of explicitly accounting for a multilayer nature of interactions in networked dynamical systems, and the fact
that such multilayer structure can greatly enhance the stability of the synchronization solution. Yet, it is evident
that considering commutative layers is too strong an assumption to adequately represent the general case encountered
in real world multilayer graphs.

When explicitly dealing with non commutative layers, the boxed term in Eqs.~(\ref{variation}) no longer vanishes.
Thus, Eqs.~(\ref{variation}) do not reduce anymore to a set of variational equations, and one must examine in detail
the effects of this extra term on the coupling of the different directions of stability for the synchronous solution.
To do so, there are some issues that need to be properly dealt with.

The first is that the boxed term in Eqs.~(\ref{variation}) is explicitly proportional
to the time derivatives of the eigenvectors of the network Laplacian, and therefore
any evolution involving instantaneous jumps from a layer to another of a multilayer
structure would result in a divergence. As a consequence, one must consider a quasi-static
evolution that depends on two different time scales: first, the network
structure is kept constant from time $t=0$ to $t=t_0$, period over which it is described
by one given layer of the network; then, from $t=t_0$ to $t=t_1$, a smooth switching
process takes place, bringing the network from one layer to another.

Second, the switching process should be carefully constructed.
Indeed, one needs to find a smooth evolution that, starting from
a generic layer whose Laplacian is $L_0$, leads to another generic
layer whose Laplacian is $L_1$, through a set of intermediate
configurations that all maintain the basic conditions for the
existence of the synchronized solution. In practice, this means
that the switching process should occur through a path of connecting
matrices, each displaying the zero-row sum condition, each being
diagonalizable, and each sharing the same eigenvector corresponding
to the null eigenvalue.

To achieve this, we first consider the matrices $A$ and $B$, consisting
of the eigenvectors of $L_0$ and $L_1$, respectively, and describe how
to transform one into the other by means of a proper rotation, and in particular
a proper rotation around a fixed axis. Then, we use this framework to find
a transformation between $L_0$ and $L_1$.

\paragraph{Eigenvector matrices}
Let $A$ and $B$ be two $N\times N$ orthogonal matrices.
In our case, we know that the vectors of $A$ and $B$ correspond
to the eigenvectors of two network Laplacians. We also
know that at least one vector $a$ is common between $A$
and $B$. This vector $a$ corresponds to the null eigenvalue
of the Laplacian, and thus to the synchronization manifold.

We want to find a one-parameter transformation group $G_s$ such that:
\begin{itemize}
 \item $G_0A=A$;
 \item $G_sA$ contains the vector $a$ for all $0\leqslant s\leqslant 1$;
 \item $G_1A=B$.
\end{itemize}

\paragraph{Existence of a solution}\label{ex}
To show that it is always possible to find a solution, first notice
that, in general, the transformation $O$ from $A$ to $B$ is a rotation,
because the vectors of both matrices form orthonormal bases of the
same space. To find $O$, one can simply solve the linear system of
$N^2$ equations in $N^2$ variables $OA=B$. As we are free of choosing
the set of coordinates, we put ourselves in the basis defined by the
matrix $A$. In this basis, $A\equiv\mathds{1}$, where $\mathds{1}$
is the identity.

Now, without loss of generality, assume that the conserved
vector $a$ is the first vector of $A$, $\left(1,0,\dotsc,0\right)^\mathrm{T}$.
Then, the transformation matrix $O$ will have the form
\begin{equation}\label{tmat} O=
\begin{pmatrix}
 1      & 0      & 0      & \dotsm & 0\\
 0      & O_{22} & O_{23} & \dotsm & O_{2N}\\
 0      & O_{32} & O_{33} & \dotsm & O_{3N}\\
 \vdots & \vdots & \vdots & \ddots & \vdots\\
 0      & O_{N2} & O_{N3} & \dotsm & O_{NN}
\end{pmatrix}\:.
\end{equation}
Notice that, $O$ is a unitary matrix, and therefore the absolute value of  $|O|$ is equal to 1.
 In general $O$ is an element of the orthogonal group $O\left(N\right)$, which
means that it is a proper rotation, if its determinant is $1$, or a composition
of rotations and reflections, if its determinant is $-1$. However, as the form
of $O$ in Eq.~(\ref{tmat}) indicates, its determinant equals the determinant of
the minor $O'$ obtained by removing from $O$ the first row and the first column.
Then, we only need to find a solution to the problem in $N-1$ dimensions. We
will henceforth use primes when referring to objects in $N-1$ dimensions.

From these considerations, we have $G'_0=\mathds{1}'$
and $G'_1=O'$. But $O'$ itself is trivially an orthogonal
matrix, belonging to $O\left(N-1\right)$. Thus, our problem
is equivalent to determining the possibility of finding
a path between the identity and $O'$.

At this point,
we have two cases.

\textbf{Case 1: $\mathbf{\left|O'\right|=1}$}. If the determinant
of $O'$ is positive, then $O'$ belongs to the special orthogonal
group $SO\left(N-1\right)$. In this case, describing a solution is straightforward.
In fact, $SO\left(N-1\right)$ is the identity component of the orthogonal
group. Moreover, as the orthogonal group is a Lie group, $SO\left(N-1\right)$
is not only connected, but also path-connected. This means that for
every orthogonal $\left(N-1\right)\times\left(N-1\right)$ matrix,
there is a continuum of orthogonal matrices of the same dimension
connecting it to the identity. Each point along this path corresponds
to an orthogonal matrix that can be embedded in $SO\left(N\right)$
by adding a $1$ in the top left corner. Since every embedded matrix
along the continuum will keep the vector corresponding to the synchronization
manifold invariant, a parametrization of the path from the identity
to $O$ will provide a direct solution to the original problem.

\textbf{Case 2: $\mathbf{\left|O'\right|=-1}$}. If the determinant
of $O'$ is negative, then $O'$ belongs to $O\left(N-1\right)\setminus SO\left(N-1\right)$.
The same considerations as for Case~1 apply. However, while $O\left(N-1\right)\setminus SO\left(N-1\right)$
is also a connected topological space, the identity does not belong
to it. This means that there is no path that can connect the identity
to $O'$, and we cannot solve the problem the same way.

Yet, in our particular case, we do not really care about the labeling
of the vectors. In other words, provided that we keep the vector $a$ constant,
we can very well swap two vectors in the basis given by the matrix $A$.
This literally just means creating a new matrix $C$ by swapping two rows
in $A$. Of course, this imposes a swap of the corresponding columns in the
transformation matrix $O$ as well. But we also know that swapping two rows
(or columns) in a matrix changes the sign of its determinant. This means
that the new matrix $O'$ does not belong to $O\left(N-1\right)\setminus SO\left(N-1\right)$,
but rather to $SO\left(N-1\right)$. Thus, it is path-connected to the identity,
and all the procedure of Case~1 can be applied. The only consequence of
the swap is that the implicit order of the eigenvalues will be changed.
As this order is irrelevant for our purposes, the problem can always be
solved. To illustrate this, consider the following simple example in $\mathbf{\mathbb{R}^3}$.

Suppose we want to go
from the basis on the left in Fig.~\ref{Fig1} (the canonical basis of $\mathbb{R}^3$)
to the one in the middle. Then, the transformation matrix is
\begin{equation*}
 \bar O=
\begin{pmatrix}
1 & 0 & 0\\
0 & 1 & 0\\
0 & 0 & -1
\end{pmatrix}.
\end{equation*}
The determinant of $\bar O$ is $-1$, so there is no continuous way to go from
the identity to $\bar O$. However, suppose we relabel the $y$ and $z$ vectors
in our original basis, obtaining the basis in the right panel of Fig.~\ref{Fig1}.
Then, the transformation matrix will be
\begin{equation*}
 O=
\begin{pmatrix}
1 & 0  & 0\\
0 & 0  & 1\\
0 & -1 & 0
\end{pmatrix}.
\end{equation*}
This time, its determinant is 1. Thus $O\in SO\left(3\right)$, or, equivalently,
$O'\in SO\left(2\right)$. This means we can find a path that connects the identity
to $O$. This path can be readily turned into a family of transformations:
\begin{equation*}
 G_s =
\begin{pmatrix}
1 & 0                                & 0\\
0 & \cos\left(-\frac{\pi}{2}s\right) & -\sin\left(-\frac{\pi}{2}s\right)\\
0 & \sin\left(-\frac{\pi}{2}s\right) & \cos\left(-\frac{\pi}{2}s\right)
\end{pmatrix}.
\end{equation*}

Note that for every value of the parameter $s$ between 0 and 1, the determinant
of $G_s$ equals 1. Also, for any value of $s$, the vector $x$ remains always
constant. However, the main point here is that while $\bar O$ leaves $y$ unchanged
and reflects $z$ across the origin, $O$ instead rotates the ``new $y$'' onto
the position previously occupied by $y$, and the ``new $z$'' onto the position
opposite to that previously occupied by $z$. In other words, the final arrangement
of the original transformation is preserved, but this time the transformation
is a proper rotation. As we are only interested in getting the final result
right, and not in the order of the eigenvectors, we can disregard this difference.

\begin{figure}
\centering
\includegraphics[width=0.75\textwidth]{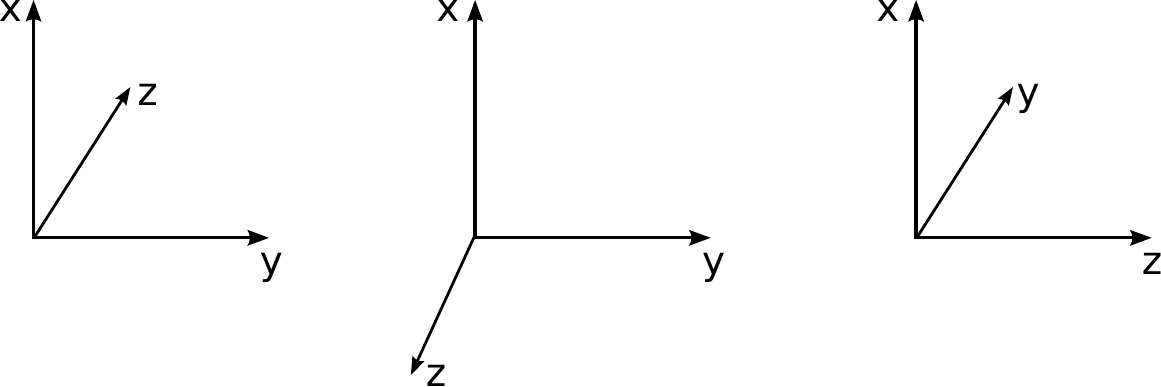}
\caption{\label{Fig1} Example of label swapping. Going from the basis
on the left to the one in the middle while keeping $x$ constant necessarily
involves a reflection. However, going from the basis on the right to
the one in the middle can be accomplished via a proper rotation.}
\end{figure}

\subsubsection{Building a solution for a generic layer switching}
Given a specific $A$ and $B$, cast in the form discussed above,
we build an explicit solution by factoring the transformation $O'$
into rotations and reflections in mutually orthogonal subspaces.
Before describing the actual procedure, we recall two useful results.
Proving these results involves constructions we will use in the
solution and provides further insight into our original problem.
Thus, we present the proofs in the following.

\paragraph{Existence of invariant subspaces}
First, any orthogonal operator $X$ in a normed space
over $\mathbb{R}$ induces an invariant subspace. This
subspace is either 1-dimensional or 2-dimensional. To
see this, given $X$, define the operator $U$ that acts
on $a+\mathrm ib$ as $U\left(a+\mathrm ib\right)=Xa+\mathrm iXb$.
As $U$ is a unitary operator, its spectrum lies on the
unit circle. Then, one can find a non-vanishing eigenvector
$x$ with eigenvalue $\lambda$. In general, $x=x_R+\mathrm ix_I$,
and $\lambda=\mathrm e^{\mathrm i\vartheta}=\lambda_R+\mathrm i\lambda_I$.
Then, we have two cases.

\textbf{Case 1: $\mathbf\lambda$ is real}. In this case, $Ux=Xx_R+\mathrm iXx_I=\pm x_R\pm\mathrm ix_I$.
Then, $Xx_R=\pm x_R$ and $Xx_I=\pm x_I$. As $x\neq\vec0$, at least one between $x_R$ and $x_I$ cannot
vanish. Without loss of generality, let the non vanishing component be $x_R$. Then, all vectors proportional
to $x_R$ are mapped by $X$ into vectors proportional to $x_R$. Thus, the 1-dimensional subspace of
the vectors proportional to $x_R$ is invariant under $X$.

\textbf{Case 2: $\mathbf\lambda$ is complex}. In this case, $x_R$ and $x_I$ are linearly independent.
We prove this by contradiction. First note that $\bar x\equiv x_R-\mathrm ix_I$ is an eigenvector of $U$ with
eigenvalue $\bar\lambda=\mathrm e^{-\mathrm i\vartheta}=\lambda_R-\mathrm i\lambda_I$, and that $x_R=x-\mathrm ix_I=\bar x+\mathrm ix_I$.

Now assume that it were $x_R=\alpha x_I$. Then one could write
\begin{equation*}
 \alpha Ux_I = Ux_R = U\left(x-\mathrm ix_I\right) = \lambda x-\mathrm iUx_I\:,
\end{equation*}
hence
\begin{equation*}
 \left(\alpha+\mathrm i\right)Ux_I = \left(\alpha+\mathrm i\right)Xx_I = \lambda x = \lambda\left(\alpha x_I+\mathrm ix_I\right) = \left(\alpha+i\right)\lambda x_I\:,
\end{equation*}
from which it would follow that
\begin{equation}\label{lone}
 Xx_I=\lambda x_I.
\end{equation}

Similarly, one could write
\begin{equation*}
 \alpha Ux_I = Ux_R = U\left(\bar x+\mathrm ix_I\right) = \bar\lambda\bar x+\mathrm iUx_I\:,
\end{equation*}
hence
\begin{equation*}
 \left(\alpha-\mathrm i\right)Ux_I = \left(\alpha-\mathrm i\right)Xx_I = \bar\lambda\bar x = \bar\lambda\left(\alpha x_I-\mathrm ix_I\right) = \left(\alpha-\mathrm i\right)\bar\lambda x_I\:,
\end{equation*}
from which it would follow that
\begin{equation}\label{ltwo}
 Xx_I=\bar\lambda x_I.
\end{equation}

Thus, Eqs.~(\ref{lone}) and~(\ref{ltwo}) would imply that $\lambda=\bar\lambda$.
But this directly contradicts the hypothesis that $\lambda\notin\mathbb{R}$.

Having proved that $x_R$ and $x_I$ are linearly independent, apply $U$ to $x$:
\begin{equation*}
 Ux = Xx_R+\mathrm iXx_I=\lambda x=\lambda_Rx_R-\lambda_Ix_I+\mathrm i\left(\lambda_Ix_R+\lambda_Rx_I\right)\:,
\end{equation*}
hence
\begin{equation}\label{xxr}
 Xx_R = \lambda_Rx_R-\lambda_Ix_I
\end{equation}
and
\begin{equation}\label{xxi}
 Xx_I = \lambda_Ix_R+\lambda_Rx_I\:.
\end{equation}

The two equations above imply that
\begin{multline*}
 X\left(\alpha x_R+\beta x_I\right) = \alpha\lambda_Rx_R-\alpha\lambda_Ix_I+\beta\lambda_Ix_R+\beta\lambda_Rx_I=\\
\left(\alpha\lambda_R+\beta\lambda_I\right)x_R+\left(\beta\lambda_R-\alpha\lambda_I\right)x_I=\gamma x_R+\delta x_I\:,
\end{multline*}
where $\gamma=\alpha\lambda_R+\beta\lambda_I$ and $\delta=\beta\lambda_R-\alpha\lambda_I$.
This means that any linear combination of $x_R$ and $x_I$ is mapped by $X$ into a linear combination
of the same vectors $x_R$ and $x_I$. As we have proved above that they are linearly independent,
this implies that the 2-dimensional subspace determined by the basis $\left\lbrace x_R,x_I\right\rbrace$
is invariant under application of $X$.

\paragraph{Orthogonality of the eigenvector components}
We show now that the components $x_R$ and $x_I$ of the eigenvector
$x$ used in Case~2 of the previous proof are not only linearly independent,
but actually orthogonal. To see this, compute the inner product between
$x_R$ and $x_I$, exploiting the fact that the operator $X$ preserves
the inner product, since it is orthogonal:
\begin{equation}\label{barbatrucco}
 \la x_R, x_I\ra = \la Xx_R, Xx_I\ra = \la\lambda_Rx_R - \lambda_Ix_I, \lambda_Ix_R + \lambda_Rx_I\ra,
\end{equation}
where we have used Eqs.~(\ref{xxr}) and~(\ref{xxi}).

Then it is
\begin{multline*}
 \la x_R, x_I\ra = \lambda_R\lambda_I\left\|x_R\right\|^2 + \lambda_R^2\la x_R, x_I\ra - \lambda_I^2\la x_I, x_R\ra - \lambda_R\lambda_I\left\|x_I\right\|^2 =\\ \lambda_R\lambda_I\left(\left\|x_R\right\|^2-\left\|x_I\right\|^2\right)+\left(\lambda_R^2-\lambda_I^2\right)\la x_R, x_I\ra\:,
\end{multline*}
hence
\begin{equation}\label{inprod}
\begin{split}
 \la x_R, x_I\ra &= \frac{\lambda_R\lambda_I}{1-\lambda_R^2+\lambda_I^2}\left(\left\|x_R\right\|^2-\left\|x_I\right\|^2\right)\\
 &= \frac{\lambda_R\lambda_I}{2\lambda_I^2}\left(\left\|x_R\right\|^2-\left\|x_I\right\|^2\right)\\
 &= \frac{\lambda_R}{2\lambda_I}\left(\left\|x_R\right\|^2-\left\|x_I\right\|^2\right)\:,
\end{split}
\end{equation}
where we have used $\lambda_R^2+\lambda_I^2=1$.

Similarly, applying the same property of $X$, and using Eq.~(\ref{xxr}), the squared norm of $x_R$ is
\begin{equation*}
\begin{split}
 \left\|x_R\right\|^2 &= \la x_R, x_R\ra\\
&= \la\lambda_Rx_R - \lambda_Ix_I, \lambda_Rx_R - \lambda_Ix_I\ra\\
&= \lambda_R^2\left\|x_R\right\|^2 - \lambda_R\lambda_I\la x_R, x_I\ra - \lambda_R\lambda_I\la x_I, x_R\ra + \lambda_I^2\left\|x_I\right\|^2\\
&= \lambda_R^2\left\|x_R\right\|^2 - 2\lambda_R\lambda_I\la x_R, x_I\ra + \lambda_I^2\left\|x_I\right\|^2\\
&= \lambda_R^2\left\|x_R\right\|^2 + \left(1-\lambda_R^2\right)\left\|x_I\right\|^2 - 2\lambda_R\lambda_I\la x_R, x_I\ra\\
&= \lambda_R^2\left(\left\|x_R\right\|^2-\left\|x_I\right\|^2\right) + \left\|x_I\right\|^2 - 2\lambda_R\lambda_I\la x_R, x_I\ra\:.
\end{split}
\end{equation*}
Then, it is
\begin{equation*}
\left\|x_R\right\|^2-\left\|x_I\right\|^2 = \lambda_R^2\left(\left\|x_R\right\|^2-\left\|x_I\right\|^2\right) - 2\lambda_R\lambda_I\la x_R, x_I\ra\:,
\end{equation*}
and, using again $\lambda_R^2+\lambda_I^2=1$,
\begin{equation}\label{norms}
 \left\|x_R\right\|^2-\left\|x_I\right\|^2 = -\frac{2\lambda_R}{\lambda_I}\la x_R, x_I\ra\:.
\end{equation}

Substituting Eq.~(\ref{norms}) into Eq.~(\ref{inprod}) yields
\begin{equation*}
 \la x_R, x_I\ra = -\frac{\lambda_R^2}{\lambda_I^2}\la x_R, x_I\ra\:.
\end{equation*}

Now, we know by hypothesis that $\lambda_I\neq 0$. Then, if $\lambda_R\neq 0$
it must be $\la x_R, x_I\ra=0$. Alternatively, if $\lambda_R=0$, then it
is $\lambda_I=\pm1$. In this case, from Eq.~(\ref{barbatrucco}), it follows
that
\begin{equation*}
 \la x_R, x_I\ra = \la \mp x_I, \pm x_R\ra = -\la x_I,x_R\ra\Rightarrow\la x_R,x_I\ra=0\:.
\end{equation*}
Either way, $x_R$ and $x_I$ are orthogonal.

\paragraph{The solution}
Using the two results just shown, we can describe
an algorithmic procedure to find the solution to our problem. The
initial situation is that we have two orthonormal bases $A$ and
$B$ of $\mathbb{R}^N$. Then, we can build a transformation
of $A$ into $B$ with the following steps:

\begin{enumerate}
 \item Express the problem in the basis $A$. In this basis, the transformation matrix
$O$ between $A=\mathds{1}$ and $B$ takes the form given by Eq.~(\ref{tmat}).
 \item Consider the operator $O'$ obtained from $O$ by removing the first row and the first column. Let $d$ be its dimension.
 \item Build the operator $U$ that acts on $a+\mathrm ib$ as $U\left(a+\mathrm ib\right)=O'a+\mathrm iO'b$,
where $a$ and $b$ belong to $\mathbb{R}^d$.
 \item Find an eigenvector $x=x_R+\mathrm ix_I$ of $U$, with eigenvalue $\lambda$.
 \item Normalize $x_R$ and $x_I$.
 \item If $\lambda\in\mathbb{R}$ \begin{enumerate}
    \item Pick the non-vanishing component between $x_R$ and $x_I$. If both are non-zero, choose one of them randomly. Without loss of generality, assume this is $x_R$.
    \item Create $d-1$ other orthonormal vectors, all orthogonal to $x_R$, and arrange all these vectors so that $x_R$ is the last of them. This set of vectors is an orthonormal basis $C$ of $\mathbb{R}^d$.
    \item Change the basis of the $d$-dimensional sub-problem to $C$. In this basis, all the elements in the last row and in the last column of $O'$ will be 0, except the last one, which will be $\pm 1$.
    \item If $d>1$, consider a new operator $O'$ obtained from the old $O'$ by removing the last row and the last column. Let $d$ be its dimension, and restart from step 3. Otherwise stop.
    \end{enumerate}
 \item If $\lambda\notin\mathbb{R}$ \begin{enumerate}
    \item Create $d-2$ other orthonormal vectors, all orthogonal to $x_R$ and $x_I$, and arrange all these vectors so that $x_R$ and $x_I$ are the first two of them. This set of vectors is an orthonormal basis $C$ of $\mathbb{R}^d$.
    \item Change the basis of the $d$-dimensional sub-problem to $C$. In this basis, all the elements in the first two rows and in the first two columns of $O'$ will be 0, except the first two.
    \item If $d>2$, consider a new operator $O'$ obtained from the old $O'$ by removing the first two rows and the first two columns. Let $d$ be its dimension, and restart from step 3. Otherwise stop.
\end{enumerate}
\end{enumerate}

The steps described above effectively solve the problem of finding the basis transformation by divide-and-conquer.
At each repetition, 1 or 2 dimensions are eliminated from the problem. All the subsequent
changes of bases do not affect the elements of the transformation matrix determined in
the previous steps, because they act on subspaces that are orthogonal to those that have been already eliminated,
which include the top left element in $A$ and $B$, corresponding to the synchronization
manifold. Thus, one can reconstruct the transformation matrix $O$ piece by piece. At the
end of the procedure, $O$ has a block-diagonal form

\begin{equation*}O=
\begin{pmatrix}
 1      & 0      & 0      & 0      & 0      & \dotsm & 0           & 0\\
 0      & O_{22} & O_{23} & 0      & 0      & \dotsm & 0           & 0\\
 0      & O_{32} & O_{33} & 0      & 0      & \dotsm & 0           & 0\\
 0      & 0      & 0      & O_{44} & O_{45} & \dotsm & 0           & 0\\
 0      & 0      & 0      & O_{54} & O_{55} & \dotsm & 0           & 0\\
 \vdots & \vdots & \vdots & \vdots & \vdots & \ddots & \vdots      & \vdots\\
 0      & 0      & 0      & 0      & 0      & \dotsm & O_{N-1 N-1} & O_{N-1 N}\\
 0      & 0      & 0      & 0      & 0      & \dotsm & O_{N N-1}   & O_{N N}\\
\end{pmatrix}\:,
\end{equation*}
where the blocks correspond to the action of the orthogonal operator on the invariant
subspaces. If the invariant subspace is 1-dimensional, then its block is a single $\pm1$
element. Conversely, if the subspace is 2-dimensional, then its block is either a rotation
or a reflection. This means that it is either

\begin{equation*}
\begin{pmatrix}
 \cos\alpha & -\sin\alpha\\
 \sin\alpha & \cos\alpha\\
\end{pmatrix}
\end{equation*}
or
\begin{equation*}
\begin{pmatrix}
 \pm 1 & 0\\
 0     & \mp 1\\
\end{pmatrix}\:.
\end{equation*}

At this point, one can proceed as follows. First, permute the basis vectors
from the second onwards so that they correspond, in order, first to all the
actual rotation blocks, then to the $-1$ elements, and finally to the $+1$
elements. It's easy to see that this can always be done. In fact, if we call
our change-of-basis matrix $T$, we can determine the new form of the transformation
matrix, $O_N$, via the following logical chain:
\begin{equation*}
OKA=KB\Rightarrow O_NTKA=TKB\Rightarrow T^{-1}O_NTKA=KB\Rightarrow T^{-1}O_NT=O\Rightarrow O_N=TOT^{-1}\:,
\end{equation*}
where $K$ is the change-of-basis matrix corresponding to the composition of
all changes of bases done in steps~6 or~7 above.

As the matrix $T$ that swaps
the position of the basis vectors is a permutation matrix, it follows that
$O_N$ has exactly the required form
\begin{equation}\label{almost}
O=
\begin{psmallmatrix}
 1 & 0 & 0 & \dotsm & 0 & 0 & 0 & 0 & 0 & \dotsm & 0 & 0\\
 0 & \cos\vartheta_1 & -\sin\vartheta_1 & \dotsm & 0 & 0 & 0 & 0 & 0 & \dotsm & 0 & 0\\
 0 & \sin\vartheta_1 & \cos\vartheta_1 & \dotsm & 0 & 0 & 0 & 0 & 0 & \dotsm & 0 & 0\\
 \scriptscriptstyle \vdots & \scriptscriptstyle \vdots & \scriptscriptstyle \vdots & \scriptscriptstyle \ddots & 0 & 0 & 0 & 0 & 0 & \dotsm & 0 & 0\\
 0 & 0 & 0 & 0 & \cos\vartheta_k & -\sin\vartheta_k & 0 & 0 & 0 & \dotsm & 0 & 0\\
 0 & 0 & 0 & 0 & \sin\vartheta_k & \cos\vartheta_k & 0 & 0 & 0 & \dotsm & 0 & 0\\
 0 & 0 & 0 & 0 & 0 & 0 & -1 & 0 & 0 & \dotsm & 0 & 0\\
 0 & 0 & 0 & 0 & 0 & 0 & 0 & -1 & 0 & \dotsm & 0 & 0\\
 \scriptscriptstyle \vdots & \scriptscriptstyle \vdots & \scriptscriptstyle \vdots & \scriptscriptstyle \vdots & \scriptscriptstyle \vdots & \scriptscriptstyle \vdots & \scriptscriptstyle \vdots & \scriptscriptstyle \vdots & \scriptscriptstyle \vdots & \scriptscriptstyle \ddots & \scriptscriptstyle \vdots & \scriptscriptstyle \vdots\\
 0 & 0 & 0 & 0 & 0 & 0 & 0 & 0 & 0 & \dotsm & 1 & 0\\
 0 & 0 & 0 & 0 & 0 & 0 & 0 & 0 & 0 & \dotsm & 0 & 1\\
\end{psmallmatrix}\:.
\end{equation}

Having obtained the form in Eq.~(\ref{almost}), there is still the possibility
that the determinant of $O$ is $-1$. However, as we saw, in this case we can
approach the problem by relabeling two vectors of the original basis, which
induces a swap of the corresponding columns of $O$. To perform this, first note
that, if $\left|O\right|=-1$, then the number of $-1$ elements in $O$ must be
odd. Then, there are three possible cases.

\textbf{Case 1: $O'$ has at least one $+1$ element}.
In this case, permute once more the basis vectors, swapping
those corresponding to the first $-1$ and the first $+1$ elements
in $O'$. Then, the first block after the ``$\sin$--$\cos$''
blocks is $\begin{pmatrix} 1&0\\0&-1\end{pmatrix}$. Now, relabel
the two corresponding vectors, as in the $\mathbb{R}^3$ example,
swapping the relevant rows in the starting matrix. This makes
the block in $O'$ become
\begin{equation*}
 \begin{pmatrix}
  0 & 1\\
  -1 & 0
 \end{pmatrix}
=\begin{pmatrix}
  \cos\left(-\frac{\pi}{2}\right) & -\sin\left(-\frac{\pi}{2}\right)\\
  \sin\left(-\frac{\pi}{2}\right) & \cos\left(-\frac{\pi}{2}\right)
 \end{pmatrix}\:.
\end{equation*}

\textbf{Case 2: $O'$ has no $+1$ elements and only one $-1$ element}.
In this case, the basis vectors to swap are those corresponding to the
first two vectors in the last $3\times 3$ block of $O'$. This will become
\begin{equation*}
 M = \begin{pmatrix}
      -\sin\vartheta_k & \cos\vartheta_k & 0\\
      \cos\vartheta_k  & \sin\vartheta_k & 0\\
      0 & 0 & -1
     \end{pmatrix}\:.
\end{equation*}

Now, change once more the basis of the problem. In particular, leave all
the basis vectors unchanged, but map the last three into the eigenvectors
of $M$. This is easily accomplished, as the matrix of eigenvectors of
$M$ is
\begin{equation*}
 V = \begin{pmatrix}
      0 & \left(-\sec\vartheta_k-\tan\vartheta_k\right)\sin\left(\frac{\pi}{4}-\frac{\vartheta_k}{2}\right) & \left(\sec\vartheta_k-\tan\vartheta_k\right)\sin\left(\frac{\vartheta_k}{2}+\frac{\pi}{4}\right)\\
      0 & 1 & 1\\
      1 & 0 & 0
     \end{pmatrix}\:.
\end{equation*}
Then, the new form of $M$ is
\begin{equation*}
 M' = V^\mathrm{T}MV = \begin{pmatrix}
                        -1 & 0  & 0\\
			0  & -1 & 0\\
			0  & 0  & 1
                       \end{pmatrix}\:.
\end{equation*}

\textbf{Case 3: $O'$ has no $+1$ elements and at least three $-1$ elements}.
Similar to the previous case, swap the basis vectors corresponding to the first
two after the ``$\sin$--$\cos$'' blocks. Their $3\times 3$ block is now
\begin{equation*}
M = \begin{pmatrix}
  0  & -1 & 0\\
  -1 & 0  & 0\\
  0  & 0  & -1
 \end{pmatrix}\:.
\end{equation*}
Next, do a change of basis as described in Case~2. The eigenvector matrix of $M$ is
\begin{equation*}
 V = \begin{pmatrix}
      0 & \frac{1}{\sqrt{2}} & -\frac{1}{\sqrt{2}}\\
      0 & \frac{1}{\sqrt{2}} & \frac{1}{\sqrt{2}}\\
      1 & 0 & 0
     \end{pmatrix}\:,
\end{equation*}
so after the basis change the new form of $M$ is once more
\begin{equation*}
 M' = V^\mathrm{T}MV = \begin{pmatrix}
                        -1 & 0  & 0\\
			0  & -1 & 0\\
			0  & 0  & 1
                       \end{pmatrix}\:.
\end{equation*}

Whether the determinant of $O$ was $+1$ to start with, or it was $-1$
and taken care of as one of the cases above, there are now either no
$-1$ elements in $O$, or an even number of them. Then, if there are any,
take every two subsequent $-1$ elements and change their diagonal blocks
into
\begin{equation*}
\begin{pmatrix}
 \cos\pi & -\sin\pi\\
 \sin\pi & \cos\pi\\
\end{pmatrix}\:.
\end{equation*}

This yields the final general form for the transformation matrix:
\begin{equation}\label{TADA}
O=
\begin{psmallmatrix}
 1 & 0 & 0 & \dotsm & 0 & 0 & 0 & 0 & 0 & 0 & 0 & \dotsm & 0 & 0\\
 0 & \cos\vartheta_1 & -\sin\vartheta_1 & \dotsm & 0 & 0 & 0 & 0 & 0 & 0 & 0 & \dotsm & 0 & 0\\
 0 & \sin\vartheta_1 & \cos\vartheta_1 & \dotsm & 0 & 0 & 0 & 0 & 0 & 0 & 0 & \dotsm & 0 & 0\\
 \scriptscriptstyle \vdots & \scriptscriptstyle \vdots & \scriptscriptstyle \vdots & \scriptscriptstyle \ddots & 0 & 0 & 0 & 0 & 0 & 0 & 0 & \dotsm & 0 & 0\\
 0 & 0 & 0 & 0 & \cos\vartheta_k & -\sin\vartheta_k & 0 & 0 & 0 & 0 & 0 & \dotsm & 0 & 0\\
 0 & 0 & 0 & 0 & \sin\vartheta_k & \cos\vartheta_k & 0 & 0 & 0 & 0 & 0 & \dotsm & 0 & 0\\
 0 & 0 & 0 & 0 & 0 & 0 & \cos\left(-\frac{\pi}{2}\right) & -\sin\left(-\frac{\pi}{2}\right) & 0 & 0 & 0 & \dotsm & 0 & 0\\
 0 & 0 & 0 & 0 & 0 & 0 & \sin\left(-\frac{\pi}{2}\right) & \cos\left(-\frac{\pi}{2}\right) & 0 & 0 & 0 & \dotsm & 0 & 0\\
 0 & 0 & 0 & 0 & 0 & 0 & 0 & 0 & \cos\pi & -\sin\pi & 0 & \dotsm & 0 & 0\\
 0 & 0 & 0 & 0 & 0 & 0 & 0 & 0 & \sin\pi & \cos\pi & 0 & \dotsm & 0 & 0\\
 \scriptscriptstyle \vdots & \scriptscriptstyle \vdots & \scriptscriptstyle \vdots & \scriptscriptstyle \vdots & \scriptscriptstyle \vdots & \scriptscriptstyle \vdots & \scriptscriptstyle \vdots & \scriptscriptstyle \vdots & \scriptscriptstyle \vdots & \scriptscriptstyle \vdots & \scriptscriptstyle \vdots & \scriptscriptstyle \ddots & \scriptscriptstyle \vdots & \scriptscriptstyle \vdots\\
 0 & 0 & 0 & 0 & 0 & 0 & 0 & 0 & 0 & 0 & 0 & \dotsm & 1 & 0\\
 0 & 0 & 0 & 0 & 0 & 0 & 0 & 0 & 0 & 0 & 0 & \dotsm & 0 & 1\\
\end{psmallmatrix}\:.
\end{equation}

Finally, introducing a parameter $s\in\left[0,1\right]$, we get the required transformation:
\begin{equation}
G_s=
\begin{psmallmatrix}
 1 & 0 & 0 & \dotsm & 0 & 0 & 0 & 0 & 0 & 0 & 0 & \dotsm & 0 & 0\\
 0 & \cos\left(\vartheta_1s\right) & -\sin\left(\vartheta_1s\right) & \dotsm & 0 & 0 & 0 & 0 & 0 & 0 & 0 & \dotsm & 0 & 0\\
 0 & \sin\left(\vartheta_1s\right) & \cos\left(\vartheta_1s\right) & \dotsm & 0 & 0 & 0 & 0 & 0 & 0 & 0 & \dotsm & 0 & 0\\
 \scriptscriptstyle \vdots & \scriptscriptstyle \vdots & \scriptscriptstyle \vdots & \scriptscriptstyle \ddots & 0 & 0 & 0 & 0 & 0 & 0 & 0 & \dotsm & 0 & 0\\
 0 & 0 & 0 & 0 & \cos\left(\vartheta_ks\right) & -\sin\left(\vartheta_ks\right) & 0 & 0 & 0 & 0 & 0 & \dotsm & 0 & 0\\
 0 & 0 & 0 & 0 & \sin\left(\vartheta_ks\right) & \cos\left(\vartheta_ks\right) & 0 & 0 & 0 & 0 & 0 & \dotsm & 0 & 0\\
 0 & 0 & 0 & 0 & 0 & 0 & \cos\left(-\frac{\pi}{2}s\right) & -\sin\left(-\frac{\pi}{2}s\right) & 0 & 0 & 0 & \dotsm & 0 & 0\\
 0 & 0 & 0 & 0 & 0 & 0 & \sin\left(-\frac{\pi}{2}s\right) & \cos\left(-\frac{\pi}{2}s\right) & 0 & 0 & 0 & \dotsm & 0 & 0\\
 0 & 0 & 0 & 0 & 0 & 0 & 0 & 0 & \cos\left(\pi s\right) & -\sin\left(\pi s\right) & 0 & \dotsm & 0 & 0\\
 0 & 0 & 0 & 0 & 0 & 0 & 0 & 0 & \sin\left(\pi s\right) & \cos\left(\pi s\right) & 0 & \dotsm & 0 & 0\\
 \scriptscriptstyle \vdots & \scriptscriptstyle \vdots & \scriptscriptstyle \vdots & \scriptscriptstyle \vdots & \scriptscriptstyle \vdots & \scriptscriptstyle \vdots & \scriptscriptstyle \vdots & \scriptscriptstyle \vdots & \scriptscriptstyle \vdots & \scriptscriptstyle \vdots & \scriptscriptstyle \vdots & \scriptscriptstyle \ddots & \scriptscriptstyle \vdots & \scriptscriptstyle \vdots\\
 0 & 0 & 0 & 0 & 0 & 0 & 0 & 0 & 0 & 0 & 0 & \dotsm & 1 & 0\\
 0 & 0 & 0 & 0 & 0 & 0 & 0 & 0 & 0 & 0 & 0 & \dotsm & 0 & 1\\
\end{psmallmatrix}\:.
\end{equation}

Notice that when $s=0$, $G_s=G_0=\mathds{1}$, and when $s=1$, $G_s=G_1=O$.
Also, for every value of $s$, the determinant of $G_s$ is always 1, and the
first vector is kept constant, which means that $G_s$ always describes
a proper rotation around the axis defined by the eigenvector corresponding
to the synchronization manifold. Moreover, note that the first vector has always
been left untouched by all the possible basis transformations. Thus, as $s$
is varied continuously between~0 and~1, the application of $G_s$ sends $A$
into $B$ smoothly and continuously, as we need.

\paragraph{Laplacian transformation}
To describe how to obtain a transformation
between the two Laplacians $L_0$ and $L_1$,
let us first summarize what we have shown.

We started from the matrices $A$ and $B$, consisting
of the eigenvectors of $L_0$ and $L_1$. These two matrices
are orthonormal bases of $\mathbb{R}^N$. We showed that
we can always find a series of basis changes casting
the problem in a reference frame where we can explicitly
construct a 1-parameter transformation group $G_s$ that
transforms $A$ into $B$ with continuity via a rotation
around an axis determined by the eigenvector corresponding
to the synchronization manifold.

In formulae, the group
$G_s$ is such that $G_0=\mathds 1$ and $G_1RA=RB$, where
$R$ is the change-of-basis matrix resulting from the
composition of all basis changes done in steps~6 and~7
of the method, the permutation of the vectors to obtain
$O$, and the eventual final adjustment in case of negative
determinant. To simplify the formalism, in the following
we let $B_0\equiv A$ and $B_1\equiv B$.

Then, since $B_0$ and $B_1$ are matrices of eigenvectors,
it is
\begin{align*}
 B_0^{\mathrm T}L_0B_0 &= D_0\\
 B_1^{\mathrm T}L_1B_1 &= D_1\:,
\end{align*}
where $D_0$ and $D_1$ are diagonal matrices
whose elements are the eigenvalues of
$L_0$ and $L_1$.

However, as $G_s$ is a group
of orthogonal transformation, we can write
that
\begin{equation*}
 G_sRB_0=RB_s\quad\forall0\leqslant s\leqslant 1\:,
\end{equation*}
where $B_s$ is a basis of $\mathbb{R}^N$.
Multiplying the equation above by $R^{\mathrm T}$
on the left, we get
\begin{equation}\label{BTdef}
 R^{\mathrm T}G_sRB_0 = R^{\mathrm T}RB_s = B_s\:,
\end{equation}
where we have used the fact that $R^{\mathrm T}=R^{-1}$.

Now, note that all the basis changes done to build
$G$ are between orthonormal bases. Therefore, their
composition $R$ is an isometry, as is any $G_s$, since
they are all proper rigid rotations. Thus, any $B_s$
still defines an orthonormal basis of $\mathbb{R}^N$.

Then, we can say that the Laplacian
for the parameter $s$ is given by the matrix
$L_s$ that solves the equation
\begin{equation}\label{BTDT}
 B_s^{\mathrm T}L_sB_s=D_s\:,
\end{equation}
where $D_s$ is a diagonal matrix whose
elements are the eigenvalues of $L_s$,
and $B_s$ consists of the eigenvectors
of $L_s$. However, the equation above
has two unknowns, namely $L_s$ and $D_s$.

To solve this last problem we impose
a linear evolution of the eigenvalues
of the Laplacians. Thus, for all
$1\leqslant i\leqslant N$,
\begin{equation}\label{valueevol}
 \lambda_i^{(s)} = \left(1-s\right)\lambda_i^{(0)}+s\lambda_i^{(1)}\:,
\end{equation}
where $\lambda_i^{(0)}$ and $\lambda_i^{(1)}$
are the $i^\mathrm{th}$ eigenvalues of $L_0$ and $L_1$, respectively.
Then, multiplying Eq.~(\ref{BTDT}) on the right by $B_s^{\mathrm T}$,
it is
\begin{equation*}
 B_s^{\mathrm T}L_sB_sB_s^{\mathrm T}=D_sB_s^{\mathrm T}\:,
\end{equation*}
hence
\begin{equation*}
 B_s^{\mathrm T}L_s=D_sB_s^{\mathrm T}\:.
\end{equation*}
Multiplying this on the left by $B_s$, we get
\begin{equation*}
 B_sB_s^{\mathrm T}L_s=B_sD_sB_s^{\mathrm T}\:,
\end{equation*}
hence
\begin{equation*}
 L_s=B_sD_sB_s^{\mathrm T}\:.
\end{equation*}
Substituting Eq.~(\ref{BTdef}) into the equation above, we get
\begin{equation}\label{LvsD}
 L_s = R^\mathrm{T}G_sRB_0D_s\left(R^\mathrm{T}G_sRB_0\right)^\mathrm{T} = R^\mathrm{T}G_sRB_0D_sB_0^\mathrm{T}R^\mathrm{T}G_s^\mathrm{T}R\:.
\end{equation}

In Eq.~(\ref{LvsD}), $B_0$ is known; $R$ and $G_s$ have been explicitly
built, so they are known as well; $D_s$ is completely determined by Eq.~(\ref{valueevol}).
Thus, Eq.~(\ref{LvsD}) defines the Laplacian for any given value of the parameter
$0\leqslant s\leqslant 1$. The evolution of the eigenvalues follows a simple
linear form, imposed by Eq.~(\ref{valueevol}), and the evolution of the eigenvectors
is given by Eq.~(\ref{BTdef}).

As a consistency check, it is straightforward to treat the left-hand side
of Eq.~(\ref{LvsD}) as an unknown and solve it for either $s=0$ or $s=1$, recovering
$L_0$ and $L_1$, respectively. Similarly, it is easy to see that for any
$s$ the determinant of $L_s$ is the product of the eigenvalues defined by
Eq.~(\ref{valueevol}).

With this result, we can finally describe the evolution via non-commutative layers.
We use the setting introduced in Sec.~\ref{NCL}: at time $t=0$, the network
starts in a layer whose Laplacian is $L_0$, in which it remains up to time
$t_0$; then, it switches to a layer whose Laplacian is $L_1$, completing the switch at time $t_1$.
What is needed is to compute the boxed term in Eqs.~(\ref{variation}) for all
times $t$.

It is clear that, from $t=0$ to $t=t_0$, the term vanishes.
To compute its value during the switch, first note that the
$i^\mathrm{th}$ eigenvector at time $t$ is the $i^\mathrm{th}$
column of $B_s$, where $s\equiv\frac{t-t_0}{t_1-t_0}$. But then, using
Eq.~(\ref{BTdef}), it is straightforward to see that the $k^\mathrm{th}$
element of the $i^\mathrm{th}$ eigenvector is
\begin{equation*}
 \left(\mathbf{v}_i\right)_k = \left(B_s\right)_{ki} = \left(R^\mathrm{T}G_sRB_0\right)_{ki} = \sum_{r=1}^N\sum_{q=1}^N\sum_{x=1}^N R_{rk}\left(G_s\right)_{rq}R_{qx}\left(B_0\right)_{xi}\:.
\end{equation*}

Notice that in the equation above the only term that depends
on time is $\left(G_s\right)_{rq}$, since $R$ is just a change-of-basis
matrix, and $B_0$ is the matrix of eigenvectors of $L_0$ at
time $t=0$. Therefore, it is
\begin{equation*}
 \ddt\left(\mathbf{v}_i\right)_k = \sum_{r=1}^N\sum_{q=1}^N\sum_{x=1}^N R_{rk}R_{qx}\left(B_0\right)_{xi}\frac{1}{t_1-t_0}\dds\left(G_s\right)_{rq}\:.
\end{equation*}

This allows us to write a final expression for the extra
term that accounts for the time variation of the eigenvectors.
In a fully explicit form, this is
\begin{multline}\label{NCbox1}
 -\sum_{i=1}^N\mathbf{v}_j^\mathrm{T}\left(t\right)\cdot\ddt\mathbf{v}_i\left(t\right)\eta_i =\\
 -\frac{1}{t_1-t_0}\sum_{i=1}^N\sum_{k=1}^N\left[\sum_{r=1}^N\sum_{q=1}^N\sum_{x=1}^N R_{rk}\left(G_s\right)_{rq}R_{qx}\left(B_0\right)_{xj}\right]\left[\sum_{r=1}^N\sum_{q=1}^N\sum_{x=1}^N R_{rk}R_{qx}\left(B_0\right)_{xi}\dds\left(G_s\right)_{rq}\right]\eta_i\:.
\end{multline}

However, for all practical purposes, one does not need
to use Eq.~(\ref{NCbox1}) directly. In fact, considering
that most elements of $G_s$ are~0, it is quite simple
to compute and store $B_s$ in a symbolic form. Similarly,
most of the terms $\dds\left(G_s\right)_{rq}$
are~0. In fact, they vanish in all the following cases:
\begin{itemize}
 \item $r=1$ or $q=1$ (the eigenvector corresponding to the synchronization
manifold, and axis of rotation of the basis of eigenvectors);
 \item if the $\left(rq\right)$ element is outside the tridiagonal, i.e. if $\left|r-q\right|>1$;
 \item if $r>2b+1$ or $q>2b+1$, where $b$ is the number of ``$\sin$--$\cos$'' blocks in $G_q$.
\end{itemize}
Also, for all other cases $\dds\left(G_s\right)_{rq}$, is proportional to a sine or a cosine. Thus, we can define
$\dot{G_s}$ to be the matrix whose $\left(rq\right)$ element
is $\frac{1}{t_1-t_0}\dds\left(G_s\right)_{rq}$;
also, we can define $\dot{B_s}\equiv R^{\mathrm T}\dot{G_s}RB_0$.
Again, $\dot{B_s}$ can be easily computed and stored in
a symbolic form. Then, Eq.~(\ref{NCbox1}) becomes
\begin{equation}\label{NCbox2}
 -\sum_{i=1}^N\mathbf{v}_j^\mathrm{T}\left(t\right)\cdot\ddt\mathbf{v}_i\left(t\right)\eta_i = -\sum_{i=1}^N\sum_{k=1}^N\left(B_s\right)_{kj}\left(\dot{B_s}\right)_{ki}\eta_i\:.
\end{equation}
Once more, Eq.~(\ref{NCbox2}) can be computed
symbolically, and evaluated at any particular time $t$, when needed.

\subsection{Synchronization in multilayer networks with coexisting layers}
A completely different scenario is that of networking dynamical units with coexisting layers.
In this case, indeed, the different layers determine \textit{as a whole} the overall coupling
structure through which the system components interact, and which in general cannot be simply
represented in terms of a single time-independent coupling matrix.

The current state of the art on the matter includes only a few direct studies,
pointing however to some important considerations and results. In the following,
we first summarize the available literature following the categorization given
in Section~2, in the hope to point out promising directions to those researchers
who, in the near future, will try to approach such a challenging issue. After
that, we concentrate on a specific case, involving a two-layer network in a master-slave
configuration, in view of the relevance of this situation that extends the very
concept of synchronization and provides practical tools to target the dynamics
of complex networks toward any desired collective behavior (not necessarily
synchronous).

\subsubsection{Hypernetworks}
In this framework, one considers a set of $N$ identical dynamical systems
linked through the connections of $M$ different layers, that is, all
the connections that correspond to the same type of coupling form a
layer, and the systems are connected by more than one layer.

Refs.~\cite{Irving2012IrenePRE,Sorrentino2012IreneNJP}
focused on the necessary and sufficient conditions for the
stability of the synchronous solution in a hypernetwork.
The main conclusion is that, in general, it is not possible
to reduce the problem to a simpler monolayer formalism.
However, for $M=2$, such a reduction is indeed possible
in three cases of interest: \textit i) if the two associated
Laplacians commute; \textit{ii}) if one of the two layers
is unweighted and fully connected; \textit{iii}) if one
of the two networks has an adjacency matrix of the form
$A_{ij}=a_j$ with $i,j=1,\dotsc,N$. The major result is
the definition of a class of networks such that if one of
the two coupling layers satisfies condition \textit{iii)},
the reduction can be obtained independently of the structural
form of the other network layer. The Authors of Refs.~\cite{Irving2012IrenePRE,Sorrentino2012IreneNJP}
also discuss some generalization of their results to the
case of hypernetworks consisting of more than two layers,
and present an example of a network of neurons connected
by both electrical gap-junctions and chemical synapses.

\subsubsection{Interconnected/interdependent networks and networks of networks}
Other studies dealt with the problem of seeking connection strategies
between interconnected networks (different layers of an overall network)
to achieve complete synchronization. In this framework, each of the
interconnected layers is a network made of elements of similar or different nature.

Ref.~\cite{aguirre2014Irene} is a combination of experimental,
numerical and analytical studies in the case of only two layers. The Authors
find that connector nodes (those at the ends of interlayer links), play a
crucial role for the setting of the synchronous behavior of the whole system.
In particular, the main result is that connecting the hubs of both networks
is the best strategy to achieve complete synchronization.

In Ref.~\cite{martinhernandez2013IreneArXiv} the same problem is investigated
in the framework of identical layers. A phase transition in the network synchronizability
is described when the number of interlinks increases. The exact location of the
transition depends exclusively on the algebraic connectivity of the graph models,
and it is always observed regardless of the interconnected graphs. Furthermore,
by means of computer simulations and analytical methods, Ref.~\cite{Wang2006IrenePRE}
investigates the effects of the degree distribution in determining mutual synchronization
of neural networks made of two identical layers. The main claim of Ref.~\cite{Wang2006IrenePRE}
is that a SF topology is more effective in inducing synchronization, or
in preventing the system from being synchronized, provided that the interlayer
links involve the nodes with larger degrees.

Refs.~\cite{Huang2006IrenePRL} and~\cite{Rad2012IrenePRL} studied synchronization
in a complex clustered network (or network of clusters) when two clusters interact
via random connections, showing that the interplay between interconnections and
intraconnections plays a key role in the global synchronizability, as well as in
determining different synchronization scenarios. For instance, Ref.~\cite{Huang2006IrenePRL}
demonstrated that the network has the strongest synchronizability only when the
number of inter and intraconnections matches perfectly, whereas any mismatch
weakens or even destroys the synchronous state. However, Ref.~\cite{Rad2012IrenePRL}
pointed out that in a modular network of phase oscillators there exists an optimal
balance condition between the number and weight of inter and intralinks that
maximizes at the same time the network's segregation and integration processes.

Furthermore, in networks of interacting populations of phase oscillators,
the onset of coherent collective behavior was studied in Ref.~\cite{Barreto2008IrenePRE}.
A typical biological application of the model is the description of interacting populations
of fireflies, where each population inhabits a separate tree. The model includes heterogeneity
at several levels. Each population consists of a collection of phase oscillators whose
natural frequencies are drawn at random from a given distribution. To allow for heterogeneity
at the population level, each population has its own frequency distribution. In addition,
the system is heterogeneous at the coupling level, as the model explicitly considers the
possibility to specify separately the coupling strengths between the members of different
populations. The assumption of a global (yet population-weighted) coupling allows for
an analytical determination of the critical condition for the onset of a coherent collective
behavior.

\subsubsection{Multiplex, multiplex with time delay and master-slave configurations}
Ref.~\cite{Louzada2013IreneSR} reports on the synchronization properties
of interconnected networks of oscillators with a time delay on the interactions
between networks, and analyzes the synchronization scenario as a function
of the coupling and communication lag. In particular, the Authors consider
the competition between an instantaneous intranetwork and a delayed internetwork
coupling. The major result is the discovery of a new breathing synchronization
regime, with the emergence of two groups in each network synchronized at
different frequencies. Each group has a counterpart in the opposite network,
one group being in phase and the other in anti-phase with the counterpart.
For strong coupling strengths, however, networks are internally synchronized,
in some cases with a phase shift between them.

In Ref.~\cite{Bogojeska2013IreneIEEE} the reader can find an interesting
study on the Master Stability Function for class~III systems~\cite{Boccaletti2006}.
The motivation is to investigate how the topologies of the layers influence
the synchronization criteria and rate for the multiplex network. For that
purpose, Ref.~\cite{Bogojeska2013IreneIEEE} considers different topologies,
such as random (ER), random regular (RR), small-world (SW) and scale-free (SF),
and studies a 2-layered multiplex network formed by interconnecting a layer
with ER, SW or SF structure with another layer with an RR structure. This
choice of the global topology is motivated by its correspondence to the
highest synchronizability. The links between the layers have a constant
weight $c$. The inspection of the eigenratio $\lambda_N/\lambda_2$ allows
to draw non-trivial and interesting conclusions for the synchronization
properties. In particular, it is shown that a SW-RR topology
enhances the synchronization, even though individually the SF topology has
a lower eigenratio and a better synchronizability than the SW topology.
In addition, it is shown that the ER-RR configuration displays similar
synchronization properties to the SW-RR configuration for small coupling
strengths, whereas at large coupling strengths the ER-RR configuration
leads to a synchronization scenario similar to that occurring for the SF-RR
configuration.

In Ref.~\cite{LI2007IrenePRE}, using Lyapunov stability theory,
the Authors predict theoretically that in a master-slave configuration,
complete synchronization between master and slave can be achieved
if they have identical connection topologies. This prediction is
verified numerically for a two-layer network of Lorenz systems.
Ref.~\cite{LI2007IrenePRE} also provides some numerical test for the case in which
the master and the slave layers have non-identical connection topologies. In this
case, the behavior of the system is highly non-trivial, with synchronization and
de-synchronization alternately appearing for increasing values of the coupling strength.
In this very same context, a fault-tolerant control scheme has been proposed
to guarantee synchronization when the coupling connections fail~\cite{Wang2013IreneNN}.

Other forms of synchronization, such as generalized synchronization,
have been explored in networks with unidirectional coupling in a master-slave
configuration~\cite{Sang2009IrenePRE,Wu2009IreneChaos,Wu2012IreneCNSNS,Xu2014IreneAAppAnal,Asheghan2013}
and with fractional-order dynamics \cite{Asheghan2011}.

Finally, in Ref.~\cite{Mao2012IreneAMC} a relatively simple model
is studied, consisting of a pair of bidirectional sub-networks
each with an arbitrary number of identical neurons in a ring
structure and two-way couplings between the ring sub-networks.
Different time delays are introduced in the internal connections
within each layer, and in the couplings between the layers.
The network exhibits different dynamical phenomena depending
on the parity of the number of neurons in the sub-network. Multiple
stability switches of the network equilibrium, synchronous and
asynchronous periodic oscillations, and the coexistence of the
bifurcated periodic oscillations are shown numerically.

\subsubsection{Bipartite and multilayer networks}
Concurrent synchronization or multisynchronous evolution in networks
is a dynamical regime where multiple groups of fully synchronized elements
coexist in distinct, possibly chaotic, dynamics, so that elements from
different groups are not necessarily synchronized. This kind of behavior
has been studied in relation to robotics and neuroscience, where groups
of interacting robots or neurons perform synchronous parallel coordinated
tasks.

In Refs.~\cite{Pham2007IreneNN,Sorrentino2007IrenePRE} the issue of multiple
synchronized, possibly chaotic, motions occurring in networks of groups has
been independently addressed. In the first study, the Authors tackled the stability
of an invariant subspace corresponding to multisynchronicity in unweighted networks.
In the second, the stability of the multisynchronous evolution in bipartite
networks is evaluated by introducing a MSF that decouples
the effects of the network topology from those of the dynamics on the nodes.
The MSF approach seems to be inadequate to assess the stability of the synchronous
evolution when both intragroup (intralayer) and extra-group (interlayer)
connections are allowed in the network. The Authors considered examples of general
network topologies, where links are also allowed to fall within each layer,
and reported numerical evidence that the presence of diffusive couplings amongst
nodes within the same layer can enhance the network synchronizability.

Based on Lyapunov stability theory and linear matrix inequality,
the Authors of Ref.~\cite{Sun2010IreneCogNeuro} studied ``inner''
(within each network) and ``outer'' (between two networks) synchronization
between two coupled discrete time networks with time delays. They
were able to provide sufficient conditions to assess the stability
of the synchronous solution within each network, but once again
the method used cannot guarantee the stability when both intra-group
and extra-group connections are allowed.

In Refs.~\cite{Gu2006IreneCSF,Liu2009IreneCSF,YanLing2009IreneCSF}
the synchronization of one-layer~\cite{Gu2006IreneCSF}, two-layer~\cite{Liu2009IreneCSF},
and three-layer center networks~\cite{YanLing2009IreneCSF} of coupled
maps has been studied, using linear stability analysis. Multicenter
layer networks are characterized by the presence in each
layer of nodes with special properties, called \emph{center nodes}.
Each layer is considered adjacent to the layer immediately preceding
it and the one immediately following it; the exceptions are the first
layer, which is only adjacent to the second, and the last, which is
only adjacent to the previous one. All center nodes in a layer are not
connected among them (except for the first layer in which they are globally connected) but to all the center nodes in the adjacent
layers. In addition, each center node in the last layer
has $n$ different non-center nodes which are not connected among them.
The Authors found conditions for stable synchronization that depend
on the number of center nodes and non-center nodes.

A natural framework for synchronization in multilayer
networks is that of transmission of information in neural
networks. A debate on the role of synchrony in nervous
system signaling is addressed in Ref.~\cite{Reyes2003IreneNatNeu},
where a detailed analysis of the relationship between
synchronization and rate encoding has been carried out
in models of feed-forward multilayer neuronal
networks~\cite{Reyes2003IreneNatNeu,Nowotny2003IreneBiolCyb,Ming2010IrenePRE}.
In Ref.~\cite{Ming2010IrenePRE}, a feed-forward multilayer
network of Hodgkin-Huxley neurons was constructed (with
no intralayer links), where synchronous firings can
develop gradually within the layered network, consistently
with experimental findings~\cite{Reyes2003IreneNatNeu}
and with other theoretical work~\cite{Nowotny2003IreneBiolCyb}.

An analogy between crowd synchrony and multilayer neural network architectures
is proposed in Ref.~\cite{Cohen2012IreneOptExp}. The Authors showed that a two-layer
system of many non-identical oscillators (semiconductor lasers) communicating
indirectly via a few mediators (hubs) can synchronize when the number of optically
delayed couplings to the hubs, or the strength of the coupling, is large enough.
The synchronization transition is discontinuous, but its order depends on the
detuning between the hubs. If the detuning is increased, the transition becomes
continuous, with a diverging synchronization time near the critical coupling that
does not depend on the number of hubs.

\subsubsection{Synchronization and the problem of targeting}
A very important point to stress is that, as long as the nodes are identical
dynamical systems represented by vector states of the kind $\mathbf x_i^l(t)$,
with $i=1,\dotsc,N$ indicating the node, and $l=1,\dotsc,M$ the layer, the analysis
of the synchronization properties may be one of two very different endeavors.

The first involves an extension of the usual synchronization concept
that we call \textit{intra}layer synchronization, and corresponds to
analyzing the conditions for which the solution $\lim_{t\to\infty}|\mathbf x_i^{l}(t)-\mathbf x_j^{l}(t)|=\mathbf 0$
(with $i\neq j$ and $\forall l$) is stable.

Notice that the stability conditions of such a solution
can be simply found by considering an overall modular-like
network of size $NM$, and applying to usual MSF formalism~\cite{Boccaletti2006}
to study the spectral properties of the $NM \times NM$
Laplacian matrix, which will have the form of $M$ diagonal
blocks describing the \textit{intra}layer connection structures,
plus extra blocks describing the \textit{inter}layer interactions.

The second case is much more general, as it encompasses
explicitly the first case as one of its particular instances.
It corresponds to assessing the stability conditions of
the solution $\lim_{t\to\infty}|\mathbf x_i^{l}(t)-\mathbf x_i^{l'}(t)|=\mathbf 0$
with $l\neq l'$ and $\forall i$). Such a solution, which
we call the \textit{inter}layer synchronized solution,
does not imply necessarily that the dynamical evolution
of nodes be synchronized within each layer.

In particular, when restricting the framework
to a two-layer network with unidirectional interactions
from a master layer to a slave layer, the importance
of the interlayer solution was stressed in connection
with the problem of \textit{targeting} the dynamics
of a complex network~\cite{Gutierrez2012StefanoSciRep},
i.e. finding a generic procedure to steer the
dynamics of a network toward a given, desired evolution.

In fact, in Ref.~\cite{Gutierrez2012StefanoSciRep}, the feasibility
of targeting \textit{any} goal dynamics $\mathbf g(t)$ compatible
with the natural evolution of the network was proven to be tantamount
to demonstrating that two identical layers coupled in a master-slave
scheme can identically synchronize~\cite{Pecora1990StefanoPRL,Kocarev96StefanoPRL}
when starting from different initial conditions. There, the Authors
considered two identical layers, namely a master network (MN) providing
the specific $\mathbf g(t)$ and a slave network (SN).
Starting from
different initial conditions, both layers produce a turbulent regime,
i.e. generic dynamics not showing any particular global or local order.
The Authors properly engineered a unidirectional pinning action from
the MN to the SN with the aim of ultimately synchronizing the dynamics
of the two layers.

In the following, we briefly summarize the main results of~Ref.~\cite{Gutierrez2012StefanoSciRep}.

The master network considered is composed of $N$ \textit{identical},
diffusively coupled chaotic units. Calling the vector state of the $i^{\mathrm{th}}$
node $\mathbf x^M_i\in\mathbb R^m$, one has:
\begin{equation}\label{master}
 \dot{\mathbf x}^M_i = \mathbf f(\mathbf x^M_i)-\sigma_1\sum_{j=1}^N\mathcal L_{ij}\mathbf h_1(\mathbf x^M_j)\:,
\end{equation}
where $\mathbf f(\mathbf x): \mathbb R^m\rightarrow\mathbb R^m$
and $\mathbf h_1(\mathbf x): \mathbb R^m\rightarrow \mathbb R^m$\
are a local evolution and an vectorial output function, respectively.
$\sigma_1$ is a coupling strength, and $\mathcal L_{ij}$ are the
elements of the corresponding zero row-sum Laplacian matrix of the
interaction network.  Notice that, here and in the following, $M$ indicates
the master network, and {\it not the number of layers}, which is here always 2 (the master and the slave). $\mathbf f$ and $\mathbf h_1$ are chosen to
make the network units of class~I, with a monotonically increasing
trend of the MSF associated to the intralayer
synchronous state $\mathbf x^M_1=\mathbf x^M_2=\dotsb=\mathbf x^M_N=\mathbf x_{sync}$~\cite{Boccaletti2006},
which is, therefore, always unstable, regardless of the choice of $\sigma_1$.

Let now
\begin{equation}\label{slaveuncoupled}
 \dot{\mathbf x}^S_i = \mathbf f(\mathbf x^S_i)-\sigma_1\sum_{j=1}^N\mathcal L_{ij}\mathbf h_1(\mathbf x^S_j)
\end{equation}
be the slave network (SN), obtained as a copy of the MN.

As long as both networks remain uncoupled and start from different initial
conditions, each one will sustain a different turbulent state, so that
$\mathbf X^S(t)\equiv (\mathbf x_1^S(t),\dotsc,\mathbf x_N^S(t))\neq\mathbf X^M(t)\equiv (\mathbf x_1^M(t),\dotsc,\mathbf x_N^M(t))=\mathbf g(t)$,
and the SN will never attain the desired goal dynamics.

In Ref.~\cite{Gutierrez2012StefanoSciRep}, the Authors implemented
a pinning strategy to target $\mathbf g(t)$ consisting of sequentially
establishing unidirectional links between nodes in the MN and their
copies in the SN. The dynamical evolution of the SN is therefore described by:
\begin{equation}\label{slave}
 \dot{\mathbf x}^S_i=\mathbf f(\mathbf x^{S}_i)-\sigma_2\chi_i\mathbf h_2(\mathbf x^S_i-\mathbf x^M_i)-\sigma_1\sum_{j=1}^N\mathcal L_{ij}\mathbf h_1(\mathbf x^S_j)\:,
\end{equation}
where $\mathbf h_2: \mathbb R^m\rightarrow \mathbb R^m$
is the coupling function \textit{between} MN and SN, with
the condition that $\mathbf h_2(\mathbf 0)=\mathbf 0$,
and $\chi_i$ encodes the pinning procedure, i.e. $\chi_i=1$
if there is a link from the $i^{\mathrm{th}}$ node of
the MN to the $i^{\mathrm{th}}$ node of the SN, and $0$
otherwise. Furthermore, $\sigma_2$ is the parameter ruling
the interlayer link strength.

It follows that the equation for the vector describing
the difference between the master and the slave layers'
dynamics $\delta\mathbf X=\mathbf X^S-\mathbf X^M\equiv (\delta\mathbf x_1,\dotsc,\delta\mathbf x_N)$
can be written in terms of its components $\delta\mathbf x_i$ as
\begin{equation}\label{error}
 \delta\dot{\mathbf x}_i=\mathbf f(\mathbf x^S_i)-\mathbf f(\mathbf x^M_i)-\sigma_1\sum_{j=1}^N\mathcal L_{ij}[\mathbf h_1(\mathbf x^S_j)-\mathbf h_1(\mathbf x^M_j)]-\sigma_2\chi_i\mathbf h_2(\mathbf x^S_i-\mathbf x^M_i)\:.
\end{equation}
A stable fixed point of Eqs.~(\ref{error}) at $\delta\mathbf X=\mathbf 0$
is a necessary condition for layers~(\ref{master}) and~(\ref{slave}) to
display the interlayer synchronized state $\mathbf X^M=\mathbf X^S$.

The
synchronization error is then defined as $\lim_{T\to\infty}(1/T)\int_0^T\|\delta\mathbf X(t)\|\mathrm{dt}$.

The linear stability of this solution
can be assessed rigorously via the analysis
of the linearized system for small $\delta\mathbf X$, which reads:
\begin{equation}\label{errorlinear}
 \delta\dot{\mathbf x}_i=[D\mathbf f(\mathbf x^M_i)-\sigma_2\chi_iD\mathbf h_2(\mathbf x^M_i)]\delta\mathbf x_i-\sigma_1\sum_{j=1}^N\mathcal L_{ij}D\mathbf h_1(\mathbf x^M_j)\delta\mathbf x_j\:,
\end{equation}
where $D\mathbf f$, $D\mathbf h_1$, and $D\mathbf h_2$ are the Jacobian
functions, and $\mathbf X^M=\mathbf g(t)$ is the MN state to be targeted.
This equation represents the MSF for the solution
corresponding to the identical synchronization \textit{between} MN and
SN. Each of the linear equations~(\ref{errorlinear}), solved in parallel
to the $N$ nonlinear equations for the MN~(\ref{master}), corresponds to
a set of $m$ conditional Lyapunov exponents at each pinning configuration
(PC). Therefore, each particular PC for which the largest of all such
exponents is negative makes the interlayer synchronous state $\mathbf X^M=\mathbf X^S$ stable.
\begin{figure}
\centering
\includegraphics[width=0.55\textwidth]{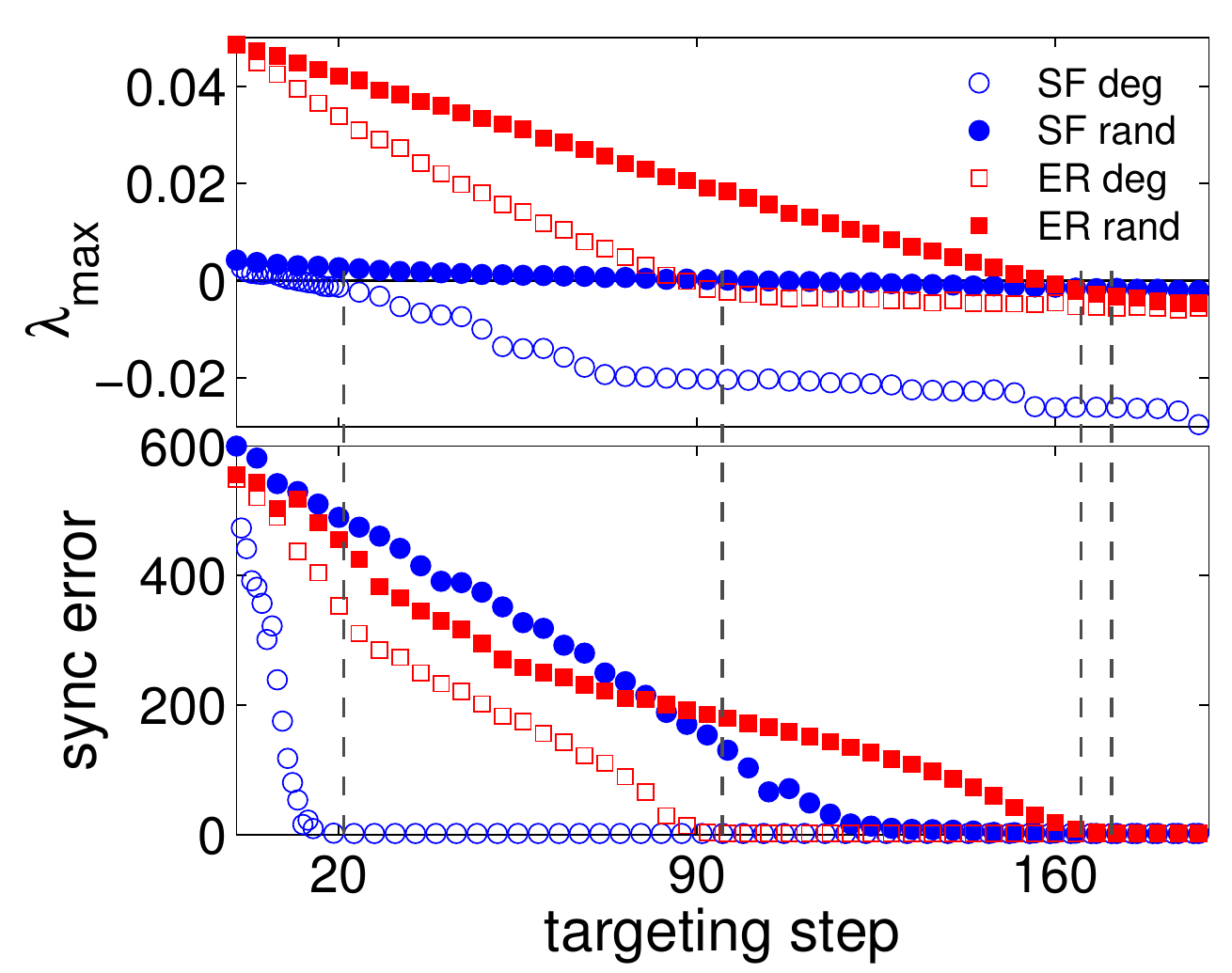}
\caption{\label{Figtargeting}(Color online). Targetability of different topologies.
Targeting scheme for ER and SF networks of $N=500$ R\"ossler oscillators.
The top panel shows $\lambda_{\max}$ as a function of the targeting
step, for degree ranking and random ranking. The zero level is highlighted
as a black horizontal line. The bottom panel shows the synchronization
error vs.\ the targeting step. The legend contains the symbol codes
for the specific topologies considered, as well as for the specific
ranking used for the targeting. All results correspond to ensemble
averages over 40 different graph realizations, and, for the random
ranking case over 40 different ranking realizations.}
\end{figure}

In Ref.~\cite{Gutierrez2012StefanoSciRep}, the validity
of such a targeting procedure was demonstrated for layers
of Rössler oscillators, configured with Erdős–Rényi (ER)~\cite{Erdos1959StefanoPM}
and Barabási-Albert SF~\cite{BA} topologies.
More precisely, a pair of identical master and slave layers
were considered, described by the following set of equations:
\begin{align*}
&\left\lbrace\begin{aligned}
\dot x_i^M &= -y_i^M - z_i^M\:,\\
\dot y_i^M &= x_i^M + 0.2 y_i^M\:,\\
\dot z_i^M &= 0.2 + z_i^M\left(x_i^M - 7.0\right) + \sigma_1 {\sum}_{j \in \mathcal{N}_i}\left(z_j^M - z_i^M\right)\:,
\end{aligned}\right.\\
&\left\lbrace\begin{aligned}
\dot x_i^S &= -y_i^S - z_i^S\:,\\
\dot y_i^S &= x_i^S + 0.2 y_i^S + \sigma_2 {\sum}_{k \in \mathcal{T}_t}(y_k^M - y_k^S)\:,\\
\dot z_i^S &= 0.2 + z_i^S\left(x_i^S - 7.0\right) + \sigma_1 {\sum}_{j \in \mathcal{N}_i}\left(z_j^S - z_i^S\right)\:,
\end{aligned}\right.
\end{align*}
where $\mathcal{N}_i$ denotes the neighborhood of the node $i$,
and $\mathcal{T}_t$ the set of pinning nodes for which a unidirectional
connection is formed from the master to the slave layers.

The equations above correspond, in the notations of Eqs.~(\ref{master}), (\ref{slaveuncoupled}) and~(\ref{slave}),
to $\mathbf x^M\equiv (x^M,y^M,z^M)$, $\mathbf x^S\equiv (x^S,y^S,z^S)$, $\mathbf h_1(\mathbf x)=(0,0, z)$,
$\mathbf h_2(\mathbf x)=(0,y,0)$ and $\mathbf f=[-y - z, x + 0.2 y, 0.2 + z (x - 7.0)]$.

By fixing $\sigma_1=\sigma_2=1$, the Authors of Ref.~\cite{Gutierrez2012StefanoSciRep}
considered both sequences of pinning nodes following directly the degree ranking, or following
a random ranking (in which the order of appearance of each node is fully random), in order
to compare between both ends of the whole spectrum of possible targeting strategies.

In Fig.~\ref{Figtargeting}, the results are shown for the first 200 targeting steps,
sufficient in all cases to attain the desired goal dynamics $\mathbf g(t)$. It is easy
to see that the maximum Lyapunov exponent $\lambda_{\max}$, obtained from Eq.~(\ref{errorlinear})
and reported in the upper panel of the figure, always crosses the zero line exactly in
correspondence of a vanishing synchronization error, reported in the lower panel of the
figure.

Furthermore, a comparative inspection of Fig.~\ref{Figtargeting}
allows to draw two important conclusions: \textit{i)} for both SF
and ER topologies, the degree ranking leads to a much better targeting,
as compared to the random scheme, and \textit{ii)} the relative improvement
in SF networks is much more evident than in the ER case, indicating
that the dynamics of heterogeneous structures, like the ones encountered
in real-world networks, are much easier to be manipulated and targeted.


\section{Applications}\label{sec:applications}

If the success of a scientific theory can be assessed by its contributions to the understanding of real-world problems, complex networks can possibly
be seen as one of the most prosperous developments in the last decades, having found applications in a plethora of different contexts \cite{costa2011analyzing}.
Nevertheless, it is usually recognized that the traditional network representation is an oversimplification
of the corresponding system; for instance, social interactions seldom develop on a single channel,
and more than one relationship can bind pairs of people. It is thus not surprising that multilayer networks
have raised a great expectation, as they should be able to better represent the topology and dynamics of real-world systems.

In the following, we present a review of works dealing with the analysis of real multilayer networks.
Most of them were introduced well before the recent development of the mathematical framework reported here,
when Erving Goffman first proposed his {\it frame analysis} social theory in 1974.
Due to this, the reader should expect a large heterogeneity in concepts and solutions.

This section is organized into three main topics: social, technological and economical networks - a synopsis
of the main references is provided in Table~\ref{table:resumeapp}.

Social networks have captured most of the attention, thanks to the data availability maintained by online communities,
and the inherent multilayer structure of social relationships.
Yet, relevant problems can also be found in the two latter topics, as for instance the analysis of transportation systems
and of trade networks. The end of this section is devoted to other applications, which have hitherto received less attention,
but whose relevance, and the relevance of the corresponding multilayer representations,
is expected to grow in the near future: this includes biomedicine, e.g. the analysis of the brain dynamics taking into account the layers of neural organization, climate (global networks created by correlation between climatological variables), ecology and psychology.

\begin{table}[!tb]
\begin{tabular}{ |l|l|l| }
\hline
\multicolumn{3}{ |c| }{Resume of topics and references} \\
\hline
Field & Topic & References \\ \hline

\multirow{4}{*}{Social}
 & Online communities & \begin{tabular}{@{}l@{}}
 Pardus: \cite{Szell2010, szell2010b, Szell2013b, klimek2013, corominas2013} \\
 Netflix: \cite{horvat2012,horvat2013} \\
 Flickr: \cite{musial2008,KaMuKa11, Brodka2010} \\
 Facebook: \cite{lewis2008, Mucha10,hashmi2012, lewis2012} \\
 Youtube: \cite{tang2012} \\
 Other online communities: \cite{ansari2011, Brodka2012, Halu13} \\
 Merging multiple communities: \cite{magnani2011,magnani2012,magnani2013,zhong2014}
 \end{tabular}
 \\

    \rule{0pt}{4ex}

 & Internet & \cite{KolBaKe05,KolBa2006, barnett2013}\\

    \rule{0pt}{4ex}

 & Citation networks & \begin{tabular}{@{}l@{}}
DBLP: \cite{cai2005, ng2011,li2012,CosRoPe13, Brodka2011, boden2013, sun2006, BerCosGi12} \\
 \end{tabular} \\

    \rule{0pt}{4ex}

 & Other social networks & \begin{tabular}{@{}l@{}}
 Scottish Community Alliance: \cite{walker2013} \\
 Politics: \cite{Mucha10, heaney2012} \\
 Terrorism: \cite{Batiston2013} \\
 Bible: \cite{duling2013} \\
 Mobile communication: \cite{zignani2014exploiting}
 \end{tabular}\\ \hline

\multirow{3}{*}{Technical}

 & Interdependent systems & \begin{tabular}{@{}l@{}}
 Power grids: \cite{Buldyrev10, BruSo12, sen2014identification} \\
 Space networks: \cite{castet2013} \\

 \end{tabular} \\

    \rule{0pt}{4ex}

 & Transportation systems & \begin{tabular}{@{}l@{}}
 Multimodal: \cite{parshani2010, halu2013} \\
 Cargo ships: \cite{kaluza2010} \\
 Air transport: \cite{CardilloSR13, CardilloEPJST13}
 \end{tabular} \\

    \rule{0pt}{4ex}

 & Other technical networks & Warfare: \cite{hauge2014selected} \\ \hline

\multirow{3}{*}{Economy}
 & Trade networks & \begin{tabular}{@{}l@{}}
 International Trade Network: \cite{Barigozzi2010,Barigozzi2011,mahutga2013} \\
 Maritime flows: \cite{ducruet2013}
 \end{tabular}\\

    \rule{0pt}{4ex}

 & Interbank market & \cite{bargigli2013multiplex}\\

    \rule{0pt}{4ex}

 & Organizational networks & \cite{amburgey2008structural, lee2011coevolution, bonacina2014multiple} \\ \hline

\multirow{4}{*}{Other applications}
 & Biomedicine & \cite{vandamme2013, hood2013, michoel2012, nicosia2014measuring, li2011,li2012b}\\

    \rule{0pt}{4ex}

 & Climate & \cite{Donges2011, feng2012} \\

    \rule{0pt}{4ex}

 & Ecology & \cite{Barret2012, mouhri2013} \\

    \rule{0pt}{4ex}

 & Psychology & \cite{varol2014connecting} \\ \hline

\hline
\end{tabular}
\caption{\label{table:resumeapp} Resume of the main application topics, and related references.}
\end{table}

\subsection{Social}

\subsubsection{Multilayer social networks: an old concept}

While the mathematical formulation of multilayer networks is very young,
the idea of studying the way social interactions develop on multiple layers dates back to the seventies.
The first of such approaches was probably introduced by Erving Goffman in 1974, along with the theory of
{\it frame analysis} \cite{goffman1974}. According to this research method,
any communication between individuals (or organizations) is constructed (or {\it framed})
in order to maximize the probability of being interpreted in a particular manner by the receiver.
Such framing may differ according to the type of relations between the involved individuals,
several of them potentially overlapping in a single communication:
thus, the output of multiple framings can be interpreted as a multilayer structure.

After this first attempt, White et al. proposed an extension of the classical work by
Nadel and Fortes \cite{nadel1957} to perform a quantitative analysis of multilayer social networks \cite{white1976}.
A simple three-layer structure is considered, where nodes represent biomedical researchers specialized
in the neural control of food and water intake; the three layers respectively represent the existence
of a bidirectional personal tie, an unidirectional awareness, and the absence of awareness \cite{griffith1973}.
Notice that this three-layer structure contains information that, in modern complex network theory,
would be synthesized into a single directed network, as the third layer is the complementary of the second.
Using this network as a test bed, they proposed a method for detecting {\it blocks}
that are coherent across the different layers: in other words, they had just introduced the first
community detection algorithm. Along this line of research, Greiger and Pattison in Ref.~\cite{breiger1986}
analyzed the multilayer social structure reported in Ref.~\cite{padgett1993},
characterized by business and marriage relationships between important families from Florence in the fifteenth century.
Authors showed how a {\it personal hierarchy}, comprising information from both layers,
can be used to create meaningful partitions of the graph.

Finally, it is worth concluding this short review on the classical social applications
by presenting the work proposed in Ref.~\cite{knipscheer1995}. Authors argued that an analysis of
the health condition in old people, e.g. their resilience to external shocks,
can be performed only when multiple networks are simultaneously considered, like {\it living arrangements} (i.e.
ties created by the residence type), family, or organizations like church or voluntary associations.

\subsubsection{Online communities}\label{online-communities}
Information and communications technology (ICT) is behind the boost experienced in the last decade by
the online communities. ICT favors users' interactions through the sharing of multiple contents,
from tastes to photos and videos, and at the same time it fulfills one of the most basic social needs:
interpersonal communication. On the other hand, all those communication exchanges are stored in a digital support:
researchers can thus find plenty of information for analyzing social behaviors,
and for representing the wide variety of human interactions under a multilayer framework.

\paragraph{The Pardus social game}
One of the most interesting online community data sets has been provided by {\it Pardus},
an online game in which more than $300.000$ players live in a virtual, futuristic universe,
exploring and interacting with others, pursuing their personal goals by means of different communication
channels ranging from instantaneous one-to-one communications, e.g. e-mail and personal messages,
while others imply a more stable relationship (friendship, cooperation, attacks, and so forth) \cite{castronova2005}.

Several papers have used this data set to explore whether particular network topological features were present.
For instance, Ref.~\cite{Szell2010} extracted a six-layers structure,
respectively representing friendship or enmity relations between pairs of players,
private messages exchange, trading activities, the presence of one-to-one aggressive acts,
and placing head money (bounties) on other players.
Different characteristics were found for each layer:
specifically, negatively connoted (i.e. aggressive) actions are characterized
by power-law topologies, while positive ones exhibit a higher clustering coefficient.
Furthermore, different layers interact in a non-trivial way:
for instance, there is a strong overlap between communication and friendship networks,
but also between communication and aggression, and between communication and friendship - see Fig.~\ref{fig:Pardus}.
Ref.~\cite{szell2010b} expands further those results,
by considering the time evolution of such a multilayer network.
The network dynamics follows some well-known processes, like {\it triadic closure} and {\it network densification}.
Furthermore, it seems that the Dunbar conjecture is confirmed \cite{dunbar1993coevolution},
as out-degrees are limited at around 150 connections.
More recently, the effect of the user's gender was tackled in Ref.~\cite{Szell2013b},
demonstrating that the structural properties of the layers are different for men and women,
especially when the interaction time scale is considered.

The problem of the triadic closure has more recently been addressed in Ref.~\cite{klimek2013},
through the analysis of a Pardus data set composed of friendship,
communication and trading relationships.
Specifically, a network growth model is presented, in which links are either added to
form a closed triangle with probability $r$, or following a standard preferential attachment mechanism
\cite{BA} with probability $1 - r$.
Such a simple model is yet able to reproduce most of the topological features
of the original multilayer network, thus suggesting that both dynamics compete in the real system.

Finally, Ref.~\cite{corominas2013} tackles the problem of detecting elite structures in the Pardus network,
i.e. subgroups of individuals that have the ability of influencing the behavior of others.
This is accomplished by means of a modified {\it k}-core algorithm. Results indicate that different
layer projections can be used to predict how players will perform socially, e.g. their leadership or wealth.

\begin{figure}
\begin{center}
  \includegraphics[width=0.9\textwidth]{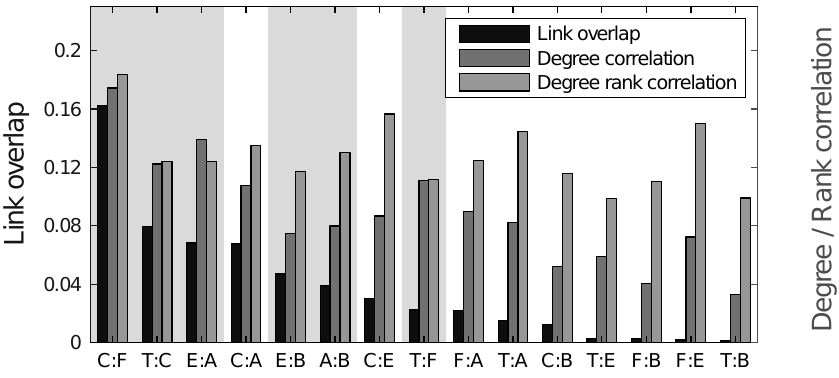}
\end{center}
\caption{Link overlap (Jaccard coefficient), degree correlation and rank correlation for all pairs of layers in the Pardus social game network. E, enmity; F, friendship; A, attack; T, trade; C, communication; and B, bounty. Pairs of equal connotation (positive-positive and negative-negative) are marked with a gray background. Reprinted figure from Ref.\cite{Szell2010}. Courtesy of S. Thurner.}
\label{fig:Pardus}
\end{figure}

\paragraph{Netflix}
Netflix, Inc. is a US-based company specialized in video rental and online streaming.
When users navigate in the website, for instance to select the next content they want to watch,
an advance recommender system \cite{lu2012} analyzes their tastes and suggests a set of videos expected
to be to their liking. In 2006, with the aim of improving this recommendation system and position it
above competitor ones,
Netflix disclosed a large data set, including millions of movie ratings assigned by users \cite{bennett2007}.

Two works \cite{horvat2012,horvat2013} have examined this data set by considering its structure as a multilayer network.
Specifically, the initial bipartite network (users can only be connected to movies)
is projected into a movie's network, i.e. where nodes represent movies,
pairwise connected when a single user rated both of them. Furthermore,
information about ratings has been used to organize links in three different layers: positive, negative and mixed ratings.
The resulting three layers have different characteristics, which are lost when the global projection is considered.
For instance, positive and negative networks display a high clustering coefficient,
indicating that users tend to like (or dislike) groups of similar movies;
on the other hand, mixed ratings create more heterogeneous structures, characterized by larger but sparser communities.

\paragraph{Flickr}
{\it Flickr} is a worldly recognized online community, specialized in the hosting and sharing of multimedia
content like photos and videos. In 2013, the popularity of this web service was such that counted more than 87
 million registered users, and 3.5 million new images uploaded daily.
Due to the large quantity of information available about interactions between users,
it is not a surprise that several studies have leveraged on Flickr to validate network analyzes,
both applied (e.g. automatic assignation of {\it tags} to photos by means of community detection algorithms \cite{shen2005}) and theoretical (as validating network growing mechanisms against real data \cite{mislove2008}).

Recently, three works have focused on the multilayer structure of the Flickr community.
While nodes always represent users, these may be linked through different types of relations.
Specifically, Authors identify 9 of them, spanning from direct user friendship, to common comments or common tag usage.

First of all, two of them \cite{musial2008,KaMuKa11} have introduced the concept of multilayer networks
in the design of improved recommendation systems, i.e. systems able to recommend new friends
or relevant contents to target users. Results indicate that users are more likely to share photos
and tags when they are already friends, or in other words, that the social layer drives the semantic one.
Also, there is a strong bidirectionality in social relations: if a user comments on the photo of another user,
the latter reciprocates by commenting a photo of the former person.
It is worth noticing that these results, and especially the relationship between social and semantic connections,
could not be inferred if all layers were collapsed
into a single-layer network.
Furthermore, time is here an essential ingredient: data is then represented as multilayer time-varying graphs.

Finally, Ref. \cite{Brodka2010} focused on the problem of identifying communities of users
across the structure.
This is accomplished by firstly generalizing the concept of {\it local clustering coefficient}
to multilayer networks (see Sec.~\ref{subsec:clustering}) averaging the metric as measured in each layer.
Secondly, the communities are detected through a growing mechanism,
in which nodes are assigned to independent communities when their respective clustering coefficient
is higher than a given threshold.

\paragraph{Facebook}
{\tt Facebook.com } is a social website where people can interact and share tastes and comments about their daily activities.
 What is uncommon about it, is its popularity: in 2013 it was recognized as the second most visited website in the world,
with $1.230$ million registered users, requiring a farm of $50$ thousand servers for ensuring its correct operation.
As for other online social networks, many works can be found in the Literature which analyze
the topology emerging from user interactions, as for instance Refs.~\cite{nazir2008,viswanath2009,schneider2009,nguyen2011};
and possibly Facebook has attracted even more scientific attention due to the large business possibilities
provided by the system.

The analysis of Facebook as a multilayer network has mainly been possible by the work presented in Ref.~\cite{lewis2008}.
In that reference, Authors have crossed online interaction information with real-world data for a group of
$1.640$ students from a private college in the northeast of the United States.
The resulting data set comprises five different layers: Facebook friendship, picture friendship (i.e.
two users sharing photos), and roommate, dormmate and groupmate relationships.
Furthermore, the data set is longitudinal, as it includes information at different time moments between
$2006$ and $2009$; finally, it includes demographic and cultural details of each user.

Taking advantage of the usefulness of such a valuable data set, several works have leveraged on it to perform
different types of analyzes, some of them focusing on its multilayer dimension.
Among others, they include the study of different types of multilayer structures
\cite{Mucha10,hashmi2012} up to dynamical processes, like peer influence \cite{lewis2012}.

\paragraph{Youtube}
{\tt Youtube.com} is a website whose purpose is video sharing, i.e. it allows users to upload personal videos
and share them with the community. In February 2014,
it was recognized as the third website in the world in terms of traffic generated.

While few works have focused on the network dimension of the website, Ref.~\cite{tang2012}
analyzed in a multilayer context the interactions among Youtube registered users.
The resulting network is composed of five layers, each one defining different relations between pairs of users:
friendship, co-contacts (i.e. when two users both add a third one as a friend),
co-subscription (when they follow the same content), co-subscribed (when two users are followed by the same people),
and when they share favorite videos. The tackled problem is the detection of communities of users,
sharing the same tastes and interests. The proposed solution is the extraction of structural features
from each layer of the network, and their integration via Principal Component Analysis \cite{abdi2010}.

\paragraph{Other online communities}
Beside these popular websites, additional (and usually less well-known) online communities
have been studied from the multilayer perspective, representing more specific types of interactions.
Here below, some of those analyzes are reported.

In the first one, Ref.~\cite{ansari2011} analyzes a Swiss online social network connecting musicians through friendship,
such that subscribers can be classified either as users (followers and fans) or as artists.
Three different layers are considered, respectively created by friendship, communication, and music download.
Notably, the three networks share a similar structure, with a clear presence of homophily and reciprocity.

Ref.~\cite{Brodka2012} studies the social network behind the web portal {\tt extradom.pl},
specialized in gathering people engaged in building their own houses in Poland.
Eleven different layers are reconstructed from user interactions, including forum groups and topic creation,
or message posting.
The interest resides in the analysis of the neighborhood of each user,
specifically through a {\it multilayered neighborhood}.
Results indicate that, while people are exposed to many different types of relationships,
they tend to focus themselves in just one or two types.

Finally, Ref.~\cite{Halu13} tackles the problem of defining node centrality (see Sec.~\ref{subsec:centrality})
by expanding the classical PageRank algorithm \cite{BrinPage1998}, and by using as a test bed the social
network of the online community at the University of California, Irvine.
Such network comprises two layers: the first one representing direct instant messaging,
the second the bipartite topology created by users and discussion groups.
As it may be expected, taking the multiplex nature of the network into account, i.e. using a
{\it multiplex PageRank}, helps uncovering the emergence of rankings of nodes that differ from the rankings
obtained from the projected topology.

\paragraph{Merging multiple social networks}
After realizing the wealth of social networks populating Internet,
and how information can straightforwardly be extracted and processed,
the reader may wonder about the possibility of merging online networks,
where nodes would represent individuals, and links representing friendship (or other relations)
between pairs of them sitting at different layers, each layer codifying a different social site.
While this is in principle feasible, the main problem resides in the identification of
common users across websites, as real names are usually concealed, and the
same user can appear under different pseudonyms, or nicknames.

One way of tackling this problem is by using information extracted from Friendfeed, a social media aggregator.
Besides performing regular activities like posting messages, it also allows users to register
their accounts in other systems, thus allowing the retrieval of the multiple accounts
associated to the same user in several social services.
Ref.~\cite{magnani2013} make use of this website to obtain information from Friendfeed, Twitter and YouTube,
creating a network of $7.628$ nodes to test different multilayer network generation algorithms.
The same data set is also the source in Refs.~\cite{magnani2011,magnani2012} for node centrality studies.

An alternative approach has been proposed by Ref.~\cite{zhong2014},
where two content-driven websites (Pinterest and Last.fm) are scanned for connections toward a generic social network,
Facebook. The work aimed at confirming social bootstrapping as a common technique used by novel websites,
that is, the act of recommending new social ties depending on the existence of that link in a different network.

\subsubsection{Internet}
In the previous part of this section, the World Wide Web has been considered as an instrument to put people in contact,
and therefore as an instrument to drive online social networks.
Yet, Internet is also the result
of social forces: for instance, the structure created by hyperlinks is partly the result of
connections between real people.
Here we present networks created by hyperlinks between web contents as a social
aspect of Internet.

Three papers addressed this problem from a multilayer point of view.
Firstly, Refs.~\cite{KolBaKe05,KolBa2006} consider a data set created by the automatic crawling of
$4.986$ unique URLs, connected between them by $122.196$ hyperlinks.
Each web page is also associated to one or more {\it anchor text terms}, i.e. the visible words of an hyperlink.
After an initial data filtering, the final network was composed of $787$ pages and $533$ terms,
each one of them defining a different layer. Upon this network, an algorithm called TOPHITS was developed,
aimed at identifying clusters of similar content web pages and at ranking web sites by taking into account the
multilayer structure of links.

Ref.~\cite{barnett2013} describes the creation of a {\it university URL-citations network},
a network generated by checking the number of hyperlinks connecting web pages of couples of universities.
While this network has a single-layer nature, complementary information has been added, which creates additional
graphs supporting (and partly explaining) the initial structure.
For instance, two universities can be connected when they share the same primary language, when they are located in the
same country, or when they are located within a given physical distance.
Results indicate that the topological structure is mainly explained by language,
USA and UK universities forming a main cluster, while the geographical distance is not relevant.

\subsubsection{Citation networks}
Since the beginning of the complex network theory, citation networks captured a lot of attention, especially as far
as the topology underlying citations between scientific papers was concerned. In this latter case,
analysis is motivated, firstly, by the relative simplicity in obtaining the network structure,
and secondly, by the interest in creating new metrics better assessing the importance of manuscripts and journals.

When moving to a multilayer setting, the main source of information comes from the DBLP
(formerly known as {\it DataBase systems and Logic Programming}, and later as {\it Digital Bibliography \& Library Project}), a computer science bibliography website hosted by Universit\"at Trier, Germany,
nowadays listing more than $2.3$ million references.

Several works have been based on such data set, starting from the seminal paper by Cai et al. about community detection \cite{cai2005}. In this network representation, nodes correspond to Authors, pairwise connected when they have published a paper in the same conference; the more than $1.000$ conferences recognized by DBLP constitute the multilayer structure of the network.
Later works have used the same network to study specific characteristics of the system,
like for instance node centrality assessment and ranking \cite{ng2011,li2012,CosRoPe13},
the discovery of shortest-paths \cite{Brodka2011}, and the identification of coherent subgraphs \cite{boden2013}.

On the other hand, the DBLP data set has also been used as a test ground for more general multilayer network analysis.
For instance, Ref.~\cite{sun2006} proposes the use of higher order tensors algebra to multilayer and time evolving networks,
i.e. an {\it offline tensor analysis} (OTA) as an extension of the celebrated {\it Principal Component Analysis} \cite{kroonenberg1980,xu2005}. Also, Ref.~\cite{BerCosGi12} proposes the extension of several well-known metrics, e.g. the degree of a node,
 as well as new ones, as the case of the {\it dimension relevance} of a node in a layer.

\subsubsection{Other social interactions}

In addition to online human interactions and citations,
other types of multilayer social interactions can be found in the Literature.
Here, we briefly report them, as an example of the variety of goals and approaches
that can be accomplished by means of the multilayer network paradigm.

As a first example, Ref.~\cite{walker2013} describes the Scottish Community Alliance,
a network composed of local communities of people, deriving from the ancient Highlanders communities.
Such a system has the structure of a {\it network of networks}, as communities are organized in larger groups
at a regional level; but it is also organized in different layers, according to their main topics:
``Community Energy Scotland'', ``Community Woodlands Association'', the ``Federation of City Farms and Community Gardens'' and the ``Scottish League of Credit Unions''.

Politics is the focus of Ref.~\cite{Mucha10} and specifically the network created by call voting:
nodes thus represent US senators, and pairs of them are connected if they shared similar voting patterns across one year.
The analysis of multiple years is thus represented as a multilayer structure.
Authors demonstrate the usefulness of such a representation by performing a community analysis,
and showing how the community structure can be used to infer information about the global political situation.
Specifically, $9$ communities are identified, which are associated to important historical situations as the Compromise of year $1850$, the beginning of the American Civil War, or the Great Depression.

Another example can be found in Ref.~\cite{Batiston2013} which analyzes
the Noordin Top Terrorist Network, composed of $78$ known terrorists active in
Indonesia \cite{roberts2011}. Each network is defined by the interactions between pairs of individuals, i.e. trust,
operational, communication and business relations.

The social network built by Saint Paul along its biblical voyage through the Aegean sea is characterized in
Ref.~\cite{duling2013}, which considers the two layers corresponding to strong and weak ties.
While the former (including close friends and family) was usually considered to be essential for spreading risky ideas
(thus including recruitment to, and spreading of the new religion),
results suggest that Saint Paul mainly achieved his mission through a large network of weak social links.

Ref.~\cite{zignani2014exploiting} is concerned about the process of social communication using
different technologies as phone calls, or short text messages (SMS).
The considered data set has been created from billing information corresponding to a large European
metropolitan area, from March 26 to May 31, 2012.
It contains more than 63 millions phone-call records, and 20 millions SMS records.
The two layers do not perfectly overlap, with a $30\%$ ($8\%$) of users using only phone calls (SMS).
Moreover, most of the links, i.e. communications between two people,
exclusively occur by means of phone calls, while the share of people exchanging only SMS is as low as $21\%$.

Finally, Ref.~\cite{heaney2012} analyzed the networking between representatives from
$168$ leading groups on national health policy located in Washington DC during year 2003.
Three overlapping networks, i.e. layers, are considered: communication, coalition overlap and issue overlap.

\subsection{Technical systems}

\subsubsection{Interdependent systems}

One of the most promising fields of application of the multilayer framework is the study of technical interdependent systems,
i.e. those sets of systems in which the correct functioning of one of them strongly depends on the status of the others.
In other words, the failure of one element does not only provoke the breakdown of similar elements,
but also affects the dynamics of other coupled systems.
This leads to an iterative process of cascading failures, as summarized before in Sec.~\ref{sec:percolation},
 that has a devastating effect on the network stability.
In a natural way, different types of systems can be represented by layers,
connected between them by {\it dependency links} \cite{Buldyrev10,parshani2011}.

Electrical power grids have been one of the first targets of such analysis, due to their high interconnectedness
with both networks of different nature, like communication networks providing synchronization between different stations
\cite{Buldyrev10}, and of similar kind, like neighboring power grids \cite{BruSo12}.
The devastating effects of such cascading failure dynamics has been firstly studied in Ref.~\cite{Buldyrev10}.
Specifically, Authors analyzed the electrical blackout that affected the major part of Italy on 28 September 2003.
The shutdown of power stations directly led to the failure of nodes in the Internet communication network,
which in turn caused further power station breakdowns.

Two years later, Brummitt et al. in Ref.~\cite{BruSo12} analyzed the interactions
between different power grids, specifically between the three main regions constituting the U.S. network - Western, Eastern, and Texas. Adding some connectivity between the different regions is beneficial, as it suppresses the largest cascade in
each system by providing alternative paths. Yet, higher levels of interconnectivity have a negative impact,
both because they open pathways for neighboring networks to inflict large cascades,
and because new interconnections increase capacity and total possible load, which fuels even larger cascades.
In the same spirit, Ref.~\cite{sen2014identification} again considered the interaction between the power grid and the
supporting communication network by focusing on the problem of detecting the most central, i.e. the most vulnerable, node.
Authors focused on the Maricopa County, the most densely populated county of Arizona, USA. The resulting network
comprised $70$ power plants, $470$ transmission lines, and $42.723$ fiber links.

Beyond power grids, it is worth noticing the completely different problem tackled in Ref.~\cite{castet2013}.
There, Authors introduced {\it space-based networks}, i.e. a technology that allows the sharing of
spacecraft on-orbit resources, such as data storage, processing, and downlink \cite{ivancic2003} -
see Fig.~\ref{fig:Spacecraft}. Each spacecraft in the network can have different subsystem composition and functionality,
and can rely on others to complement (or back-up) its equipment. Thus, in this case, each layer represents a spacecraft,
with links connecting different layers representing the possibility of sharing capabilities.
Various design insights are derived and discussed, and the capability to perform trade-space analysis with
the proposed approach for various network characteristics is indicated.

\begin{figure}
\begin{center}
\includegraphics[width=0.7\textwidth]{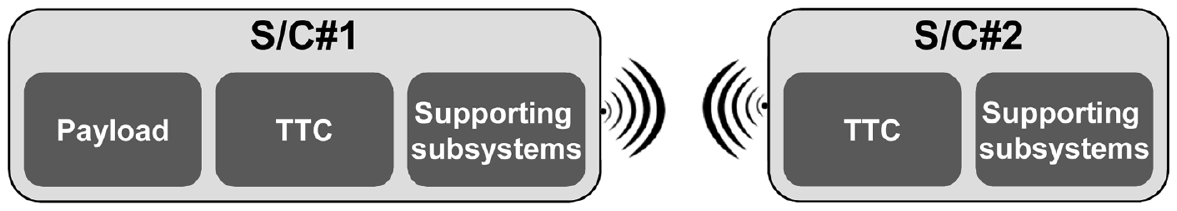}
\end{center}
\caption{Example of a space-based network. The two light gray boxes
  represent two spacecrafts, connected through a high-capacity
  communication system. Each one of them is composed of different
  modules, e.g. the Telemetry, Tracking and Command (TTC): when
  one of them fails, the corresponding spacecraft can rely on the other to backup that function, and to fulfill its mission.
 Reprinted figure from Ref.~\cite{castet2013}.}
\label{fig:Spacecraft}
\end{figure}

\subsubsection{Transportation systems}

Transportation systems are natural candidates for a multilayer network representation. Firstly because connections
between different nodes, e.g. direct flights or trains, are explicit, that is, their existence is published,
requiring little effort for the network construction. Secondly, because transportation systems are usually
interconnected, with a multilayer approach thus being the natural choice for intermodality studies. This latter aspect, i.e. the multimodal nature of transportation, has been tackled in two recent papers. In the first one,
a two-layer structure has been created by merging the world wide port and airport networks
\cite{parshani2010}, with the aim of studying the global robustness against random attacks.
While previous studies have considered the resilience of coupled systems when such a coupling was of a random nature
\cite{Buldyrev10,Parshani10}, there Authors focused on the level of similarity between layers, given for instance
the fact that well-connected airports are more connected to central ports.
Such a similarity is assessed by means of two metrics, namely, the inter degree-degree correlation and the
interclustering coefficient. Simulations executed using the topology of such a transportation system reveal that,
as the networks become more intersimilar, the system becomes significantly more robust to random failure,
as one network is able to {\it back-up} the other, when needed.

In the second publication \cite{halu2013}, a multilayer structure is used to model the Indian air and train
transportation networks, the nodes of the latter being stations, with pairs of them being connected if within
one stop distance.
Pairs of nodes belonging to different layers, i.e. airport-station pairs, are connected whenever
a direct road access between both of them is present. It is worth noticing that both networks have a very different nature.
While the airport network is small (with 78 nodes) and presents a scale-free disassortative structure,
the rail one is much larger ($6.769$ stations) and has a topology strongly constrained by the geography,
implying a narrow degree distribution and an assortative topology.
Yet, such heterogeneity enables a good performance of the global network in terms of navigability.

As for the analysis of individual transportation networks, a few examples can be found in the Literature,
in which a multilayer structure is used to model different aspects of the transportation process.
For instance, Ref.~\cite{kaluza2010} analyzed the network created by worldwide cargo ship movements;
 while the $90\%$ of world trade is carried by sea, it is noteworthy that little attention has been devoted
to this transportation mode from the field of complex networks.
The multilayer structure is given by the three most common classes of cargo ships, i.e. container ships,
bulk dry carriers and oil tankers, which usually connect different ports and travel following different patterns.
For instance, bulk dry carriers and oil tankers tend to move in a less regular manner between ports than container ships.
The understanding of the specific patterns of movement created by each type of ship, instead of analyzing the
topological projection of this multilayer network, is motivated by the problem of invasive species spreading:
bulk dry carriers and oil tankers often sail empty, thus exchanging large quantities
of ballast water and favoring the spreading process.

Finally, the air transport network \cite{zanin2013} has also been studied alone,
but taking into account the multilayer structure created by different airlines.
As demonstrated in Ref.~\cite{CardilloSR13}, the global network is the result of the interactions
between two main types of layers: those created by {\it major} airlines on the one side,
and those generated by {\it low cost} (or {\it low fares}) carriers on the other.
The former are characterized by a {\it hub-and-spoke} topology, with one central hub receiving and
dispatching all flights; the latter by a {\it point-to-point} configuration, of a more random nature.
The topological properties of the global network, as for instance its low diameter or the high clustering coefficient,
only appear when several layers of different nature are merged together.
While an analysis of the topology of connections may be relevant, at least from a theoretical point of view,
the study of the dynamics taking place on top of it is of high interest for the air transport community,
as it would allow tackling problems like the propagation of delays through the system,
or the effects of such dynamics on the movement of passengers \cite{zanin2013}.
Ref.~\cite{CardilloEPJST13} tackles the problem of passenger rescheduling when a fraction of the links, i.e. of flights,
is randomly removed. A simple dynamical model is executed using two networks: the single-layer projected network,
and the whole multilayer structure. In the latter, airlines incur in a cost whenever a passenger needs to change layer,
thus effectively reducing the flexibility of the system. Results indicate that the layered structure is
associated to a lower resilience of the system against such kind of perturbations.
In other words, the use of a projection is a simplification that results in an over-estimation of the resilience
of the air transport network, as it does not take into account the cost of moving between layers
- see Fig.~\ref{fig:ATN} for more details.
At the intersection between the topological and dynamical analysis of the air transport network,
Ref.~\cite{communicability} focused on the {\it communicability}, i.e. the number of possible routes
that two nodes can choose to communicate with each other \cite{comm1}.
By aggregating the communicability corresponding to each departure node, i.e.
considering the broadcasting of information to passengers,
and by varying the degree of interactions between different layers,
it is possible to create airport rankings that takes into account the multilayer structure or, in other words,
quantifies the effects of the airline structure on the movement of passengers.
The most central node thus varies: while the airport of Paris CdG is the most central in Europe when
layers are disconnected (main consequence of the extension of the Air France network),
this role is passed to London Stansted and Frankfurt when increasing the coupling.

\begin{figure}[t!]
\begin{center}
\includegraphics[width=0.95\textwidth]{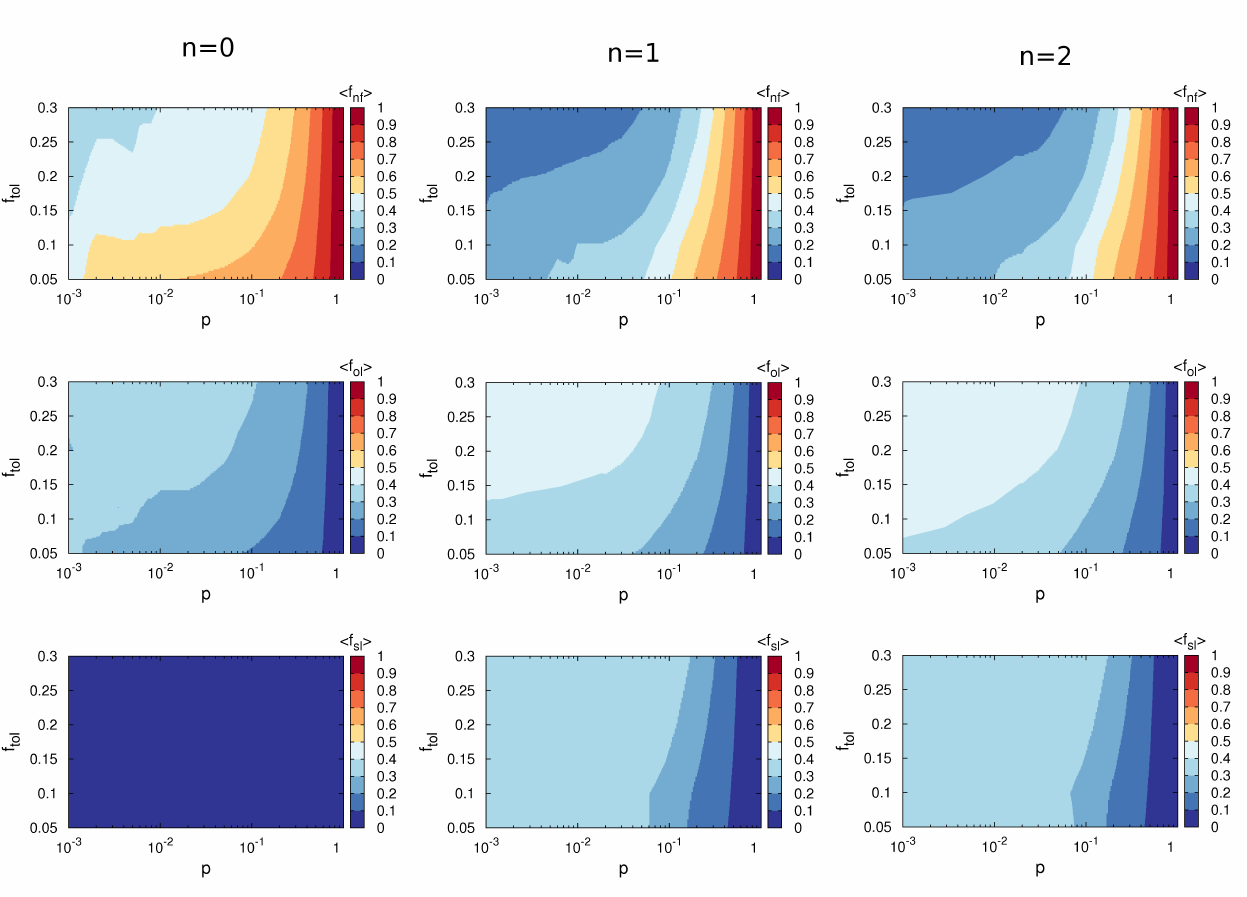}
\end{center}
\caption{(Color online). Outcome of the re-scheduling process in the
  multilayer air transport network, as a function of the probability
  of link failure, $p$, and load tolerance, $f_{tol}$. Each column
  displays (from top to bottom): the average fraction of passengers
  that cannot fly, that of those that are re-scheduled in other
  layers, and that for those re-scheduled within the original
  layer. Each column accounts for the possibility of scheduling
  passengers on paths with length $0$ (left), $1$ (center), and $2$
  (right).  Reprinted figure from Ref.~\cite{CardilloEPJST13}. With kind
permission from Springer Science and Business Media \copyright \, 2013.}
\label{fig:ATN}
\end{figure}

\subsubsection{Other technical systems}
As a final note on technological systems, it is worth noticing
Ref.~\cite{hauge2014selected}, which analyzes the problem of establishing a reliable communication network
between heterogeneous elements, and specifically between different warfighters in an army.
Thanks to the technological progress, each soldier and vehicle in the battlefield can be connected
to others by means of different communication channels, each one having specific characteristics:
being wireless or not, bandwidth, range, etc. When these different systems are considered as heterogeneous layers,
the battlefield can be represented by a single multilayer network - see Fig.~\ref{fig:Warfare}.
The challenge is then to ensure a high end-to-end Quality of Service (QoS), i.e. that two units can share information
when needed, making use of different communication channels.


\begin{figure}
\begin{center}
\includegraphics[width=0.7\textwidth]{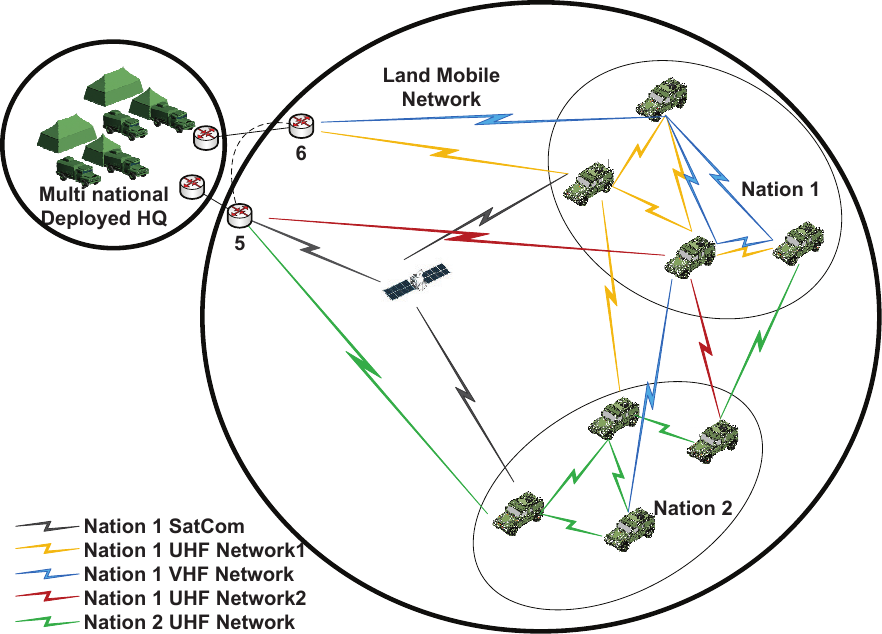}
\end{center}
\caption{(Color online). Example of an heterogeneous mobile
  communication network.
The battlefield can be seen as a set of resources,  e.g. tanks, soldiers, or even satellites,
which require to continuously transmit tactical information to the
others. Due to the different characteristics and capabilities of each resource, multiple
communication networks are deployed (as, for instance, satellite, UHF or VHF links), making the system a multilayer network.
Reprinted figure from Ref.~\cite{hauge2014selected}. Courtesy of P. Lubkowski.}
\label{fig:Warfare}
\end{figure}

\subsection{Economy}

\subsubsection{Trade networks}

The International Trade Network (ITN), also known as the World Trade Web (WTW), is defined as the graph of all
import/export relationships between world countries. In recent years, it has been extensively analyzed from the
complex network theory point of view, due to its importance: for instance, its topology can be used to assess the
likelihood that economic shocks might be transmitted between countries \cite{kali2007}.

Traditional network studies have considered the ITN at the aggregate level,
in which links represent the presence of a trade relation, irrespectively of the commodity actually traded.
Nevertheless, the introduction of the concept of multilayer network has enabled the unfolding of the aggregate ITN
such that each layer represents import and export relationships for a given commodity class, according to some standard classification scheme.

This latter approach has been conducted in Refs.~\cite{Barigozzi2010,Barigozzi2011},
by using a network composed of 162 countries (nodes) and 97 layers, evolving yearly from $1992$ to $2003$.
In the former work, Authors considered different topological metrics and how they differ across the layers:
among others, node centralities (and thus, country rankings) and connected components.
On the other hand, the latter work focused on the identification of communities,
highlighting the fact that commodity-specific communities have a degree of granularity which
is lost in the aggregate analysis.
A similar study was performed in Ref.~\cite{mahutga2013}, but relying on a different data set:
the {\it UNCOMTRADE}, collected by the United Nations, which includes information for $94$ countries starting from year $1962$.

A different approach to trade analysis has been proposed in
Ref.~\cite{ducruet2013} regarding the study of maritime flows.
Extending the large research body devoted to the study of maritime flows by means of standard complex networks
\cite{kaluza2010,keller2011,ducruet2012}, Ref.~\cite{ducruet2013} presents a model where ports (nodes)
are connected through five different layers,
representing the five main categories of commodities exchanged in maritime business:
liquid bulk (i.e. crude oil, oil products, chemicals, etc.), solid bulk (like aggregates, cement, or ores),
containers, passengers/vehicles and general cargo.
Data have been extracted from the {\it Lloyd's Voyage Records}, covering the months of October and November 2004.
An inspection of this disaggregated network suggests that most networks measures,
like centrality or clustering, exhibit strong sensitivity to the type of commodity traded in a given set of ports.
The most diversified ports are on average, the bigger and more dominant they are in the network,
while they also connect over greater physical distances than more specialized ports - see Fig. \ref{fig:Maritime}.

\begin{figure}[t!]
\begin{center}
\includegraphics[width=0.9\textwidth]{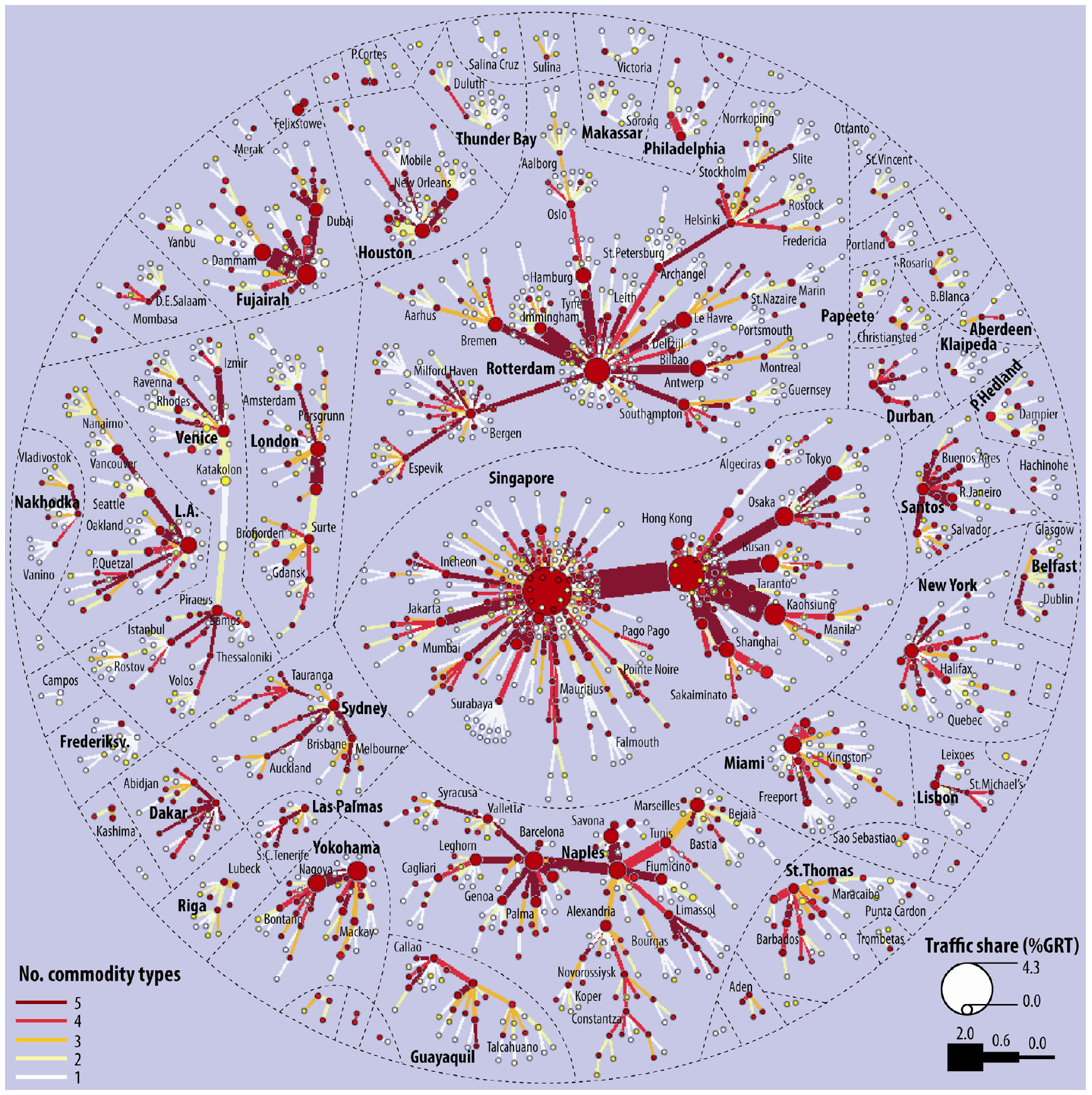}
\end{center}
\caption{(Color online). Commodity diversity and nodal regions in the
  maritime flow network.
Nodes represent ports, pairwise connected by links when a commodity is transported
between them. Different commodities are considered, thus making this
graph a multilayer network. In order to represent this heterogeneity, links are
colored according to the number of commodity types traveling between
ports. Furthermore, both nodes and links sizes represent the corresponding traffic
share. Reprinted figure from Ref.~\cite{ducruet2013}. \copyright\, 2013 with
permission from Elsevier.}
\label{fig:Maritime}
\end{figure}

\subsubsection{Interbank market}

The 2008 financial crisis has raised the interest of the scientific community
toward the understanding of the complex system emerging from the interaction between financial institutions and markets
\cite{boss2004network,soramaki2007topology,gai2010contagion}.
Most contributions have focused on the interbank market, where nodes represent financial institutions,
and links credit relations between two counterparties. Networks are usually directed and weighted,
such that directionality identifies the borrower and the lender, and the weight of the link represents the loan amount.

Ref.~\cite{bargigli2013multiplex} is hitherto the only work analyzing this network from a multilayer point of view.
Authors utilize a unique database of supervisory reports of Italian banks to the
{\it Banca d’Italia}, including all bilateral exposures of all Italian banks,
broken down by maturity and by the secured and unsecured nature of the contract.
Each layer in the network represents a different type of exposure, namely:
unsecured overnight, unsecured short-term (less than 12 months),
unsecured long-term (more than 12 months), secured short-term and secured long-term.
The analysis of these five layers shows a high heterogeneity between them,
with the topology of the overall network largely reflecting the one of the unsecured overnight layer.
This layer is especially important from the policy-making point of view,
as it is the focus of monetary policy operations in several jurisdictions.
Notably, such layer is characterized by a high persistence of links over time,
such that, if two banks are connected at a given time, they will have a high probability of being connected
in the future. This may have important consequences in the stability of the system, in terms of perturbations contagion.

\subsubsection{Organizational networks}

A third active field of research in economy is the study of
{\it organizational networks}, i.e. the study of how different organizations interact,
by exchanging information, money, or other resources,
with the aim of collectively produce a good or a service.
While traditionally such interactions had a hierarchical structure,
in recent years more complex topologies have emerged,
in which the power of the decision making process is spread among many participants
\cite{camarinha2005collaborative}. Once again, complex networks stand for as the most suitable
instrument for the analysis of the resulting
interaction structures \cite{dooley2007managing,durugbo2011modelling}.
While the importance of the multilayer nature of these networks was recognized almost three decades ago
\cite{granovetter1985economic,galaskiewicz1998nonprofit}, only recently this problem has been mathematically tackled.

One of the first works analyzing organizational networks as multilayer structures
is Ref.~\cite{amburgey2008structural}. There, Authors analyzed more than $6.000$
{\it Research \& Development} and more than $6.500$ {\it Management \& Development} alliances
involving $1.000$ biotech firms in the United States.
The time window considered covers over 30 years, in which data were available quarterly,
thus enabling the construction of a 120-layers temporal network.
The underlying hypothesis is that the actions of a given firm can be better understood
if one takes into account changes in the network structure,
like the firm position (centrality).
The main result suggests that biotech firms act following a process of preferential attachment, i.e.
organizations are more likely to form ties with organizations of similar institutional and structural status.

The multiplex paradigm has also been used to study
the interactions between organizations using Information and Communications Technology (ICT)
firms for development goals, i.e. the use of ICTs as a tool for empowering less developed countries
\cite{lee2011coevolution}. Specifically, two nodes can be connected when the corresponding organizations
collaborate in implementation or knowledge-sharing projects, thus creating a two-layers network.
In Ref.~\cite{lee2011coevolution}, the studied data set comprised 323 projects between years 1997 and 2005, 211
of which were implementation projects and 112 knowledge-sharing projects, for a total of 752 organizations.
Results indicate that the existence  of a link in one layer is influenced by several factors:
the presence of the corresponding link in the other layer,
the presence of a common third-party, or the centrality of connected nodes, among others.

Within organizational networks analysis, of special interest has been {\it Corporate Governance},
defined as the study of the interrelations between ownership and control of a firm.
Stockholders invest their money in a company whose strategic decisions are taken by the top management:
as both of them are embedded in a network (for instance, a single investor may hold shares of multiple companies),
the resulting structure can be seen as a multilayer network composed of two layers,
respectively ownership and control. Thus, a multilayer analysis may give a unique perspective on the relevance
of management rather than ownership in the company’s policy.
Such an approach has been addressed in Ref.~\cite{bonacina2014multiple},
focusing on the shareholding and board structures in the year 2013 for 273 quoted companies on the Italian Stock Market.
The main claim is that corporate governance cannot be understood unless ownership and management
are taken simultaneously, as the two layers are characterized by a small overlap:
while the core of shareholding is composed of banks and financial institutions,
the most central nodes in the management layer are large industrial companies.

\subsection{Other applications}

\subsubsection{Biomedicine}

The introduction of the complex network methodology in biomedicine caused a kind of a small revolution,
as it made possible a real {\it systems medicine}: not only studying the elements constituting a living being,
but analyzing how they communicate and interact in a systemic way.
Two are the main fields where the network approach has been successful.
On the one side, the characterization of how the main constituents of the cell are organized:
specifically, genes \cite{barabasi2004,barabasi2011}, proteins \cite{jeong2001,mutation}
and metabolites \cite{guimera2005}. On the other side, and from a  higher level perspective,
one may find all the works in neuroscience that deals with the complexity of the brain as a network,
both anatomically \cite{sporns2011} and functionally \cite{bullmore2009}.

Up to now, these networks, e.g. the ones created by interactions between genes, proteins and metabolites,
 have been considered in an independent fashion. Nevertheless, they are clearly not independent:
genes control the production of proteins, which in turn catalyze metabolic reactions.
Malfunctions in cells are seldom due to just one element, being instead the result of multiple
interactions at different levels. It has been recently suggested that a multilayer approach, i.e.
a representation of the cell composed of different layers for each one of these elements,
may yield relevant knowledge about the normal dynamics, as well as about the appearance of
pathological conditions \cite{vandamme2013}. The final aim is called {\it personalized medicine}, i.e.
a medicine focused on the individual and proactive in nature,
which would allow a significant improvement in health care by customizing the treatment according to
each patient needs \cite{hood2013}. In spite of its importance, such multilayer approach has not yet been achieved.

While a full multilayer representation of the cell has not yet been constructed,
there are several situations in which different networks can be combined or analyzed together.
For instance, molecule interaction networks can be extracted from different organisms,
the objective being the identification of modules that have been preserved across different evolution steps.
Following this line of research, in Ref.~\cite{michoel2012} Authors found
common modules by means of a spectral clustering algorithm extending
the Perron-Frobenius theorem. Among the resulting shared modules, some examples include the
{\it Mini Chromosome Maintenance} complex, which plays an important role in eukaryotes DNA replication \cite{fletcher2003}, or
the V-ATPase, an ancient and highly preserved enzyme that performs different functions in eukaryotes cells \cite{kibak1992}.
It has also been proposed the multilayer analysis of the protein-protein interaction network
by considering the type of relation connecting pairs of them:
physical when they can bind together to form a larger complex,
and genetic if one of the two proteins regulates the other, i.e. by triggering the activation (or repression)
of the gene responsible for the production of the target protein \cite{nicosia2014measuring}.

Another example of a multilayer approach emerges when a set of data, representing the same molecules,
are extracted from different conditions,
 or are analyzed by different means. These two approaches are explored in Refs.~\cite{li2011,li2012b}.
Specifically, Ref.~\cite{li2011} works with a multilayer structure composed of 130 co-expression networks,
in which each layer represents a different experimental condition.
The aim was the identification of {\it recurrent heavy subgraphs}, i.e. subsets of densely interconnected
nodes that consistently appear in the different layers. While this is similar to a community identification
task in a single layer, the simultaneous analysis of multiple layers is expected to enhance signal-noise separation.
In Ref.~\cite{li2012b}, an initial data set of gene expressions is used to create two different networks:
firstly, a standard co-expression network, where pairs of genes are connected whether a relationship
(e.g. statistical correlation) is found between their expression levels; and secondly,
an exon co-splicing network, in which nodes represent exons and the edge weights represent correlations
between the inclusion rates of two exons across all samples in the data set.
These two networks are not independent, but indeed form a two-layers system, as each gene contains several exons.
Furthermore, transcription and splicing factors are expected to be consistent among related genes, thus giving rise to
{\it coupled modules}.
This last hypothesis has been confirmed by comparing the obtained modules with data sets of gene functions.

\subsubsection{Climate}

The last decade has witnessed an increasing interest toward the application of complex network
theory to climate science. Networks are usually reconstructed such that nodes correspond to spatial grid
points of the Earth, and links are added between pairs of nodes depending on the degree of statistical
interdependence between the corresponding pairs of anomaly time series taken from the climate data set.
The graph approach then allows a topological and dynamical analysis of the climate system over many scales, i.e.
from local properties of a single point and its influence on the whole planet,
to global network measures associated to stabilization mechanisms \cite{yamasaki2008,donges2009,steinhaeuser2012}.

Ref.~\cite{Donges2011} constructs a multilayer network, where each layer represents the Earth's atmospheric
condition at a given geopotential height; nodes are associated to pairs of coordinates (i.e. latitude and longitude),
and pairs of them are connected with a weight defined by the zero-lag linear correlation between the time series
corresponding to the two nodes. In this way, the multilayer network provides a representation of the geostrophic
wind field as well as large-scale convection processes in the troposphere and the lower stratosphere.
By analyzing its topology, and especially the centrality of nodes (by means of adapted degree,
closeness and betweenness centralities), it is possible to extract relevant knowledge about the atmosphere dynamics,
without considering fields of temperature, moisture content or other relevant climatological variables
that are not straightforward to obtain.

A different approach is proposed in Ref.~\cite{feng2012}, where a two-layer network is constructed to analyze sea-air
interactions. Nodes in the two layers respectively codify positions (pairs of latitudes and longitudes),
and links are once more weighted according to a zero-lag linear correlation between time series associated to each node:
geopotential heights for points in the air, and surface sea temperature for points in the sea.
By analyzing this network, it is possible to recover the coupled Asian monsoon and Walker circulations,
known as the {\it Indian and Pacific Ocean} ({\it GIP}) model.

\begin{figure}
\begin{center}
\includegraphics[width=\textwidth]{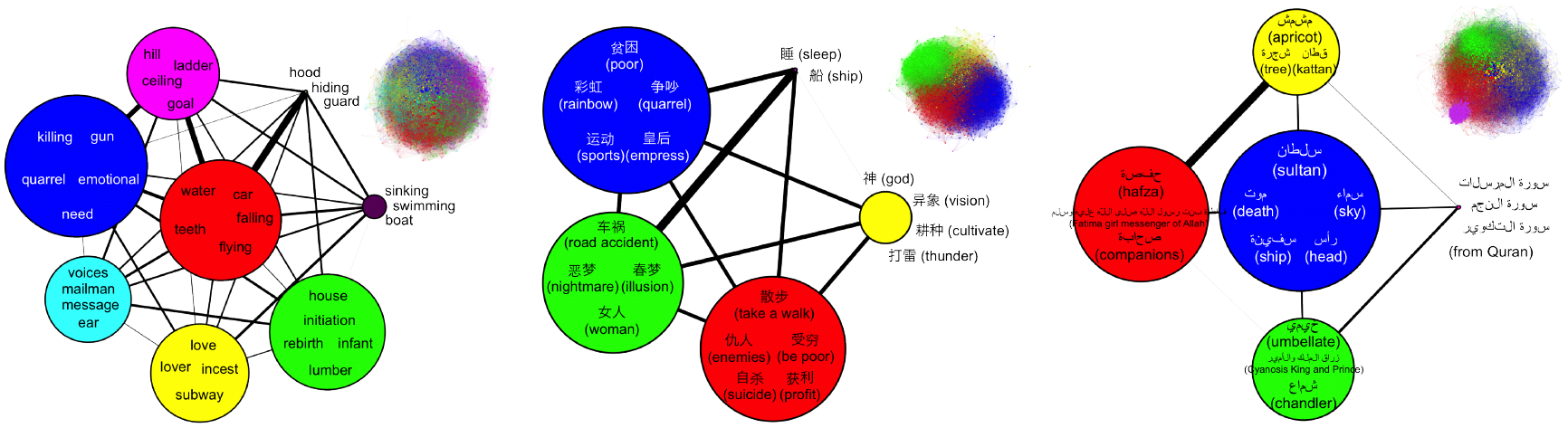}
\end{center}
\caption{(Color online). Communities in the dream interpretation network: English (left), Chinese (center), and Arabic (right). Symbols with the highest eigenvector centrality are used to label each community. The complete networks are illustrated in the upper right corners of each figure, with colors corresponding to the communities (colors are not intended to match across different networks). Reprinted figure from Ref.~\cite{varol2014connecting}. }
\label{fig:Dreams}
\end{figure}

\subsubsection{Ecology}

In this subsection, we concentrate on two paperworks that analyzed, from a multilayer point of view,
the interactions between animals and the structure of an aquifer system, respectively.

In the first one, the interaction network of a group of 12 female baboons
in the De Hoop Nature Reserve, South Africa,
is considered \cite{Barret2012}. Specifically, three different types of interactions were recorded along 10 years,
by means of human observations: agonistic, spatial and grooming interactions.
This remarkable data set has then been used to build a multilayer structure, in which each layer represents
each one of those social aspects. Beyond the complete, or {\it control}, network,
two more have been reconstructed, by monitoring the social changes resulting from the death of two subjects,
specifically a dominant and a low-ranked female. This allowed Authors to compare the topological
properties of the three networks, specifically their clustering coefficients, and analyze how social ties are
modified by the two events. The loss of the dominant female resulted in an instability in the dominance hierarchy,
which was compensated for by adjustments in the spatial association network.
On the other hand, the loss of the low-ranked female had little effect on the social network.
This suggests that dominance serves to regulate the interactions of all group members.

A completely different application is proposed in Ref.~\cite{mouhri2013}, where the aquifer system of the
Orageval catchment in France is studied. In order to correctly describe the structure of this system,
a two-layer network representation has been reconstructed, where the two layers respectively map the Oligocene
and the Eocene geological layers. After geophysical and drilling investigations, aimed at assessing the general
structure of the aquifer system, electrical resistivity tomography was used to reconstruct the
connectivity patterns at different spatial resolutions. When combined with temperature and hydraulic head information,
this multilayer structure provides the researcher with valuable information about the stream-aquifer interface.

\subsubsection{Psychology}

While psychology has not traditionally contemplated complex networks as a tool for analyzing human behavior,
it is worth noticing the study proposed in Ref.~\cite{varol2014connecting}. In this work,
the symbolism of dreams and their interpretation, i.e. oneirology \cite{de1982dreams}, is studied by means of networks:
nodes represent dream symbols, pairwise connected according to the overlap in their meaning (i.e. the similarity
in their corresponding interpretation documents). A different network is created for each considered culture
(English, Chinese and Arabic), for then constructing a network of networks by connecting common symbols.
Findings suggest that sentiments connected with a given interpretation can be associated with communities
in the network, and that communities can be themselves clustered into three main categories of positive,
negative, and neutral dream symbols - thus revealing a complex multiscale structure - see Fig. \ref{fig:Dreams}.

\section{Conclusions and open questions}

After having revisited the main theory and applications of multilayer networks, we feel our duty to draw a few concluding remarks, and pinpoint questions that remain still open to future progresses, at the same time  addressing the reader to some
issues and problems that we consider of key relevance, and that we believe it is more than likely that would attract soon the attention of scientists in the area.

In Section \ref{sec:structuremultilayer} we started with trying to establish a common notation (and some basic definitions) for multilayer networks, this way  joining forces that have been spent in the last years to describe the structure of several complex systems. The main inspiration was trying to prove that the different terminologies reported so far could be, in fact, encompassed within a single framework. However, there is still a lot of work ahead in order to set a proper and comprehensive mathematical formalism, and possibly a long way to go. Among the different open problems to be solved, we highlight the following three: {\it i)} the need of setting up other metric concepts that could possibly affect relevant parameters of the systems, such as the betweenness, the vulnerability and the efficiency amongst others; {\it ii)}  the need of gathering a better knowledge and understanding on the mathematical relationships among the values of centrality parameters when measured on each single network's layer and when, instead, measured on the whole multilayer structure; {\it iii)} the need of paying effort for extending other quantifications pertinent to the dynamics of single-layer networks, such as, for instance, those regarding controllability and observability.

In Section \ref{sec:generativemodels}, we followed up by reviewing the different models that have been proposed for generating multilayer networks endowed with the topological structures observed in real world systems. Possibly, this is the part of state of the art that still lies in its infancy. Definitely, and much more than being confident of, this issue is of the utmost importance. Retrieving the mechanisms at the basis of what is ubiquitously seen in social, technological and biological environments will certainly contribute to uncover the intimate physical processes that ultimately determine the way networked systems shape their interactions. Multilayer and multiplex networks encode indeed, significant more information in their structure with respect to their single-layers taken in isolation. A direct way to extract information from these structures is to look at correlations. For multiplex networks, these can be represented by degree correlations between different layers, or overlap of the links, or correlations in the activities of the nodes. For networks of networks, the degrees of freedom are even more pronounced since the interlinks between the nodes in different layers can be assigned using a large variety of rules.
In our Section \ref{sec:generativemodels}, we have tried to give an overview of generative models using both a growing multilayer network framework, and the framework of an ensemble of networks generating unbiased null models.
Growing multiplex network models can be used to generate multiplexes with positive correlation in the degrees of the nodes, when the preferential attachment is linear, or tunable (positive or negative) degree correlation when the preferential attachment is non linear.
On the other side, multiplex network ensembles using multilinks can be used to generate multiplex networks with tunable levels of overlap of the links incident to each node, that cannot be generated using the configuration model in each layer of the multiplex.

At variance, in Section \ref{sec:percolation} we attempted to review what are probably the most studied problems in multilayer networks, i.e. the issues connected to the overall network's resilience (or fragility) to external perturbations, and to percolation. Such processes, indeed, once defined on multilayer networks, display many relevant and unexpected properties that cannot be simply reported to the case of a single layer. As soon as interdependencies between the nodes in different layers are taken into account, one of the most important questions is to characterize their robustness properties, i.e. their response to random damage. In systems like complex infrastructures, we observe that an increased level of interdependencies can be responsible for an increased fragility of the system that might be affected by cascades of failure events. 
On the other hand, in a network with interdependencies between the nodes of different layers, the percolation transition can be discontinuous, indicating a discontinuous emergence/disruption of the mutually connected component. At the same time, the response to damage of these networks generates {\it eventually} a disrupting avalanche of failures events. The investigation of the effects of multiplexity into the robustness of networks has revealed that when the networks are uncorrelated and each node has at least one interdependency link, the transition is {\it always} expected to be discontinuous and characterized by avalanches of failures.

The nature (and structural characteristic) of the interlink might also affect the robustness properties of the network. For instance, in the context of networks of networks, if any node is linked to its replica node and there is a fixed supernetwork, the size of the mutually connected component is independent of the structure of the supernetwork as long  as it is connected, while in the case of nodes randomly linked to other nodes in other layers,  the structure of the supernetwork matters.Moreover, if the nodes are only connected to their replica ones, either all the layers percolate, or none does. In the case in which the nodes in one layer are linked to random nodes in other layers, one has that layers with higher superdegrees are more fragile than layers with lower superdegrees.
Moreover, in the presence of partial interdependence or overlap of the
links, the transition can become continuous, and the response to damage less disruptive. This opens the challenging problem of characterizing the general structural conditions that can enhance resilience and robustness of the entire system. While for infrastructure and man made networks it is plausible to assume that the design of the system has not taken into account the role of interdependencies and has therefore resulted in fragile structures, in biological system such as the cell (or the brain)  we are more inclined to believe that evolution must have driven the system toward a relatively robust configuration, at least finding a tradeoff between robustness and adaptability of the entire system.

In Sec.~\ref{sec:spreading} we have shown that spreading processes in multilayer networks constitute possibly one of the best suited dynamical frameworks to unveil the effects that the coupling between networks has on their functioning. For instance, linear diffusion being the most simple dynamical setup allows to monitor the transition between two limiting regimes (one in which the networks behave independently and another corresponding to a coordinated functioning) with the transition between these two phases appearing to be abrupt according to the spectrum of the Laplacian. The use of more refined diffusion processes such as random walkers allows to go one step further and device tools for ranking the importance of the nodes, and monitoring it as a function of the coupling between networks. The study of even more realistic diffusion processes, such as those mimicking the transport of information packets, on interconnected networks has been also addressed, although more results on this line are certainly to be expected. In this regard, the transition to congestion in information networks is of key importance.

In addition to the report on diffusion processes, Sec.~\ref{sec:spreading} paid attention to the problem of disease spreading on multilayer networks, where several related disease dynamics propagate on separated network layers, or both social information and
infection information simultaneously diffuse and affect each other. Compared with the case of single-layer networks, this new
framework can greatly change the threshold of disease outbreak, which is usually shown in the form of a phase transition
between the healthy and epidemic phases. More research regarding this important issue is needed since SIR and SIS models seem
to behave differently. In particular, SIR dynamics allows system states in which the epidemic regime only occurs in one
of the networks (while not affecting the others) whereas for the SIS dynamics the propagation in one network automatically
induces the epidemic in all the remaining networks.
Besides the prediction of the epidemic threshold, we also emphasized studies that take advantage of the multilayer
benchmark to couple the spreading dynamics with the development of voluntary prevention measures to control and
eradicate the infection risk. Along this research line, there is still room for many proposals considering interconnected networks.
As it is known, social dynamics, including information propagation and opinion formation, plays a non-trivial role in the risk
perception under an epidemic threat. Apart from studying the emergence of individual contention measures under disease outbreaks,
the same approach seems to be adequate to explore dynamical prevention measures in the Internet, so to control,
in a dynamical way, the attack of viruses.

As we have seen in Sec.~\ref{sec:spreading}, evolutionary games are another fascinating field of research that can benefit from the multilayer framework.
At variance with games in single-layer networks, evolutionary games on different network layers can be coupled
via either the utility function or the strategy of the nodes. We summarized the main influence of these kinds of inter-layer
coupling on the maintenance of collective cooperation behavior both in pair-wise games (such as the Prisoner's Dilemma)
and $n$-person ones (such as the Public Goods Game). The reviewed studies clearly point out that the different
coupling between networks enriches the enhancement of cooperative behaviors provided by spatial reciprocity.
In spite of the great achievements, there are still unexplored problems related to evolutionary games that merit further attention.
The most obvious issue is extending to multilayer networks other types of games like the Ultimatum game
\cite{guth1982experimental,bolton1995anonymity}, the Rock-Scissor-Paper game \cite{peltomaki2008three},
the Naming game {\cite{baronchelli2004,baronchelli2006topology}} and the Collective-risk social dilemma \cite{milinski2008collective}.
These games usually focus on the emergence of other collective states such as, fairness, species diversity, cyclical dominance, language evolution and prevention of dangerous climate change. Within the novel multilayer framework, one can get a  better understanding on the impact that the interaction among different networks plays on the emergence of collective states. For instance,
during the campaign of mitigating climate change, the mutual impact of different local systems is crucial.
If the total contribution of these systems is less than the expected target, it may cause a cascade of reducing investment leading to the tragedy of the commons \cite{hardin:1968}.
Experimental research is another unexplored field of evolutionary games on multilayer networks. Previous experiments on single-layer networks have unveiled that heterogeneous topology does not promote cooperation \cite{gracia2012heterogeneous},
which is inconsistent with theoretical predictions \cite{santosprl}. Naturally, if the multilayer interaction is involved,
how cooperation trait varies will become of particular importance.

In section \ref{sec:synchro} we have given a rather complete overview of the current results about
synchronization in multilayer networked systems. Starting
from the simplest case of only two layers, we have first reported on the enhancement of synchronization when the topological
evolution of a network happens through a sequence of commutative layers.
We have then demonstrated how one can in fact obtain a fully
general treatment, without the need to impose any constraint
on the layers and their mutual relations.  Other specific
cases we considered are coexisting layers, hypergraphs, interconnected
networks, multiplex networks and bipartite graphs. Finally,
we discussed recent results on the problem of targeting dynamical
trajectories of a system.
The new approach we described for general network evolution
opens several possibilities for research areas in the near
future. In particular, it should be noted that by the application
of Eq.~(\ref{valueevol}) and Eq.~(\ref{NCbox2}) one can effectively specify the evolution of the
Laplacian eigenvalues and eigenvectors during the switch from one layer to another.
Thus, one can impose any kind of transient behavior in the
network. This method can be exploited to achieve finer control
on time-evolving complex networks, with potential applications
to many areas of social impact.

Further, the role of symmetry on the dynamical properties
of multilayer networks is currently completely unexplored.
Studies exist describing the importance of topological
symmetries for synchronization, but they all deal
with single-layer graphs. The addition of new layers
could lead to substantially different results, which
need to be investigated. Notice that the two cases
of temporal layers and spatial layers are likely
to involve different mathematics, with potentially
different conclusions.

Another very relevant subject of investigation that will certainly attract attention is
the study of how a multilayer distribution of links can spontaneously emerge as a consequence of
adaptation of the interactions between constituents of a system.
In Refs.~\cite{Assenza2011BoccaSR,Gutierrez2011BoccaPRL,Avalos2012BoccaPRE}
it was indeed shown that, when considering a single-layer network of phase oscillators and when the interaction weights are
themselves dynamical variables evolving by means of differential equations explicitly coupled with the state
of the network's nodes, the asymptotic emergence of a synchronous state is associated with a structural self-organization into
a modular and scale-free topology. The extension of this formalism to the case of multilayer interactions between oscillators is in progress,
and promises to unveil relevant information on how a multilayer structure of connections emerges spontaneously in connection with
the setting of specific collective dynamics.

Finally, in Sec.~\ref{sec:applications} we have reviewed the current literature on applications of the
multilayer formalism and representation to several cases of relevance in social sciences, technological systems and economy.
We firmly believe that in the future biology and biomedicine will greatly take advantage of a multilayer representation,
although these two fields of research have until now lag behind.
This is mainly due to the difficulty associated with the recording of real data in living organisms,
especially with the high resolution needed to discriminate the physical layers.
This is, for instance, the case of the human brain: while the human cortex is composed of six layers,
each one processing information in a distinctive way, actual recording technologies
(e.g. EEG or MEG) can only detect the activity of the top-most layers. When functional representations
(i.e. representing the activity of the brain in terms of information processing) will be merged with
structural ones (e.g. the connectome, that is, the structure of fibers connecting different parts of the brain),
 the result will be a complex multilayer network, describing both how and why information moves.
Beyond the brain, it is commonly accepted that cells are composed of different networks interacting
between them: from the RNA, mRNA, up to proteins and metabolites. While these networks have been deeply studied
in separated and isolated fashion, many pathological conditions generate from the wrong
interaction of the different networks (layers). Adopting the multilayer paradigm will then provide an elegant mathematical way of
representing these cross-interactions, paving the way for a deeper comprehension of the structure and function of cells.

\section*{Acknowledgements}
\addcontentsline{toc}{section}{Acknowledgements}

We would like to acknowledge gratefully all
colleagues with whom we maintained (and are currently maintaining) interactions and discussions on the topic
covered in our report.

In particular, we would like to thank J.A. Almendral, A. Amann, R. Amritkar, K. Anand, R.F.S. Andrade, F.T. Arecchi, A.
Arenas, S. Assenza, V. Avalos-Gayt\'an, A. Ayali, N. Azimi-Tafreshi
% ...

The uncountable number of stimulating and inspired discussions with them (and their sharing with us of
unpublished results on the subject) are largely responsible for having spurred and encouraged us to provide the present account on
this rapidly growing field of research.

At the same time, we also would like to gratefully acknowledge all the discussions we had (in meetings, Conferences, Workshops, and personal visits) with other colleagues who are not mentioned in the above list (and, for that, we really apologize), which equally inspired our efforts, and opened up our minds in a way that contributed, eventually and substantially, to the realization of the present survey.

This work was partly supported by the Spanish MINECO under projects FIS2011-25167, FIS2012-38949-C03-01, and FIS2012-38266-C02-01; by the European FET project MULTIPLEX ``Foundational research on multilevel complex networks and systems'' (Reference Number 317532); by the Comunidad de Aragon (FENOL group); the Brazilian CNPq through the PVE project of the Ciencia Sem Fronteiras.
C.D.G. acknowledges support by EINS, Network of Excellence in Internet Science, via the European Commission's FP7 under Communications Networks, Content and Technologies, grant No.~288021; J.G.G is supported by MINECO through the Ramon y Cajal program. Z.W. acknowledges the National Natural Science Foundation of
China (Grant No. 11005047).

\section*{Note added in proof}
\addcontentsline{toc}{section}{Note added in proof}
After editing the present Report, we learned about the publication of
other recent and important contributions which relate to subjects and
arguments treated in our Sections. In the following, we provide a list of these additional References that the
reader may find useful to consult, together with the number of the Section they
refer to.

\bigskip

\noindent Section 2:\\
M. De Domenico, and V. Nicosia, ArXiv e-prints (2014) 1405.0425v1;
I. Laut, C. R\"ath, L. W\"orner, V. Nosenko, S. K. Zhdanov,
J. Schablinski, D. Block, H.M. Thomas, G. E. Morfill, Phys. Rev. E 89
(2014) 023104; R. J. S\'{a}nchez-Garc\'{\i}a, E. Cozzo, Y. Moreno,
Phys. Rev. E 89 (2014) 052815; M. De Domenico, M. ~A. Porter,
A. Arenas, ArXiv e-prints (2014) 1405.0843; C. W. Loe, H. J. Jensen,
ArXiv e-prints (2014) 1406.2205; A. Sol\'e-Ribalta, M. De Domenico,
S. G\'omez, A. Arenas, WebSci '14 Proceedings of the 2014 ACM conference on Web science
 (2014) 149-155.

\bigskip

\noindent Section 4:\\
F. Tan, Y. Xia, ArXiv e-prints (2014) 1405.4342; Z. Su, L. Li,
H. Peng, J. Kurths, J. Xiao, Y. Yang, Sci. Rep.  4 (2014) 5413;
L. Daqing, J. Yinan, K. Rui, S. Havlin, Sci. Rep. 4 (2014) 5381; 
B. Kotnis, ArXiv e-prints (2014) 1406.2106.

\bigskip

\noindent Section 5:\\
 E. Valdano, L. Ferreri, C. Poletto, V. Colizza, ArXiv
e-print (2014) 1406.4815; F. Bagnoli, E. Massaro, ArXiv e-prints
(2014) 1405.5348; W. Wang, M. Tang, H. Yang, Y. Do, Y. C. Lai,
G. W. Lee, Sci. Rep. 4 (2014) 57; 
B. Wang, Z. Pei, L. Wang, ArXiv e-prints (2014) 1405.1573; 
B. Sun, B. Leonard, P. Ronhovde, Z. Nussinov, ArXiv e-prints (2014)
1406.7282;  J. Sanz, Ch.-Y. Xia, S. Meloni, Y. Moreno, ArXiv e-prints
(2014) 1402.4523; M. Salehi, R. Sharma, M. Marzolla, D. Montesi, P. Sivari, M. Magnani, ArXiv
e-prints (2014) 1405.4329; C. Granell, S. Gomez, A. Arenas, ArXiv
e-prints (2014) 1405.4480; A. Chmiel, P. Klimek, S Thurner, ArXiv e-prints (2014) 1405.3801

\bigskip

\noindent Section 6:\\
L. M. Pecora, F. Sorrentino, A. M. Hagerstrom, T. E. Murphy,
R. Roy, Nat. Comm. 5 (2014) 4079; 
V. Nicosia, P. S. Skardal, V. Latora, A. Arenas, ArXiv e-prints (2014)
1405.5855v1; M. Asllani, D. M. Busiello, T. Carletti, D. Fanelli,
G. Planchon, ArXiv e-prints (2014) 1406.6401.

\bigskip

\noindent Section 7:\\
D. Hristova, M. Musolesi, C. Mascolo, Proceedings of the 8th AAAI
International Conference on Weblogs and Social Media (ICWSM'14), Ann Arbor, Michigan, USA.

\bigskip

Furthermore, a part of the material quoted in our presentation was taken
from the ArXiv public repository, and we warn the reader that it is very likely
that those Manuscripts could have appeared meanwhile on peer-reviewed Journals.
For instance, we ourselves were made aware of publication of some of those
sources, that we gladly report here below:\\

\noindent 
Our Ref. \cite{Domenico13} was published as
Proc. Nat. Acad. Sci. U.S.A 111 (2014) 8351.\\
Our Ref. \cite{BD2} was published as Phys. Rev. E 89 (2014) 062814.\\
Our Ref. \cite{tan2014} was published as Phys. Rev. E 89 (6) (2014)
062813.\\
Our Ref. \cite{sahnehcomp} was published as Phys. Rev. E 39 (2014) 062817.\\
Our Ref. \cite{Buono} was published as PLoS ONE 9 (3) (2014)
e92200. \\
Our Ref. \cite{masuda2014ZhenArx} was published as Phys. Rev. E 90 (2014) 012802. \\
Our Ref. \cite{diakonova2014ZhenWiv} was published as Phys. Rev. E 89
(6) (2014) 062818.

\phantomsection
\section*{References}
\addcontentsline{toc}{section}{References}

\bibliography{references}

\end{document}